\documentclass{aa}  

\usepackage{graphicx}
\usepackage{longtable}
\usepackage[varg]{txfonts}
\usepackage{natbib}
\bibpunct[]{(}{)}{;}{a}{}{,}

\begin{document}

\title{Star formation in Chamaeleon I and III:\\
a molecular line study of the starless core population}
\author{A. E. Tsitali\inst{1} \and A. Belloche\inst{1} \and R. T. Garrod\inst{2} \and B. Parise\inst{1,3} \and K. M. Menten\inst{1}}

\institute{Max-Planck-Institut f\"{u}r Radioastronomie, Auf dem H\"{u}gel 69, 53121, Bonn, Germany \and Center for Radiophysics and Space Research, Cornell University, Ithaca, NY 14853-6801, USA \and School of Physics and Astronomy, Cardiff University, Queen's Buildings, The Parade, Cardiff CF24 3AA, UK }

\date{Received 31 October 2013; accepted 31 October 2014}

\abstract
% context heading (optional)
{The Chamaeleon dark molecular clouds are excellent nearby targets for low-mass star formation studies. Even though they belong to the same cloud complex, Cha I and II are actively forming stars while Cha III shows no sign of ongoing star formation.} 
% aims heading (mandatory)
{We aim to determine the driving factors that have led to the very different levels of star formation activity in Cha I and III and examine the dynamical state and possible evolution of the starless cores within them.}
% methods heading (mandatory)
{Observations were performed in various molecular transitions with the APEX and Mopra telescopes. We examine the kinematics of the starless cores in the clouds through a virial analysis, a search for contraction motions, and velocity gradients. The chemical differences in the two clouds are explored through their fractional molecular abundances, derived from a non-LTE analysis, and comparison to predictions of chemical models. }
% results heading (mandatory)
{Five cores are gravitationally bound in Cha I and one in Cha III. 
The so-called infall signature indicating contraction motions is
seen toward 8--17 cores in Cha I and 2--5 cores in Cha III, which leads to a range of 13--28\% of the cores in Cha I and 10--25\% of the cores in Cha III that are contracting and may become prestellar. There is no significant difference in the turbulence level in the two clouds. Future dynamical interactions between the cores will not be dynamically significant in either Cha I or III, but the subregion Cha I North may experience collisions between cores within $\sim 0.7$~Myr. Turbulence dissipation in the cores of both clouds is seen in the high-density tracers N$_2$H$^+$ 1--0 and HC$_3$N 10--9 which have lower non-thermal velocity dispersions compared to C$^{17}$O 2--1, C$^{18}$O 2--1, and C$^{34}$S 2--1. Evidence of depletion in the Cha I core interiors is seen in the abundance distributions of the latter three molecules. The median fractional abundance of C$^{18}$O is lower in Cha III than Cha I by a factor of $\sim 2$. The median abundances of most molecules (except methanol) in the Cha III cores lie at the lower end of the values found in the Cha I cores. A difference in chemistry is thus seen. Chemical models suitable for the Cha I and III cores are used to constrain the effectiveness of the HC$_3$N to N$_2$H$^+$ abundance ratio as an evolutionary indicator. Both contraction and static chemical models indicate that this ratio is a good evolutionary indicator in the prestellar phase for both gravitationally bound and unbound cores. In the framework of these models, we find that the cores in Cha III and the southern part of Cha I are in a similar evolutionary stage and are less chemically evolved than the central region of Cha I. }
% conclusions heading (optional), leave it empty if necessary 
{The measured HC$_3$N/N$_2$H$^+$ abundance ratio and the evidence for contraction motions seen towards the Cha III starless cores suggest that Cha III is younger than Cha I Centre and that some of its cores may form stars in the future if contraction does not cease. The cores in Cha I South may on the other hand be transient structures.}


\maketitle

\section{Introduction}
\label{sec:intro}

The Chamaeleon cloud complex is a prime target for studying the earliest phases of low-mass star formation. The molecular clouds Chamaeleon I and III contain a relatively large population of starless condensations at nearby distances.
Only a few of these starless cores show signs of being gravitationally bound \citep{belloche11a, belloche11b}. One of our aims is to explore the dynamics and kinematics of these cores in more detail, and determine whether a significant fraction of the cores will become prestellar, or whether the majority are merely transient structures that will disperse as a result of turbulence. Throughout this paper, we use the definition that a starless core is a core without an embedded protostellar object. Prestellar cores are defined as the subset of starless cores that are gravitationally bound and will therefore likely form stars \citep{andre00, difrancesco07}. 
Understanding which evolutionary path the condensations will 
undertake is important to assess the dynamical state and 
the likely future evolution of the clouds themselves.

Studying the initial phase of star formation can help establish the link between the stellar initial mass function (IMF) and the core mass function (CMF) \citep[e.g.,][]{andre07, andre09, motte98, johnstone00}. Large and clustered core populations can also be used to shed light into the dominant dynamical processes taking place at the prestellar phase. On the one hand, in the competitive accretion scenario the dynamical interactions between the condensations play an important role in the subsequent core evolution and emergent IMF \citep[e.g.,][]{bate03, bonnell04, bonnell98}. On the other hand, the process of gravoturbulent cloud fragmentation suggests that stellar masses are determined at the prestellar phase and dynamical interactions with other cores and 
their surroundings are not significant 
\citep[e.g.,][]{padoan02,myers98, nakano98}.

Chamaeleon I (Cha I) is a nearby, low-mass star forming cloud located 
at a distance of $\sim$150 pc \citep{whittet97, knudehog98}. The Chamaeleon cloud complex consists of Cha I, II, and III. Despite their close proximity 
to each other (all within $\sim 16$~pc in projection) they differ 
greatly in their level of star formation activity. Cha I has the highest level of star formation activity with more than 200 young stellar objects (YSOs) produced so far \citep{luhman08}. A gas mass of $\sim950$~M$_{\odot}$ was derived from $^{12}$CO observations, as well as a diameter of 8 pc \citep{boulanger98}. It also has the highest dense gas fraction in the Chamaelon cloud complex \citep{mizuno99}. \citet{belloche11a} performed a deep, unbiased dust continuum survey of Cha I at 870~$\mu$m with the Large APEX Bolometer Camera \citep[LABOCA; ][]{siringo09} at the Atacama Pathfinder Experiment telescope (APEX{\footnote{The Atacama Pathfinder Experiment telescope (APEX) is a collaboration between the Max-Planck Institut f\"{u}r Radioastronomie, the European Southern Observatory, and the Onsala Space Observatory.}}). They identified 59 starless cores, 21 YSOs, 1 first hydrostatic core (FHSC) candidate, and a network of five filaments where most of the cores are located. They found that at most $\sim 17$\% of the starless cores are prestellar based on a simple thermal analysis and concluded that the rate of star formation in Cha I is decreasing. 
Based on the YSO population, \citet{luhman07} had similarly proposed 
that the star formation activity in Cha I is continuing but gradually 
declining. \citet{haikala05} identified 71 C$^{18}$O 1--0 clumps in Cha I, 
but there is no one-to-one correspondence with the 870 $\mu$m continuum 
sources. 
The CMF of the starless cores in Cha I seems to be in good agreement with the shape of the IMF at the high-mass end \citep{belloche11a}, as was found for various other molecular clouds \citep[Aquila, Taurus, Pipe nebula, Serpens \& Perseus, Ophiuchus;][, respectively]{andre10, sadavoy10, alves07, enoch08, motte98}, but it ceases to be consistent at the low-mass end of the spectrum. 

Chamaeleon III (Cha III) is also located at a distance of 
$\sim 150$ pc \citep{whittet97, knudehog98}. Although it belongs to the same cloud complex as Cha I and has a $^{12}$CO mass of $\sim1400$ M$_{\odot}$ and a diameter of 10 pc \citep{boulanger98}, it contains a lower fraction of dense gas \citep[][]{mizuno99} and no YSO \citep[e.g.,][]{luhman08}. \citet{belloche11b} identified 29 starless condensations in Cha III through an unbiased 870~$\mu$m dust continuum survey with LABOCA at APEX, performed with the same sensitivity as for Cha I. They found that only two sources out of the 29 starless cores are prestellar based on the Bonnor-Ebert critical mass criterion. A total of 38 clumps were previously detected in C$^{18}$O~1--0 with the Swedish-ESO Submillimetre Telescope (SEST) \citep{gahm02}. As in Cha I, no one-to-one correspondence is seen between the C$^{18}$O 1--0 clumps and the detected 870~$\mu$m sources \citep[][]{belloche11b}.
Unlike for Cha I, no prominent filamentary structure is seen in the 870~$\mu$m dust continuum \citep{belloche11b}, but the core distribution within Cha III seems to trace out an underlying filamentary structure. The level of turbulence is of the same order in both Cha I and III based on C$^{18}$O~1--0 \citep{mizuno99} and CO~1--0 \citep{mizuno01, boulanger98} data. 

The striking difference in the number of YSOs in Cha I and III is bewildering given their proximity and the similar densities of their starless core population. A scenario that could explain why Cha III contains no YSOs, is that the cloud is at a younger evolutionary stage than Cha I and has not had enough time to produce protostars.
To address this and other issues, we have undertaken a molecular line survey of the starless core population in Cha I and III, which aims to complement the results obtained from the dust continuum surveys of \citet{belloche11a, belloche11b}. We start by presenting the observational details of our data in Sect.~\ref{sec:observations} and the basic results in Sect.~\ref{sec:definitions}. We then proceed to the analysis of the kinematics (velocity distribution, contraction motions, turbulence) of the cores in Sect.~\ref{sec:results_kinematics}. Section~\ref{sec:results_abundances} presents the analysis of the molecular abundances of the cores which are compared to predictions of chemical models. Finally, we discuss the implications of our results in Sect.~\ref{sec:discussion} and give a summary of the conclusions in Sect.~\ref{sec:conclusions}.

\section{Observations}
\label{sec:observations}

The observations were performed toward the starless condensations identified in the continuum surveys of \citet{belloche11a, belloche11b}. The data 
was reduced with the CLASS software\footnote{see http://www.iram.fr/IRAMFR/GILDAS.}. 

\subsection{Chamaeleon I data}
\label{sec:chaI_data}

The 60 cores detected in continuum emission were observed with the APEX telescope in 2010 May and July in the following molecular lines: C$^{18}$O~2--1, $^{13}$CO~2--1, C$^{17}$O~2--1, H$_2$CO~3$_{1,2}$--2$_{1,1}$, HCN~3--2, HCO$^+$~3--2,  and N$_2$H$^+$~3--2. Table~\ref{table:chaI_lines} lists the 
transitions observed for each core in our sample. The cores are numbered by decreasing 870 $\mu$m peak flux density \citep[see][]{belloche11a}.
All 60 cores were observed in C$^{18}$O~2--1. $^{13}$CO~2--1 was observed in parallel to C$^{18}$O~2--1. C$^{17}$O~2--1 was observed toward the 33 cores with the strongest C$^{18}$O~2--1 emission so that the transition pair can be used to determine line opacities. H$_2$CO~3$_{1,2}$--2$_{1,1}$ was observed in parallel to C$^{17}$O~2--1. 
Owing to observing time constraints only the brightest 30 cores were targeted in HCO$^+$~3--2 and HCN~3--2, which were observed in parallel. N$_2$H$^+$ 3--2 was observed only toward the cores Cha1-C1 -- C34 which had N$_2$H$^+$ 1--0 Mopra detections (see below).

Table~\ref{table:obs_details_apex} gives information about the frequencies, resolutions, telescope receivers, forward/main beam efficiencies, system temperatures, and rms noise values for all observed transitions. The coordinates of the cores can be found in Table 2 of \citet{belloche11a}. The observations were conducted in position-switching mode with a reference position for the CO isotopologues at $\alpha_{\rm J2000}$=11$^{h}$24$^{m}$33$^{s}$.5, $\delta_{\rm J2000}$=-77$^{\circ}$01$^{\prime}$56.8$^{\prime\prime}$, which was free of emission in $^{13}$CO 2--1 with an rms of $\sim 60$~mK for the spectral resolution given in Table~\ref{table:obs_details_apex}. A second reference position at $\alpha_{\rm J2000}$=11$^{h}$05$^{m}$23$^{s}$.7, $\delta_{\rm J2000}$=-77$^{\circ}$11$^{\prime}$02.2$^{\prime\prime}$ was used for all other transitions and was also free of emission at all frequencies with an rms of $\sim 50$ mK in N$_2$H$^+$ 3--2 (spectral resolution listed in Table~\ref{table:obs_details_apex}). 

A subset of 57 cores from the continuum sample was observed with the Mopra telescope in 2010 June. The zoom mode of the high resolution spectrometer MOPS was used, with the receiver tuned at 94554 MHz with 12 zoom windows of 138 MHz selected within the 8 GHz bandwidth. The subset consists of cores Cha1-C1 to Cha1-C56 and core Cha1-C60. All cores were observed in various molecular transitions from which the following were detected: HNC~1--0, HC$_3$N~10--9, N$_2$H$^+$~1--0, C$_4$H~10$_{10}$--9$_{9}$, C$^{34}$S~2--1, CH$_3$OH~2$_{1}$--1$_{1}$~E, CH$_3$OH~2$_{0}$--1$_{0}$~A$^{+}$, and CS~2--1. Information about the parameters of the Mopra observations for each transition (frequency, angular resolution, main beam efficiency, channel spacing, system temperature, rms) are given in Table~\ref{table:obs_details_mopra}. Position-switching observations were performed with a reference position at $\alpha_{\rm J2000}$=11$^{h}$05$^{m}$23$^{s}$.7, $\delta_{\rm J2000}$=-77$^{\circ}$11$^{\prime}$02.2$^{\prime\prime}$ (free of emission with an rms of 40 -- 53 mK for the transitions and spectral resolutions given in Table~\ref{table:obs_details_mopra}). 
IRAM 04191+1522 and Oph A SM1N were used as calibration 
sources. The pointing was checked every $\sim1$ hour . \citet{tsitali13} calculated a value of 0.34 for the main beam efficiency of the telescope through a detailed calibration analysis.

The northern part of Cha I was mapped in CO 3--2 with APEX in 2006 June with the on-the-fly mode. The mapped region has a size of $6.9^{\arcmin} \times 5.4^{\arcmin}$, centred on the equatorial offset position (60$^{\arcsec}$, 60$^{\arcsec}$) with respect to Cha-MMS2 at $\alpha_{\rm J2000}$=11$^{h}$09$^{m}$57$^{s}$.9, $\delta_{\rm J2000}$=-76$^{\circ}$35$^{\prime}$00.1$^{\arcsec}$. The rms of the cube at a region that is free of emission is $\sim 0.6$~K at a spectral resolution of 122 kHz. The reference position ($\alpha_{\rm J2000}$=11$^{h}$23$^{m}$01$^{s}$.39, $\delta_{\rm J2000}$=-76$^{\circ}$08$^{\prime}$53.4$^{\arcsec}$) was also checked to be free of CO 3--2 emission with an rms of $\sim 0.14$ K at a spectral resolution of 122 kHz. The pointing was checked every $\sim 1$ h on 07454-7112 and the focus was optimized on Saturn. Table~\ref{table:obs_details_apex} gives more information about the CO 3--2 observational setup. 

\vspace*{-2ex}
\begin{table}
\begin{center}
\caption{Molecular transitions observed with APEX for the cores in Cha~I.} 
\vspace*{-3ex}
\hfill{}
\begin{tabular}{lcccc} 
\hline
\hline
Core    & \multicolumn{4}{c}{Molecular Transition} \\
         & C$^{18}$O & C$^{17}$O         & HCO$^+$       & N$_2$H$^+$  \\ 
         & 2--1 \tablefootmark{a}  & 2--1 \tablefootmark{b}  & 3--2 \tablefootmark{c}  & 3--2   \\ \hline

Cha1-C1    & y\tablefootmark{d} & y & y                    & y            \\
Cha1-C2    & y   & y        & y                            & y            \\
Cha1-C3    & y   & y        & y                            & y            \\
Cha1-C4    & y   & y        & y                           & y            \\
Cha1-C5    & y   & y        & y                          & --            \\
Cha1-C6    & y   & --      & y                           & --            \\
Cha1-C7    & y   & y       & y                           & --            \\ 
Cha1-C8    & y   & --      & y                           & y             \\           
Cha1-C9    & y   & y       & y                           & --            \\
Cha1-C10   & y   & y       & y                           & y             \\
Cha1-C11   & y   & y       & y                           & y             \\
Cha1-C12   & y   & y       & y                           & --            \\
Cha1-C13   & y   & --      & y                           & --            \\      
Cha1-C14   & y   & --      & y                           & y            \\
Cha1-C15   & y   & y       & y                           & --            \\
Cha1-C16   & y   & --      & y                           & --            \\
Cha1-C17   & y   & y       & y                           & --            \\
Cha1-C18   & y   & y       & y                           & --            \\
Cha1-C19   & y   & y       & y                           & y             \\
Cha1-C20   & y   & --      & y                           & --            \\
Cha1-C21   & y   & y       & y                           & y             \\
Cha1-C22   & y   & --      & y                           & --            \\
Cha1-C23   & y   & --      & y                           & --            \\
Cha1-C24   & y   & --      & y                           & --            \\
Cha1-C25   & y   & --      & y                           & --            \\
Cha1-C26   & y   & y       & y                           & --            \\
Cha1-C27   & y   & y       & y                           & --            \\
Cha1-C28   & y   & --      & y                           & --            \\
Cha1-C29   & y   & y       & y                           & y             \\
Cha1-C30   & y   & y       & y                           & y             \\
Cha1-C31   & y   & --      & --                          & y             \\
Cha1-C32   & y   & --      & --                          & --            \\
Cha1-C33   & y   & --      & --                          & y             \\
Cha1-C34   & y   & y       & --                          & y             \\
Cha1-C35   & y   & y       & --                          & --            \\
Cha1-C36   & y   & y       & --                          & --            \\
Cha1-C37   & y   & --      & --                          & --            \\
Cha1-C38   & y   & y       & --                          & --            \\
Cha1-C39   & y   & y       & --                          & --            \\
Cha1-C40   & y   & --      & --                         & --            \\
Cha1-C41   & y   & y       & --                          & --            \\
Cha1-C42   & y   & --      & --                          & --            \\
Cha1-C43   & y   & y       & --                          & --            \\
Cha1-C44   & y   & --      & --                          & --            \\
Cha1-C45   & y   & --      & --                          & --            \\
Cha1-C46   & y   & --      & --                          & --            \\
Cha1-C47   & y   & --      & --                          & --            \\
Cha1-C48   & y   & --      & --                         & --            \\
Cha1-C49   & y   & y       & --                          & --            \\
Cha1-C50   & y   & --      & --                         & --            \\
Cha1-C51   & y   & --      & --                          & --            \\
Cha1-C52   & y   & --      & --                         & --            \\
Cha1-C53   & y   & y       & --                         & --            \\
Cha1-C54   & y   & y       & --                         & --            \\
Cha1-C55   & y   & --      & --                         & --            \\
Cha1-C56   & y   & --      & --                         & --            \\
Cha1-C57   & y   & y       & --                         & --            \\
Cha1-C58   & y   & y       & --                         & --            \\
Cha1-C59   & y   & y       & --                         & --            \\
Cha1-C60   & y   & y       & --                         & --            \\
\emph{Total} \tablefootmark{e} & 60  & 33  & 30         & 15            \\
\hline
\end{tabular}
\hfill{}
\label{table:chaI_lines}
\end{center}
\vspace*{-3ex}
\tablefoot{
All cores but Cha1-C1 are starless. Cha1-C1 corresponds to Cha-MMS1, which is a candidate first hydrostatic core \citep[see ][]{belloche06,tsitali13}.
\tablefoottext{a}{$^{13}$CO 2--1 was observed in parallel to C$^{18}$O 2--1.}
\tablefoottext{b}{H$_2$CO 3$_{1,2}$--2$_{2,1}$ was observed in parallel to C$^{17}$O 2--1.}
\tablefoottext{c}{HCN 3--2 was observed in parallel to HCO$^+$ 3--2.}
\tablefoottext{d}{Core was (y) or was not (--) observed in the respective transition.}
\tablefoottext{e}{Total number of cores observed.}
}
\end{table}

\begin{table*}
\begin{center}
\caption{Parameters of Cha I and III observations with APEX.} 
\vspace*{-1ex}
\hfill{}
\begin{tabular}{@{\extracolsep{-7pt}}lllllllllllll} 
\hline
\hline
Transition & $f$\tablefootmark{a}         & ${\sigma_{\nu}}$\tablefootmark{b} & $HPBW$\tablefootmark{c}     & Receiver & Mixing & Backend & $\delta f$\tablefootmark{d} & $\delta V$\tablefootmark{e} &  ${N_{\mathrm{cores}}}$\tablefootmark{f}    & ${B_{\mathrm{eff}}}$\tablefootmark{g}        & $T_{\mathrm{sys}}$\tablefootmark{h}   & rms\tablefootmark{i}  \\
           & {\scriptsize (MHz)} & {\scriptsize(kHz)} & {\scriptsize($^{\prime\prime}$)} &   & &  & {\scriptsize (kHz)}   & {\scriptsize (km~s$^{-1}$)}  &    &   {\scriptsize(\%)} & {\scriptsize(K)} & {\scriptsize(mK)}  \\ \hline
\multicolumn{13}{c}{\underline{Cha I}} \\
C$^{18}$O~2--1                    & 219560.3541  &  1.5  & 27.7 & APEX-1 & SSB   & FFTS1 & 122 & 0.17 & 60  & 75 & 187--208 & 62--192 \\
$^{13}$CO~2--1  \tablefootmark{j} & 220398.6842  &  0.1  & 27.6 & APEX-1 & SSB   & FFTS1 & 122 & 0.17 & 60  & 75 & 174--198 & 52--160 \\
C$^{17}$O~2--1 \tablefootmark{j}  & 224714.1870  &  80   & 27.1 & APEX-1 & SSB   & FFTS1 & 122 & 0.16 & 33  & 75 & 216--359 & 65--122 \\
H$_2$CO~3$_{1,2}$--2$_{1,1}$       & 225697.7750  &  10   & 27.0 & APEX-1 & SSB   & FFTS1 & 122 & 0.16 & 33  & 75 & 198--334 & 52--95 \\
HCN~3--2                         & 265886.4339  &  0.4  & 22.9 & APEX-2 & SSB   & FFTS1  & 122 & 0.14 & 30  & 74 & 141--269 & 28--81 \\
HCO$^+$~3--2                     & 267557.6259  &  1.1  & 22.8 & APEX-2 & SSB   & FFTS1  & 122 & 0.14 & 30  & 74 & 176--344 & 35--94 \\
N$_2$H$^+$~3--2  \tablefootmark{j} & 279511.7843  & 6.3 & 21.8 & APEX-2   & SSB & FFTS1 & 122  & 0.13 & 10  & 60 & 160--290 & 38--55 \\
N$_2$H$^+$~3--2  \tablefootmark{j} & 279511.7843  & 6.3 & 21.8 & FLASH345 & 2SB & XFFTS & 76   & 0.08 & 5  & 60 & 172--192 & 33--63 \\ 
CO~3--2                       & 345795.9899  &  0.5  & 17.6 & APEX-2A & DSB     & FFTS1 & 122 & 0.11 & -- \tablefootmark{k} &  73 & 160--235 & 600 \\
\multicolumn{13}{c}{\underline{Cha III}} \\
C$^{18}$O 2--1                & 219560.3541  & 1.5   & 27.7 & APEX-1 &  SSB   & XFFTS2 & 76  & 0.10 & 29 &  75 & 205--261 & 98--144 \\
C$^{17}$O 2--1                & 224714.1870  & 80    & 27.1 & APEX-1 &  SSB   & XFFTS2 & 76  & 0.10 & 2  &  75 & 218--220 & 69--72 \\
\hline
\end{tabular}
\hfill{}
\label{table:obs_details_apex}
\end{center}
\vspace*{-1ex}
\tablefoot{The forward efficiency for all transitions is 95\%.
\tablefoottext{a}{Rest frequency taken from the Cologne Database for Molecular Spectroscopy \citep[CDMS;][]{muller05}.}
\tablefoottext{b}{Frequency uncertainty taken from the CDMS.}
\tablefoottext{c}{Angular resolution.}
\tablefoottext{d}{Channel spacing in frequency.} 
\tablefoottext{e}{Channel spacing in velocity.} 
\tablefoottext{f}{Number of observed cores.}
\tablefoottext{g}{Main-beam efficiency.}
\tablefoottext{h}{System temperature.}
\tablefoottext{i}{rms sensitivity in $T_a^\star$ scale.}
\tablefoottext{j}{Transition with hyperfine structure.}
\tablefoottext{k}{Map of part of Cha I North.}
}
\end{table*}

\begin{table*}
\begin{center}
\caption{Parameters of Cha I and III observations with Mopra.} 
\vspace*{-1ex}
\hfill{}
\begin{tabular}{llllllll} 
\hline
\hline
Transition & $f$\tablefootmark{a}         & ${\sigma_{\nu}}$\tablefootmark{b} & $HPBW$\tablefootmark{c}  & $\delta V$\tablefootmark{d} &  ${N_{\mathrm{cores}}}$\tablefootmark{e}         & $T_{\mathrm{sys}}$\tablefootmark{f}   & rms\tablefootmark{g}  \\
           & {\scriptsize (MHz)} & {\scriptsize(kHz)} & {\scriptsize($^{\prime\prime}$)}    & {\scriptsize (km~s$^{-1}$)}  &     & {\scriptsize(K)} & {\scriptsize(mK)}  \\ \hline
\multicolumn{8}{c}{\underline{Cha I}} \\
HNC~1--0                                       &     90663.5680   & 40.0  &  38.0  & 0.11 & 57  & 155--218 & 15--40  \\
HC$_3$N~10$_{11}$--9$_{10}$ \tablefootmark{h}   &    90979.0024   & 1.0   &  37.9  & 0.11 & 57 &  155--218 & 15--39 \\
CH$_3$CN~5--4 \tablefootmark{i}                &    91987.0876   & 0.1   &  37.4  & 0.11 & 57 &  155--218 & 15--39 \\
$^{13}$CS~2--1                                 &    92494.3080   & 50.0  &  37.2  & 0.11 & 57 &  155--218 & 16--41 \\
N$_2$H$^+$~1$_{2,3}$--0$_{1,2}$  \tablefootmark{h} &    93173.7642   & 2.4   &  37.0  & 0.11 & 57 &  160--228 & 27--50 \\
C$_4$H~10$_{11}$--9$_{10}$                &    95150.3971   & 31.0   &  36.2  & 0.11 & 57 &  162--232 & 13--46 \\
CH$_3$OH~8$_{0}$--7$_{1}$ A$^+$  \tablefootmark{i} &    95169.4630   & 10.0  &  36.2  & 0.11 & 57 &  162--232 & 15--43 \\
C$_4$H~10$_{10}$--9$_{9}$                  &    95188.9481   & 30.0   &  36.2  & 0.11 & 57 &  162--232 & 14--46 \\
C$^{34}$S~2--1                                 &    96412.9495   & 2.2   &  35.7  & 0.11 & 57 &  162--232 & 17--45 \\
CH$_3$OH~2$_{1}$--1$_{1}$ E                    &    96739.3620   & 5.0   &  35.6  & 0.10 & 57 &  172--248 & 29--43 \\
CH$_3$OH~2$_{0}$--1$_{0}$ A$^+$                &    96741.3750   & 5.0   &  35.6  & 0.10 & 57 &  172--248 & 29--47 \\
C$^{33}$S~2--1 \tablefootmark{i}              &    97172.0639   & 0.2   &  35.4  & 0.10 & 57 &  172--248 & 16--47 \\
OCS~8--7 \tablefootmark{i}                    &    97301.2085   & 0.1   &  35.4  & 0.10 & 57 &  172--248 & 17--49 \\
CS~2--1                                       &    97980.9533   & 2.3   &  35.2  & 0.10 & 57 &  172--248 & 15--45 \\

\multicolumn{8}{c}{\underline{Cha III}} \\
H$^{13}$CO$^+$~1$_{2,2}$--0$_{1,1}$ \tablefootmark{h} & 86754.3004 & 3.9 & 39.7  & 0.12 & 20  & 177--200 & 35--55 \\
HN$^{13}$C~1$_{2,3,3}$--0$_{1,2,2}$ \tablefootmark{h} & 87090.8298 & 3.8 & 39.5  & 0.12 & 20  & 177--200 & 35--56 \\
HNCO~4$_{0,4,5}$--3$_{0,3,4}$ \tablefootmark{h}      & 87925.2178 & 0.3 & 39.2  & 0.12 & 20  & 175--198 & 36--71 \\
HCN~1--0                                           & 88631.6022 & 0.1  & 38.9  & 0.11 & 20  & 175--264 & 30--58 \\  
HCO$^+$~1--0                                      & 89188.5247 & 4.1  & 38.6  & 0.11 & 20  & 167--189 & 40--63 \\
HNC~1--0                                          & 90663.5680 & 40.0 & 38.0  & 0.11 & 20  & 167--264 & 29--59 \\
HC$_3$N~10$_{11}$--9$_{10}$ \tablefootmark{h}      & 90979.0024 & 1.0  & 37.9  & 0.11 & 20  & 167--264 & 28--53 \\ 
$^{13}$CS 2--1 \tablefootmark{i}                   & 92494.3080 & 50.0 & 37.2  & 0.11 & 20  & 200--264  & 28--42 \\
N$_2$H$^+$~1$_{2,3}$--0$_{1,2}$ \tablefootmark{h}   & 93173.7642 & 2.4  & 37.0  & 0.11 & 20 & 208--274 & 32--45 \\
C$_4$H~10$_{11}$--9$_{10}$ \tablefootmark{i}  & 95150.3971 & 31.0 & 36.2   & 0.11 & 20 & 210--279 & 30--47 \\
C$_4$H~10$_{10}$--9$_{9}$ \tablefootmark{i}    & 95188.9481 & 30.0 & 36.2  & 0.11 & 20  & 210--279 & 30--47 \\
C$^{34}$S 2--1  \tablefootmark{i}                  & 96412.9495 & 2.2  & 35.7  & 0.11 & 20  & 209--280 & 33--47  \\
CH$_3$OH~2$_{1}$--1$_{1}$ E                      & 96739.3620 & 5.0  & 35.6  & 0.10 & 20  & 227--300 & 35--49 \\
CH$_3$OH~2$_{0}$--1$_{0}$ A$^+$                      & 96741.3750 & 5.0  & 35.6  & 0.10 & 20  & 227--300 & 36--50 \\
C$^{33}$S 2--1 \tablefootmark{i}                  & 97172.0639 & 0.2  & 35.4  & 0.10 & 20  & 225--300 & 32--47 \\
CS~2--1                                           & 97980.9533 & 2.3  & 35.2  & 0.10 & 20  & 227--300 & 38--47 \\
\hline
\end{tabular}
\hfill{}
\label{table:obs_details_mopra}
\end{center}
\vspace*{-1ex}
\tablefoot{The main-beam efficiency for all transitions is 34\% \citep[see][]{tsitali13} and the channel spacing in frequency is 33.7 kHz. 
\tablefoottext{a}{Rest frequency taken from the CDMS.}
\tablefoottext{b}{Frequency uncertainty taken from the CDMS.}
\tablefoottext{c}{Angular resolution.}
\tablefoottext{d}{Channel spacing in velocity.} 
\tablefoottext{e}{Number of observed cores.}
\tablefoottext{f}{System temperature.}
\tablefoottext{g}{rms sensitivity in $T_a^\star$ scale.}
\tablefoottext{h}{Transition with hyperfine structure.}
\tablefoottext{i}{There were no detections in these transitions.}
}
\end{table*}

\subsection{Chamaeleon III data}
\label{chaIII_data}

Twenty starless cores in Cha III were observed with the Mopra telescope in 2012 May. The sample was chosen by selecting the cores with the highest peak flux density in 870 $\mu$m continuum emission \citep{belloche11b}. The receiver was tuned at two different frequencies, 94554~MHz and 87190~MHz. Single position observations were performed in various molecular transitions from which we detected the following: H$^{13}$CO$^+$~1--0, HN$^{13}$C~1--0, HNCO~4$_{0,4}$--3$_{0,3}$, HCN~1--0, HCO$^+$~1--0, HNC~1--0, HC$_3$N~10--9, CH$_3$OH~2$_{1}$--1$_{1}$~E, CH$_3$OH~2$_{0}$--1$_{0}$~A$^+$, N$_2$H$^+$~1--0, CS~2--1. We list the observation parameters for all the detected transitions and some of the non-detected transitions in Table~\ref{table:obs_details_mopra}. The coordinates of the observed cores are given in Table 2 of \citet{belloche11b}. The observations were performed in position-switching mode with the reference position at $\alpha_{\rm J2000}$=13$^{h}$00$^{m}$00$^{s}$, $\delta_{\rm J2000}$=-79$^{\circ}$40$^{\prime}$00$^{\prime\prime}$. The reference position was free of emission at all frequencies with a maximum rms of 56 mK for the spectral resolution given in Table~\ref{table:obs_details_mopra}. IRAM 04191+1522 and Cha1-C1 were used as calibration sources.

We observed all 29 continuum sources in C$^{18}$O~2--1 (and $^{13}$CO 2--1 in parallel) and two sources (Cha3-C1 and Cha3-C4) in C$^{17}$O~2--1 with the APEX telescope in 2013 June. Table~\ref{table:obs_details_apex} gives the respective observational parameters of the two transitions. The observations were done in position-switching mode with the same reference position at $\alpha_{\rm J2000}$=13$^{h}$00$^{m}$00$^{s}$, $\delta_{\rm J2000}$=-79$^{\circ}$40$^{\prime}$00$^{\prime\prime}$ (free of emission in C$^{18}$O 2--1 and $^{13}$CO 2--1 with an rms of $\sim 100$ mK). The pointing was checked every 1 -- 1.5 h on 07454-7112. 

\subsection{Complementary C$^{18}$O 1--0 data}

We used the C$^{18}$O~1--0 emission maps of Cha I and III observed with the Swedish-ESO Submillimetre Telescope (SEST) from \citet{haikala05} and \citet{gahm02}, respectively. Details about the observational setup can be found in these papers. 

\section{Results}
\label{sec:definitions}

We use a detection threshold of $3\sigma$ in peak temperature for all spectra used in the analysis. The cores' systemic velocities and the linewidths of the detected transitions were extracted from single Gaussian or hyperfine-structure fits to the spectra. The fitting was performed in CLASS using the methods ``GAUSS'' or ``HFS'', respectively. The transitions to which we 
applied fits are listed in Table~\ref{table:detections_chaI_chaIII}. 
For the lines that do not have hyperfine structure we apply Gaussian fits as long as their shape is symmetric and not self-absorbed. The number of cores with two velocity components in their spectra is listed as $N_{\rm 2vel}$. We define a core with two velocity components as a core with spectra having two emission peaks in both optically thick \emph{and} optically thin transitions in order to exclude spectra that might be self-absorbed. The results of the ``GAUSS'' and ``HFS'' fits in CLASS are given in Tables~\ref{table:physical_parameters_chaI_2} and ~\ref{table:physical_parameters_chaIII_2} for Cha I and III, respectively. Transitions that are thought to be optically thick and in particular those that are self-absorbed toward most cores (HNC~1--0, $^{13}$CO~2--1, and CS 2--1) are not shown in Tables~\ref{table:physical_parameters_chaI_2} and~\ref{table:physical_parameters_chaIII_2} and are not used to derive kinematical properties or column densities (Sect.~\ref{sec:results_abundances}).

In the following sections we also use a composite C$^{18}$O and 
C$^{17}$O sample when discussing non-thermal velocity dispersions in 
the clouds. The C$^{18}$O 2--1 transition was observed and detected toward all 
sixty cores in Cha I so it is a very convenient starting point for a kinematic 
analysis.
However, we calculated the opacity of the C$^{18}$O~2--1 line for the sample of 32 cores that also have C$^{17}$O~2--1 detections and found that only 50\% of the cores are optically thin in the former line with an opacity $\tau \le 1$ (see Sect~\ref{sec:opacity_c18o}). There is therefore a large fraction of cores that are somewhat optically thick in C$^{18}$O 2--1. The 32 cores observed in C$^{17}$O~2--1 were those with the strongest C$^{18}$O~2--1 emission. Since C$^{17}$O 2--1 is optically thin toward these cores, the C$^{17}$O spectra are 
expected to be more reliable in deriving linewidths toward these 32 cores. 
The mean ratio of the C$^{18}$O to C$^{17}$O 2--1 linewidths for the 
sample of cores that are optically thick in C$^{18}$O 2--1 is 
$1.19 \pm 0.16$. For the sample of cores optically thin in 
C$^{18}$O 2--1, the mean ratio is $1.16 \pm 0.29$, slightly lower than for the
optically thick sample, but the difference is not significant given the large
dispersions. Even so, we create a composite sample that consists of the 32 
cores observed in C$^{17}$O and the remaining 28 cores observed only in 
C$^{18}$O. Using 
this composite sample, we gain the advantage of having the \emph{full} set of 
cores, while reducing possible biases introduced by the higher 
optical depth of C$^{18}$O.

However, we do not use the composite sample for the centroid velocity 
analysis presented in Sect.~\ref{sec:cen_velocities} because we noticed a 
systematic difference between the systemic velocities derived from C$^{18}$O 
and C$^{17}$O~2--1. For the Cha~I sample observed in both transitions, we find 
an average velocity difference $V_{18} - V_{17}$ of $0.17 \pm 0.04$ km~s$^{-1}$. 
This difference likely comes from either inaccurate spectroscopic predictions 
or an inaccurate frequency/velocity calibration of the observed spectra. We 
tested new spectroscopic predictions for C$^{17}$O that take only laboratory 
measurements made in Bologna into account (H.~S.~P. M\"uller, priv. comm.), 
but this only reduces the discrepancy by 0.04 km~s$^{-1}$. Therefore, the 
problem is still unsolved.

\begin{figure}
%\centerline{\includegraphics[height=8.3cm,angle=0]{Figs/vel_components_ab.eps}\hspace*{3mm}\includegraphics[height=8.3cm,angle=0]{Figs/vel_components2_ab.eps}}
\centerline{\includegraphics[height=8.3cm,angle=0]{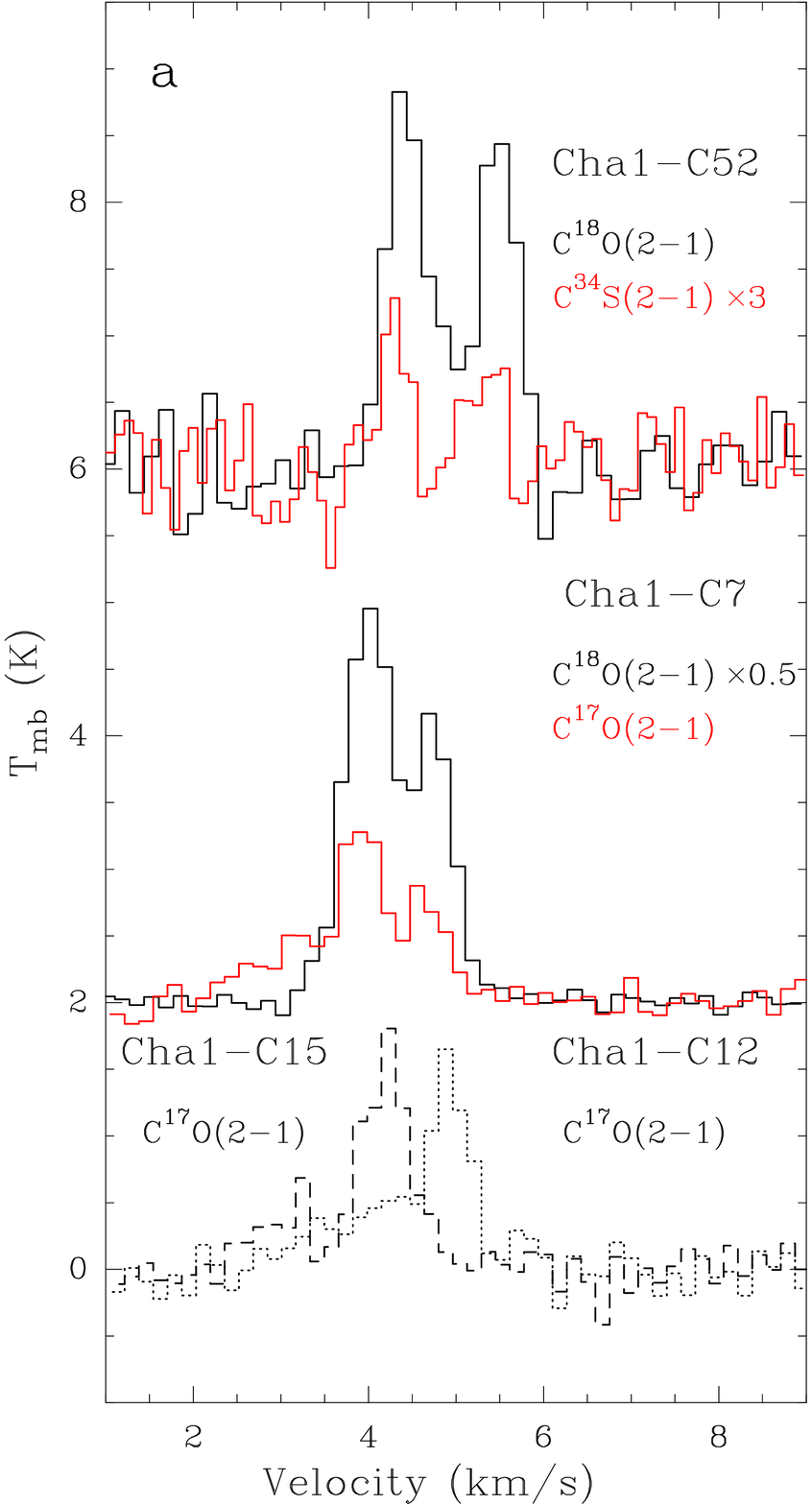}\hspace*{3mm}\includegraphics[height=8.3cm,angle=0]{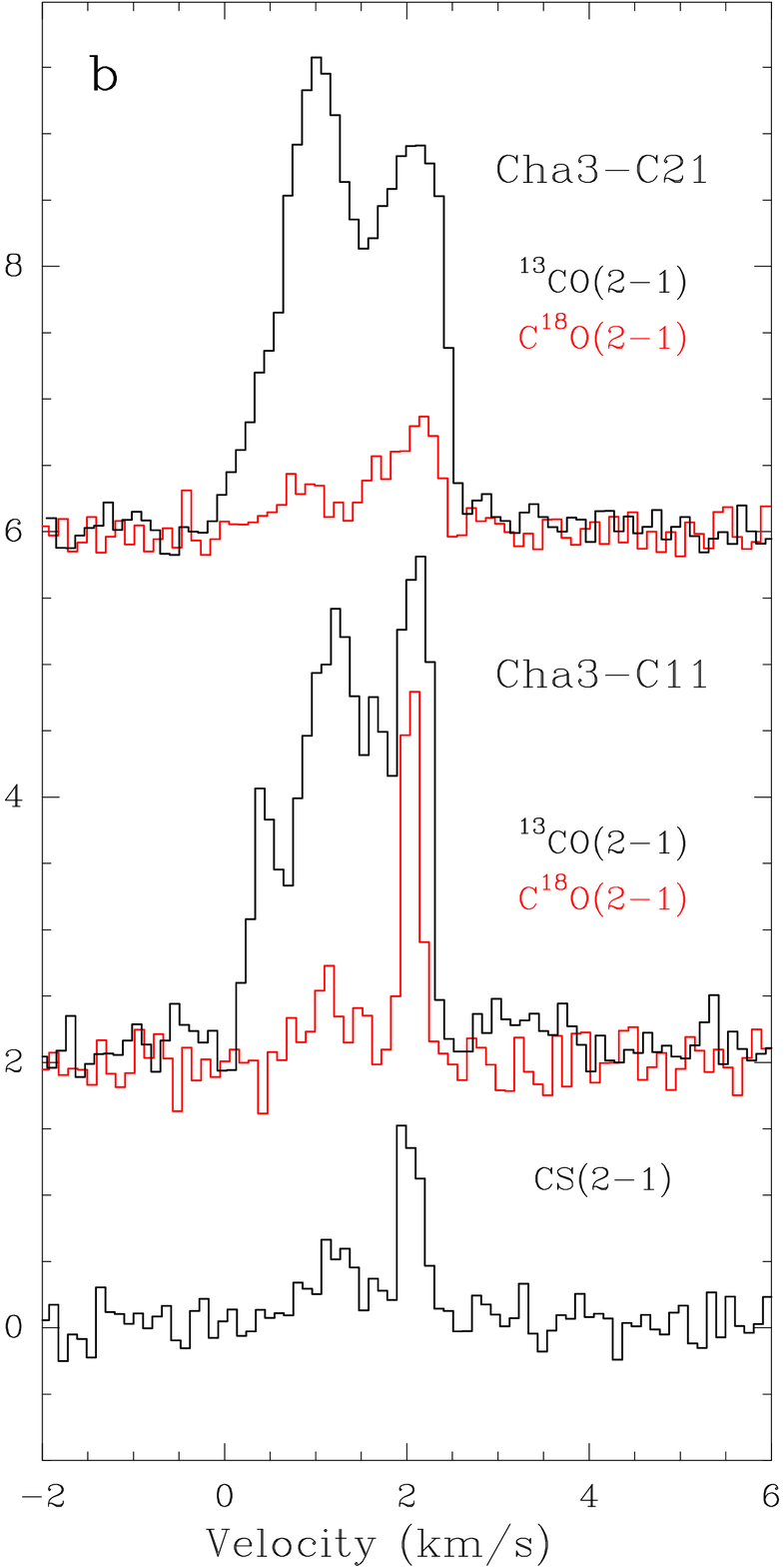}}
 \caption[]{Examples of spectra showing two emission peaks at different velocities that are not due to self-absorption, in Cha~I (\textbf{a}) and in Cha~III (\textbf{b}). \textbf{a} Cha1-C7 and Cha1-C52 have two velocity components in C$^{18}$O 2--1 (black) and in the optically thin C$^{17}$O 2--1 and C$^{34}$S 2--1 (red), respectively. Cha1-C15 (dashed) and Cha1-C12 (dotted) are located on either side of Cha1-C7 (see Fig~\ref{fig:2vel}a). They both have only one emission peak that roughly coincides with \emph{either} the lower velocity or the higher velocity peak 
of Cha1-C7. 
\textbf{b} Cha3-C11 and Cha3-C21 have two velocity components in
$^{13}$CO 2--1 (black) and in the optically thin C$^{18}$O 2-1 (red), as well 
as in CS 2--1 for Cha3-C11.
\label{fig:vel_components}}
\end{figure}

\subsection{Multiple velocity components}
\label{sec:multiple_components}

Spectra with two emission peaks at different velocities are seen in the transitions C$^{18}$O 2--1, C$^{17}$O 2--1, and C$^{34}$S 2--1 toward some dense structures in the northern and central parts of Cha~I 
and in the central part of Cha~III (see four examples in
Fig.~\ref{fig:vel_components}). Spectra with two velocity components that are not due to self-absorption might arise as projection effects if two physically unconnected regions coincide along the line of sight. Alternatively they 
could be an indication that different structures in the same region are interacting with one another. We used the C$^{18}$O 1--0 maps of \citet{haikala05} and \citet{gahm02} alongside our data to better understand the physical origin of these features. 

\begin{figure*}
\begin{center}
\begin{tabular}{ccc}
\hspace*{-1ex}
\includegraphics[width=40mm,angle=0]{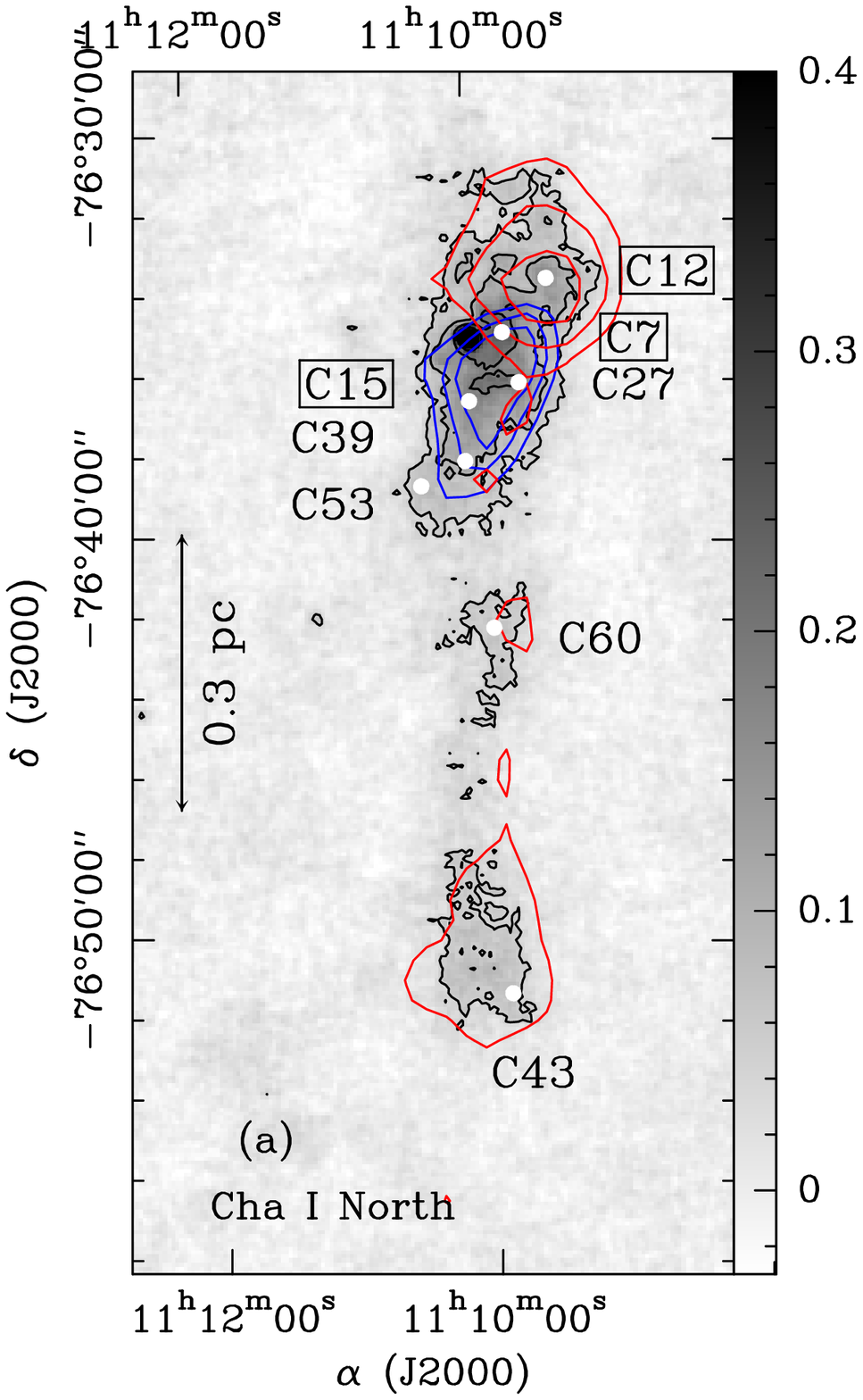} &
\hspace*{-1ex}
\includegraphics[width=85mm,angle=0]{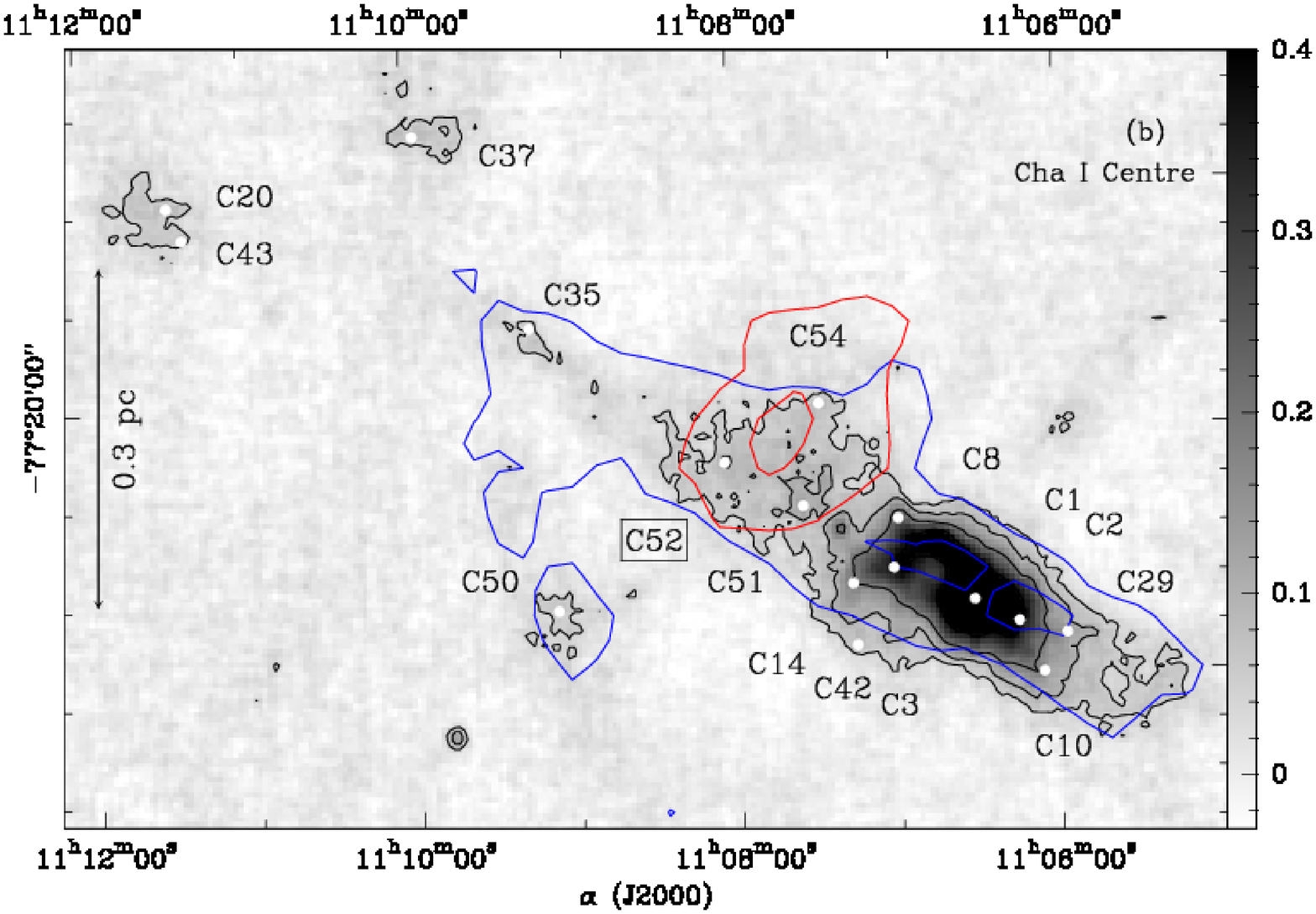} &
\hspace*{-1ex}
\includegraphics[width=50mm,angle=0]{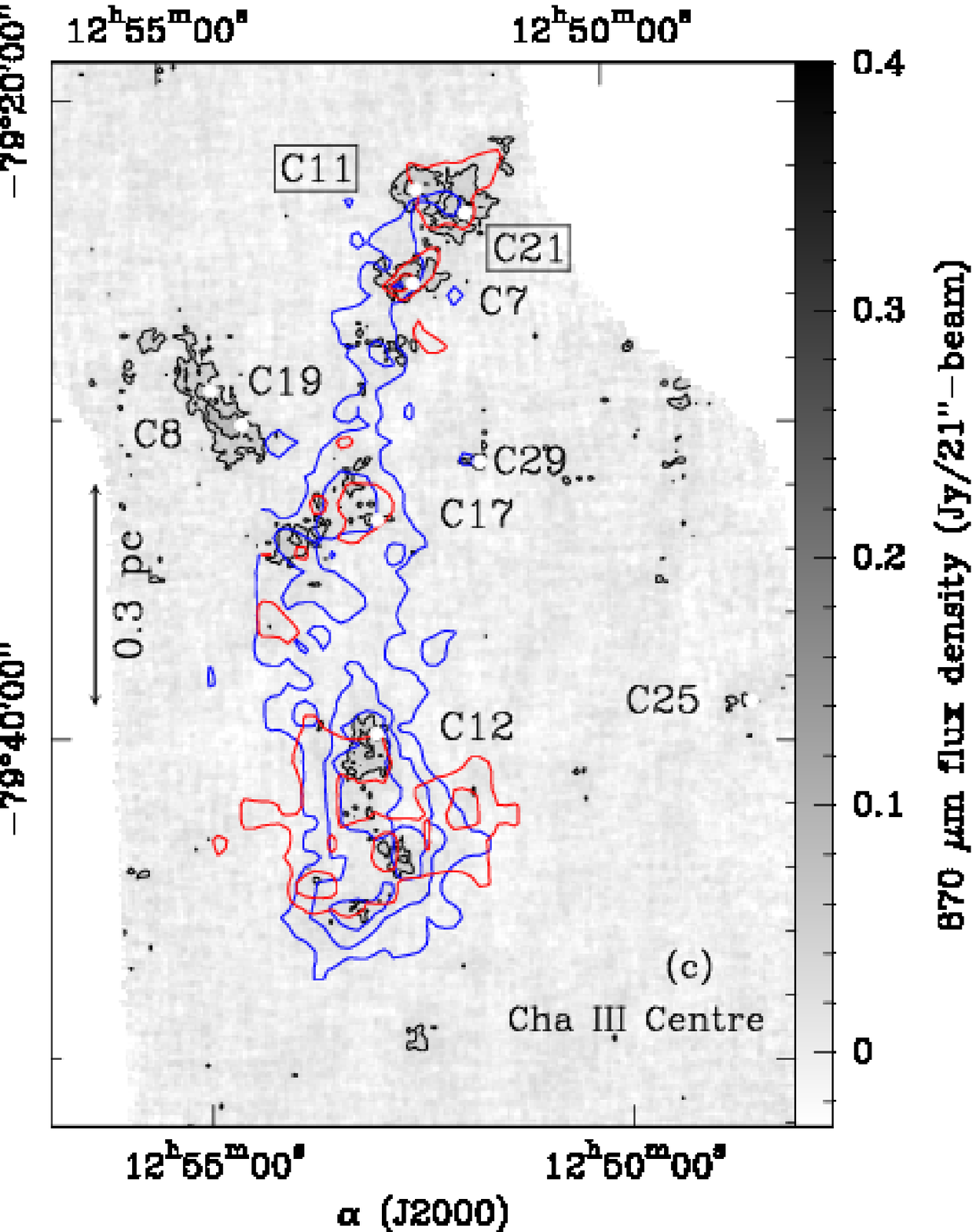} \\
\end{tabular}
 \caption[]{C$^{18}$O 1--0 blueshifted and redshifted emissions in (a) Cha I North, (b) Cha I Centre, and (c) Cha III Centre overlaid on the 870 $\mu$m dust continuum maps of Cha I and III \citep{belloche11a, belloche11b}. The C$^{18}$O 1--0 data are from 
\citet{haikala05} for Cha~I and \citet{gahm02} for Cha~III.
The continuum contour levels (black) correspond to $a$, $2a$, $4a$, $8a$, $16a$, $32a$, with $a = 48$~mJy/$21\arcsec$-beam ($4\sigma$) for (a) and (b), while $a = 34.5$ mJy/$21\arcsec$-beam ($3\sigma$) for (c). The blueshifted (subscript $b$) and redshifted (subscript $r$) C$^{18}$O 1--0 contours correspond to the levels (a) $3\sigma_b$ ($3\sigma_r$), $4\sigma_b$ ($4\sigma_r$), $5\sigma_b$ ($5\sigma_r$) with $\sigma_b = 0.17$ K km s$^{-1}$ and $\sigma_r = 0.3$ K km s$^{-1}$, (b) $3\sigma_b$ ($3\sigma_r$), $4.5\sigma_b$ ($5\sigma_r$) with $\sigma_b = 0.33$ K km s$^{-1}$ and $\sigma_r = 0.22$ K km s$^{-1}$, and (c) $4\sigma_b$ ($4\sigma_r$), $6\sigma_b$ ($6\sigma_r$), $8\sigma_b$ with $\sigma_b = 0.08$ K km s$^{-1}$ and $\sigma_r = 0.09$ K km s$^{-1}$.
The velocity ranges for the blueshifted and redshifted integrated intensity emissions are $v_b = $ 3.5 -- 4.3 km s$^{-1}$ and $v_r =$ 4.3 -- 5.5 km s$^{-1}$ for (a), $v_b = $ 3.8 -- 4.8 km s$^{-1}$ and $v_r =$ 4.8 -- 6.0 km s$^{-1}$ for (b), and $v_b = $ 0.5 -- 1.5  km s$^{-1}$ and $v_r =$ 1.5 -- 2.5 km s$^{-1}$ for (c). The white dots show the location of the continuum cores in each region. 
The sources with spectra displayed in Fig.~\ref{fig:vel_components} are highlighted with a box around their name.
\label{fig:2vel}}
\end{center}
\end{figure*}

The spectrum of Cha1-C7, located in Cha I North, peaks at the velocities $\sim 4.0$ and $\sim 4.8$ km s$^{-1}$ in C$^{18}$O 2--1, C$^{17}$O 2--1, and CS 2--1 (see Fig.~\ref{fig:vel_components}a). The fact that we see the second velocity peak in the optically thin transition C$^{17}$O 2--1 excludes the possibility that it is due to self-absorption. Cha1-C7 is located at the northern elongated structure as seen in the 870 $\mu$m dust continuum (Fig.~\ref{fig:2vel}a). Cores Cha1-C12 and Cha1-C15 lie on either side of Cha1-C7 but only have one component each, the former roughly corresponding to the higher velocity component of Cha1-C7 and the latter to the lower velocity one (Fig.~\ref{fig:vel_components}a). We produced integrated intensity maps for the velocity ranges corresponding to each C$^{18}$O 2--1 emission peak using the 1--0 map of \citet{haikala05}. The intensity contours of both lower velocity and higher velocity emissions are overlaid on the $870$ $\mu m$ continuum in Fig.~\ref{fig:2vel}a. The lower velocity and higher velocity integrated intensity emissions peak at different locations on the continuum map. The lower velocity emission seems to follow the peak of the dust continuum emission. The higher velocity emission, however, peaks close to the position of Cha1-C12. Both velocity components overlap along the line of sight and Cha1-C7 lies in the overlap region which is the reason why its spectra show two velocity components.

A YSO, Cha-MMS2, is located very close to Cha1-C7 and drives a 
bipolar outflow \citep[see, e.g.,][]{bally06, reipurth96, mattila89}. We examined the possibility that the second, higher velocity component that we see in our spectra coincides with the redshifted outflow component. We use our CO 3--2 map of the Cha-MMS2 region (see Sect.~\ref{sec:chaI_data}) and find that the outflow does not contribute to the higher velocity emission in the velocity range  $ 4.5 < v < 5.5$ km s$^{-1}$.

Two velocity components are seen toward three cores located in (or close to) the central elongated structrure of Cha I Centre: Cha1-C50, C52 
(see Fig.~\ref{fig:vel_components}a), and C54. Cha1-C52 and C54 are located at the northern edge of the filament as seen in the dust continuum (Fig.~\ref{fig:2vel}b) and C50 lies close to that part of the structure but is not embedded in it. The lower velocity component peaks close to $\sim4.5$ km s$^{-1}$ for all three cores and the higher velocity component peaks at $\sim5.2$--5.5 km s$^{-1}$. The integrated intensity contours of the lower and higher velocity C$^{18}$O 1--0 emissions are shown in Fig.~\ref{fig:2vel}b overlaid on the dust continuum emission. The lower velocity emission closely follows the dust continuum emission, while the higher velocity component is less related to the continuum structure. Its emission seems to trace a direction perpendicular to the filamentary structure traced by the dust continuum and the lower velocity C$^{18}$O~1--0 emission (see Sect.~\ref{sec:discussion_2vel} for further discussion). 

Two sources in Cha III have C$^{18}$O 2--1 spectra with two emission peaks at different velocities, Cha3-C11 and Cha3-C21, which are located adjacent to 
each other (see Figs.~\ref{fig:vel_components}b and \ref{fig:2vel}c).  
The brightest peak of Cha3-C11 is at $v \sim 2$~km s$^{-1}$ and the weakest peak is at $v \sim 1.2$~km~s$^{-1}$. The two emission peaks are also seen at the same velocities in the spectra of $^{13}$CO 2--1 and CS 2--1. The C$^{18}$O 2--1 and $^{13}$CO 2--1 spectra of Cha3-C21 peak at $\sim2$~km~s$^{-1}$ and $\sim0.9$~km~s$^{-1}$, with the former being the strongest in emission in C$^{18}$O 2--1 (the opposite is true for $^{13}$CO 2--1). Cha3-C11 and Cha3-C21 are located in the overlap region of two velocity components that are revealed in the C$^{18}$O 1--0 map shown in Fig.~\ref{fig:2vel}c. The velocities of these two C$^{18}$O 1--0 components match the velocities derived from our C$^{18}$O 2--1, 
$^{13}$CO 2--1, and CS 2--1 spectra.

In the spectra that have two emission peaks in either Cha I or Cha III, we select the strongest emission peak as the main velocity component, and the weakest peak as the \emph{second} velocity component. For the cores Cha1-C52 and Cha1-C54 we define the higher velocity peak as the second velocity component since we have an indication that it is not related to emission arising from the central elongated structure  seen in the dust continuum. 
Finally, Fig.~\ref{fig:2vel}c suggests that the 870~$\mu$m dust continuum emission of Cha3-C11 and Cha3-C21 is closely related to the higher velocity component, which supports our decision to take the higher velocity component as the main component for these two cores.

\section{Analysis: kinematics}
\label{sec:results_kinematics}

\begin{table}
\caption{List of transitions to which Gaussian or hyperfine-structure fits were applied.} 
\begin{tabular}{@{\extracolsep{-7pt}}llllll}
\hline\hline
Transition &  $N_{\rm obs}$\tablefootmark{a} & $N_{\rm det}$\tablefootmark{b} & $N_{\rm 2vel}$\tablefootmark{c}  & $N_{\rm gb}$\tablefootmark{d} &  $N_{\rm vir}$\tablefootmark{e}   \\ 
           &                                                   &                                &                                &                                    &                                 \\ \hline
\multicolumn{6}{c}{\underline{Chamaeleon I}}                      \\ 
HNC 1--0   \tablefootmark{f}      & 57   & 46 & -- & -- & --   \\
HC$_3$N 10--9   \tablefootmark{g} & 57   & 18 &  0 & 4\tablefootmark{h} & 1   \\
N$_2$H$^+$ 1--0 \tablefootmark{g} & 57   & 19 &  0 & 4\tablefootmark{h} & 1   \\
C$_4$H 10$_{11}$--9$_{10}$          & 57  & 10 &  0 & 3\tablefootmark{i} & 1   \\
C$_4$H 10$_{10}$--9$_{9}$      & 57  &  8 &  0 & 3\tablefootmark{j} & 1   \\
C$^{34}$S 2--1                      & 57 & 23 &  1 & 1 & 1   \\
CH$_3$OH 2$_{1}$--1$_{1}$ E         & 57 & 31 &  0 & 3\tablefootmark{k} & 1   \\
CH$_3$OH 2$_{0}$--1$_{0}$ A$^+$     & 57 & 41 &  0 & 2\tablefootmark{l} & 1   \\
CS 2--1 \tablefootmark{f}         & 57 & 56 & -- & -- & --   \\
C$^{18}$O 2--1                     & 60 & 60 &  4 & 1 & 1   \\
C$^{17}$O 2--1 \tablefootmark{g}   & 33 & 32 &  1 & 3\tablefootmark{j} & 1   \\
H$_2$CO 3$_{1,2}$--2$_{1,1}$        & 33 & 8  &  0 & 2\tablefootmark{l} & 0   \\
N$_2$H$^+$ 3--2                    & 15 & 5  &  0 & 4\tablefootmark{h} & 1   \\

\multicolumn{6}{c}{\underline{Chamaeleon III}}                     \\ 
H$^{13}$CO$^+$ 1--0                & 20  &  5 &   0 & 0 & 0    \\
HNC 1--0  \tablefootmark{f}       & 20  & 17 &  -- & -- & --    \\
HC$_3$N 10--9  \tablefootmark{g}   & 20  &  3 &   0 & 1 & 0    \\
N$_2$H$^+$ 1--0  \tablefootmark{g} & 20  &  2 &   0 & 1 & 0    \\
C$^{34}$S 2--1                     & 20  &  2 &   0 & 0 & 0    \\
CH$_3$OH 2$_{1}$--1$_{1}$ E        & 20  & 10 &   0 & 1 & 0    \\
CH$_3$OH 2$_{0}$--1$_{0}$ A$^+$    & 20  & 15 &   0 & 1 & 0    \\
CS 2--1  \tablefootmark{f}        & 20  & 20 &  -- & -- & --   \\
C$^{18}$O 2--1                    & 29  & 29 &  2 & 1 & 0    \\
C$^{17}$O 2--1  \tablefootmark{g} &  2  &  2 &  0 & 1 & 0    \\

\hline
\end{tabular}
\label{table:detections_chaI_chaIII}
\vspace*{-1ex}
\tablefoot{
\tablefoottext{a}{Number of observed cores.}
\tablefoottext{b}{Number of cores with $3\sigma$ detections in terms of the peak temperature that were used for the fitting (hyperfine structure fitting or symmetric, Gaussian lines).}
\tablefoottext{c}{Number of cores showing (and fitted by) two velocity components in their spectra.}
\tablefoottext{d}{Number of gravitationally bound cores assuming a temperature of 10 K to derive the non-thermal velocity dispersion.}
\tablefoottext{e}{Number of virialized cores.}
\tablefoottext{f}{CS 2--1 is optically thick in both Cha I and III toward all the cores detected in C$^{34}$S 2--1 (see Sect.~\ref{sec:opacity_cs}). The majority of cores detected in HNC 1--0 have asymmetric, non-Gaussian like spectra which might be a sign of optical thickness. We therefore do not fit these two transitions. }
\tablefoottext{g}{Hyperfine-structure fit.}
\tablefoottext{h}{Bound cores: Cha1-C1, C2, C3, C4.} 
\tablefoottext{i}{Bound cores: Cha1-C1, C3, C4.}
\tablefoottext{j}{Bound cores: Cha1-C1, C2, C3.}
\tablefoottext{k}{Bound cores: Cha1-C1, C2, C5}
\tablefoottext{l}{Bound cores: Cha1-C1, C2.}
}
\end{table}

\subsection{Turbulence}
\label{sec:turbulence}

\subsubsection{Virial analysis}
\label{sec:vir_analysis}

\begin{figure*}
\begin{center}
\begin{tabular}{cc}
\hspace*{-8ex}
\includegraphics[width=105mm,angle=0]{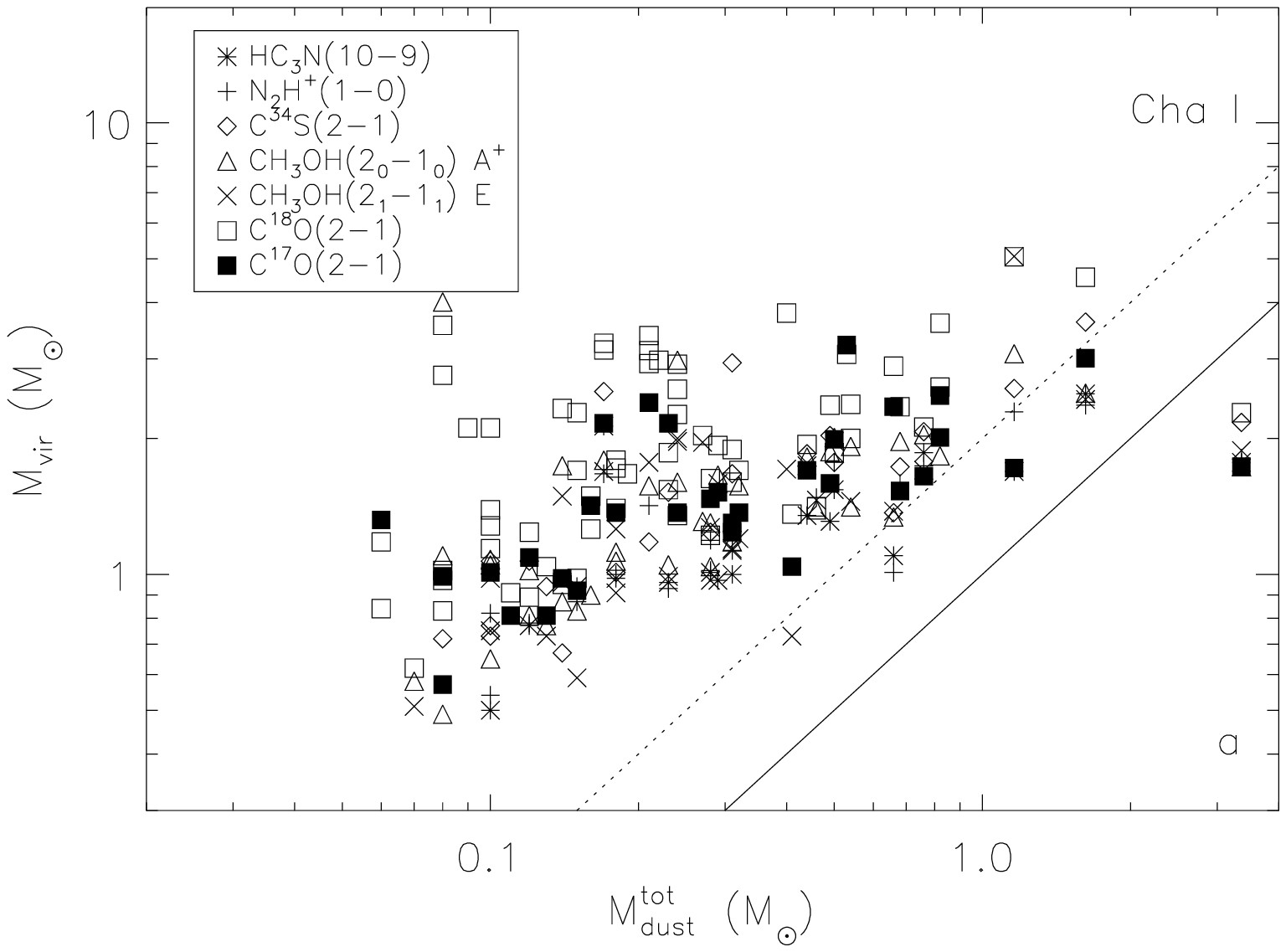} &
\hspace*{-13ex}
\includegraphics[width=105mm,angle=0]{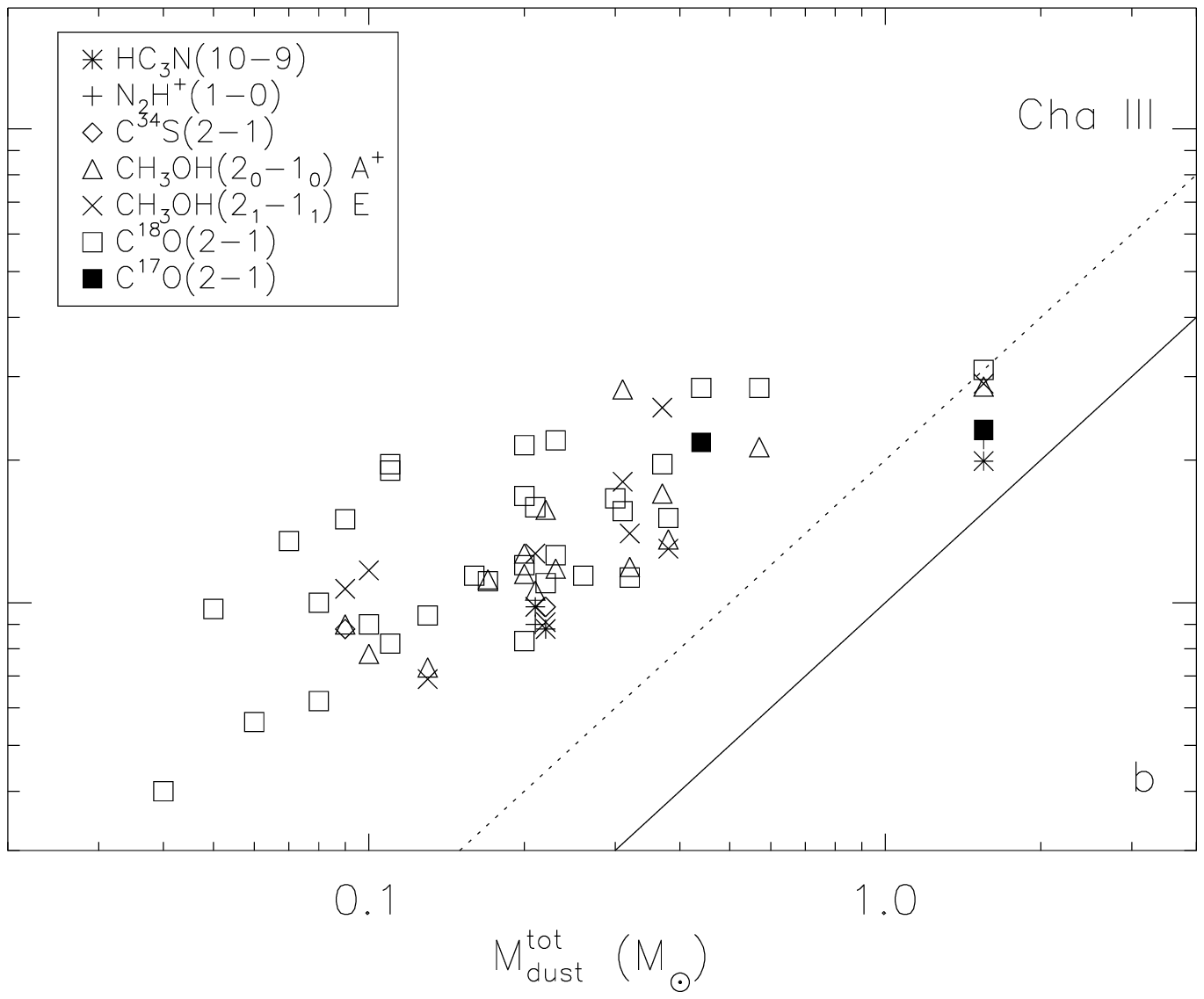} \\
\hspace*{-8ex}
\includegraphics[width=105mm,angle=0]{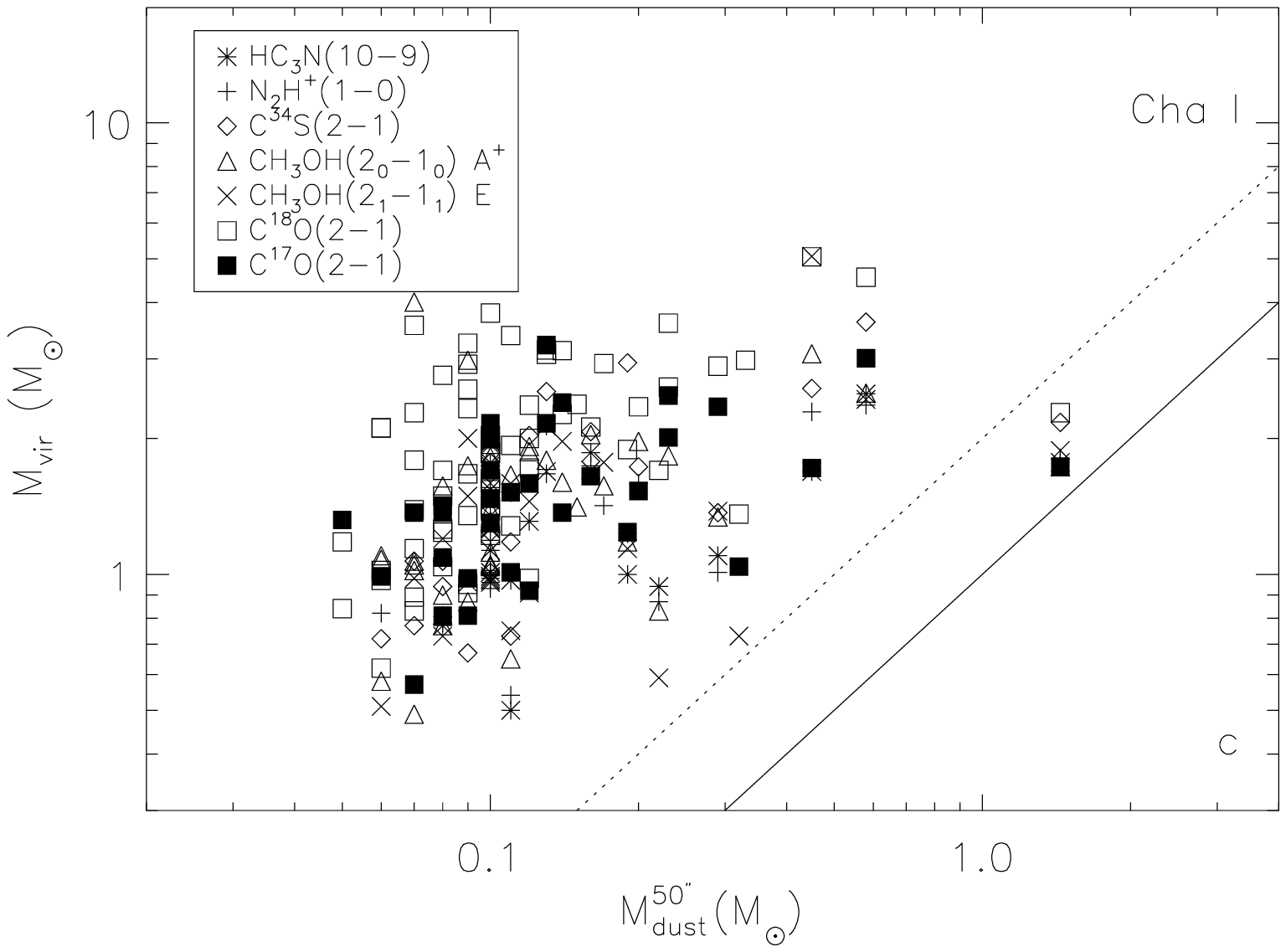} &
\hspace*{-13ex}
\includegraphics[width=105mm,angle=0]{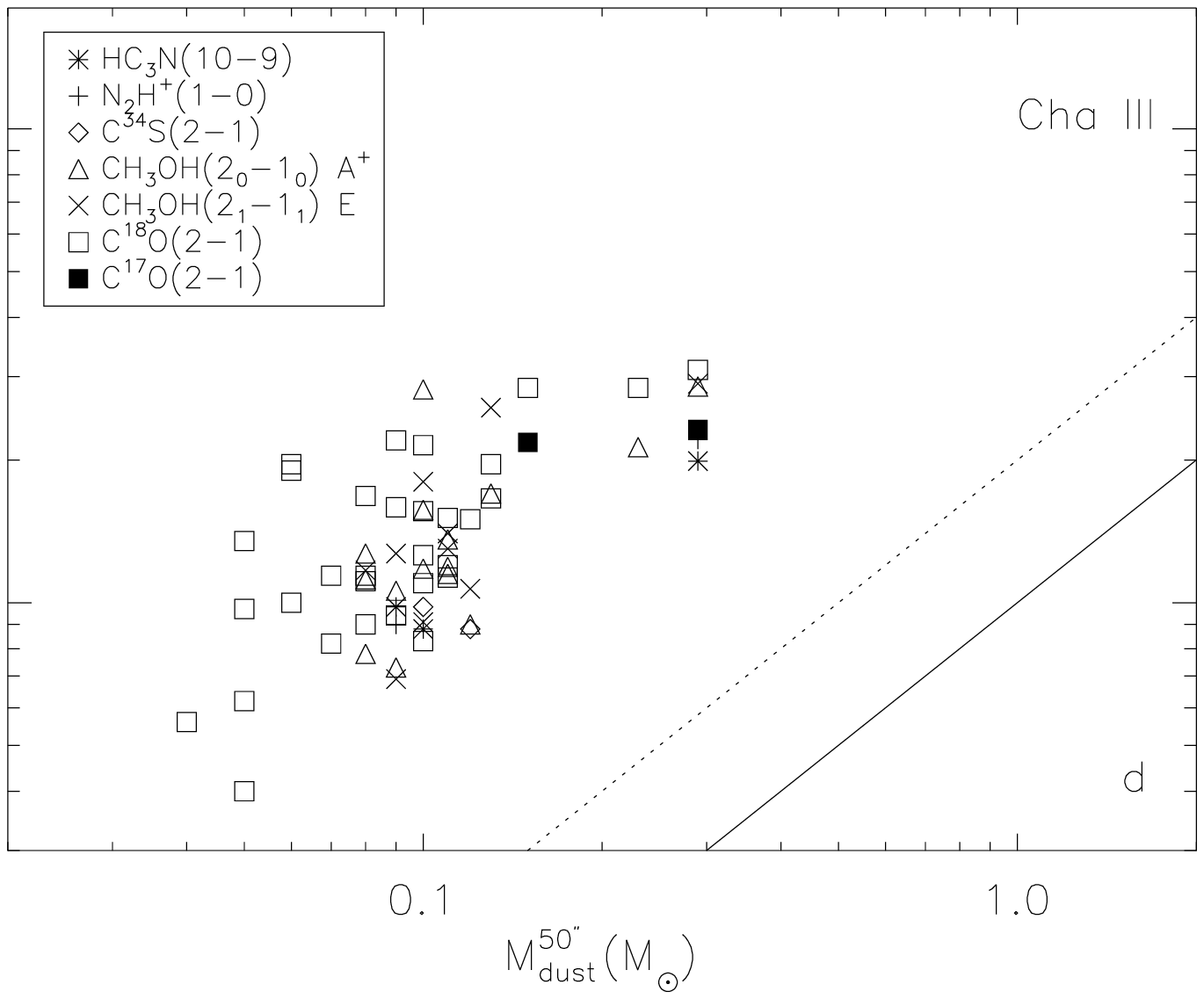} \\
\end{tabular}
 \caption[]{Mass derived from dust emission versus virial mass of the Cha I (a, c) and Cha III (b, d) cores computed for the transitions HC$_3$N 10--9, N$_2$H$^+$ 1--0, C$^{34}$S 2--1, CH$_3$OH 2$_{0}$--1$_{0}$ A$^+$, CH$_3$OH 2$_{1}$--1$_{1}$ E, C$^{18}$O 2--1, and C$^{17}$O 2--1. Panels a and b show the total mass of the cores while panels c and d show the mass obtained for a circular aperture of diameter 50$^{\prime\prime}$. The solid line defines the region in which the cores are virialized ($\frac{M_{\rm vir}}{M_{\rm dust}} \le 1$). The cores located
to the right of the dotted line are considered to be gravitationally bound ($\frac{M_{\rm vir}}{M_{\rm dust}} \le 2 $). \label{fig:mvir_plot}}
\end{center}
\end{figure*}

If the external pressure is negligible, the virial mass of a spherically symmetric core with a density distribution proportional to r$^{-2}$ can be estimated through the expression
\begin{equation}
M_{\rm vir} \sim 3\times\frac{R{\sigma_{\rm tot}}^2}{G}, 
\label{eq:virialmass}
\end{equation}
where the radius $R$ is given by the square root of the product of the major and minor $FWHM$ axes as measured in the dust maps \citep{belloche11a, belloche11b}, $\sqrt{FWHM_{\rm maj} \times FWHM_{\rm min}}$,  and
\begin{equation}
\sigma_{\rm tot} = \sqrt{( {\sigma_{\rm obs}}^2 - {\sigma_{\rm th,mol}}^2 + {\sigma_{\rm th,mean}}^2)},
\end{equation}
with $\sigma_{\rm obs}$ the measured velocity dispersion computed as
\begin{equation}
\sigma_{\rm obs} = \frac{FWHM_{\rm obs}}{\sqrt{8\ \ln 2}},
\end{equation}
where $FWHM_{\rm obs}$ is the observed spectral linewidth. The outer (radii $\ge 20000$ AU) gas kinetic temperature of the densest core in Cha I, 
Cha1-C1 (Cha-MMS1) 
was constrained to 9 K through radiative transfer modelling \citep{tsitali13}. Less dense cores may have a slightly higher kinetic temperature, and we assume $T = 10$ K as an approximation.
The mean thermal velocity dispersions are computed as
\begin{equation}
\sigma_{\rm th,mean} = \sqrt{ \frac{k_{\rm B}T}{\mu m_{\rm H}} },
\end{equation}
\begin{equation}
\sigma_{\rm th,mol} = \sqrt{ \frac{k_{\rm B}T}{m_{\rm mol}} },
\end{equation}
where $k_{\rm B}$ is the Boltzmann constant and $T$ the kinetic temperature in K. The molecular weight of the mean particle is $\mu m_{\rm H}$ ($\mu = 2.37$), $m_{\rm mol} = \mu_{\rm mol} m_H$ is the molecular mass of the tracer, and $m_{\rm H}$ the proton mass. Equation~\ref{eq:virialmass} therefore becomes
\begin{equation}
M_{\rm vir} = 3 \times \frac{R}{G} \left( {\sigma_{\rm obs}}^2 - \frac{k_{\rm B}T}{m_{\rm mol}} + \frac{k_{\rm B}T}{\mu m_{\rm H}} \right).
\end{equation}
A core is considered to be gravitationally bound if its mass corresponds to at least half its virial mass and it is in gravitational virial equilibrium if its mass and virial mass are equal \citep[e.g.,][]{bertoldi92}. To calculate the respective mass to virial mass ratios, we use the masses derived by \citet{belloche11a, belloche11b} through dust continuum observations at 870 $\mu$m. We calculate the mass ratios both for the total dust mass derived from a Gaussian fit (Figs.~\ref{fig:mvir_plot}a and ~\ref{fig:mvir_plot}b) and the dust mass derived for 
a circular aperture of diameter 50$^{\prime\prime}$ (Figs.~\ref{fig:mvir_plot}c and ~\ref{fig:mvir_plot}d), which corresponds to typical mean densities of $\sim 0.5$~--~$1 \times 10^5$~cm$^{-3}$ \citep{belloche11a, belloche11b}. It is not straightforward to decide which dust mass is the most appropriate to compare to the virial masses because the lines trace material along the line of sight, the extent of which cannot be known without a detailed radiative transfer modelling.

Table~\ref{table:detections_chaI_chaIII} and Fig~\ref{fig:mvir_plot}a show that at most five cores are gravitationally bound (Cha1-C1, C2, C3, C4, C5)  
and one is virialized in Cha I when taking the \emph{total} dust mass into account. The virialized core is Cha1-C1 (also known as Cha-MMS1), which is a first hydrostatic core candidate \citep{belloche06, belloche11a, tsitali13}.  
On the other hand, no single molecular line indicates that any core in
Cha~III is virialized and we find that only Cha3-C1 is gravitationally bound 
(see Tables~\ref{table:detections_chaI_chaIII} and ~\ref{table:physical_parameters_chaIII_2}, and Fig.~\ref{fig:mvir_plot}b). Repeating the calculations for the dust mass derived from a 50$^{\prime\prime}$ aperture we find that only Cha1-C1 is gravitationally bound in Cha I and no core is gravitationally bound in Cha III (Figs.~\ref{fig:mvir_plot}c and d). From now on, we will only use the \emph{total} dust mass of the cores and consider the numbers of gravitationally bound and virialized cores as \emph{upper limits}.

\subsubsection{Comparison between tracers}
\label{sec:tracers_comparison}

\begin{table*}
\caption{Comparison of systemic velocities, linewidths, and non-thermal velocity dispersions for selected pairs of transitions for Cha I and III.} 
\begin{tabular}{lllllll}
\hline\hline
 Transitions   & $N_{\rm cores}$\tablefootmark{a} & $v_{\rm LSR,mean}$\tablefootmark{b} & $FWHM_{\rm mean}$\tablefootmark{c} & $\sigma_{\rm nth,mean}$\tablefootmark{d} & $[\frac{\sigma_{\rm nth1}}{\sigma_{\rm nth2}}]_{\rm mean}$\tablefootmark{e} & $[\frac{\sigma_{\rm nth}}{\sigma_{\rm th,mean}}]_{\rm mean}$\tablefootmark{f}  \\ 
              &                                  & \scriptsize{ (km s$^{-1}$) }     &  \scriptsize{ (km s$^{-1}$) } & \scriptsize{ (km s$^{-1}$) } & & \\ \hline

\multicolumn{7}{c}{\underline{Chamaeleon I}} \\
 N$_2$H$^+$ 1--0 & 14 &  4.72$\pm$0.36 &  0.36$\pm$0.11 &  0.14$\pm$0.05 &  0.75$^{+0.18}_{-0.19}$ &  0.75$\pm$0.28 \\
 C$^{34}$S 2--1 & 14 &  4.74$\pm$0.33 &  0.47$\pm$0.16 &  0.19$\pm$0.07 &  --                  &  1.04$\pm$0.38 \\ \hline

 N$_2$H$^+$ 1--0 & 12 &  4.68$\pm$0.35 &  0.38$\pm$0.11 &  0.15$\pm$0.05 &  0.85$^{+0.22}_{-0.17}$ &  0.80$\pm$0.26 \\
 C$^{17}$O 2--1 & 12 &  4.63$\pm$0.30 &  0.46$\pm$0.12 &  0.19$\pm$0.05 &  -- &  0.99$\pm$0.29 \\\hline
 
 N$_2$H$^+$ 1--0                      & 19 &  4.65$\pm$0.34 &  0.37$\pm$0.13 &  0.14$\pm$0.06 &  0.77$^{+0.30}_{-0.30}$ &  0.77$\pm$0.31 \\
 C$^{18}$O/C$^{17}$O \tablefootmark{g} & 19 &  4.67$\pm$0.29 &  0.50$\pm$0.20 &  0.20$\pm$0.09 &  --                   &  1.09$\pm$0.47 \\ \hline

 N$_2$H$^+$ 1--0  & 19 &  4.65$\pm$0.34 &  0.37$\pm$0.13 &  0.14$\pm$0.06 &  0.87$^{+0.17}_{-0.11}$ &  0.77$\pm$0.31 \\
 CH$_3$OH 2$_{0}$--1$_{0}$ A$^+$ & 19 &  4.66$\pm$0.32 &  0.41$\pm$0.13 &  0.16$\pm$0.06 &  -- &  0.88$\pm$0.31 \\\hline

 HC$_3$N 10--9 & 17 &  4.89$\pm$0.40 &  0.30$\pm$0.14 &  0.12$\pm$0.06 &  0.63$^{+0.20}_{-0.22}$ &  0.64$\pm$0.32 \\
 C$^{34}$S 2--1 & 17 &  4.83$\pm$0.35 &  0.46$\pm$0.15 &  0.19$\pm$0.07 &  -- &  1.01$\pm$0.35 \\\hline

 HC$_3$N 10--9  & 15 &  4.87$\pm$0.40 &  0.31$\pm$0.14 &  0.12$\pm$0.06 &  0.71$^{+0.28}_{-0.20}$ &  0.66$\pm$0.33 \\
 C$^{17}$O 2--1 & 15 &  4.73$\pm$0.33 &  0.43$\pm$0.12 &  0.17$\pm$0.05 &  --                   &  0.93$\pm$0.29 \\ \hline

 HC$_3$N 10--9 & 18 &  4.87$\pm$0.39 &  0.31$\pm$0.15 &  0.13$\pm$0.07 &  0.68$^{+0.31}_{-0.17}$ &  0.68$\pm$0.35 \\
 C$^{18}$O/C$^{17}$O \tablefootmark{g} & 18 &  4.75$\pm$0.33 &  0.46$\pm$0.16 &  0.19$\pm$0.07 &  -- &  1.00$\pm$0.37 \\ \hline

HC$_3$N 10--9                  & 18 &  4.87$\pm$0.39 &  0.32$\pm$0.15 &  0.13$\pm$0.07 &  0.75$^{+0.25}_{-0.19}$ &  0.68$\pm$0.35 \\
CH$_3$OH 2$_{0}$--1$_{0}$ A$^+$ & 18 &  4.78$\pm$0.36 &  0.41$\pm$0.11 &  0.16$\pm$0.05 &  -- &  0.88$\pm$0.26 \\\hline

HC$_3$N 10--9  & 14 &  4.75$\pm$0.36 &  0.35$\pm$0.15 &  0.14$\pm$0.07 &  0.96$^{+0.17}_{-0.22}$ &  0.76$\pm$0.36 \\
 N$_2$H$^+$ 1--0 & 14 &  4.69$\pm$0.36 &  0.38$\pm$0.13 &  0.15$\pm$0.06 &  --                  &  0.80$\pm$0.31 \\ \hline
 
C$^{17}$O 2--1                   & 24 &  4.68$\pm$0.37 &  0.41$\pm$0.11 &  0.17$\pm$0.05 &  1.09$^{+0.31}_{-0.30}$ &  0.89$\pm$0.27 \\
 CH$_3$OH 2$_{0}$--1$_{0}$ A$^+$ & 24 &  4.76$\pm$0.34 &  0.39$\pm$0.11 &  0.16$\pm$0.05 &  --              &  0.84$\pm$0.26 \\ \hline
 
C$^{34}$S 2--1 & 18 &  4.86$\pm$0.35 &  0.45$\pm$0.15 &  0.19$\pm$0.06 &  1.14$^{+0.31}_{-0.35}$ &  1.00$\pm$0.34 \\
 C$^{17}$O 2--1 & 18 &  4.78$\pm$0.32 &  0.42$\pm$0.12 &  0.17$\pm$0.05 &  --                  &  0.91$\pm$0.28 \\ \hline
 
CH$_3$OH 2$_{0}$--1$_{0}$ A$^+$ & 28 &  4.72$\pm$0.31 &  0.41$\pm$0.13 &  0.17$\pm$0.06 &  1.11$^{+0.28}_{-0.24}$ &  0.88$\pm$0.32 \\
 CH$_3$OH 2$_{1}$--1$_{1}$ E    & 28 &  4.72$\pm$0.32 &  0.40$\pm$0.16 &  0.16$\pm$0.07 &  --                  &  0.86$\pm$0.39 \\ \hline
 
C$^{18}$O/C$^{17}$O \tablefootmark{g} & 41 &  4.67$\pm$0.33 &  0.45$\pm$0.17 &  0.18$\pm$0.07 &  1.18$^{+0.45}_{-0.30}$ &  0.98$\pm$0.40 \\
 CH$_3$OH 2$_{0}$--1$_{0}$ A$^+$      & 41 &  4.70$\pm$0.32 &  0.41$\pm$0.18 &  0.17$\pm$0.08 &  --                    &  0.89$\pm$0.42 \\ \hline

 C$^{18}$O/C$^{17}$O  \tablefootmark{g} & 22 &  4.77$\pm$0.35 &  0.45$\pm$0.13 &  0.18$\pm$0.06 &  1.18$^{+0.35}_{-0.48}$ &  0.97$\pm$0.32 \\
 C$^{34}$S 2--1 & 22 &  4.84$\pm$0.36 &  0.43$\pm$0.15 &  0.18$\pm$0.06 &  --                  &  0.94$\pm$0.34 \\ \hline

\multicolumn{7}{c}{\underline{Chamaeleon III}} \\

HC$_3$N 10--9 & 3 &  1.51$\pm$0.04 &  0.23$\pm$0.04 &  0.09$\pm$0.02 &  0.54$\pm$0.04 &  0.48$\pm$0.10 \\
C$^{18}$O 2--1 & 3 &  1.40$\pm$0.13 &  0.42$\pm$0.08 &  0.17$\pm$0.04 &  -- &  0.91$\pm$0.19 \\ \hline

HC$_3$N 10--9  &  3 &  1.51$\pm$0.04 &  0.23$\pm$0.04 &  0.09$\pm$0.02 &  0.60$\pm$0.19 &  0.48$\pm$0.10 \\
CH$_3$OH 2$_{0}$--1$_{0}$ A$^+$  &  3 &  1.43$\pm$0.13 &  0.40$\pm$0.08 &  0.16$\pm$0.04 &  -- &  0.86$\pm$0.19 \\ \hline

C$^{18}$O 2--1   & 15 &  1.41$\pm$0.16 &  0.43$\pm$0.10 &  0.18$\pm$0.05 &  1.15$^{+0.21}_{-0.27}$ &  0.94$\pm$0.25 \\
CH$_3$OH 2$_{0}$--1$_{0}$ A$^+$  & 15 &  1.41$\pm$0.18 &  0.39$\pm$0.09 &  0.16$\pm$0.04 &  -- &  0.85$\pm$0.22 \\ \hline

CH$_3$OH 2$_{0}$--1$_{0}$ A$^+$  & 10 &  1.38$\pm$0.15 &  0.38$\pm$0.11 &  0.15$\pm$0.05 &  1.09$^{+0.40}_{-0.34}$ &  0.83$\pm$0.26 \\
CH$_3$OH 2$_{1}$--1$_{1}$ E  & 10 &  1.37$\pm$0.15 &  0.39$\pm$0.12 &  0.16$\pm$0.06 &  -- &  0.85$\pm$0.30 \\ \hline

C$^{18}$O 2--1        &  5 &  1.30$\pm$0.07 &  0.46$\pm$0.07 &  0.19$\pm$0.03 &  1.04$\pm$0.16 &  1.00$\pm$0.17 \\
H$^{13}$CO$^+$ 1--0   &  5 &  1.33$\pm$0.07 &  0.46$\pm$0.10 &  0.19$\pm$0.05 &  -- &  0.99$\pm$0.25 \\ \hline

H$^{13}$CO$^+$ 1--0  &  5 &  1.33$\pm$0.07 &  0.46$\pm$0.10 &  0.19$\pm$0.05 &  1.35$\pm$0.36 &  0.99$\pm$0.25 \\
CH$_3$OH 2$_{0}$--1$_{0}$ A$^+$  &  5 &  1.29$\pm$0.09 &  0.36$\pm$0.08 &  0.14$\pm$0.04 &  -- &  0.76$\pm$0.20 \\  \hline

\end{tabular}
\label{table:samplescompare_chaI}
\vspace*{-1ex}
\tablefoot{
All uncertainties are dispersions around the mean of each distribution apart from the uncertainty of the ratio of the non-thermal velocity dispersions. The latter is the dispersion above or below the mean computed by taking the difference between the largest value within 68\% of the population (above or below the mean) and the mean value itself.  
\tablefoottext{a}{Number of cores with detections in both tracers.}
\tablefoottext{b}{Average systemic velocities. }
\tablefoottext{c}{Average $FWHM$ linewidths.}
\tablefoottext{d}{Average non-thermal velocity dispersion, calculated for $T = 10$~K. }
\tablefoottext{e}{Mean ratio of the non-thermal dispersions of each pair of tracers. The transition listed first is denoted with the subscript 1 and the second with the subscript 2.  }
\tablefoottext{f}{Average ratio of the non-thermal to the mean thermal velocity dispersions.}
\tablefoottext{g}{Composite C$^{18}$O and C$^{17}$O sample (see Sect.~\ref{sec:definitions}).}
}
\end{table*}

We compare the non-thermal velocity dispersions derived for a pair of transitions by using only the cores in which they are both detected. Table~\ref{table:samplescompare_chaI} lists the average systemic velocities, linewidths, non-thermal velocity dispersions, the ratio of the non-thermal velocity dispersions, and the mean ratio of the non-thermal to the mean thermal velocity dispersion for selected pairs of transitions in Cha I and III. 

%% Cha I
N$_2$H$^+$~1--0 has systematically lower (by $\sim20$--30\%) non-thermal velocity dispersions when compared to either C$^{17}$O~2--1, C$^{34}$S~2--1, and the composite C$^{18}$O/C$^{17}$O sample in Cha I.  
HC$_3$N~10--9 has a similar behaviour to N$_2$H$^+$~1--0, with non-thermal dispersions lower than C$^{17}$O~2--1, C$^{34}$S~2--1, and the composite C$^{18}$O/C$^{17}$O by $\sim30$--40\%. The non-thermal dispersion ratio of the last two transitions is around unity. The two methanol species have similar non-thermal velocity dispersions. CH$_3$OH~2$_{0}$--1$_{0}$ has higher non-thermal velocity dispersions than both N$_2$H$^+$ 1--0 and HC$_3$N 10--9 by $\sim10$\% and $\sim20$\%, respectively. It has, however, a similar non-thermal dispersion to C$^{17}$O 2--1.
%% Cha III
HC$_3$N~10--9 similarly has a lower non-thermal dispersion in Cha III compared to C$^{18}$O~2--1 by $\sim50$\% and CH$_3$OH~2$_0$--1$_0$~A$^+$ by $\sim40$\%. The methanol CH$_3$OH~2$_{0}$--1$_{0}$~A$^+$ and CH$_3$OH 2$_{1}$--1$_{1}$~E transitions have similar non-thermal dispersions in both Cha I and III.

Overall, both high- and low-density tracers show subsonic to transonic turbulence in Cha I and III, with their non-thermal velocity dispersions being on average lower than or equal to the mean thermal velocity dispersion.

\subsection{Contraction motions}
\label{sec:infall_signature}

\begin{table}
\caption{``Infall'' and opposite signatures.} 
\vspace*{-1ex}
\hfill{}
\begin{tabular}{@{\extracolsep{-3pt}}llllc}
\hline\hline
Source & Line$_{\rm thick}$\tablefootmark{a}  & Line$_{\rm thin}$\tablefootmark{a} & $T_{\rm blue}$/$T_{\rm red}$\tablefootmark{b} & $\delta V$\tablefootmark{c}    \\ \hline 
\multicolumn{5}{c}{\underline{``Infall'' signature - Chamaeleon I}} \\
Cha1-C1 & CS 2--1         & C$^{34}$S 2--1 &     2.09$\pm$0.09 &      -0.41$\pm$0.10\\ 
Cha1-C2 & CS 2--1         & C$^{34}$S 2--1 &    1.99$\pm$0.13 &      -0.29$\pm$0.12\\
Cha1-C3 & CS 2--1         & C$^{34}$S 2--1 &      2.59$\pm$0.23 &      -0.41$\pm$0.28\\
Cha1-C8 & CS 2--1         & N$_2$H$^+$ 1--0 &    2.10$\pm$0.35 &      -0.63$\pm$0.53\\ 
Cha1-C9 & CS 2--1         & C$^{34}$S 2--1 &     1.58$\pm$0.26 &      -0.28$\pm$0.26\\
Cha1-C10 & CS 2--1        & C$^{34}$S 2--1 &     1.39$\pm$0.10 &      -0.25$\pm$0.17\\  
Cha1-C11 & HNC 1--0       & N$_2$H$^+$ 1--0 &     1.65$\pm$0.11 &      -0.80$\pm$0.58\\  
Cha1-C14 & CS 2--1        & CH$_3$OH \tablefootmark{d} &  1.97$\pm$0.28 &  -0.50$\pm$0.39\\
Cha1-C18 & HNC 1--0      & HC$_3$N 10--9 &       1.67$\pm$0.28 &      -0.91$\pm$0.60\\
Cha1-C19 & HNC 1--0      & C$^{34}$S 2--1 &       2.15$\pm$0.10 &      -0.79$\pm$0.19\\
Cha1-C21 & CS 2--1       & C$^{34}$S 2--1 &      1.25$\pm$0.09 &      -0.64$\pm$0.35\\
Cha1-C27 & $^{13}$CO 2--1 & C$^{18}$O 2--1 &      3.05$\pm$0.08 &      -0.27$\pm$0.05\\
Cha1-C29 & CS 2--1        & C$^{34}$S 2--1 &     2.61$\pm$0.29 &      -0.36$\pm$0.19\\
Cha1-C33 & HNC 1--0       & C$^{34}$S 2--1 &      1.90$\pm$0.08 &      -0.78$\pm$0.31\\
Cha1-C34 & CS 2--1        & C$^{34}$S 2--1 &      2.05$\pm$0.18 &      -0.42$\pm$0.22\\
Cha1-C35 & HNC 1--0       & C$^{34}$S 2--1 &     4.47$\pm$0.44 &      -0.62$\pm$0.27\\
Cha1-C41 & CS 2--1       & C$^{34}$S 2--1 &      1.52$\pm$0.13 &      -0.30$\pm$0.21\\
Cha1-C46 & $^{13}$CO 2--1 & C$^{18}$O 2--1 &     1.36$\pm$0.13 &      -0.36$\pm$0.19\\
\multicolumn{5}{c}{\underline{Opposite Signature - Chamaeleon I}} \\
Cha1-C6 & $^{13}$CO 2--1 & C$^{18}$O 2--1    &       0.85$\pm$0.04 &       0.73$\pm$0.12\\    
Cha1-C12 & $^{13}$CO 2--1 & C$^{18}$O 2--1  &       0.72$\pm$0.05 &       0.43$\pm$0.06\\    
Cha1-C17 & CS 2--1        & C$^{34}$S 2--1              & 0.84$\pm$0.07 &       0.29$\pm$0.16\\ 
Cha1-C21 & HNC 1--0 & C$^{34}$S 2--1 &       0.68$\pm$0.06 &       1.33$\pm$0.62\\
Cha1-C22 & $^{13}$CO 2--1 & CH$_3$OH \tablefootmark{d} &   0.76$\pm$0.06 &       0.82$\pm$0.39\\
Cha1-C24 & $^{13}$CO 2--1 & C$^{18}$O 2--1                &  0.78$\pm$0.06 &       1.46$\pm$0.27\\
Cha1-C30 & CS 2--1        & CH$_3$OH \tablefootmark{d} & 0.33$\pm$0.10 &       0.93$\pm$0.43\\  
Cha1-C31 & CS 2--1        & CH$_3$OH\tablefootmark{d} & 0.58$\pm$0.17 &       0.69$\pm$0.40\\ 
\multicolumn{5}{c}{\underline{``Infall'' signature - Chamaeleon III}} \\
Cha3-C1 & HNC 1--0       & CH$_3$OH \tablefootmark{d} &  2.00$\pm$0.25 &      -0.40$\pm$0.16\\ 
Cha3-C9 & $^{13}$CO 2--1 & C$^{18}$O 2--1                &  1.38$\pm$0.09 &      -0.31$\pm$0.12\\ 
Cha3-C10 & $^{13}$CO 2--1 & C$^{18}$O 2--1               & 1.33$\pm$0.09 &      -0.34$\pm$0.08\\
Cha3-C13 & HNC 1--0      & C$^{18}$O 2--1                & 2.21$\pm$0.75 &      -0.55$\pm$0.28\\ 
Cha3-C20 & $^{13}$CO 2--1 & C$^{18}$O 2--1                & 1.22$\pm$0.06 &      -0.34$\pm$0.10\\ 
\hline
\end{tabular}
\hfill{}
\label{table:infall_table}
\tablefoot{
\tablefoottext{a}{Optically thick and thin transitions shown in Figs.~\ref{fig:infall_chaI}, ~\ref{fig:inversesig}, and ~\ref{fig:infall_chaIII}.}
\tablefoottext{b}{Ratio of the blue and red peak temperatures of the optically thick spectra and 1$\sigma$ uncertainty.}
\tablefoottext{c}{Dimensionless parameter defined in Equation~\ref{eq:dv}. The uncertainty corresponds to the 3$\sigma$ error. }
\tablefoottext{d}{CH$_3$OH 2$_{0}$--1$_{0}$ A$^+$.}
}
\end{table}

\begin{figure*}
\begin{center}
\begin{tabular}{c}
\hspace*{-5ex}
\includegraphics[width=185mm,angle=0]{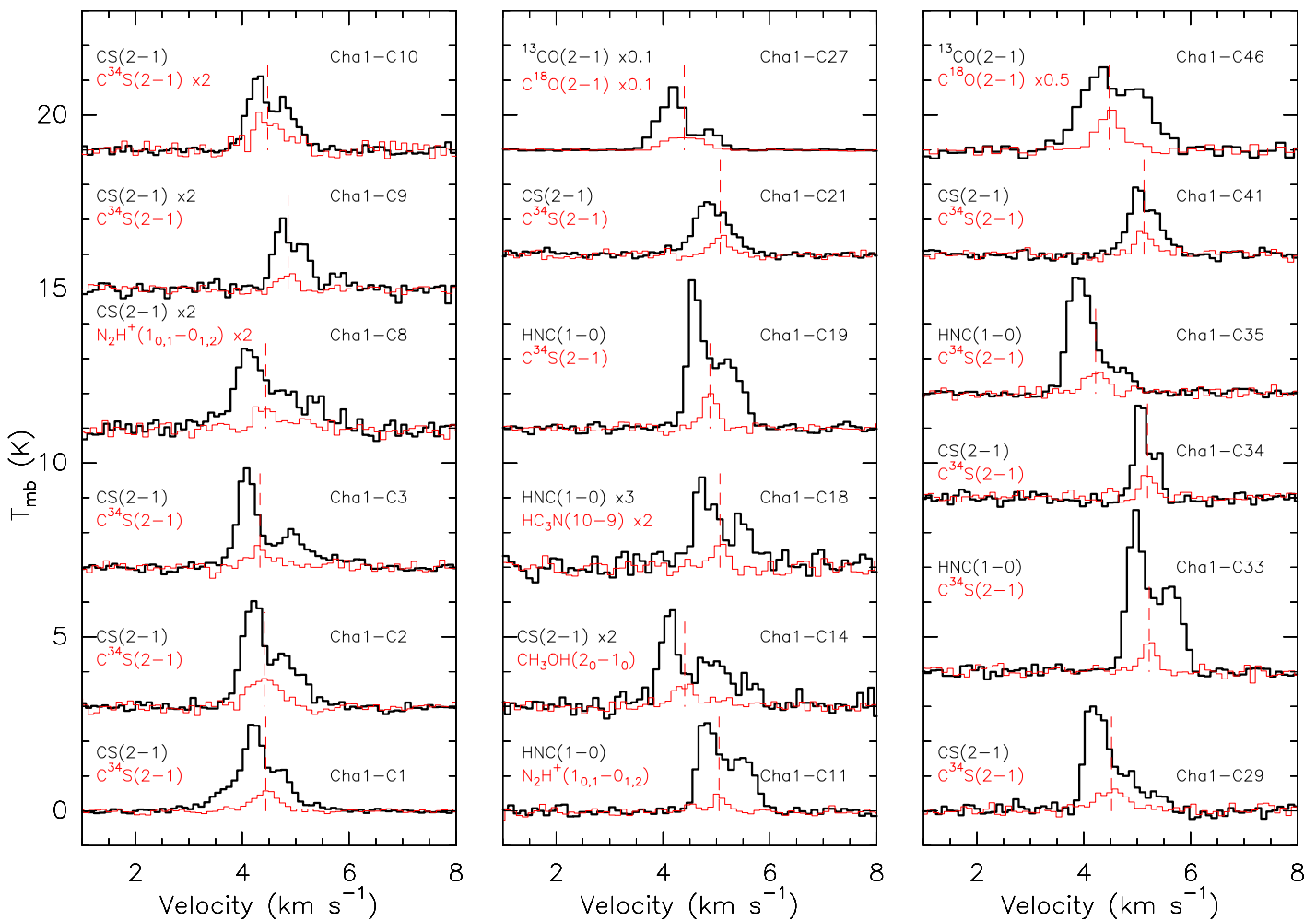} 
\end{tabular}
 \caption[]{``Infall'' signature toward cores in Cha I. The optically thick, self-absorbed transitions are shown with black thick lines and the optically thin transitions with red thin lines. The dashed red line shows the systemic velocity of each core derived from a Gaussian or hyperfine-structure fit to the optically thin line shown in red. The name of the cores is given on the right-hand side of each transition depicted. The factor by which each spectrum was multiplied is given next to the name of the transitions and it corresponds to 1 if not specified.  \label{fig:infall_chaI}}
\end{center}
\end{figure*}

\begin{figure}
\begin{tabular}{c}
\includegraphics[width=70mm,angle=0]{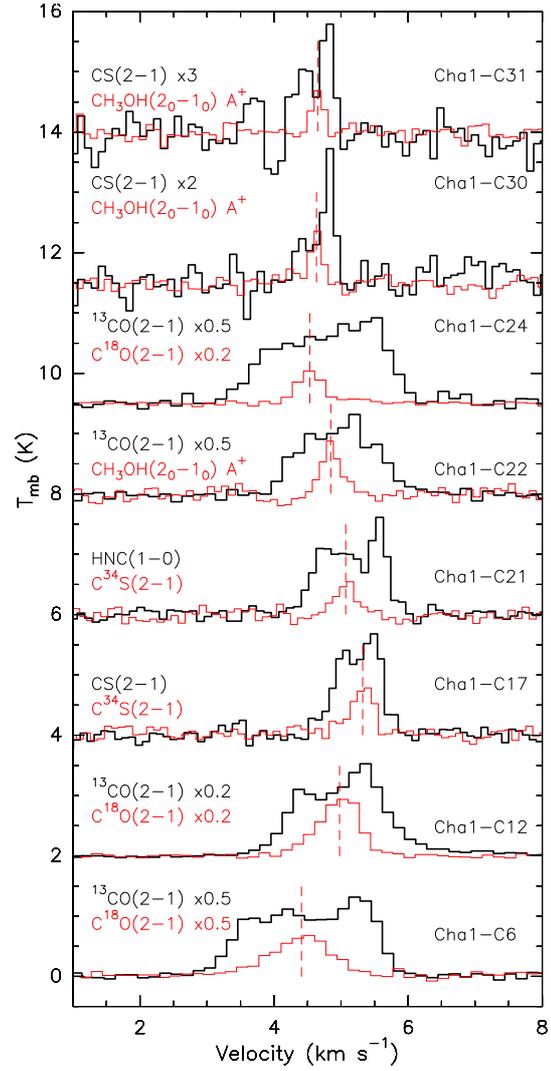} 
\end{tabular}
 \caption[]{Same as Fig.~\ref{fig:infall_chaI}, for the cores in Cha I showing the opposite signature. 
 \label{fig:inversesig}}
\end{figure}

\begin{figure}
\begin{tabular}{c} 
\hspace*{-5ex}
\includegraphics[width=70mm,angle=0]{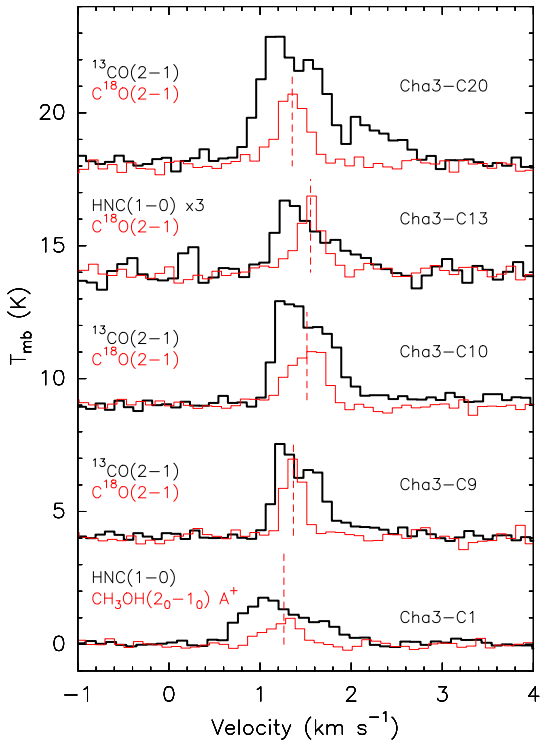}
\end{tabular}
 \caption[]{Same as Fig.~\ref{fig:infall_chaI}, but for the cores in Cha III.  \label{fig:infall_chaIII}}
\end{figure}

\begin{table}
\caption{Line shapes of optically thick tracers\tablefootmark{a} for the cores that show the ``infall'' or opposite signature in Cha I and III.} 
\begin{tabular}{@{\extracolsep{-2pt}}llll}
\hline\hline
Source  & CS 2--1  & HNC 1--0  & $^{13}$CO 2--1 \\ \hline
\multicolumn{4}{c}{\underline{Cores with ``infall'' signature in Cha I }} \\
Cha1-C1 & B, A, I  & B, A, I  & B, A, I  \\
Cha1-C2 & B, A, I  & B, A, I  & B, A, I  \\
Cha1-C3 & B, A, I  & B, A, I  & B, A, I  \\
Cha1-C8 & B, A, I  & B, A, I  & B, A, I  \\
Cha1-C9 & B, A, I  & S, A     & S, N  \\
Cha1-C10 & B, A, I & B, N, I  & B, N, I \\
Cha1-C11 & S, N    & B, A, I  & S, N \\
Cha1-C14 & B, A, I & B, A, I  & B, A, I \\
Cha1-C18 & S, N    & B, A, I  & S, N \\
Cha1-C19 & S, N    & B, A, I  & B, N \\
Cha1-C21 & B, A, I & R, A, O  & S, N \\
Cha1-C27 & S, N    & low S/N  & B, A, I    \\
Cha1-C29 & B, A, I & B, A, I  & B, A, I \\
Cha1-C33 & B, A, I & B, A, I & R, N\tablefootmark{b} \\
Cha1-C34 & B, A, I & B, A, I  & B, N, I  \\
Cha1-C35 & S, N    & B, A, I  & B, A, I  \\
Cha1-C41 & B, A, I  & S, N    & B, A, I \\
Cha1-C46 & S, N     & low S/N  & B, A, I \\ 

\multicolumn{4}{c}{\underline{Cores with ``infall'' signature in Cha III }} \\
Cha3-C1 & S, N & B, A, I &  B, A, I \\
Cha3-C9 & S, N & S, N &  B, A, I \\
Cha3-C10 & S, N & low S/N & B, N, I \\
Cha3-C13 & S, N & B, N, I & B, A \\
Cha3-C20 & S, N & S, N & B, A, I \\

\multicolumn{4}{c}{\underline{Cores with opposite signature in Cha I}} \\
Cha1-C6 & S, N   & --      & R, A, O \\
Cha1-C12 & S, N  & S, N    & R, A, O \\
Cha1-C17 & R, A, O  & R, A, O  & R, N, O \\
Cha1-C21 & B, A, I & R, A, O  & S, N \\
Cha1-C22 & low S/N & low S/N & R, A, O \\
Cha1-C24 & S, N & S, N & R, N, O \\
Cha1-C30 & R, A, O  & R, N, O  & S, N \\
Cha1-C31 & R, A, O  & R, A, O & S, N \\

\hline
\end{tabular}
\hfill{}
\label{table:lineshapes_table}
\tablefoot{
\tablefoottext{a}{Abbreviations used: B = blue asymmetry, R = red asymmetry, A = self-absorption, N = no self-absorption, S = symmetric, I = ``infall'' signature (with one of the optically thin lines), O = opposite signature.}
\tablefoottext{b}{No opposite signature with this line (i.e., optically thin line peaks underneath blue peak).}
}
\end{table}

The infall signature is an optical depth effect and is produced when the line excitation temperature increases toward the centre of a collapsing core \citep{evans03}. Assuming spherical symmetry, the infall signature of a core consists of two elements. The first one is an asymmetric profile of an optically thick line that is skewed to the blue and sometimes features a self-absorption dip (due to absorption from the outer low-excitation material). In order to exclude the possibility that the observed spectral profile is due to two separate, overlapping velocity components, an optically thin, symmetric (and single-peaked) line is required to peak at a velocity in between the red and blue portions of the optically thick line \citep{leungbrown77, walker86, zhou92, myers95, myers96}. 
Strictly speaking, the infall signature may in some cases trace early 
or large-scale contraction motions in layers that are not (yet) 
gravitationally collapsing. Therefore, we use the expression ``infall'' 
signature with quotation marks in the rest of the article to remind the reader 
that the signature may not trace proper infall motions but only contraction 
motions when gravity does not (yet) dominate the dynamical evolution of these 
layers.

The opposite signature, i.e., with the red peak emission of the self-absorbed transition stronger than the blue peak, has also been seen in other surveys of dense cores in the past \citep[e.g.,][]{gregersen97, mardones97, gregersen00, gregersen01}. The red-skewed profiles are generally not very well understood. They may arise as a result of outflowing or expanding material \citep{evans03}. In most of the aforementioned surveys, the blue profiles outnumbered the red profiles which was taken as an indication of ``infall'' motions. 

\citet{mardones97} proposed a non-dimensional parameter, $\delta V$, 
to quantify the observed asymmetry for both blue-skewed and red-skewed profiles. This parameter is
\begin{equation}
\label{eq:dv}
\delta V = \frac{V_{\rm thick} - V_{\rm thin}}{\Delta V_{\rm thin}},
\end{equation}
where $V_{\rm thick}$ and $V_{\rm thin}$ are the peak velocities of the brightest optically thick peak and the optically thin peak, respectively, and $\Delta V_{\rm thin}$ is the linewidth ($FWHM$) of the optically thin transition. This parameter is given in Table~\ref{table:infall_table} along with the peak temperature ratio of the blueshifted and redshifted peaks of the optically thick line. This way we quantify both the peak asymmetry in the optically thick spectrum and the amount by which the brightest peak of the optically thick line is blueshifted or redshifted with respect to the peak of the optically thin line. The peak velocities, linewidths, and respective uncertainties of the optically thin transitions were derived by performing Gaussian fits in CLASS. The velocity of the strongest optically thick peak ($V_{\rm thick}$) was determined by applying Gaussian fits to the spectrum after masking the channels of the weaker peak. We used the rms of the spectrum as the uncertainty of $T_{\rm blue}$ and $T_{\rm red}$, and the velocity uncertainty from the Gaussian fit for $V_{\rm thick}$. Negative values of $\delta V$ correspond to blue-skewed profiles (``infall'' signature) and positive values to red-skewed profiles (opposite signature). 

For Cha I, we look for the ``infall'' signature in pairs of optically thick/thin isotopologues in CS 2--1 and C$^{34}$S 2--1. In the case of a non-detection in C$^{34}$S 2--1 we use the detected optically thin transitions N$_2$H$^+$ 1--0, HC$_3$N 10--9, CH$_3$OH 2$_{0}$--1$_{0}$ A$^+$, or C$^{18}$O 2--1 for the cores in which it is optically thin (see Tables~\ref{table:opacity_c18o} and ~\ref{table:opacity_c18o_chaIII}). If CS 2--1 is not self-absorbed, we search for the ``infall'' signature in the optically thick tracers HNC 1--0 and $^{13}$CO 2--1. CS 2--1 was detected toward all 20 cores in Cha III, but no spectrum shows a significant self-absorption in this transition. Only two cores have detections in C$^{34}$S 2--1. We therefore use the HNC 1--0 and $^{13}$CO 2--1 as optically thick tracers for the ``infall'' signature in Cha III. $^{13}$CO 2--1 likely traces the outer core material \citep[e.g.,][]{onishi99, arce11} and any ``infall'' signature found with this transition may only correspond to large-scale contraction motions. On the other hand, C$^{18}$O and C$^{17}$O 2--1 were found to correlate very well with the dust extinction between 6 
and 15~mag in Cha I \citep{belloche11a}. These values correspond to H$_2$ column densities of $\sim4$--$9\times 10^{21}$ cm$^{-2}$, which are similar to the peak column densities of the cores in Cha I and III \citep[see][]{belloche11a, belloche11b}. These two transitions are therefore expected to trace the bulk mass of the cores (excluding the innermost parts where depletion occurs) in Cha I and Cha III very well for extinctions in the aforementioned range. In general, the ``infall'' signatures discussed in this section may not signify gravitational collapse, but instead large-scale contraction motions of material accumulating onto the cores. This accumulation of material may turn some of the cores from unbound to gravitationally bound structures in the future. Table~\ref{table:lineshapes_table} lists the line shape characteristics of the optically thick spectra of all the cores that show an ``infall'' or opposite signature in Cha I and III.
Figure~\ref{fig:appendix_spectra} shows the spectra of all 
transitions used to search for signatures of contraction motions in the sources 
listed in Table~\ref{table:lineshapes_table}.

\begin{figure}
\begin{tabular}{c}
\hspace*{-4ex}
\includegraphics[width=110mm,angle=270]{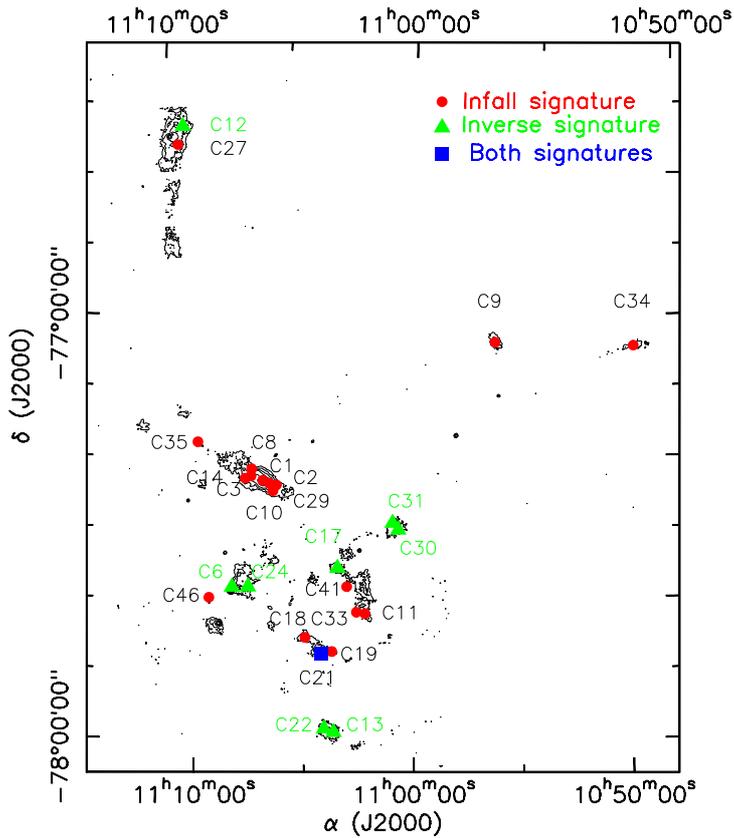} 
\end{tabular}
 \caption[]{The cores showing the ``infall'' signature in Cha I (corresponding to Fig.~\ref{fig:infall_chaI}) are plotted as red filled circles, the cores showing the opposite signature as green triangles, and the cores showing both signatures as blue squares. The contours correspond to the 870 $\mu$m dust continuum emission obtained with LABOCA \citep{belloche11a} with contour levels of $a$, $2a$, $4a$, $8a$, $16a$, $32a$, and $a = 48$~mJy/21$^{\prime\prime}$-beam (4$\sigma$).\label{fig:infallingcores}}
\end{figure}

Starting with Cha I, a total of 18 cores have a negative $\delta V$ parameter whose absolute value is greater than the 3$\sigma$ uncertainty and thus appear to be contracting based on their ``infall'' signatures (Fig.~\ref{fig:infall_chaI}). All of them have a peak temperature asymmetry 
$T_{\rm blue}/T_{\rm red}$ deviating from unity by more than 1$\sigma$ and 15 of 
them by more than 3$\sigma$, the three exception being
Cha1-C9, Cha1-C18, and Cha1-C21 (Table~\ref{table:infall_table}). Cha1-C21, however, also shows the opposite signature in HNC 1--0 (see Fig.~\ref{fig:inversesig}) and is therefore not a reliable candidate for contraction.
The total number of cores showing the ``infall'' signature is therefore 17, or 28\% of the Cha I cores. Their locations within the cloud are shown in Fig.~\ref{fig:infallingcores}. The optically thick, self-absorbed transitions CS~2--1, HNC~1--0, and $^{13}$CO~2--1, and the assumed optically thin transitions, C$^{34}$S~2--1, HC$_3$N~10--9, C$^{18}$O~2--1, N$_2$H$^+$~1$_{0,1}$--0$_{1,2}$ (isolated component), and CH$_3$OH~2$_{0}$--1$_{0}$ A$^+$, form this typical signature for the 18 cores in Cha I (Fig.~\ref{fig:infall_chaI}). Table~\ref{table:opacity_cs} shows that C$^{34}$S~2--1 is indeed optically thin for all the relevant cores (for which this transition is used) in Fig.~\ref{fig:infall_chaI}, as well as the cores discussed later showing the opposite signature. We have no similar opacity estimates for HC$_3$N~10--9 and CH$_3$OH~2$_{0}$--1$_{0}$ A$^+$, but 
these transitions have a Gaussian line shape for all cores in which 
they are detected. C$^{18}$O 2--1 is optically thin toward approximately half of the cores that have C$^{17}$O~2--1 detections (Table~\ref{table:opacity_c18o}). The core Cha1-C27 is optically thin in C$^{18}$O 2--1. We do not have an estimate of the C$^{18}$O 2--1 opacity for Cha1-C46 but the line looks symmetric and peaks in between the blue and red peaks of the $^{13}$CO 2--1 spectrum. We finally use the isolated, likely optically thin component of N$_2$H$^+$ 1--0 for only two cores, Cha1-C8 and Cha1-C11. It is interesting to note that 88\% of the cores showing the ``infall'' signature (14 out of 17; 
the exceptions being Cha1-C1, C2, and C3) do not appear to be gravitationally bound based on the virial analysis (Sect.~\ref{sec:vir_analysis}). 

The opposite signature is seen in eight cores in Cha I (Fig.~\ref{fig:inversesig}) and it corresponds to positive $\delta V$ parameters (Table~\ref{table:infall_table}). As Cha1-C21 shows both an opposite and ``infall'' signature, it is not a reliable candidate either for contraction or expansion 
motions alone. We therefore take the remaining seven cores to be showing evidence of expansion (small- or large-scale depending on the tracer).
Four of the seven cores showing the opposite signature have an asymmetric profile in $^{13}$CO 2--1. It is likely that this transition is sensitive to the ambient, low-density gas and might not be appropriate for tracing motions within the dense cores themselves. If we excluded the low-density tracer $^{13}$CO 2--1,
the actual number of opposite signature spectra could be reduced to 3 and the number of ``infall'' signature spectra to 16. 

The two signatures (``infall'' and opposite) could be produced with the same
probability by oscillations
This interpretation was proposed for the starless core B68 where both blue-skewed and red-skewed spectra were seen in CS 2--1 \citep{lada03} and HCO$^+$ 3--2 across the core \citep{ redman06}. This pattern was interpreted as an oscillation, or pulsation of the outer core layers about an equilibrium state due to external pressure perturbations \citep{lada03, redman06, ketofield05}. \citet{aguti07} also suggested pulsation as a possible physical state for the outer layers of the starless core FeSt 1-457.

If the ``infall'' and opposite signatures also trace oscillating motions, then given that we see the opposite signature in $\sim 7$ cores overall we could expect the sample of ``infalling'' cores to be contaminated by $\sim 7$ oscillating cores, which would reduce the number of infalling cores to $\sim 10$ in Cha I. The fact that we see both the ``infall'' and the opposite signature in Cha1-C21 may be an indication of such an oscillation. It is not straightforward to conclude if pulsations are occuring as most sources, apart from Cha1-C21, do not show both signatures simultaneously in different tracers and we do not have spatial information. 

\begin{table*}
\caption{Average velocity dispersions, crossing time, and collisional time for the Cha I and III cores detected in different tracers.} 
\begin{tabular}{@{\extracolsep{-2pt}}lllllllllllll}
\hline\hline
Transition      & $D_{\rm cl}$\tablefootmark{a} & $D_{\rm cyl}$\tablefootmark{b}  & $N_{\rm fit}$\tablefootmark{c} &  $R_{\rm mean}$\tablefootmark{d} & $M_{\rm mean}$\tablefootmark{e} &  $v_{\rm mean}$\tablefootmark{f}  & $\sigma_{\rm 1D}$\tablefootmark{g} & $\sigma_{\rm 3D}$\tablefootmark{h} & $v_{\rm rel,COM}$\tablefootmark{i} & $v_{\rm rel,cores}$\tablefootmark{j}  & $t_{\rm cross}$\tablefootmark{k}      & $t_{\rm coll}$\tablefootmark{l}     \\

                &  \scriptsize{(pc)}           & \scriptsize{(pc)} &                           &    \scriptsize{(AU)}           &   \scriptsize{($M_{\odot}$)}    & \scriptsize{(km s$^{-1}$)}       & \scriptsize{(km s$^{-1}$)}         & \scriptsize{(km s$^{-1}$)}            & \scriptsize{(km s$^{-1}$)}        & \scriptsize{(km s$^{-1}$)}            &         \scriptsize{(Myr)}          &               \scriptsize{(Myr)}            \\ \hline

\multicolumn{13}{c}{\underline{Cha I} \tablefootmark{m}} \\ 

C$^{18}$O 2--1                 & 4.0 & 3.0 & 60  &   6462 &      0.35 &   4.72 &     0.34 &       0.59 &       0.54  &      0.77 &    5.0--6.7 &   138  \\

CH$_3$OH 2$_{0}$--1$_{0}$ A$^+$ & 4.0 & 3.0 & 41  &   6783 &      0.42 &   4.70 &     0.32 &       0.56 &       0.51 &       0.73 &    5.3--7.0 &    179  \\

\multicolumn{13}{c}{\underline{Cha I North} \tablefootmark{n}} \\
 
C$^{18}$O 2--1                 & 0.9  & 0.15 & 7 &    7761 &      0.36 &   4.47 &      0.29 &       0.50 &       0.46 &       0.65 &    0.3--1.8 &  0.5  \\
CH$_3$OH 2$_{0}$--1$_{0}$ A$^+$ & 0.9 & 0.15  & 4 &    7755 &      0.39 &   4.61 &      0.16 &       0.27 &       0.25 &       0.36 &    0.5--3.2 &  0.9   \\

\multicolumn{13}{c}{\underline{Cha I Centre} \tablefootmark{n}} \\
        
C$^{18}$O 2--1                 & 0.9 & 0.3 & 16 &  5640 &      0.52 &  4.48 &     0.20 &       0.35 &       0.32 &       0.46 &     0.8--2.5 &  1.2   \\
CH$_3$OH 2$_{0}$--1$_{0}$ A$^+$ & 0.9 & 0.3 & 15 &  5616 &      0.54 &  4.50 &     0.29 &       0.50 &       0.46 &       0.65 &      0.6--1.8 &  1.4  \\

\multicolumn{13}{c}{\underline{Cha I South} \tablefootmark{n}} \\

C$^{18}$O 2--1                  & 1.3 & -- & 29   &   6331 &       0.26 &  4.83 &     0.32 &      0.56  &       0.51 &       0.73 &      2.3 &  13   \\
CH$_3$OH 2$_{0}$--1$_{0}$ A$^+$  & 1.3 & -- & 18   &   6806 &       0.29 &  4.85 &    0.30 &      0.51 &       0.47 &       0.67 &        2.5 &  19   \\

\multicolumn{13}{c}{\underline{Cha III \tablefootmark{m}}} \\
C$^{18}$O 2--1                 & 4.9 & 0.8 & 29 &  5969 &       0.25 &   1.50 &       0.23 &       0.40 &       0.37 &       0.52 &     2.0--12 &  36  \\ 
CH$_3$OH 2$_{0}$--1$_{0}$ A$^+$ & 4.9 & 0.8 & 15 & 6845 &       0.34 &   1.41 &       0.18 &       0.32 &       0.29 &       0.42 &      2.5--15 & 49   \\   

\multicolumn{13}{c}{\underline{Cha III North \tablefootmark{n}}} \\

C$^{18}$O 2--1                 & 0.6 & 0.3  & 8  & 6317  & 0.38 & 1.36  & 0.09 & 0.15 & 0.14 & 0.19 & 2.0--4.0 & 1.1  \\
CH$_3$OH 2$_{0}$--1$_{0}$ A$^+$ & 0.6 & 0.3  & 6  & 7276   & 0.48 & 1.32 & 0.09 & 0.15  & 0.14 & 0.20 & 1.9--3.9 & 1.1  \\

\multicolumn{13}{c}{\underline{Cha III Centre\tablefootmark{n}}} \\

C$^{18}$O 2--1                   & 0.9  & 0.5 & 9 & 5408   & 0.18 & 1.53 & 0.31 & 0.53 & 0.49 & 0.69 & 0.9--1.7 & 9.8 \\
CH$_3$OH 2$_{0}$--1$_{0}$ A$^+$  &  0.9 & 0.5 & 4 & 6911  & 0.26  & 1.37 & 0.18 & 0.31 & 0.28 & 0.40 & 1.6--2.9 & 15  \\

\multicolumn{13}{c}{\underline{Cha III South\tablefootmark{n}}} \\

C$^{18}$O 2--1                  & 1.7 & 0.5 & 11 & 6429 & 0.23 & 1.59 & 0.17 & 0.29 & 0.27 & 0.38 & 1.7--5.7 & 12  \\
CH$_3$OH 2$_{0}$--1$_{0}$ A$^+$ & 1.7 & 0.5  & 5  & 6276 & 0.23 & 1.56  & 0.19 & 0.32 & 0.30 & 0.42 & 1.5--5.1 & 28  \\ 

\hline
\end{tabular}
\label{table:dispersions_lines}
\vspace*{-1ex}
\tablefoot{
\tablefoottext{a}{Approximate diameter of the corresponding region (or height $H$ when cylindrical symmetry is assumed). }
\tablefoottext{b}{Approximate diameter of cylinder (when cylindrical symmetry is assumed).}
\tablefoottext{c}{Number of spectra that were fitted, excluding second velocity components.}
\tablefoottext{d}{Mean radius of the cores.}
\tablefoottext{e}{Mean total mass of the cores.}
\tablefoottext{f}{Mean $V_{\rm LSR}$.}
\tablefoottext{g}{1D velocity dispersion.}
\tablefoottext{h}{3D velocity dispersion.}
\tablefoottext{i}{Mean velocity relative to the centre of mass of the system.}
\tablefoottext{j}{Mean relative speed between cores.}
\tablefoottext{k}{Time needed for a core to cross the region, calculated 
with Eq.~\ref{eq:tcross} in the spherical case. In the 
cylindrical case the higher value is computed with Eq.~\ref{eq:tcross} and
the lower value with Eq.~\ref{eq:tcross2}, i.e., they correspond to the long and 
short axes of the cylinder, respectively.}
\tablefoottext{l}{Time needed for cores to collide with one another.}
\tablefoottext{m}{Results for the whole cloud.}
\tablefoottext{n}{The cores located in Cha I North, Centre, and South are shown in Fig.~\ref{fig:velgradients}, and the cores in Cha III North, Centre, and South are shown in Fig.~\ref{fig:veldistribution_chaIII}.}
}
\end{table*}

Five cores in Cha III show the classical ``infall'' signature (Fig.~\ref{fig:infall_chaIII}). The spectrum of Cha3-C1 in HNC 1--0 is not self-absorbed, but it is clearly skewed to the blue. CH$_3$OH~2$_{0}$--1$_{0}$ A$^+$ is redshifted with respect to the blue peak of HNC 1--0, thus making Cha3-C1 a probable collapsing core. The asymmetry is more pronounced for the cores Cha3-C9, Cha3-C10, Cha3-C13, and Cha3-C20 with optically thin lines peaking in between the blue and red peaks of the optically thick spectrum. Cha3-C1 is the only core that also appears to be gravitationally bound from our virial analysis (see Sect~\ref{sec:vir_analysis}). All cores showing the ``infall'' signature in Cha III have a significant (negative) $\delta V$ parameter at the 3$\sigma$ level. The $T_{\rm blue}/T_{\rm red}$ asymmetry is larger than $\sim 1.2$ in all cases.
Three of the contracting cores, however, show the ``infall'' signature in $^{13}$CO 2--1. If this transition is not a good tracer of core motions, then only two cores are valid ``infall'' candidates. 

Therefore, taking into account the observed line profiles, the parameter $\delta V$, and the ambiguity of the $^{13}$CO 2--1 profiles, the number of the ``infall'' candidates is 8--17 (with/without oscillations and without/with $^{13}$CO 2--1) cores in Cha I and $\sim 2$ -- 5 (without/with $^{13}$CO 2--1) in Cha III. The opposite signature is seen toward 3--7 (without/with $^{13}$CO 2--1) cores in Cha I. Using these ranges, 13--28\% of the core population in Cha I and 10--25\% of the observed Cha III sample (or $\sim 7$--17\% of the overall Cha III starless core population) are ``infall'' candidates.

\subsection{Centroid velocities}
\label{sec:cen_velocities}

\subsubsection{Core velocity distribution}
\label{sec:vel_distribution}

We follow the analysis in \citet{andre07} to estimate if interactions between cores are likely to be dynamically important for the evolution of Cha I and III. This analysis assumes Maxwellian, isotropic velocities. We perform the analysis for the whole sample of cores in Cha I and Cha III, and also for the subsamples of cores in Cha I North, Centre, South, and Cha III North, Centre, and South, separately. 
We use the average velocity of all cores in each transition for our calculations (Table~\ref{table:dispersions_lines}). 

\begin{figure*}
\begin{center}
\begin{tabular}{cc}
\includegraphics[width=95mm,angle=270]{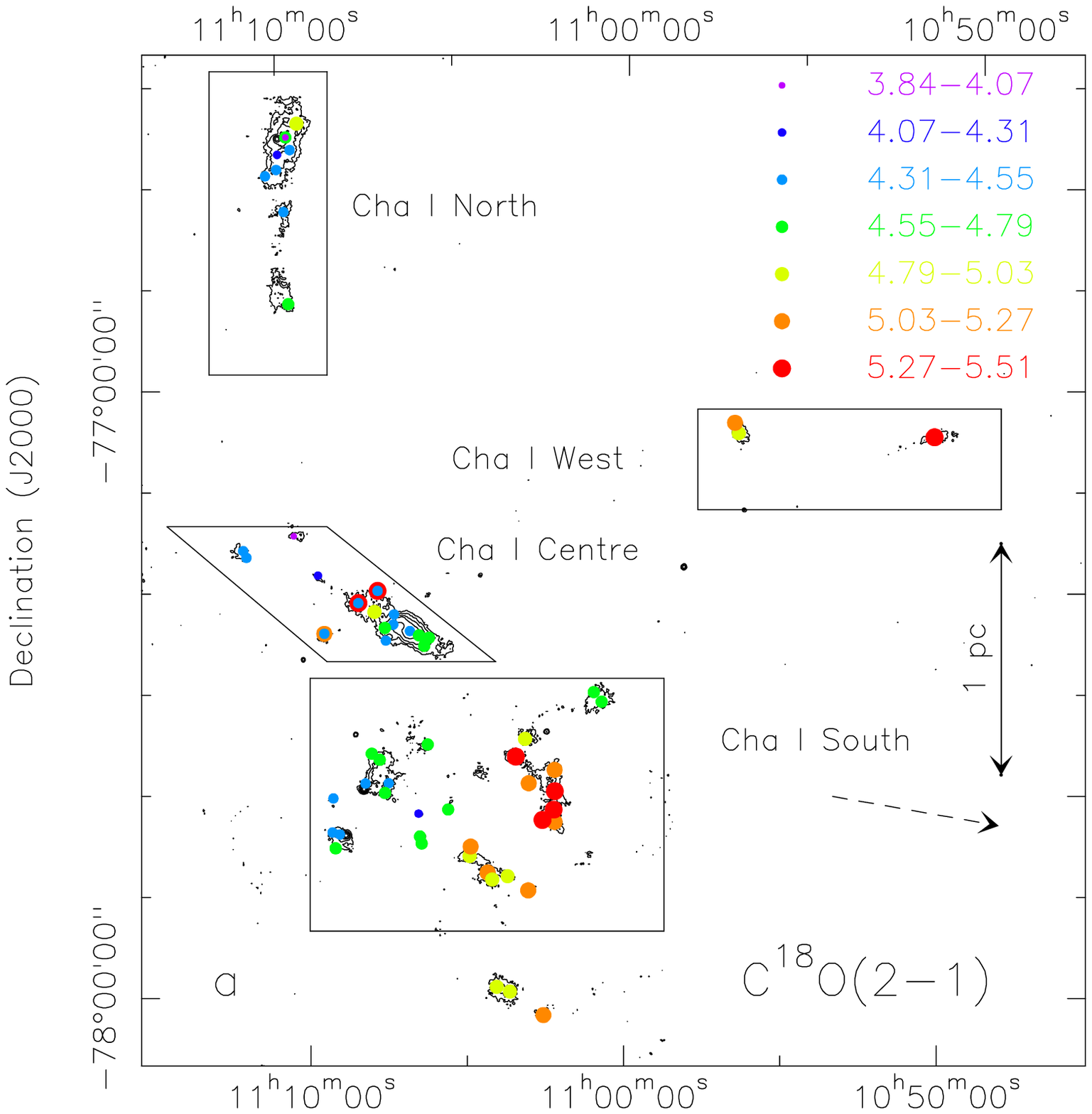} &
\includegraphics[width=95mm,angle=270]{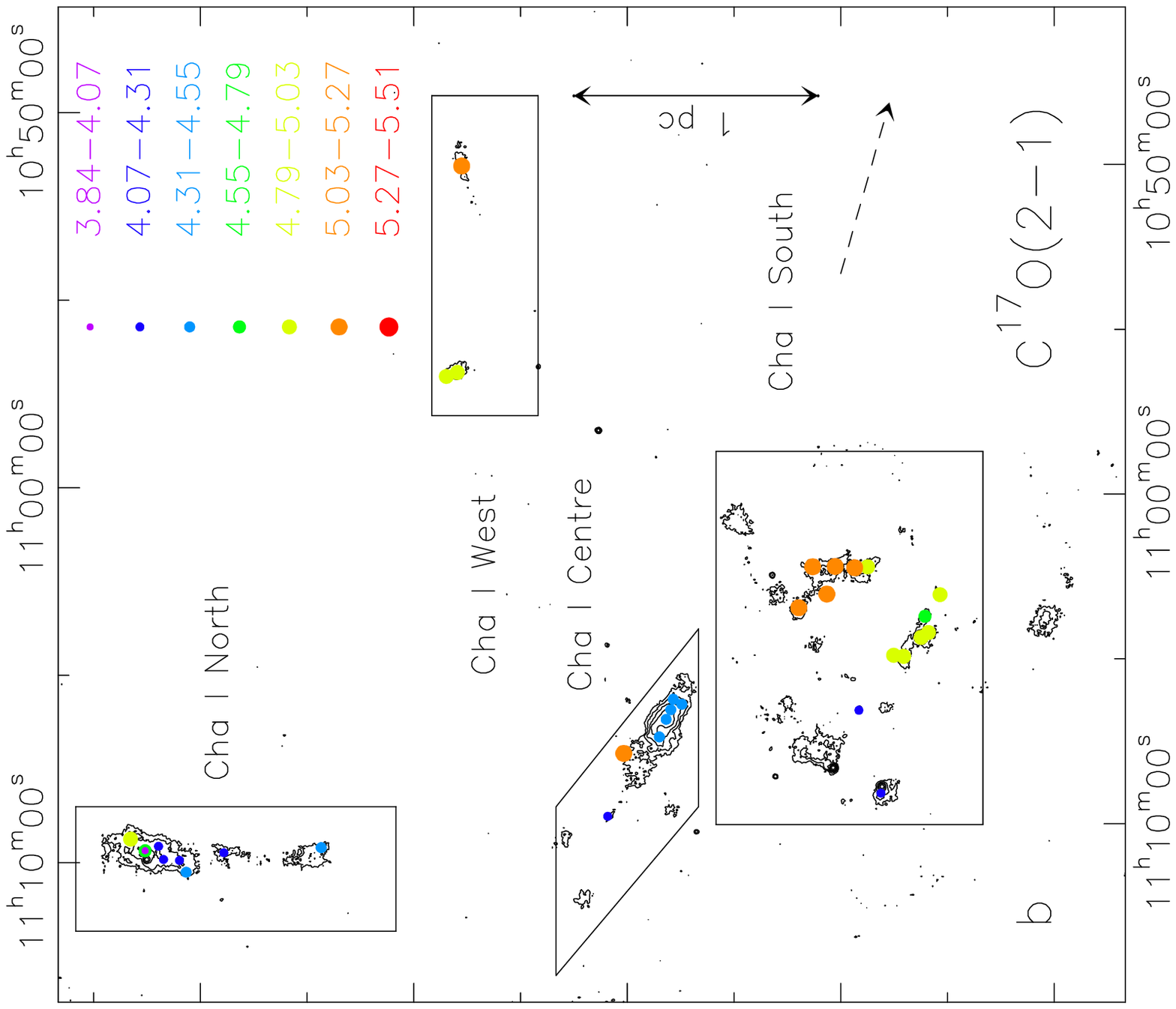} \\
\includegraphics[width=97mm,angle=270]{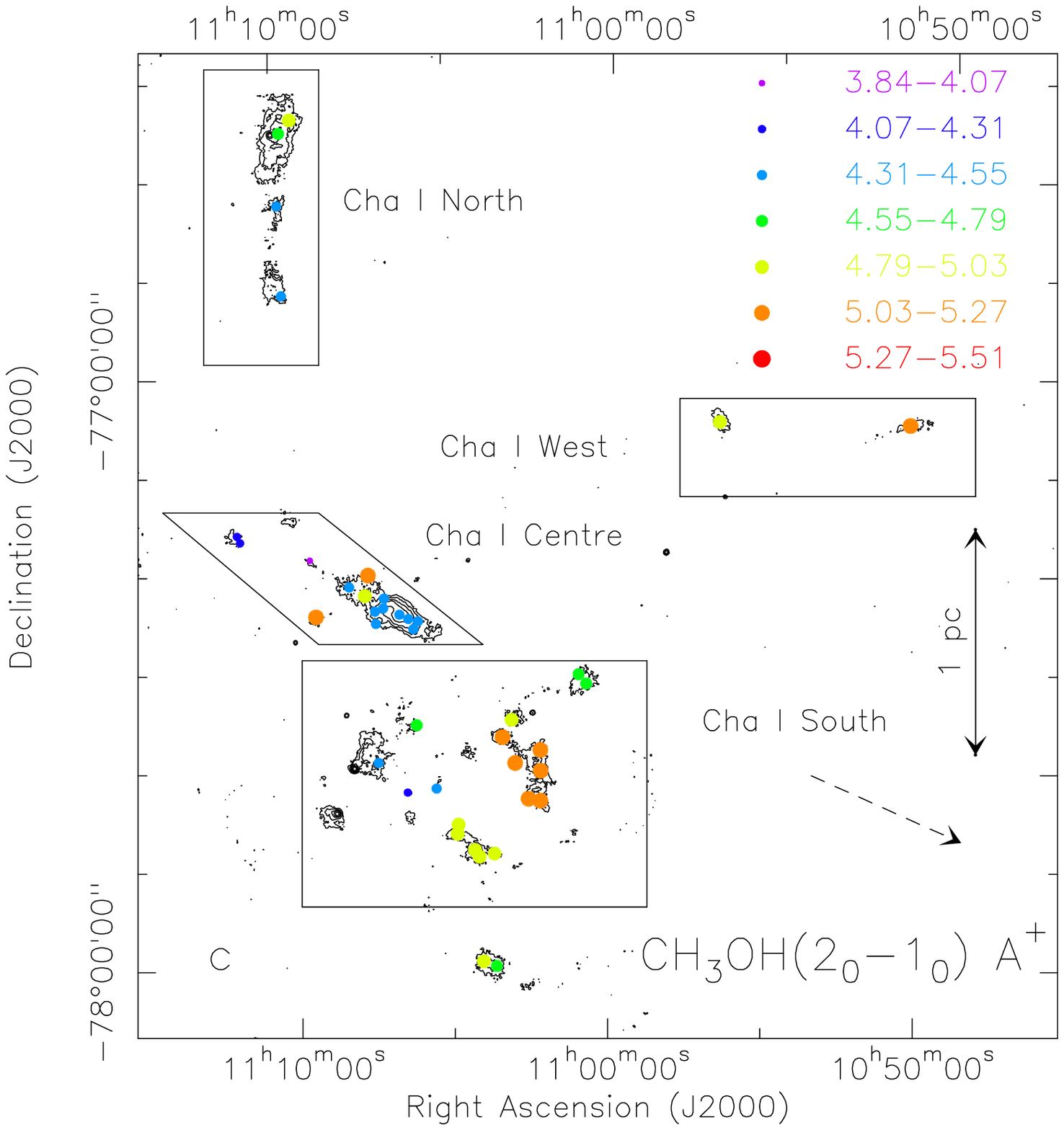} &
\includegraphics[width=97mm,angle=270]{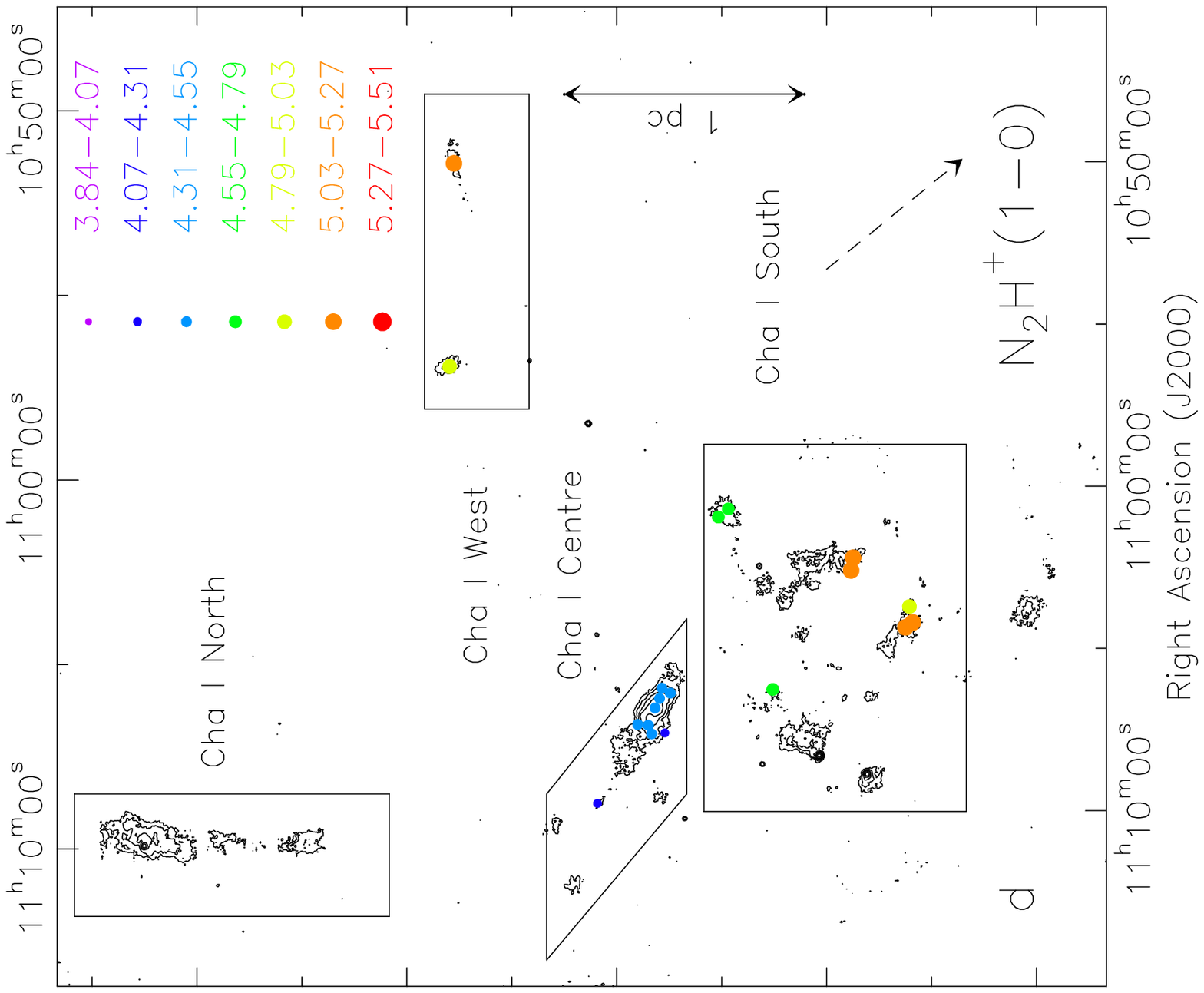}
\end{tabular}
 \caption[]{Systemic velocities of the cores in Cha I overplotted on the 870 $\mu$m continuum emission map obtained with LABOCA \citep{belloche11a} for the transitions: (a) C$^{18}$O 2--1, (b) C$^{17}$O 2--1, (c) CH$_3$OH 2$_0$--1$_0$ A$^+$, and (d) N$_2$H$^+$ 1--0. The contour levels correspond to $a$, $2a$, $4a$, $8a$, $16a$, $32a$, with $a = 48$~mJy/21$^{\prime\prime}$-beam (4$\sigma$). The core velocities were derived from Gaussian or hyperfine-structure fits to the observed spectra. The filled circles are colour-coded and increase in size to emphasize increasing velocities. Cores with two velocity components are shown with the lower velocity overplotted onto the higher velocity component. The dashed arrow indicates the position angle of the velocity gradient fit. \label{fig:velgradients}}
\end{center}
\end{figure*}

The global velocity dispersion of the cores in each tracer is given by
\begin{equation}
\sigma_{\rm 1D} = \sqrt{ \Sigma_{i=1}^N  \frac{(v_{\mathrm{core,}i} - v_{\rm mean})^2}{N} },
\end{equation}
where $N$ is the number of cores detected in the specific tracer,
$v_{{\rm core},i}$ the systemic velocity of each individual core measured with 
this tracer, and $v_{\rm mean}$ is the average systemic velocity of all cores in 
this tracer. 
The 3D velocity dispersion is
\begin{equation}
\sigma_{\rm 3D} = \sqrt{3} \times \sigma_{\rm 1D}.
\end{equation}  
We obtain the mean velocity relative to the centre of mass of the system, $v_{\rm rel,COM}$, using the relation 
\begin{equation}
v_{\rm rel,COM} = \sqrt{\frac{8}{\pi}} \times \sigma_{\rm 1D}.
\end{equation}
Finally, the relative speed between cores is computed as
\begin{equation}
v_{\rm rel,cores} = \sqrt{2} \times v_{\rm rel,COM} = \frac{4}{\sqrt{\pi}} \sigma_{\rm 1D}.
\end{equation}
The velocity dispersions are given in Table~\ref{table:dispersions_lines} for both Cha I and III as a whole, as well as the subregions of Cha I and III.  

\begin{figure*}
\begin{tabular}{cc}
\includegraphics[width=115mm,angle=270]{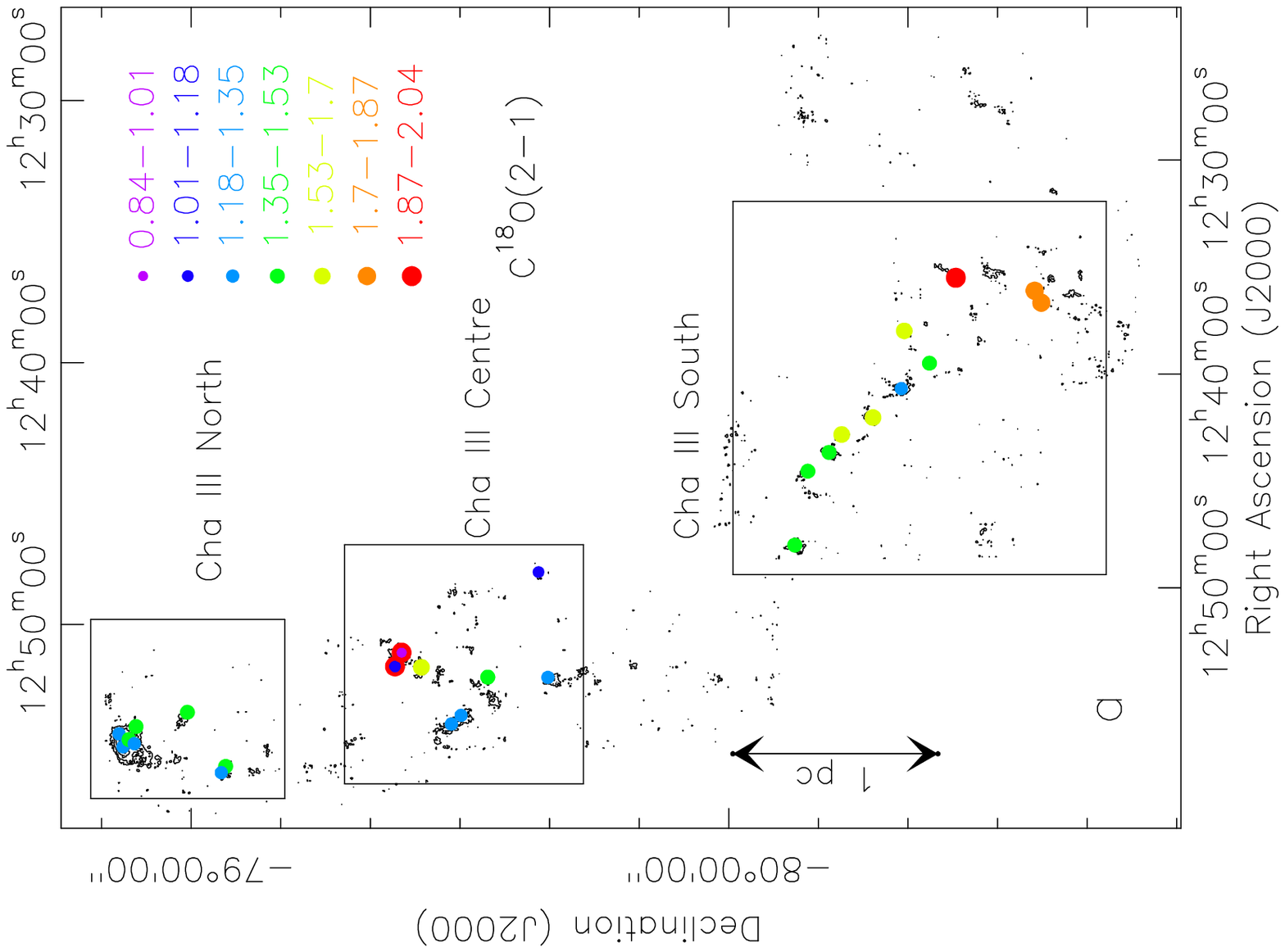} &
\hspace*{-145ex}
\includegraphics[width=115mm,angle=270]{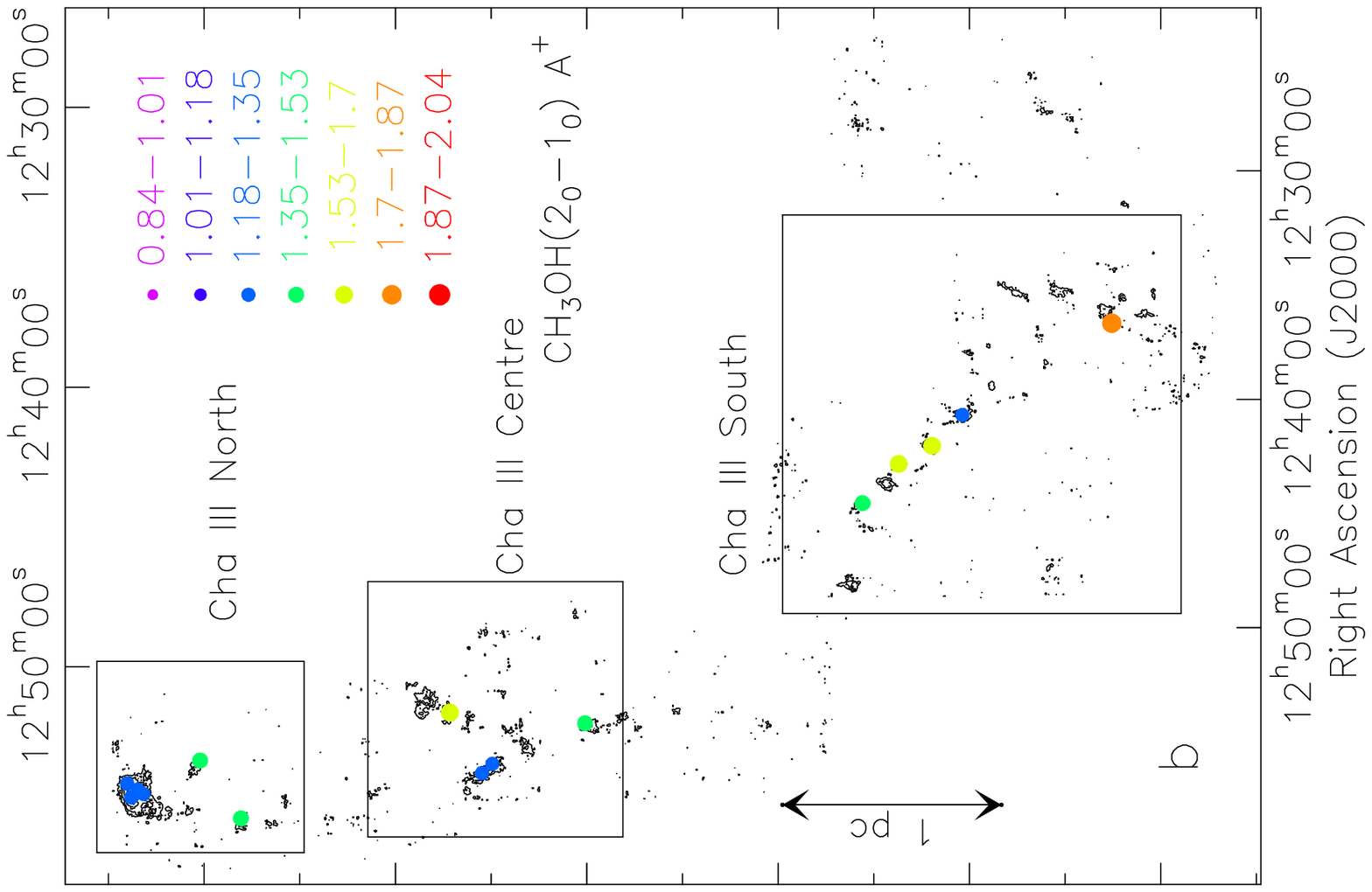} \\
\end{tabular}
 \caption[]{Same as Fig~\ref{fig:velgradients} but for Cha III and the transitions (a) C$^{18}$O 2--1 and (b) CH$_3$OH 2$_0$--1$_0$ A$^+$.  The contour levels correspond to the 870~$\mu$m continuum emission obtained with LABOCA \citep{belloche11b} at $a$, $2a$, $4a$, $6a$, with $a = 34.5$~mJy/21$^{\prime\prime}$-beam (3$\sigma$). \label{fig:veldistribution_chaIII}}
\end{figure*}

Figure~\ref{fig:velgradients} shows the colour-coded (in terms of magnitude) systemic velocities of the cores in Cha I for the molecular tracers C$^{18}$O~2--1, C$^{17}$O~2--1, CH$_3$OH~2$_{0}$--1$_{0}$ A$^+$, and N$_2$H$^+$~1--0 overlaid on the 870~$\mu$m continuum LABOCA map. Most of the cores are located in structures that appear to be filamentary in the dust continuum. We define four regions, Cha I North, Centre, South, and West, with the first three containing each at least one of the three distinct elongated structures in Cha I. The subregion Cha I West does not contain an elongated structure, but rather two small clumps with three cores altogether. 
Figure~\ref{fig:veldistribution_chaIII} overlays the systemic velocities that were derived for the Cha III cores in C$^{18}$O~2--1 and CH$_3$OH~2$_{0}$--1$_{0}$ A$^+$ on the 870~$\mu$m map of the cloud. We also define the subregions Cha III North, Centre, and South, in which the cores appear more clustered. 

We find a mean systemic velocity ($v_{\rm mean}$) of $4.72\pm 0.34$~km~s$^{-1}$ for the C$^{18}$O sample in Cha I. The uncertainty is the 1D velocity dispersion. Second velocity components are excluded in this calculation. We believe that the higher velocity peak, or second velocity component, is likely not associated with the material in the region we are studying and is either due to projection effects or due to material interacting with those regions (see Sect.~\ref{sec:multiple_components}). The mean velocities for Cha I North, Centre, and South yield $4.47\pm0.29$~km~s$^{-1}$, $4.48 \pm 0.20$~km~s$^{-1}$, and $4.83 \pm 0.32$~km~s$^{-1}$ for the C$^{18}$O sample and for the number of cores given by $N_{\rm fit}$ in Table~\ref{table:dispersions_lines}.

All 29 cores in Cha III were detected ($>3 \sigma$) in C$^{18}$O~2--1. 
We derive a mean velocity of $1.50 \pm 0.23$~km~s$^{-1}$, with the 
uncertainty being the 1D velocity dispersion. The mean systemic velocities in C$^{18}$O 2--1 for the subregions of Cha III are $1.36 \pm 0.09$~km~s$^{-1}$ (Cha III North), $1.53 \pm 0.31$~km~s$^{-1}$ (Cha III Centre), and $1.59 \pm 0.17$~km~s$^{-1}$ (Cha III South) for the number of cores given by $N_{\rm fit}$ in Table~\ref{table:dispersions_lines}.

\subsubsection{Crossing and collisional times}

To estimate the crossing and collisional times, $t_{\rm cross}$ and $t_{\rm coll}$, for the Cha I and III clouds as a whole and their subregions we assume cylindrical symmetry due to their elongated shape. The only exception is Cha I South for which we assume spherical symmetry. For a number of cores N, we take $n$ to be the core number density in the respective regions, given by $N/(\pi H R_{\rm cyl}^2)$ in the cylindrical case and $N/(\frac{4}{3}\pi R_{\rm cl}^3)$ in the spherical case. We call $D_{\rm cl}$ the largest projected distance between two cores 
in the cloud or region. In the spherical case, $D_{\rm cl}$ corresponds 
to the diameter of the sphere and the radius is $R_{\rm cl} = D_{\rm cl}/2$. In 
the cylindrical case, the height $H$ of the cylinder is assumed to be 
equal to $D_{\rm cl}$.
The diameter of the cylinder, $D_{\rm cyl}$, is estimated as the largest 
distance between two cores in the direction perpendicular to the main axis of 
the cylinder, and the radius is $R_{\rm cyl} = D_{\rm cyl}/2$.
Assuming isotropic, Maxwellian velocities we can estimate the crossing and collisional times as \citep[see ][]{binney_tremaine, andre07}
\begin{equation}
\label{eq:tcross}
t_{\rm cross} = \frac{D_{\rm cl}}{\sigma_{\rm 3D}}
\end{equation}
for the long axis of the cylinder or for the sphere in question, and 
\begin{equation}
\label{eq:tcross2}
t_{\rm cross} = \frac{D_{\rm cyl}}{\sigma_{\rm 3D}}
\end{equation}
for the short axis of the cylinder. Finally,
\begin{equation}
\frac{1}{t_{\rm coll}} = 4\sqrt{\pi} \, n \, \sigma_{\rm 1D} \, R_{\rm mean}^2 (1 + \Theta)
\end{equation} 
where
\begin{equation}
1+\Theta = 1 + \frac{G M_{\rm mean}}{\sigma_{\rm 1D}^2 R_{\rm mean}}.
\end{equation}
The mean mass ($M_{\rm mean}$), mean core radius ($R_{\rm mean}$), number of cores ($N$) used in each tracer and the resulting velocity dispersions and times are given in Table~\ref{table:dispersions_lines}. The radii $R$ of the cores are calculated as $\sqrt{FWHM_{\rm maj}\times FWHM_{\rm min}}$, as in \citet{andre07}. 
We select the C$^{18}$O 2--1 and CH$_3$OH 2$_{0}$ -- 1$_{0}$ A$^+$ 
samples for Cha I as these transitions were observed toward many 
cores. We use the same tracers for Cha III.

We compute maximum cloud crossing times, i.e., along the long axis in 
the cylindrical cases, $t_{\rm cross} \sim7$~Myr for Cha~I as a whole and 
$\sim 2.5$ Myr for each of its subregions. The time for core 
collisions is $\sim 160$~Myr for the whole Cha~I cloud, but it is
much lower 
in its subregions. We find collision times of $\sim0.7$ Myr for Cha~I 
North, $\sim 1.3$~Myr for Cha~I Centre, and $\sim16$ Myr for Cha~I South. 
Interactions between cores in Cha~I North and Cha~I Centre could 
therefore occur at a higher rate than in Cha~I South.

We perform the same calculations for the cores in Cha III and its subregions. The velocity dispersions and timescales for every transition are listed in Table~\ref{table:dispersions_lines}. We find maximum cloud crossing times 
of $\sim13$ Myr for the cloud as a whole, and $\sim4$ Myr, $\sim2.3$ 
Myr, and $\sim5.4$ Myr for Cha III North, Centre, and South, respectively. 
The collisional time is $\sim 40$~Myr for the whole cloud. It 
decreases, however, by a large amount when focusing on the Cha III 
subregions, for which we find $\sim 1$ Myr (Cha III North), $\sim10$ Myr 
(Cha III Centre), and $\sim 12$ Myr (Cha III South).

\subsubsection{Velocity gradients}
\label{sec:vel_gradients}

To search for velocity gradients in Cha I, we performed a 2D least-square fit on the systemic velocities of the cores detected in each tracer. We perform the fitting with a three-parameter function \citep{goodman93},
\begin{equation}
v_{LSR} = v_0 + a \Delta\alpha + b \Delta\beta,
\end{equation}
where $v_{LSR}$, $\Delta\alpha$, and $\Delta\beta$ are the systemic velocity of the cores, the offset of the cores in right ascension, and the offset in declination (radians). The magnitude of the linear velocity gradient (which assumes solid-body rotation) and its direction (east of north) are given by \citep{goodman93} 
\begin{equation}
\label{eq:gradient}
Grad = \frac{\sqrt{(a^2 + b^2)}}{d},
\end{equation}
\begin{equation}
\theta_{\rm Grad} = \tan^{-1}\frac{a}{b}, 
\end{equation}
where $d$ is the distance of the cloud. The derived velocity gradients and their position angles are given in Table~\ref{table:vel_gradients}. They are
shown in Fig.~\ref{fig:velgradients}.
We emphasize that the assumption of solid-body rotation only serves as a rough approximation as the velocities across the cloud vary in a more complex manner.

\begin{table}
\caption{Velocity gradients in Cha I and III.} 
\vspace*{-1ex}
\hfill{}
\begin{tabular}{lllll}
\hline \hline
Transition                   & $D$\tablefootmark{a}  & $N_{\rm fit}$\tablefootmark{b} & Gradient                          & $\theta_{\rm Grad}$ \\
                             & \scriptsize{(pc)}   &           & \scriptsize{(km~s$^{-1}$~pc$^{-1}$)} & \scriptsize{($^\circ$)} \\ \hline
\multicolumn{5}{c}{\underline{Chamaeleon I}} \\
C$^{18}$O 2--1               & 4.0  & 65       & 0.36$\pm$0.06 & -100 \\
C$^{17}$O 2--1               & 4.0  & 33       & 0.34$\pm$0.08 & -106 \\
CH$_3$OH 2$_{0}$--1$_{0}$ A$^+$  & 4.0 & 42     & 0.34$\pm$0.07 & -114 \\
N$_2$H$^{+}$ 1--0            & 4.0 & 19        & 0.63$\pm$0.06 & -141 \\
\multicolumn{5}{c}{\underline{Chamaeleon III}} \\
C$^{18}$O 2--1                & 4.9 & 29        & $<$ 0.24\tablefootmark{c}    &  --  \\
CH$_3$OH 2$_{0}$--1$_0$ A$^+$ & 4.9 & 15        & $<$ 0.28\tablefootmark{c}   &  --  \\ 
\hline
\end{tabular}
\hfill{}
\label{table:vel_gradients}
\tablefoot{
\tablefoottext{a}{Projected distance between most distant cores.}
\tablefoottext{b}{Number of Gaussian and hyperfine-structure fits used to compute the velocity gradients.}
\tablefoottext{c}{The upper limit corresponds to three times the uncertainty obtained from Equation~\ref{eq:gradient}.}
}
\end{table}

The higher velocities of the cores in the south-western part of Cha I (Fig.~\ref{fig:velgradients}) are mostly responsible for the velocity gradient in Cha I. All tracers give consistent results, the higher value derived for N$_2$H$^+$ 1--0 being due to its non-detection in Cha I North. We tested the latter effect by computing the velocity gradients in C$^{18}$O 2--1, C$^{17}$O 2--1, CH$_3$OH 2$_{0}$--1$_0$ A$^+$ without taking the cores in Cha I North into account. We indeed found higher gradients (i.e., $\sim0.5$--$0.6$ km~s$^{-1}$~pc$^{-1}$) as in N$_2$H$^+$ 1--0. We find no significant velocity gradient in Cha III, but we list the CH$_3$OH 2$_{0}$--1$_0$ A$^+$  and C$^{18}$O 2--1 upper limits 
in Table~\ref{table:vel_gradients}.

\section{Analysis: Molecular abundances}
\label{sec:results_abundances}

\subsection{Observational molecular abundances}
\label{sec:observed_abundances}

We used the offline-version of the non-LTE radiative transfer code RADEX \citep{vanderTak07} to calculate the column densities of C$^{18}$O, C$^{17}$O,
CH$_3$OH, C$^{34}$S, HC$_3$N, and N$_2$H$^+$, applying the escape probability 
method for a uniform sphere. We assumed a kinetic temperature of 10 K for all observed transitions except for C$^{18}$O and C$^{17}$O 2--1 for which we used 12 K. The initial assumption of 10 K for C$^{18}$O and C$^{17}$O 2--1 gave rise to inconsistent (unrealistically high) column densities for a few cores. In addition, the dust temperature in the cold cores of the Chamaeleon clouds was found to be around 12 K \citep[see][]{toth00}. These two transitions may trace more diffuse material than the other higher-density tracers, which is likely to be at a higher kinetic temperature. RADEX calculates the column density and excitation temperature of a line, given the spectrum's main-beam peak temperature, linewidth, and core density. The core densities for Cha I and III are taken from the 870 $\mu$m dust continuum surveys of \citet{belloche11a} and \citet{belloche11b}, respectively. For each core, we interpolate between the 
peak density and the 
density computed from the flux contained in a 50$^{\arcsec}$ aperture to derive 
the mean density in the beam ($HPBW$) corresponding to each transition (listed in Tables~\ref{table:obs_details_apex} and ~\ref{table:obs_details_mopra}). Finally, the molecular datafiles are taken from the LAMDA database\footnote{http://home.strw.leidenuniv.nl/~moldata/}.

We have no sign that HC$_3$N 10--9, C$^{34}$S 2--1, CH$_3$OH 2$_{0}$--1$_{0}$ A$^+$, CH$_3$OH 2$_{1}$--1$_{1}$ E, and C$^{17}$O 2--1 might be optically thick judging from the symmetric, Gaussian-like spectra (see Sects.~\ref{sec:opacity_c18o} and ~\ref{sec:opacity_cs} for estimates of C$^{17}$O and C$^{34}$S opacities). C$^{18}$O 2--1 is slightly optically thick for some cores in Cha I (Sect.~\ref{sec:opacity_c18o}). We do not expect the cores in Cha III to be optically thick in C$^{18}$O 2--1 as the brightest core, Cha3-C1, is only at the limit of being optically thick. Cha3-C4 is optically thin in C$^{18}$O 2--1 (see Sect.~\ref{sec:opacity_c18o}). 

\begin{table*}
\caption{Average and median molecular abundances in Cha I and III.} 
\vspace*{-1ex}
\hfill{}
\begin{tabular}{lllllllllll} 
\hline
\hline
Molecule & Trans.\tablefootmark{a}  &  & \multicolumn{3}{c}{Cha I } & & \multicolumn{3}{c}{Cha III} & Ratio\tablefootmark{e} \\
\cline{4-6}
\cline{8-10}
        &           &  & Abundance\tablefootmark{b}  & Dispersion\tablefootmark{c}   & $N_{\rm cores}$\tablefootmark{d} &    & Abundance\tablefootmark{b}  & Dispersion\tablefootmark{c}  & $N_{\rm cores}$\tablefootmark{d}  & $\frac{X_{\rm ChaI,med}}{X_{\rm ChaIII,med}}$ \\ \hline

C$^{18}$O\tablefootmark{f}  & 2--1   &   Average  &   4.2$\times 10^{-7}$ &   1.3$\times 10^{-6}$   & 60 &  & 1.4$\times 10^{-7}$ &   4.9$\times 10^{-8}$ & 29 & 1.6  \\
                            & &  Median &   2.2$\times 10^{-7}$             & & &                         & 1.3$\times 10^{-7}$              & &  & \\  

C$^{17}$O  & 2--1              &   Average  &  5.3$\times 10^{-8}$  &   2.7$\times 10^{-8}$   & 32 &  & 2.7$\times 10^{-8}$ &   3.4$\times 10^{-9}$ &  2 & 1.8 \\
                           & &  Median &   4.9$\times 10^{-8}$              & & &                         & 2.7$\times 10^{-8}$              & & & \\  

CH$_3$OH &  2$_{0}$--1$_{0}$ A$^+$ &  Average &   1.3$\times 10^{-9}$  &   7.4$\times 10^{-10}$  & 41 & & 9.4$\times 10^{-10}$ & 2.9$\times 10^{-10}$  & 15 & 1.2 \\
                           &  &  Median  &   1.1$\times 10^{-9}$            & & &                        & 9.6$\times 10^{-10}$             & & & \\
 
CH$_3$OH &  2$_{1}$--1$_{1}$ E &  Average &    1.2$\times 10^{-9}$  & 6.4$\times 10^{-10}$    & 32 & & 1.1$\times 10^{-9}$  & 2.8$\times 10^{-10}$ & 10 & 1.1 \\
                           &  &  Median  &   1.2$\times 10^{-9}$             & & &                         & 1.1$\times 10^{-9}$             & & & \\  

C$^{34}$S &  2--1              &  Average &   3.0$\times 10^{-10}$ &   2.0$\times 10^{-10}$ & 23 & & 1.8$\times 10^{-10}$  & 3.4$\times 10^{-11}$ & 2 & 1.4 \\
                            & &  Median  &   2.5$\times 10^{-10}$            & & &                        & 1.8$\times 10^{-10}$            & & & \\

HC$_3$N &  10--9   &  Average &   9.0$\times 10^{-10}$ &  1.0$\times 10^{-9}$   & 18 &  &  4.8$\times 10^{-10}$ &   4.7$\times 10^{-10}$ & 3 & 2.3 \\
                           & &  Median  &   4.7$\times 10^{-10}$            & &  &                        &  1.8$\times 10^{-10}$   & & & \\  

N$_2$H$^+$ &  1--0            &   Average &   2.6$\times 10^{-10}$ &   1.1$\times 10^{-10}$  & 19 & & 8.4$\times 10^{-11}$ &   4.1$\times 10^{-11}$ & 2 & 3.1 \\
                            & &  Median  &   2.6$\times 10^{-10}$            & & &                    & 8.4$\times 10^{-10}$            & & & \\
\hline
\end{tabular}
\hfill{}
\label{table:abundances}
\vspace*{-1ex}
\tablefoot{
\tablefoottext{a}{Transition used to calculate molecular abundances.}
\tablefoottext{b}{Average or median abundance relative to H$_2$.}
\tablefoottext{c}{Standard deviation of abundance values shown in Fig.~\ref{fig:abundances}.}
\tablefoottext{d}{Number of cores used when calculating the mean and median abundances.}
\tablefoottext{e}{Ratio of median molecular abundance in Cha I to the median abundance in Cha III.}
\tablefoottext{f}{C$^{18}$O 2--1 is slightly optically thick for some cores in Cha I and its abundance should therefore be considered with caution.}
}
\vspace*{-1ex}
\end{table*}  

For HC$_3$N 10--9, C$^{17}$O 2--1, and N$_2$H$^+$ 1--0, which have a hyperfine structure, we apply hyperfine structure fits with CLASS to derive their linewidths. There are no molecular datafiles on LAMDA that consider the hyperfine structure of HC$_3$N 10--9 and C$^{17}$O 2--1. As approximately $\sim 30$\% of the C$^{17}$O 2--1 flux is missing from the central components, we apply a correction to the column density of $1/0.7$, i.e., $\sim1.43$. No correction is applied for HC$_3$N 10--9.
We use the strongest, single-frequency component N$_2$H$^+$~1$_{2,3}$--0$_{1,2}$ along with its hyperfine structure datafile from LAMDA to derive its column density with RADEX. The four densest cores in Cha I, C1-C4, are optically thick in this transition. Therefore, for these four cores we use the weakest, optically thin component N$_2$H$^+$~1$_{1,0}$--0$_{1,1}$ to derive their column densities.

The excitation temperatures we derive with RADEX show that C$^{18}$O and C$^{17}$O 2--1 are thermalized at $T_{\rm k} = 12$ K. The two CH$_3$OH species are also nearly thermalized with excitation temperatures around 8--9.5 K at a kinetic temperature of 10 K. C$^{34}$S 2--1 and N$_2$H$^+$ 1--0 both have excitation temperatures of $\sim4$--7 K in both clouds, while HC$_3$N 10--9 has slightly higher excitation temperatures of $\sim6$--8.5 K (excluding the two densest cores in Cha I).

\begin{figure*}
\begin{tabular}{cc}
\vspace*{-8ex}
\includegraphics[width=93.5mm,angle=0]{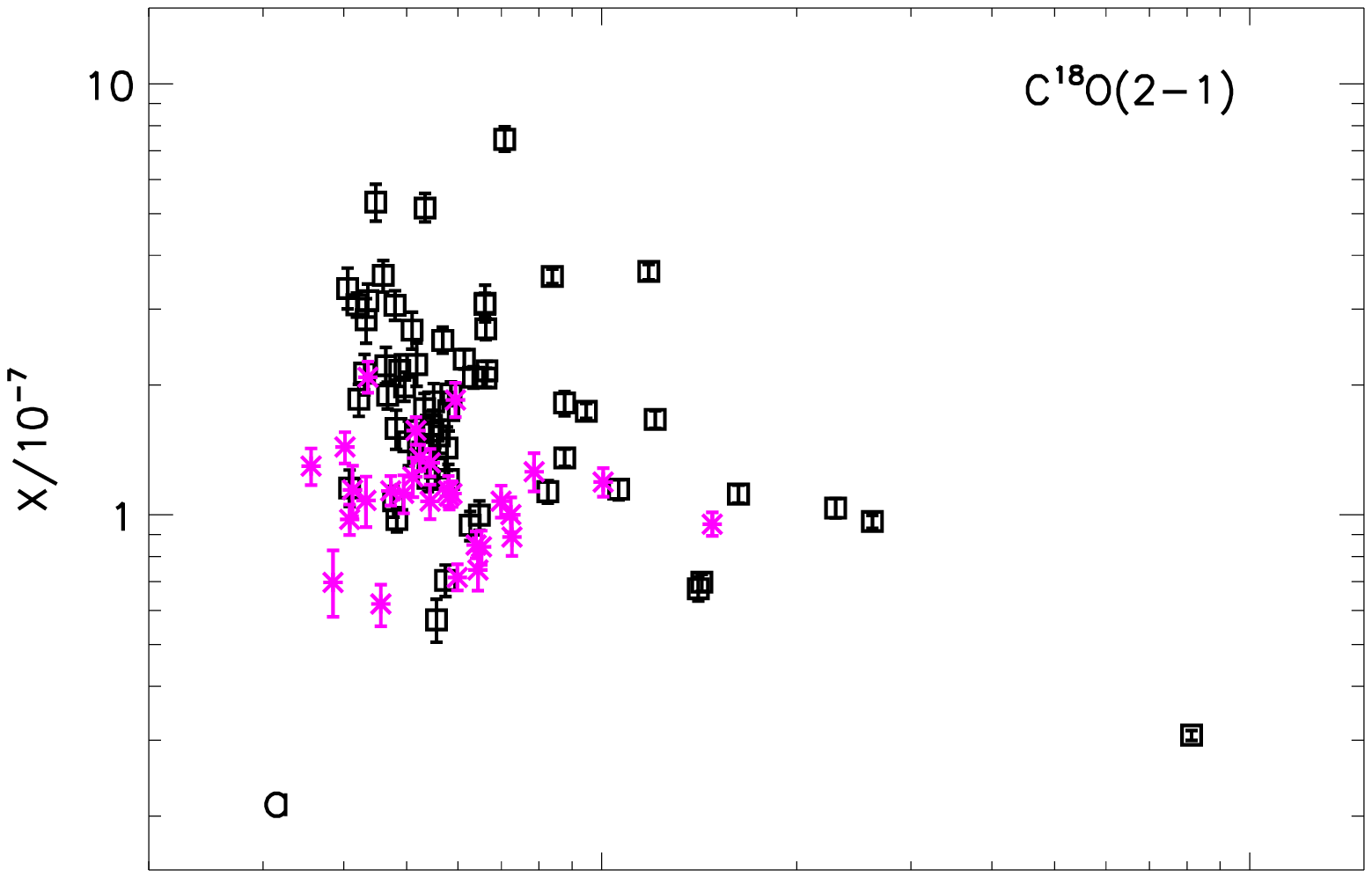} &
\hspace*{-7ex}
\includegraphics[width=93.5mm,angle=0]{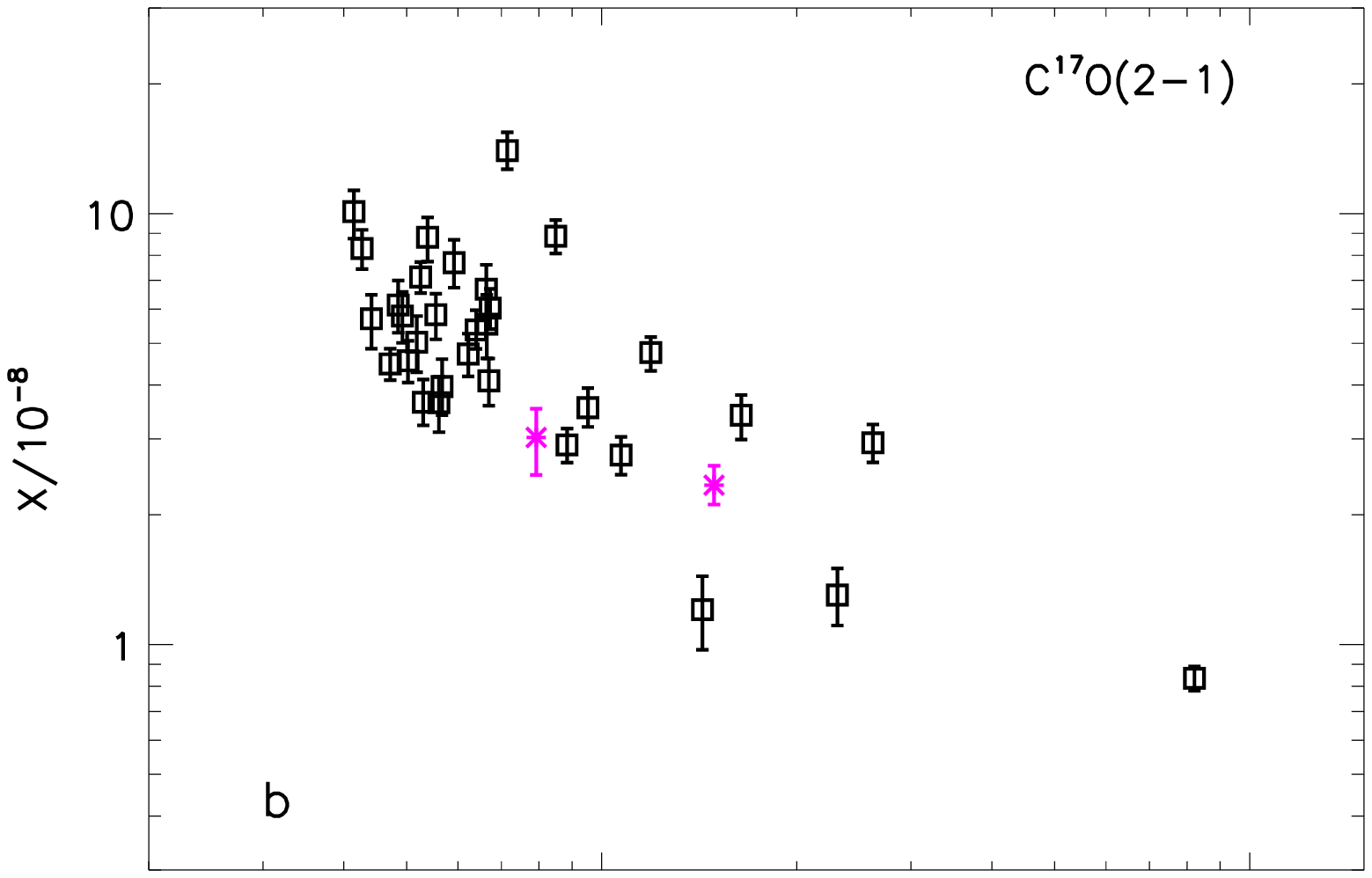}  \\
\vspace*{-8ex}
\includegraphics[width=93.5mm,angle=0]{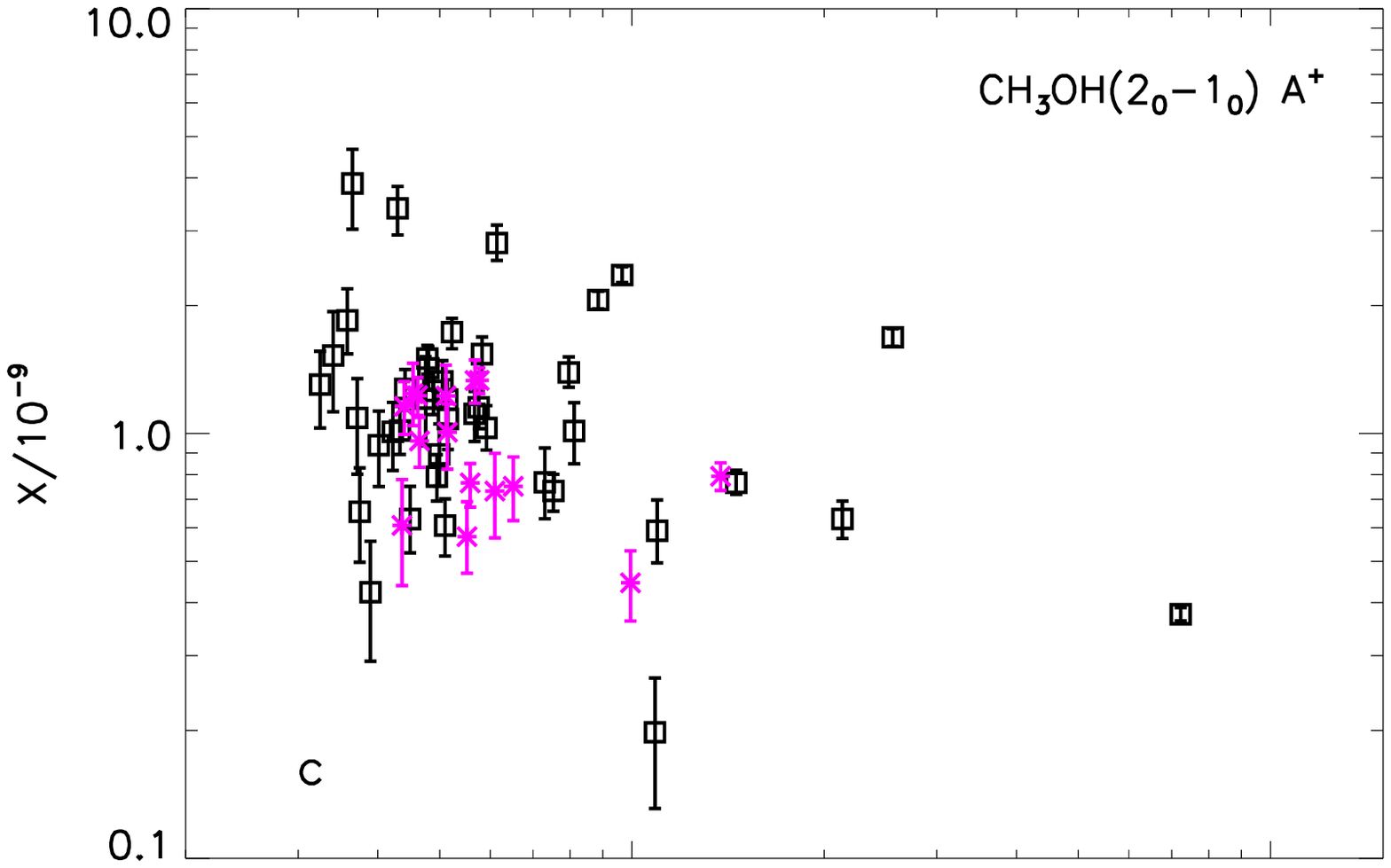} &
\hspace*{-7ex}
\includegraphics[width=93.5mm,angle=0]{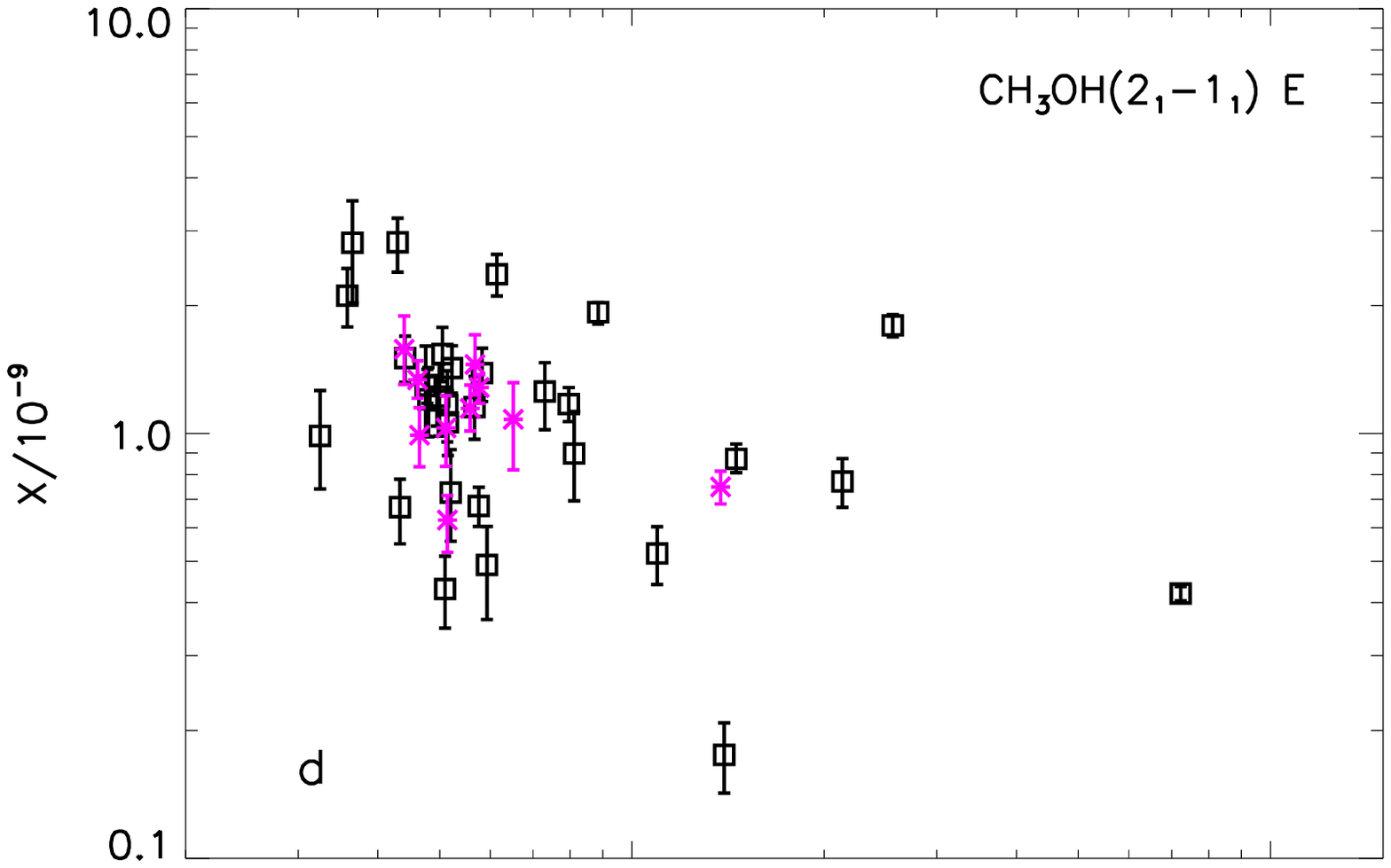} \\
\vspace*{-8ex}
\includegraphics[width=93.5mm,angle=0]{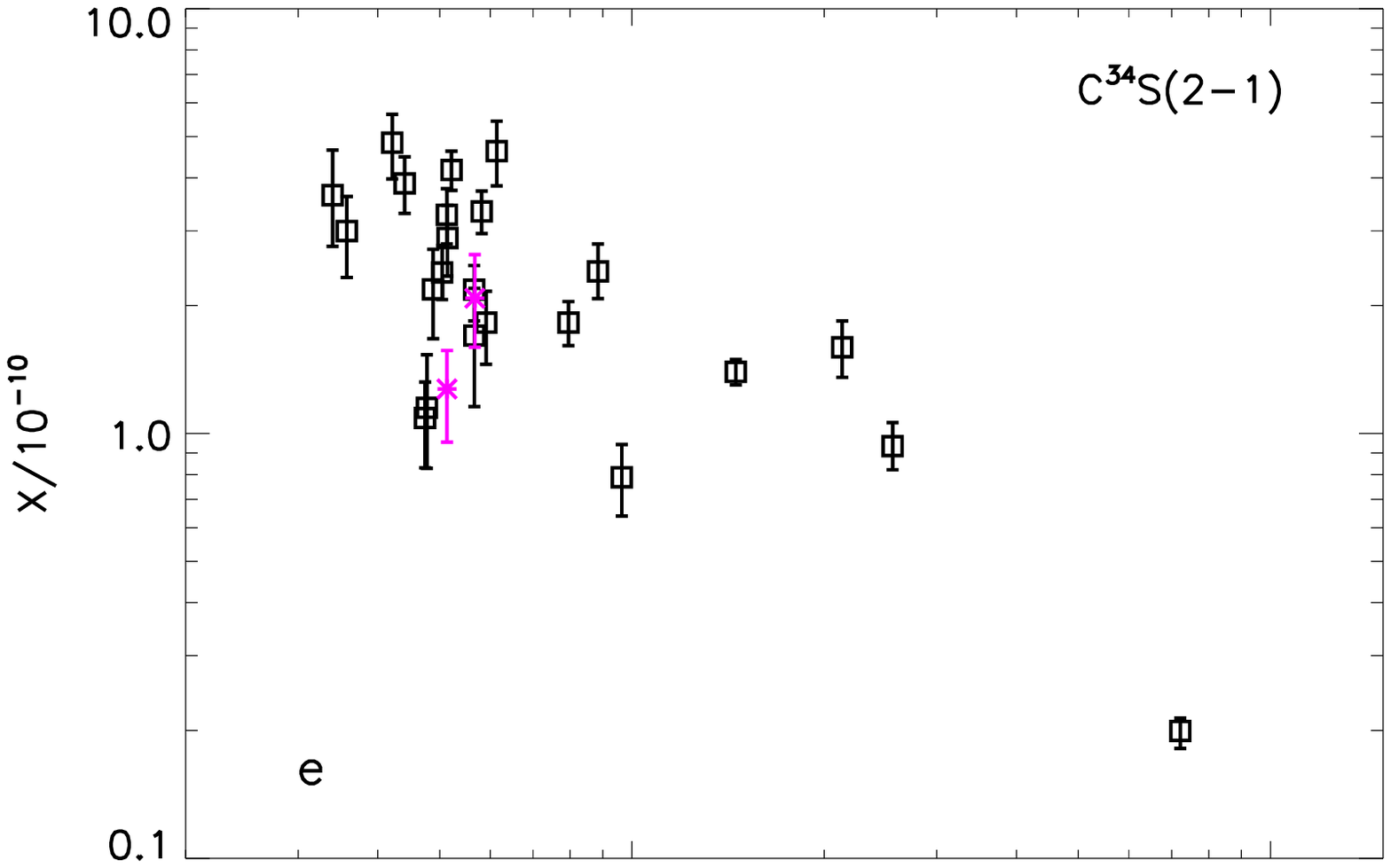} &
\hspace*{-7ex}
\includegraphics[width=93.5mm,angle=0]{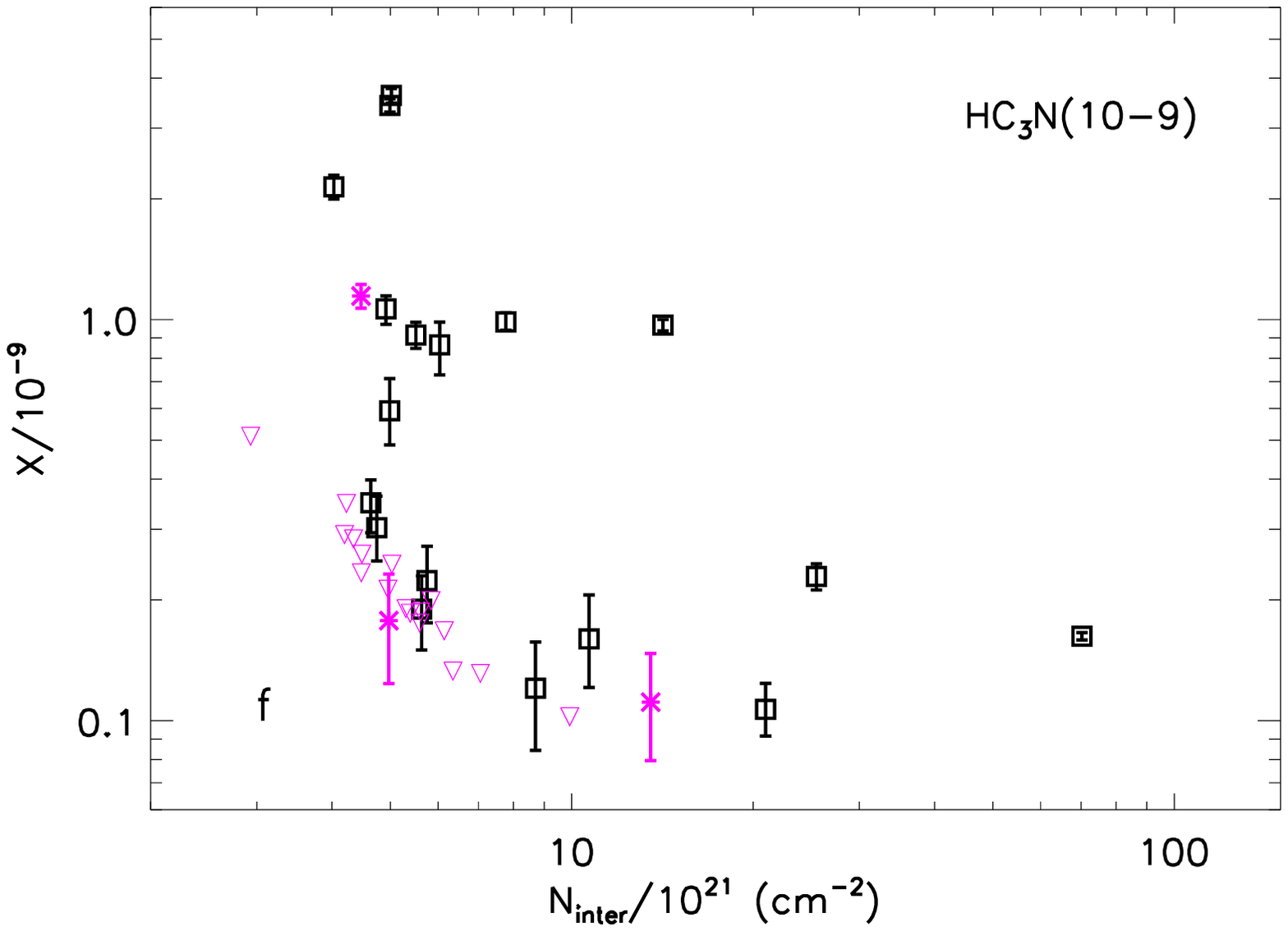} \\
\includegraphics[width=93.5mm,angle=0]{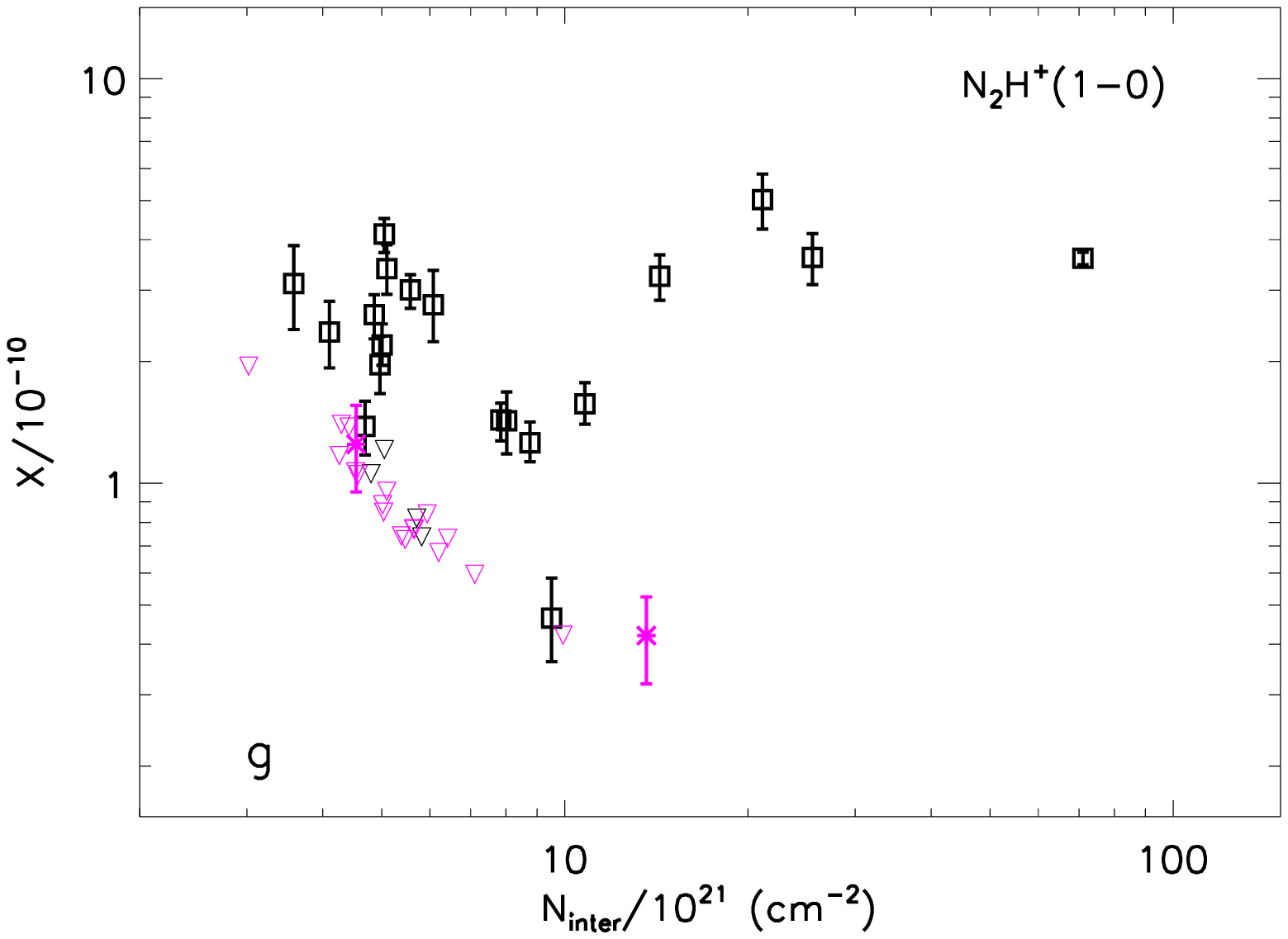} &
\end{tabular}
\vspace*{-3ex}
 \caption[]{Molecular abundances relative to H$_2$ against H$_2$ column densities interpolated within an aperture of diameter equal to the $HPBW$ of each transition. The abundances of the Cha I and III cores are shown as black squares and pink asterisks, respectively. Upper limits (3$\sigma$ uncertainties) are shown as pink and black downward triangles for Cha III and Cha I, respectively. The N$_2$H$^+$ abundance of the four densest cores in panel (g) was derived from the observed peak temperature of the weakest hyperfine component 
(1$_{1,0}$--0$_{1,1}$) while the strongest component 
(1$_{2,3}$--0$_{1,2}$) was used for all other cores
(see Sect.~\ref{sec:observed_abundances}). \label{fig:abundances}}
\end{figure*}

We compute the abundances relative to H$_2$ averaged along the line of sight 
using the H$_2$ column densities of the cores derived from the dust 
emission by \citet{belloche11a, belloche11b}. We use the interpolated column 
densities, using the same method as described above for the interpolated 
densities.
We refer to the interpolated H$_2$ column density as $N_{\rm inter}$. 
The average and median abundances relative to H$_2$ and the abundance 
dispersion derived from C$^{18}$O~2--1, C$^{17}$O~2--1, 
CH$_3$OH~2$_{0}$--1$_{0}$~A$^+$, CH$_3$OH~2$_{1}$--1$_{1}$~E, C$^{34}$S~2--1, 
HC$_3$N~10--9, and N$_2$H$^+$ 1--0
are listed in Table~\ref{table:abundances} for both Cha I and III. Figure~\ref{fig:abundances} plots the individual core abundances derived for each tracer against their interpolated H$_2$ column densities.

Most of the Cha III molecular abundances lie at the \emph{lower end} of the range of abundances in Cha I, apart from the core abundances of the two methanol species (Fig.~\ref{fig:abundances}). The fact that only a few 
cores in Cha~III are detected in each molecule makes it difficult to conduct any reliable statistical comparisons between the two clouds, especially in C$^{17}$O, HC$_3$N, C$^{34}$S, and  N$_2$H$^+$. Nevertheless, it is apparent that C$^{18}$O is on average more abundant in Cha I than in Cha III by a factor of $\sim 1.6$ (see Table~\ref{table:abundances}). This difference in molecular abundances can also be seen in Fig.~\ref{fig:abundances}.

\begin{figure}
\hspace*{-3ex}
\includegraphics[width=95mm,angle=0]{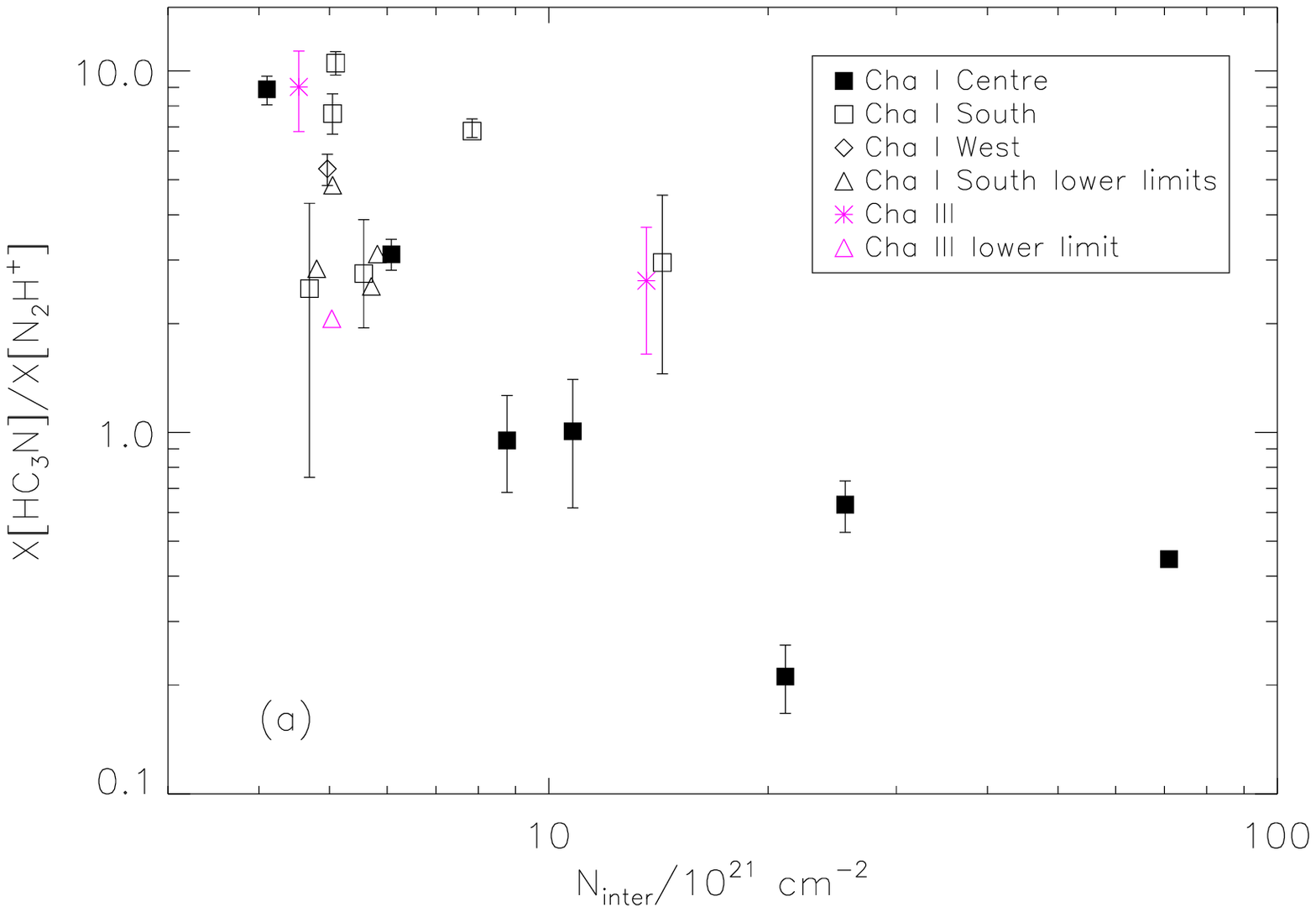}  \\[-2ex]
\includegraphics[width=95mm,angle=270]{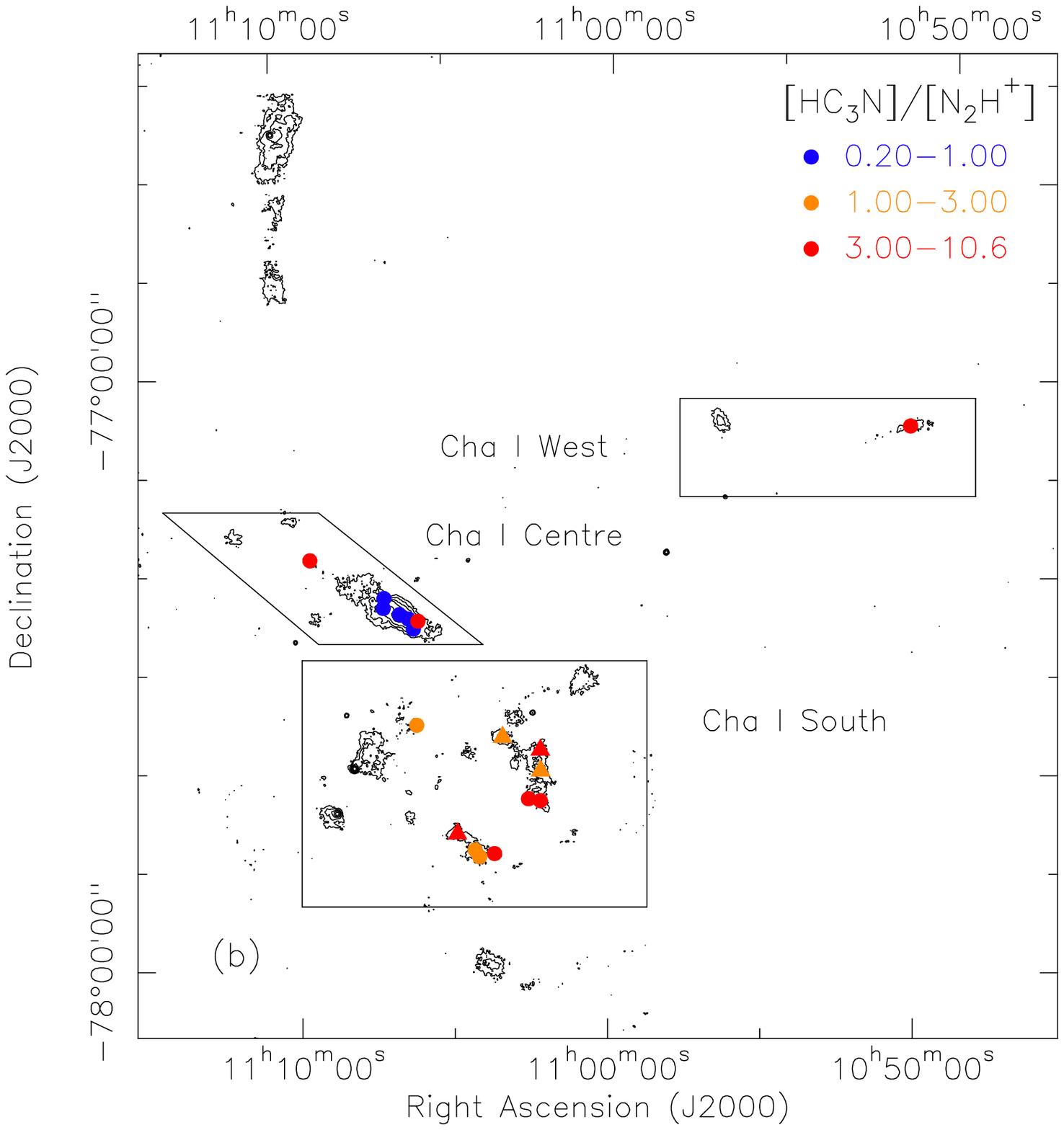} 
\caption[]{(a) Observed HC$_3$N/N$_2$H$^+$ abundance ratio as a function of H$_2$ column density for the cores in Cha I Centre (black, filled squares), Cha I South (black, empty squares), Cha I West (black diamond), Cha I South lower limits (black, empty triangles), Cha III (pink asterisks), and Cha III lower limits (pink, empty triangle). (b) Observed HC$_3$N/N$_2$H$^+$ abundance ratios for Cha I (circles) and lower limits (triangles) overplotted on the 870 $\mu m$ dust continuum emission obtained with LABOCA \citep{belloche11a}. The contour levels correspond to $a$, $2a$, $4a$, $8a$, $16a$, $32a$, with $a = 48$~mJy/21$^{\prime\prime}$-beam (4$\sigma$).  \label{fig:observed_ratio}}
\end{figure}

\citet{benedettini12} found that the ratio of HC$_3$N/N$_2$H$^+$ varies between the (prestellar and protostellar) cores in the Lupus 1, 3, and 4 molecular clouds. They conclude that this ratio decreases as a dense core or a protostar evolves, both observationally in their core sample and in their chemical model predictions. The HC$_3$N/N$_2$H$^+$ ratio for the cores in Cha I and III is shown in Fig.~\ref{fig:observed_ratio}a. Figure~\ref{fig:observed_ratio}b overplots the observed HC$_3$N/N$_2$H$^+$ ratio in Cha I on the 870 $\mu m$ continuum map. Two cores in Cha III have detections in both molecules (Cha3-C1, C15) and fourteen in Cha I. We calculate \emph{lower limits} for one additional core in Cha III (Cha3-C13) and four cores in Cha I (Cha1-C17, C18, C26, C38) that have a HC$_3$N detection and a N$_2$H$^+$ upper limit. The cores located in Cha I South and West have on average a higher HC$_3$N/N$_2$H$^+$ ratio than the ones in Cha I Centre, while the cores in Cha III have values similar to Cha I South (Figs.~\ref{fig:observed_ratio}a,b).

\subsection{Depletion}

The distributions of the abundances of C$^{18}$O, C$^{17}$O, CH$_3$OH, and C$^{34}$S suggest that these molecules are affected by depletion in the core interiors in Cha I (Fig.~\ref{fig:abundances}). The abundances of these  molecules appear to systematically decrease for cores of higher column density. C$^{17}$O and C$^{34}$S in particular show this abundance decrease with a narrower dispersion at each column density compared to C$^{18}$O and CH$_3$OH that have a larger spread of abundances. For instance, the abundances of C$^{17}$O are of the order $\sim 3$--20$\times 10^{-8}$ at a column density of $\sim 5 \times 10^{21}$ cm$^{-2}$ and $\sim 1$--3$\times 10^{-8}$ at a column density of $\sim 15 \times 10^{21}$ cm$^{-2}$. There is also a hint of HC$_3$N depletion at higher densities (Fig.~\ref{fig:abundances}f), but for this molecule the trend of abundance decrease as a function of increasing density is more uncertain than for C$^{18}$O, C$^{17}$O, CH$_3$OH, and C$^{34}$S.
The sample of cores in Cha III is not large enough to draw a conclusion about depletion within the cores.

\subsection{Comparison to predictions of chemical models}
\label{sec:chemical_models}

Both physically-static and core-contraction chemical models have been run for a selection of gas densities.
 
The static models adopt free-particle densities of $1 \times 10^5$, $3 \times 10^5$, and $1 \times 10^6$ cm$^{-3}$ corresponding to total hydrogen number densities of $n_{\rm H} = 1.7 \times 10^5$, $5.1 \times 10^5$, and $1.7 \times 10^6$ cm$^{-3}$ (see Appendix~\ref{sec:density_conversions}). The adopted densities are comparable to the interpolated (for 37$^{\arcsec}$) averaged Mopra free-particle densities of the Cha I and III cores (Cha I; $n_{\rm inter} \sim 7.4 \times 10^4$-- $1.7 \times 10^6$~cm$^{-3}$, Cha III;  $n_{\rm inter} \sim 7.3 \times 10^4$-- $3.3 \times 10^5$~cm$^{-3}$).

 Visual extinctions are sufficient to render unimportant those processes that depend on the external UV radiation field. The chemical model (MAGICKAL) and reaction network are those presented by \citet{garrod13}, which employ a fully-coupled gas-phase and dust grain-surface chemistry. Because of the low dust temperatures assumed here (8 K), only surface chemistry is considered; chemistry within the bulk ices is switched off. All material except hydrogen begins in atomic/ionic form in the gas phase, as shown in Table 1 of \citet{garrod13}. A gas temperature of 9 K is assumed.
We expect lower-density cores to have slightly higher temperatures (e.g., $\sim 10$--12 K). Somewhat higher temperatures do not significantly affect the model predictions ($T = 12$ K has been tested).

\begin{figure*}
\hspace*{3ex}
\begin{tabular}{cc}
\vspace*{-5ex}
\hspace*{-7ex}
\includegraphics[width=95mm,angle=0]{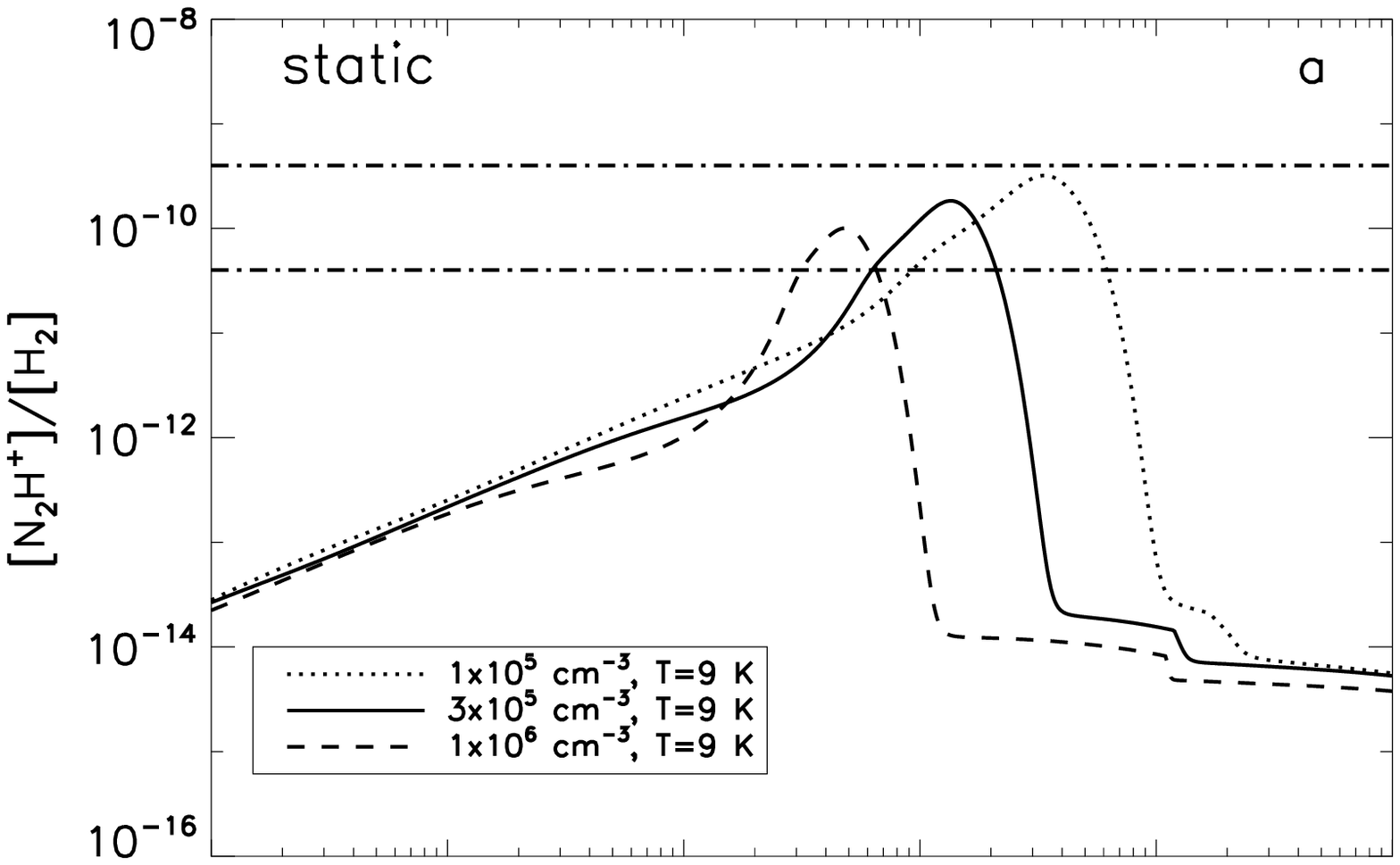} &
\hspace*{-7ex}
\includegraphics[width=95mm,angle=0]{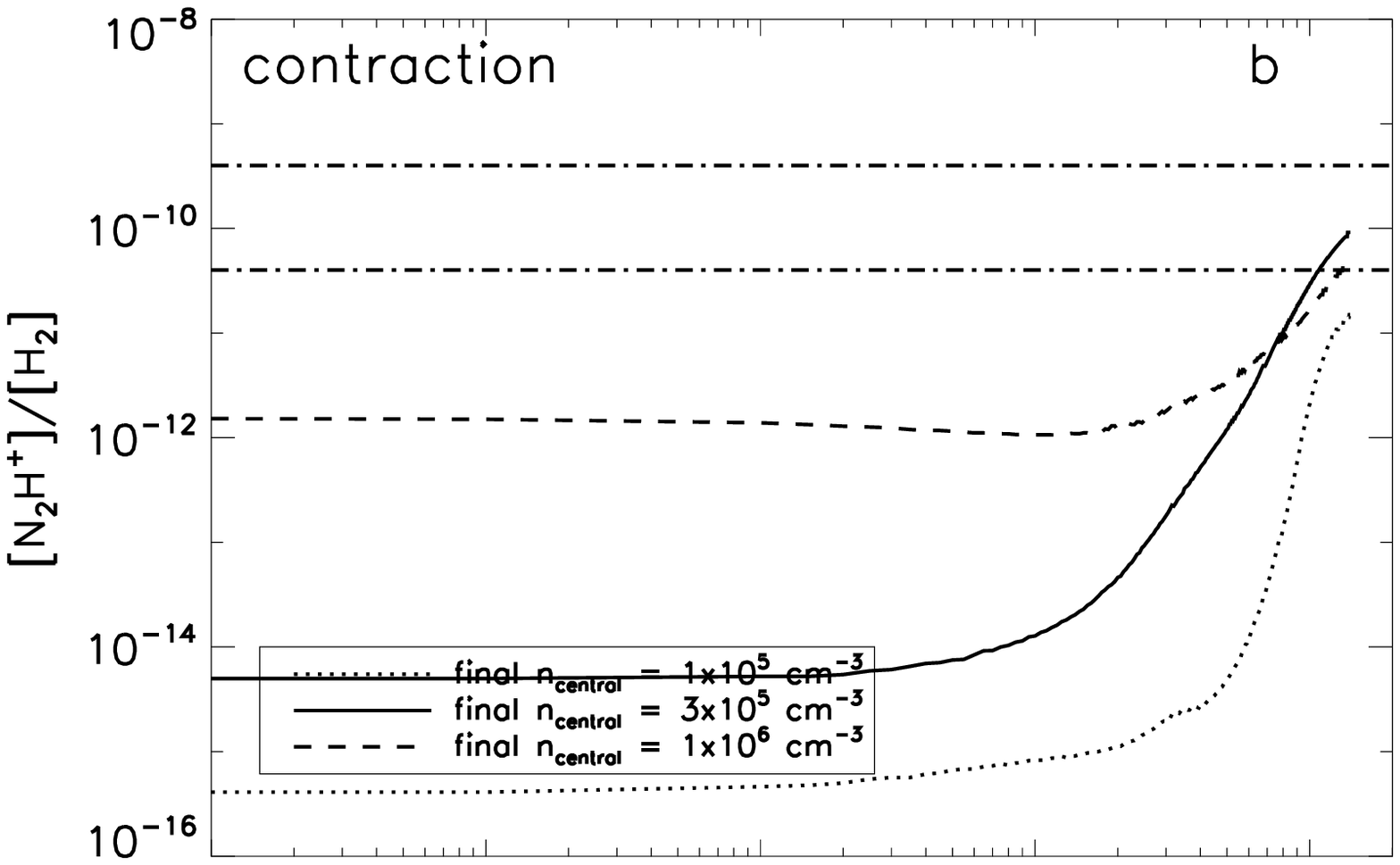} \\

\vspace*{-5ex}
\hspace*{-7ex}
\includegraphics[width=95mm,angle=0]{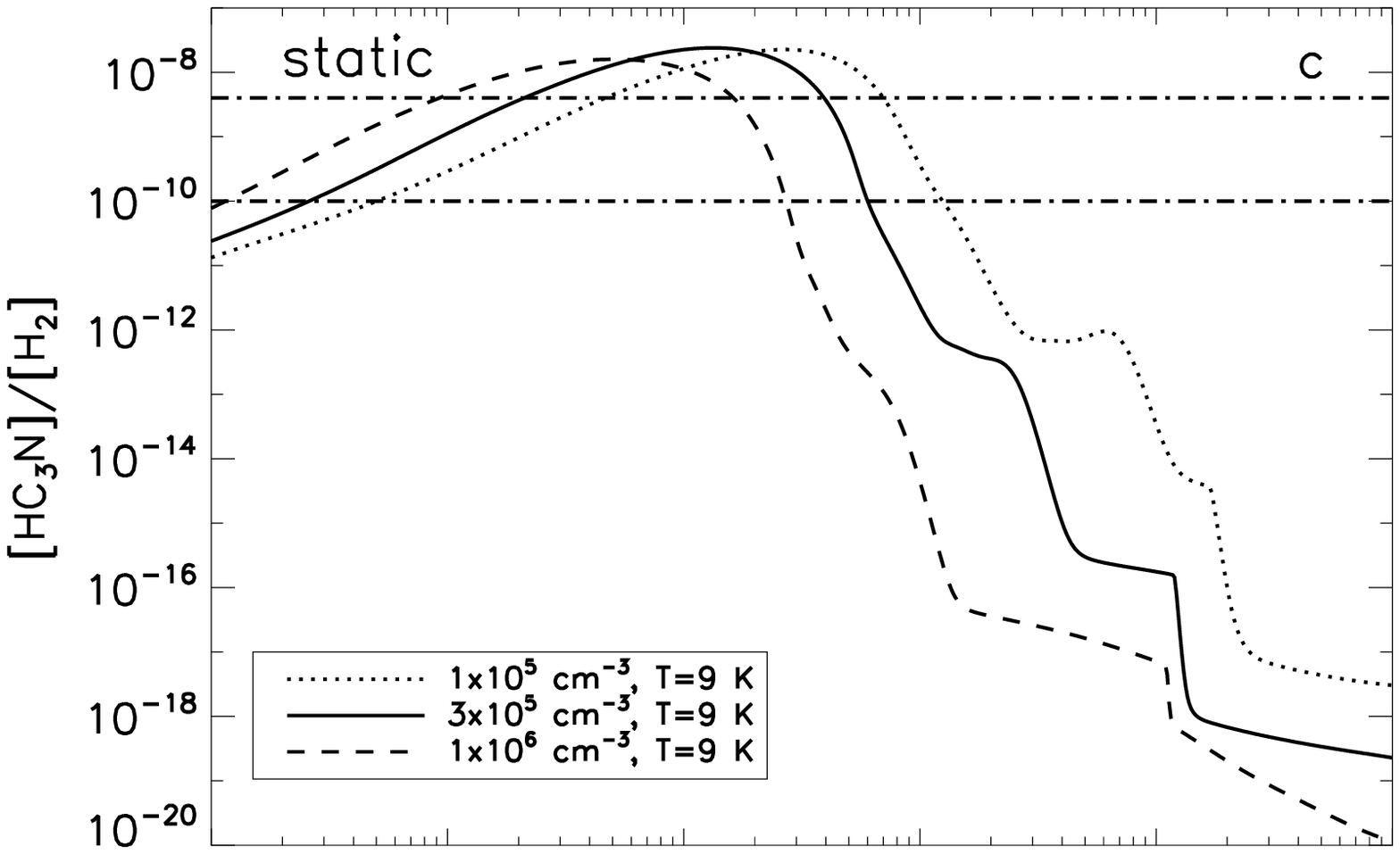} &
\hspace*{-7ex}
\includegraphics[width=95mm,angle=0]{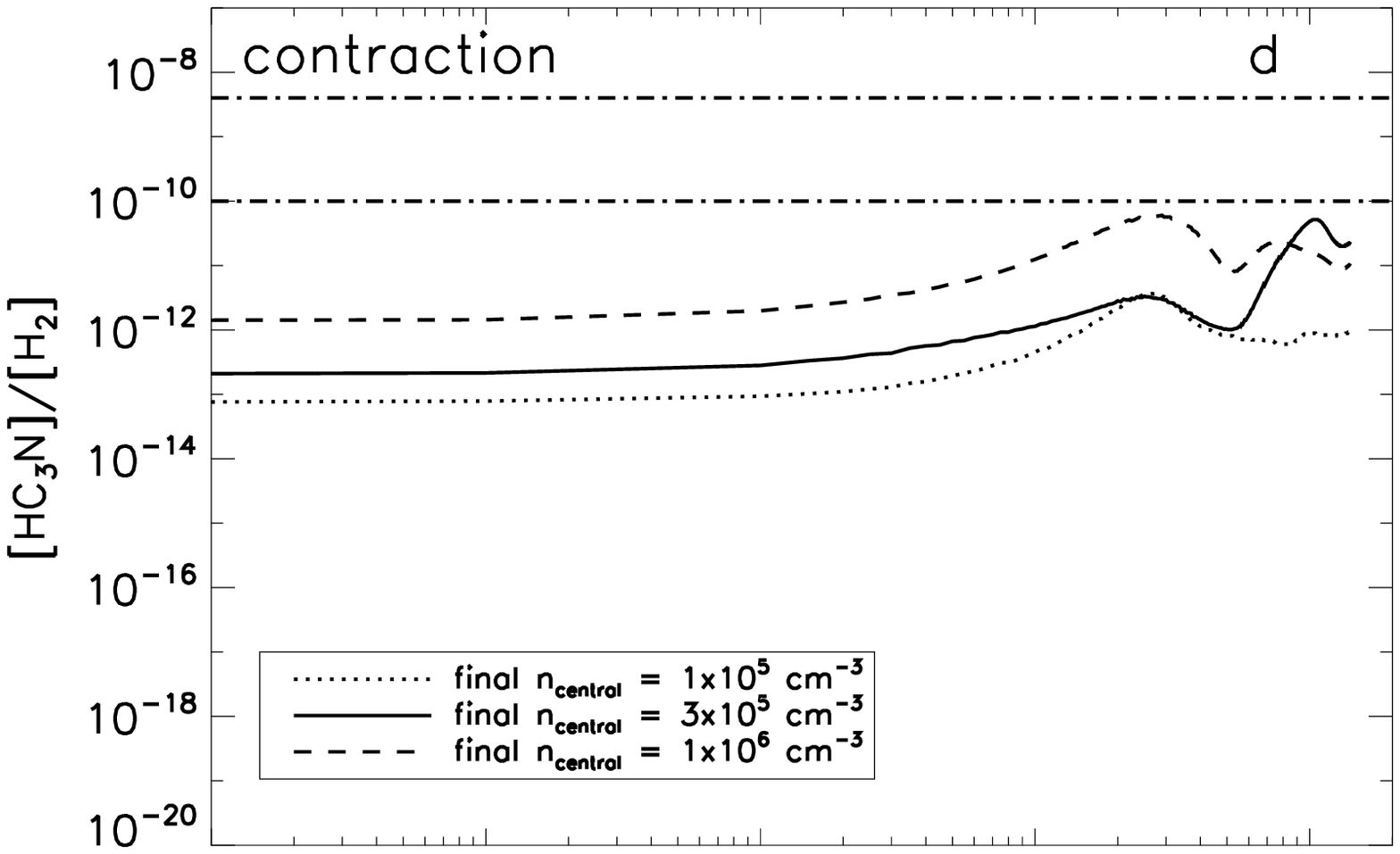} \\

\hspace*{-7ex}
\includegraphics[width=95mm,angle=0]{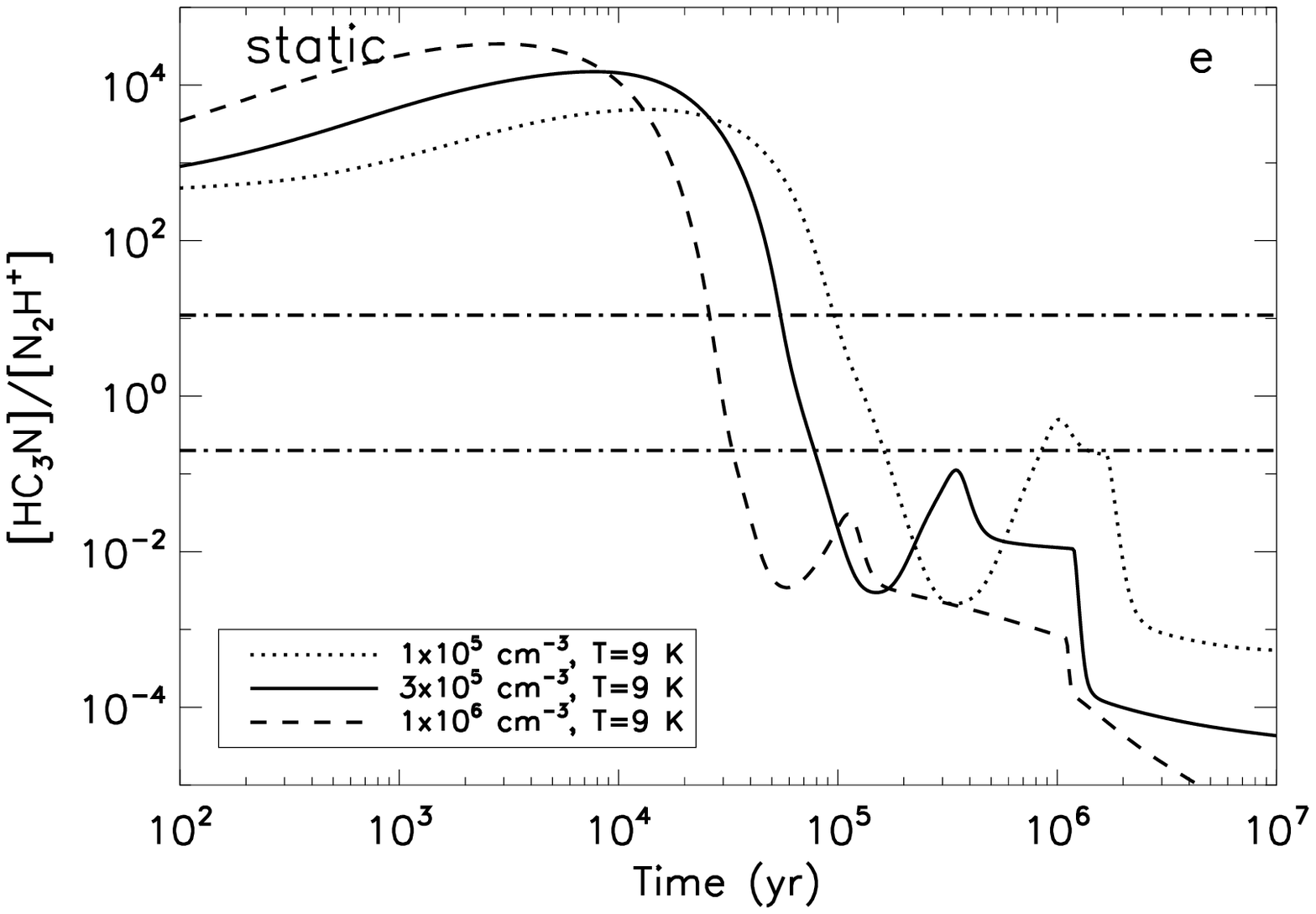} &
\hspace*{-7ex}
\includegraphics[width=95mm,angle=0]{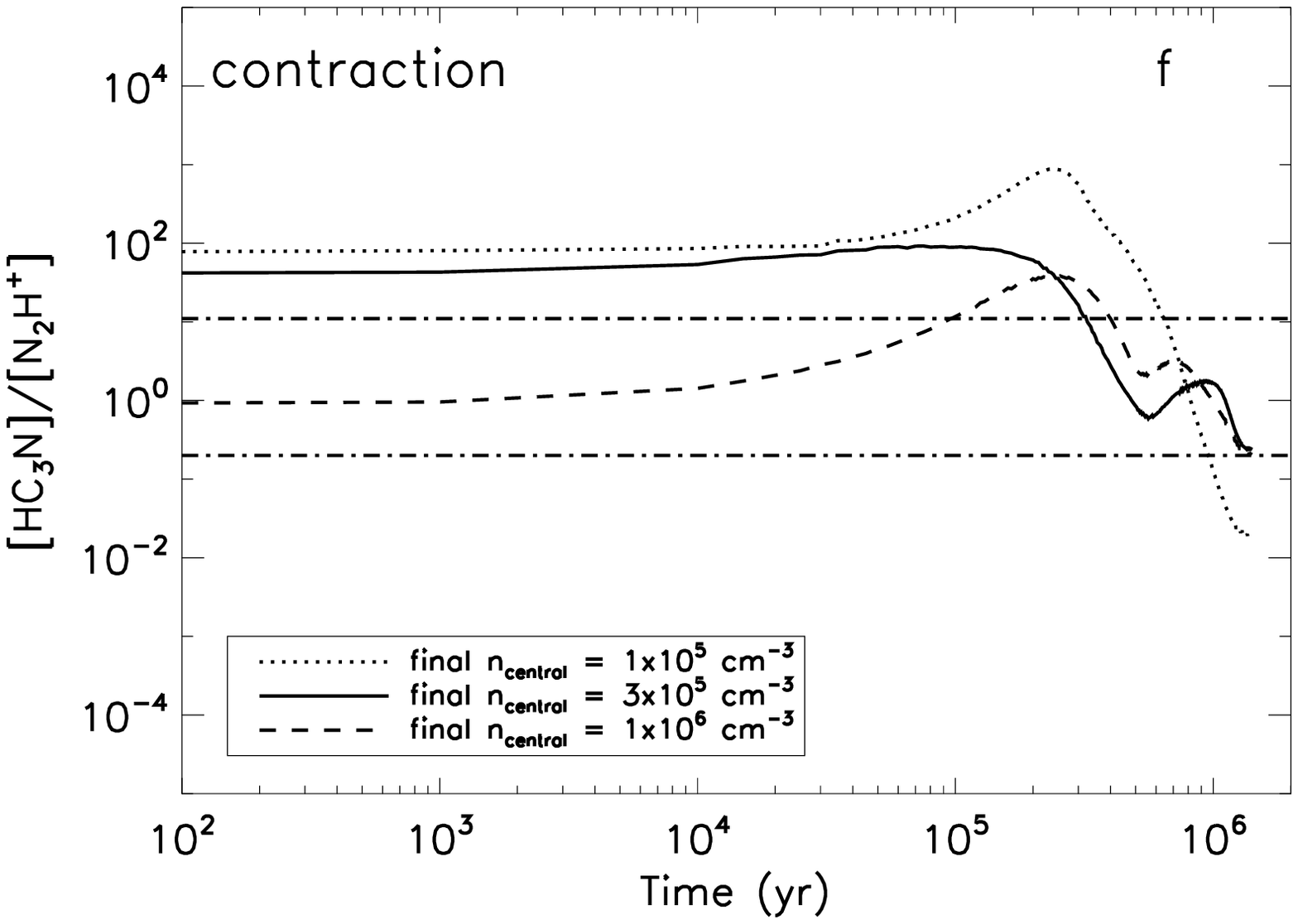} \\
\end{tabular}
 \caption[]{Evolution of the N$_2$H$^+$ (a, b) and HC$_3$N (c, d) abundances, and the HC$_3$N to N$_2$H$^+$ abundance ratio (e, f) as a function of time for free-particle densities of $1\times10^{5}$~cm$^{-3}$ (dotted), $3\times10^{5}$~cm$^{-3}$ (solid), and $1\times10^{6}$~cm$^{-3}$ (dashed line) and a kinetic temperature of 9~K. 
Each column shows predictions from one specific model (static, contraction). The predictions from the contraction models correspond to abundances derived from integrated (along all depth points) column densities \citep[see][]{garrod06a, garrod06b}. 
The horizontal dot-dashed lines indicate the \emph{range} of observed values for the N$_2$H$^+$ (top row), HC$_3$N abundances (middle row), and the HC$_3$N to N$_2$H$^+$ ratio (bottom row). 
The maximum H$_2$ column densities for the contraction models are achieved at 1 Myr and they are $\sim4 \times 10^{21}$~cm$^{-2}$, $\sim1.3 \times 10^{22}$~cm$^{-2}$, and $\sim4 \times 10^{22}$~cm$^{-2}$, respectively. \label{fig:model_ratio}}
\end{figure*}

\subsubsection{Static models}
\label{sec:static_models}

The evolution of the N$_2$H$^+$ and HC$_3$N fractional abundances, and the HC$_3$N/N$_2$H$^+$ abundance ratio for the static models are shown in Figs.~\ref{fig:model_ratio}a, c, and e. 
After an initial peak, the HC$_3$N to N$_2$H$^+$ abundance ratio decreases from a value of $\sim 10^3$--$10^4$ down to $\sim 10^{-3}$ (Fig.~\ref{fig:model_ratio}e).
After it reaches a minimum value it increases once more by a factor of $\sim 10$ (highest density) to 100 (lowest density model) toward the end of the core evolution. The range of observed values of the HC$_3$N/N$_2$H$^+$ abundance ratio ($\sim 0.2$ -- $11$) is also shown in Fig.~\ref{fig:model_ratio}e. 
For the densities modelled in this section, the observed values match the model predictions in a phase during which the HC$_3$N/N$_2$H$^+$ abundance ratio decreases with increasing time.

The static models therefore appear to support the hypothesis that a greater HC$_3$N/N$_2$H$^+$ abundance ratio is associated with early-time chemistry. However, these models include no treatment of the time-dependent condensation of a core, nor account for the chemistry in regions outside of the central density peak. Furthermore, the initial chemical abundances correspond to diffuse, atomic conditions that are not appropriate for the higher densities observed in the cores. Since the early formation of HC$_3$N appears to be dependent on the availability of atomic carbon, which is not fully locked up in CO at early times, the consideration of static models alone is not sufficient to support the interpretation of the observed trend. 

\subsubsection{Contraction models}
\label{sec:contraction_models}

To remedy these shortcomings, a comparison is made with chemical models of cloud cores investigated by \citet{garrod05} and later publications. These models trace the chemical evolution at 12 depth points arranged along a radius through a cloud core, as it condenses from diffuse to dense conditions over a period of 1 Myr, but without the re-expansion of the core that takes place after 1 Myr in the standard models of \citet{garrod05}. Here, the physical conditions are kept steady after 1 Myr. The initial elemental abundances are the same for both the static and the core contraction models.
Initial total hydrogen densities follow a Gaussian profile; values range from 300 to 1000 cm$^{-3}$ between the outermost point and the central core position, with visual extinctions ranging from 0.65 to 1.7 mag. This profile is held steady for an arbitrary period of time until a chemical steady-state is achieved (no freeze-out onto dust grains is active during this time). This produces an initial chemical composition that is appropriate to the physical conditions.

During the main period of physical evolution of the core, the density follows a Gaussian time-dependence. This process is described in detail by \citet{garrod05}. Free-particle peak central densities of $1 \times 10^5$, $3 \times 10^5$, and $1 \times 10^6$ cm$^{-3}$ with respective H$_2$ peak edge-to-edge column densities of $\sim 4 \times 10^{21}$, $\sim 1.3 \times 10^{22}$, and $\sim 4 \times 10^{22}$ cm$^{-2}$ are achieved after 1 Myr. The equivalent initial, edge-to-edge visual extinctions achieved are $\sim 1$, $\sim 2$, $\sim 5$ mag, while the final values are $\sim4$, $\sim 14$, and $\sim 43$ mag \citep[conversion factor of $9.4\times 10^{20}$ cm$^{-2}$ mag$^{-1}$; ][]{bohlin78}. The densities and visual extinctions at each modelled depth point vary with time. As points achieve visual extinctions greater than 2.5 mag, the freeze-out of gas-phase material begins. No surface chemistry is explicitly modelled in this treatment; accreted atoms are assumed to be fully hydrogenated, and remain on the grain surfaces (until the visual extinction again falls below the threshold value of 2.5 mag). 

These models allow column densities to be calculated along lines of sight passing through each depth point. Values calculated in this way take account of both the chemical abundances and gas densities specific to each point. Figures in \citet{garrod06a, garrod06b} show column densities calculated in this way, as a function of time during the physical evolution of the core. Species that are strongly dependent on carbon for their formation are found to peak prior to the attainment of maximum core density, while other species, such as N$_2$H$^+$, are found to trace the density more closely.

Figures~\ref{fig:model_ratio}b and d show the predicted peak column densities 
of N$_2$H$^+$ and HC$_3$N for the contraction models, normalized 
to the total H$_2$ column density to provide a comparison with the 
single-point fractional-abundance results.
The range of the observed HC$_3$N to N$_2$H$^+$ column density ratio
is in agreement with predicted values (Fig.~\ref{fig:model_ratio}f), 
and shows the same trend in its time-dependent behaviour as is found in the 
case of the static models.

The observed H$_2$ column densities for the Mopra beam of 37$^{\arcsec}$ (for N$_2$H$^+$ 1--0) are found after interpolating between the peak H$_2$ column density (for $HPBW$ 21.2$^{\arcsec}$, $\sim3200$ AU) and the column density within an aperture of 50$^{\arcsec}$. They are $3.3\times10^{21}$--$1.4\times10^{22}$~cm$^{-2}$ for the cores in Cha III and $3.6\times10^{21}$--$7.1\times10^{22}$~cm$^{-2}$ in Cha I (or $3.6\times10^{21}$--$1.4\times10^{22}$~cm$^{-2}$ excluding the three densest cores) \citep{belloche11a, belloche11b}. The $HPBW$ of HC$_3$N 10--9 is similar to N$_2$H$^+$ 1--0 (37.9$^{\arcsec}$). The observed H$_2$ column densities averaged over the Mopra beam are therefore comparable to the model values. The visual extinctions derived from 2MASS range from 5.1--19 mag for the Cha I cores with an average extinction of $\sim10$~mag, and $\sim2$--9 mag for the Cha III cores, with an average of $\sim5$~mag \citep{belloche11a, belloche11b}. The cores in Cha III and the low $A_V$ cores in Cha I are therefore expected to be better described by the two lower-density, contraction models with final edge-to-edge extinctions of 4 and 14 mag, while the model with the final $A_V$ of $\sim 43$ mag is better suited for the Cha I cores with visual extinctions approaching 20 mag. 

The peak, predicted N$_2$H$^+$ and HC$_3$N abundances for the contraction models shown in Figs.~\ref{fig:model_ratio}b and d are slightly lower (factor of $\sim1$--5) than the equivalent observed abundances. The range of the observed HC$_3$N to N$_2$H$^+$ abundance ratio is, on the other hand, in good agreement with the predicted values. The (downward) discrepancies in the individual abundances predicted for each molecule may result from the true density profiles of the cores being less sharp than the Gaussian density profile assumed in the contraction models. A more centrally-flattened density profile, with higher gas densities in the wings, would result in a somewhat larger region of molecule production. This could plausibly produce N$_2$H$^+$ and HC$_3$N column densities that are a few times greater, while maintaining the same overall H$_2$ column density, which would better match the observational fractional abundances. Nevertheless, given the agreement of the models and observations regarding the HC$_3$N to N$_2$H$^+$ abundance ratio, this discrepancy is small enough that it is unlikely to affect the qualitative results.

While the hypothesized cause of the trend in the HC$_3$N/N$_2$H$^+$ abundance ratio appears to agree with the models, i.e., that a lower ratio is indicative of a greater degree of physical evolution, it is unlikely that a more specific determination of core age may be made without a more specific model for individual cores in the cloud. 

Finally, the models presented here for peak densities of $\sim1 \times 10^5$--$1 \times 10^6$ cm$^{-3}$ represent densities that match those of both prestellar (gravitationally bound) and currently unbound cores in Cha I and III. Our predictions in Fig.~\ref{fig:model_ratio} therefore show that the HC$_3$N/N$_2$H$^+$ abundance ratio cannot be used to differentiate between gravitationally bound and unbound cores. Instead, it only points to an early or late stage of chemical evolution.

\subsection{Comparison to other clouds}
 
Starting with the low-density tracers (in terms of critical density), the mean abundances of C$^{18}$O and C$^{17}$O in Cha I relative to H$_2$ are $\sim 4.2 \times 10^{-7}$ and $\sim 5.3 \times 10^{-8}$, respectively. We derive lower abundances in Cha III by a factor of $\sim 3$ and $\sim 2$, respectively. \citet{frerking82} estimated a C$^{18}$O abundance of 1.7$\times$10$^{-7}$ within cloud interiors in $\rho$ Ophiuchus and Taurus. \citet{miettinen11} derived C$^{17}$O abundances ranging from $4 \times 10^{-8}$ to $1.6 \times 10^{-7}$ toward seven clumps in massive infrared dark clouds (IRCDs). \citet{friberg88} observed methanol toward three dark clouds. They found CH$_3$OH abundances of $\sim 2 \times 10^{-9}$ for TMC1 and L134N, and $\sim 0.5$ -- $1 \times 10^{-9}$ for the Class 0 object B335 and for a kinetic temperature of 10 K. The observed C$^{18}$O, C$^{17}$O, and CH$_3$OH abundances toward the aforementioned clouds are all very similar to the mean (and median) abundances we derive for both Cha I and Cha III (see Table~\ref{table:abundances}).

We now discuss the abundances of the high-density tracers. \citet{vasyunina11} report an average N$_2$H$^+$ abundance of $7.7 \times 10^{-10}$ for a sample of low-mass IRDCs. The average N$_2$H$^+$ abundances we derive are \emph{lower} by a factor of $\sim3$ in Cha I and $\sim 7$ in Cha III. The low-mass IRDC abundance average for HC$_3$N is $\sim 1.5 \times 10^{-9}$ \citep{vasyunina11}, which is higher than the mean abundance we find in Cha III by a factor of $\sim 3$. Molecular abundances of $\sim 5.8$ -- $17 \times 10^{-11}$ were found for C$^{34}$S in the IC5146 dark cloud \citep{bergin01}. Our reported C$^{34}$S abundances in Cha I and III are similar to the high-end value. We therefore see a significant abundance difference between Cha (I and III) and other low-mass dark clouds only in N$_2$H$^+$ and in HC$_3$N for Cha III, but the statistics are low in the last case. These molecules are both \emph{less} abundant in Cha I and III than in other low-mass clouds.

\section{Discussion}
\label{sec:discussion}

\subsection{Turbulence in Cha I and III}
\label{sec:discussion_turbulence}

In order to directly compare the linewidths and non-thermal velocity dispersions of the transitions detected in both clouds, we choose to only compare core samples of similar peak column densities. This way we compare cores of similar physical properties. For this reason we exclude the cores Cha1-C1, Cha1-C2, and Cha1-C3, whose observed peak column densities are higher by a factor of $1.5$--$6$ than the maximum peak column densities in Cha III. We also \emph{omit} the second velocity components (see definition in Sect.~\ref{sec:definitions}) for this comparison. We assume the \emph{same kinetic temperature} of 10 K for both clouds, which is plausible since both clouds have similar dust temperatures \citep[see][]{toth00} and mean densities \citep{belloche11a, belloche11b}.

\begin{figure}
\begin{tabular}{c}
\hspace*{-6ex}
\includegraphics[width=100mm,angle=0]{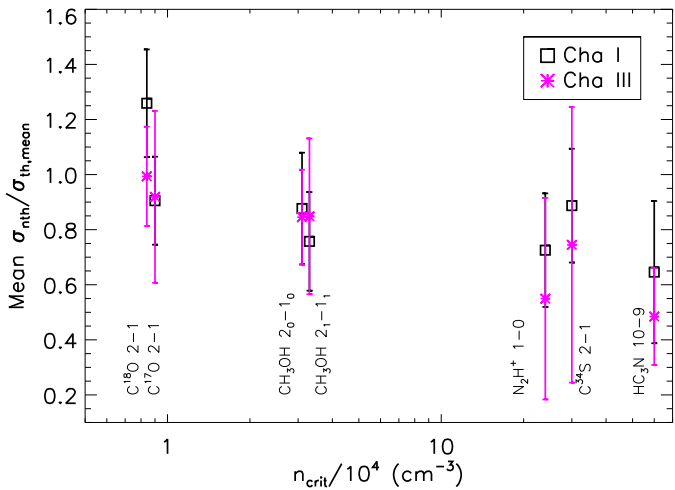} \\
\end{tabular}
 \caption[]{Mean ratio of non-thermal to thermal (of the mean particle) velocity dispersions against the critical density for the transitions listed in Table~\ref{table:turbulence} for the Cha I (black squares) and III (pink asterisks) clouds. The errorbars show the dispersions from the mean. \label{fig:turbulence}}
\end{figure}

A direct comparison of the non-thermal to thermal velocity dispersion ratios of the transitions C$^{34}$S 2--1, CH$_3$OH 2$_{0}$--1$_{0}$ A$^+$, CH$_3$OH 2$_{1}$--1$_{1}$ E, N$_2$H$^+$ 1--0, C$^{18}$O 2--1, C$^{17}$O 2--1, and HC$_3$N 10--9 shows that they are similar in both clouds (see Table~\ref{table:turbulence} and Fig.~\ref{fig:turbulence}). This ratio is larger in C$^{18}$O 2--1 in Cha I, but as the line is affected by optical depth effects (see Sect.~\ref{sec:opacity_c18o}) the difference between the two clouds is most likely not significant. Hence, there is no significant difference in the turbulence level within Cha I and III that could account for the difference in their star formation activities.

\begin{table*}
\caption{Linewidths and non-thermal velocity dispersions in Cha I and III.} 
\vspace*{-1ex}
\hfill{}
\begin{tabular}{@{\extracolsep{-2pt}}llllllllllll} 
\hline
\hline
Line & \multicolumn{5}{c}{Cha I \tablefootmark{a}} & & \multicolumn{5}{c}{Cha III} \\
\cline{2-6}
\cline{8-12}
    &   $FWHM$\tablefootmark{b} & $FWHM$\tablefootmark{c} & $[\frac{\sigma_{\rm nth}}{\sigma_{\rm th,mean}}]$\tablefootmark{d} & $[\frac{\sigma_{\rm nth}}{\sigma_{\rm th,mean}}]$\tablefootmark{e}   & $N$\tablefootmark{f} & & $FWHM$\tablefootmark{b} & $FWHM$\tablefootmark{c} & $[\frac{\sigma_{\rm nth}}{\sigma_{\rm th,mean}}]$\tablefootmark{d} & $[\frac{\sigma_{\rm nth}}{\sigma_{\rm th,mean}}]$\tablefootmark{e} & $N$\tablefootmark{f} \\ 
   &  \small{average}                        & \small{median}                     &          \small{average}                                                              & \small{median}                                            &     & & \small{average}    &   \small{median}   & \small{average}                                             & \small{median}                                                    & \\
    &   \scriptsize{(km s$^{-1}$)} & \scriptsize{(km s$^{-1}$)}  &  &   &     &   & \scriptsize{(km s$^{-1}$)} & \scriptsize{(km s$^{-1}$)} & & &  \\ 
\hline
C$^{18}$O 2--1 \tablefootmark{g}  &  0.57$\pm$0.07 & 0.52 & 1.26$\pm$0.07 & 1.15 & 57  &  &    0.45$\pm$0.03  &  0.44 & 0.97$\pm$0.06 & 0.96 & 29 \\
C$^{17}$O 2--1 \tablefootmark{h}  &   0.42$\pm$0.06 & 0.40 & 0.91$\pm$0.05 & 0.86 & 29  &  &    0.42$\pm$0.04 & 0.42 & 0.92$\pm$0.10   & 0.92 & 2 \\ 
CH$_3$OH 2$_{1}$--1$_{1}$ E        &  0.36$\pm$0.06 & 0.36 & 0.76$\pm$0.06 & 0.74 & 28  & &     0.39$\pm$0.04  &  0.40 & 0.85$\pm$0.09 & 0.86 & 10 \\
CH$_3$OH 2$_{0}$--1$_{0}$ A$^+$    &  0.41$\pm$0.06 & 0.36 & 0.88$\pm$0.07 & 0.77 & 38  & &     0.39$\pm$0.02  &  0.40 & 0.85$\pm$0.06 & 0.87  & 15 \\
N$_2$H$^+$ 1--0 \tablefootmark{h} &  0.35$\pm$0.07 & 0.32 & 0.73$\pm$0.07 & 0.64 & 16  & &     0.27$\pm$0.05  &  0.27 & 0.55$\pm$0.12 & 0.55 &  2 \\
C$^{34}$S 2--1                     &  0.40$\pm$0.07  & 0.37 & 0.89$\pm$0.07 & 0.81 & 20  & &    0.34$\pm$0.07 &  0.34 & 0.74$\pm$0.17 & 0.74  & 2 \\
HC$_3$N 10--9 \tablefootmark{h}   &  0.30$\pm$0.06 & 0.27 & 0.65$\pm$0.09 & 0.55 & 15  &  &    0.23$\pm$0.02 & 0.23 & 0.48$\pm$0.06 & 0.48 & 3 \\
\hline
\end{tabular}
\hfill{}
\label{table:turbulence}
\vspace*{-1ex}
\tablefoot{
\tablefoottext{a}{Average values exclude cores Cha1-C1, C2, and C3 as we only compare cores with the same range of peak column densities in both clouds. The errors are dispersions from the mean value.}
\tablefoottext{b}{Average $FWHM$ linewidths over all cores in each transition.}
\tablefoottext{c}{Median $FWHM$ linewidth in each transition. }
\tablefoottext{d}{Average ratio of the non-thermal-to-mean-thermal velocity dispersions for the core sample in each transition. }
\tablefoottext{e}{Median non-thermal-to-mean-thermal velocity dispersion ratio. }
\tablefoottext{f}{Number of cores used for the calculation. }
\tablefoottext{g}{A fraction of the C$^{18}$O 2--1 spectra in Cha I are marginally optically thick and may therefore be slightly broadened (see Sect.~\ref{sec:opacity_c18o}). }
\tablefoottext{h}{Intrinsic linewidths ($FWHM$) derived from a seven-component hyperfine-structure fit.}
}

\vspace*{-1ex}
\end{table*}       

\subsubsection{Turbulence dissipation}

C$^{17}$O 2--1 and C$^{18}$O 2--1 have critical densities of 
$\sim 9 \times 10^3$~cm$^{-3}$ at 10 K (see Table~\ref{table:crit_dens}). In 
addition, C$^{17}$O and C$^{18}$O are usually depleted in the interiors of cold 
starless cores. Therefore these transitions preferentially trace low-density 
material. Transitions with higher critical densities, such as N$_2$H$^+$ 1--0 
and HC$_3$N 10--9, are expected to trace core interiors. The dense, inner parts of the cores are expected to be regions of minimum turbulence in contrast to the less dense, outer parts of the cores, which are less shielded from the ambient turbulent motions \citep{goodman98, larson81}. The low-density transitions likely trace these outer, and more turbulent core regions. The larger non-thermal dispersions that we find in C$^{17}$O 2--1 and C$^{18}$O 2--1 compared to HC$_3$N and N$_2$H$^+$ in both clouds are consistent with the picture of turbulence dissipation in the core interiors in Cha I and III. CH$_3$OH~2$_0$--1$_{0}$~A$^+$ has a similar critical density to C$^{17}$O, C$^{18}$O, within a factor of $\sim3$, as well as a similar abundance distribution (see Fig.~\ref{fig:abundances}). We also find that CH$_3$OH~2$_0$--1$_{0}$~A$^+$, C$^{17}$O, and C$^{18}$O have similar non-thermal velocity dispersions in both Cha I and III, therefore indicating that these transitions likely trace similar regions. 

N$_2$H$^+$ 1--0 and HC$_3$N 10--9 are both high-density tracers with critical densities of $\sim 2.4 \times 10^5$~cm$^{-3}$ and $\sim 6.0 \times 10^5$~cm$^{-3}$, respectively (Table~\ref{table:crit_dens}). Both transitions have non-thermal velocity dispersions comparable to each other in both Cha I and III (Sect.~\ref{sec:tracers_comparison}). This is expected if they both trace regions of similar density. The non-thermal velocity dispersions of these transitions are also found to be systematically smaller than the non-thermal dispersions of C$^{17}$O 2--1 and C$^{18}$O 2--1 in both Cha I and III. We obtain a N$_2$H$^+$ 1--0 to C$^{34}$S 2--1 mean non-thermal velocity dispersion ratio of $0.75_{-0.19}^{+0.18}$ in Cha I. C$^{34}$S 2--1 is excited at densities similar to N$_2$H$^+$ 1--0 and these two transitions would therefore be expected to have similar non-thermal velocity dispersions if they both have similar abundance profiles. The higher C$^{34}$S 2--1 non-thermal dispersions is a sign of C$^{34}$S depletion in the core interiors, making this molecule sensitive to the outer, more turbulent core regions. We also find evidence of C$^{34}$S depletion in the distribution of core abundances in Cha I as a function of their column density (Fig.~\ref{fig:abundances}e). The fact that C$^{34}$S 2--1 and C$^{17}$O 2--1 have a mean non-thermal dispersion ratio close to unity is an additional hint that C$^{34}$S 2--1 traces similar lower-density regions to C$^{17}$O 2--1 and that N$_2$H$^+$ 1--0 and HC$_3$N 10--9 trace the quiescent inner core regions.  

Overall, all tracers excluding C$^{18}$O 2--1 have subsonic non-thermal dispersions, indicating that the core interiors in Cha I and III are indeed quiescent, as found in other clouds \citep{goodman93}.

\subsection{Core interactions in Cha I and III}
\label{sec:discussion_interactions}

The free-fall time for spherically symmetric gravitational collapse is given by
\begin{equation}
t_{\rm ff} = \sqrt{\frac{3\pi}{32 G \rho_{\rm mean}}},
\end{equation}
where $\rho_{\rm mean}$ is the mean mass density of the cores. We estimate the core densities as $\rho$=$n_{\rm mean} \times \mu m_{\rm H}$, with $n_{\rm mean}$ taken from \citet{belloche11b} for Cha III, and computed using the total mass and a radius of $\sqrt{FWHM_{\rm maj} \times FWHM_{\rm min}}$ for the cores in Cha I \citep[total mass and radius from ][]{belloche11a}. We use a mean molecular weight of $\mu \sim 2.37$. We find \emph{average} free-fall times of 0.18~Myr for both Cha I and III. 

The expected lifetimes and evolution of dense cores largely depend on the theoretical model used to describe the gravitational collapse of the cloud \citep[see][ for a review on core lifetimes]{wardthompson07}. Collapse models governed by ambipolar diffusion tend to predict core lifetimes longer than the free-fall time, which increases the more the initial core conditions depart from the magnetically supercritical state \citep{shu87, mouschovias91, myers_khersonsky95}. The collapse proceeds faster when the core is initially nearly supercritical \citep[e.g.,][]{nakano98} and if turbulence is dynamically important \citep{ciolekbasu01, fatuzzo02, vazquezsemadeni05}. In models of initially magnetically subcritical, turbulent clouds the dense cores produced have lifetimes that range from $\sim1.5$ to $\sim 10$ times the free-fall time \citep[e.g.,][]{nakamura05}. Models of magnetically supercritical, turbulent cloud collapse also predict core lifetimes of the order of a few times the free-fall time \citep{galvan07, vazquezsemadeni05} compared to the turbulence dominated star formation scenario that supports faster evolution timescales of the order of one free-fall time \citep{elmegreen00, wardthompson07, kirk05, ballesteros99, vazquezsemadeni00, maclow04, ballesteros07}. \citet{galvan07} find mean core lifetimes of the order of $\sim$6t$_{\rm ff}$ in their numerical simulations of magnetically supercritical, turbulent, and isothermal clouds. 

Observationally, there are various estimates of the core lifetimes, some of which are $\sim$0.5 Myr \citep{evans09}, $\sim$0.3--1.6 Myr \citep{leemyers99}, $\sim$0.4 Myr \citep{onishi98,onishi02}, and $\sim$0.3 Myr \citep{kirk05}. \citet{wardthompson07} conclude that most observational estimates yield core lifetimes of $\sim$2--5 free-fall times for densities of $\sim$10$^4$--10$^5$ cm$^{-3}$.

If we assume core lifetimes of the order of $3$--$6$ times the free-fall time as suggested by both the observational evidence \citep[e.g.,][]{wardthompson07} and collapse models \citep[e.g.,][]{galvan07}, we obtain a dynamical lifetime of $\sim0.5$--1~Myr for Cha I and III with dispersions of $\sim0.2$--$0.3$~Myr. 
A comparison of these lifetime estimates to the collisional times we 
obtained in Table~\ref{table:dispersions_lines} for Cha I and III as a whole 
and their subregions shows that, with the exception of 
Cha~I North ($t_{\rm coll} \sim 0.7$~Myr), the collisional times are
longer than the dynamical lifetimes, and in most cases even much longer.
The cores in Cha I North might therefore be the only ones to experience 
dynamical interactions in the future, thus increasing the probability of 
competitive accretion in this region. We note, however, that the cores 
in Cha~I Centre and Cha~III North have collisional times only slightly longer 
than 1~Myr, so they may also experience dynamical interactions in the future.

The collision times we compute for the \emph{subregions} of Cha I and III are of the same order as the ones found in \citet{andre07}, while the deviations are large when we compare these values for the Cha I and III clouds as a whole. \citet{andre07} estimated a collision time of $\sim 1$ -- 10 Myr between the 57 starless cores in Ophiuchus and a dynamical core lifetime of $\sim0.02$ -- $0.5$ Myr (assuming $3 \times t_{\rm ff}$ and using the mean core density). They concluded that interactions between the condensations in Ophiuchus cannot happen within their lifetimes. \citet{sadavoy12} studied nine dense cores in Perseus B1-E and found that the cores have near-neighbour separations less than one Jeans length and interactions might occur on a timescale of $\sim1$~Myr, which makes competitive accretion plausible for that region.

We conclude that, as is the case for Ophiuchus, dynamical interactions between cores are unlikely to be dynamically significant in either Cha~I, Cha~III, and the majority of their subregions. Competitive accretion may only affect the future dynamical state of the cores in Cha~I North and, to a lesser extent,
Cha~I Centre and Cha~III North.

\subsection{Dynamical state of starless cores}
\label{sec:discussion_dynamical}

Our virial equilibrium analysis suggests that up to five cores are gravitationally bound in Cha I (Cha1-C1, Cha1-C2, Cha1-C3, Cha1-C4, Cha1-C5) and only one is virialized (Cha1-C1, also known as Cha-MMS1). Cha1-C1 is virialized in most detected transitions (see Table~\ref{table:detections_chaI_chaIII}). Therefore, $\sim8$\% of the whole sample at most, show evidence of \emph{already} being gravitationally bound in Cha I. 

Nevertheless, we found that $13$\% -- $28$\% of the total core population in Cha I are either already prestellar or likely to become gravitationally bound and therefore prestellar in the future based on the ``infall'' signature (see Sect.~\ref{sec:infall_signature}). For the number of cores not gravitationally bound, the ``infall'' signature is interpreted as contraction motions, which may turn the currently unbound cores into bound ones (prestellar). The upper limit corresponds to the full sample of cores showing the ``infall'' signature, while the lower limit reflects a more conservative number after excluding the number of ``infall'' signatures that may only trace large-scale contraction motions (as traced by $^{13}$CO 2--1) or pulsations.
\citet{anathpindika13} also stress that inward motions are not a sufficient criterion for gravitational boundedness. \citet{belloche11a} found that $\sim17$\% of the starless cores are \emph{currently} prestellar based on their Bonnor-Ebert analysis and \citet{belloche11b} estimated that up to $\sim50$\% might become unstable in the future based on theoretical predictions. We therefore find that the fraction of the cores that might become prestellar based on \emph{observational evidence} and the ``infall'' signature ($\sim 13$--28\%) is in agreement with the $\sim50$\% upper limit of \citet{belloche11b}.%the fraction 

The estimated fraction of prestellar cores in Aquila amounts to $\sim60$\% \citep{andre10}, while it is as high as $\sim70$\% in Ophiuchus \citep{motte98}. \citet{curtis11} found that most of the dense cores in Perseus seem to be close to the critical virial ratio, thus implying that they are gravitationally bound, thus prestellar \citep[see also][]{enoch08}. 
 \citet{belloche11a} suggested that Cha I is experiencing the end of its star formation activity. Overall, the data show that Cha I is experiencing significantly lower star formation activity compared to other nearby clouds and our kinematical analysis so far suggests that only up to $\sim28$\% of the Cha I starless cores might become prestellar in the future. \citet{ortiz12} also found that almost all the starless cores in the Vela-D molecular cloud are not virialized or gravitationally bound. Overall, they estimated that approximately 30\% of the whole core population in Vela-D is gravitationally bound, which is comparable to the number of cores in Cha I that will likely become unstable in the future. 

%Cha III 
We find that no core is virialized in Cha III and only one, Cha3-C1, is gravitationally bound based on our virial analysis. We also find that 2--5 cores are likely to become prestellar based on the ``infall'' signature, which corresponds to 10--25\% of the observed sample (see Sect.~\ref{sec:infall_signature}). \citet{belloche11b} concluded that two cores in Cha III are gravitationally unstable (Cha3-C1 and Cha3-C2) from their Bonnor-Ebert analysis.

The virial analysis of Cha I and Cha III confirms the result previously suggested that Cha I is in general more active than Cha III \citep{belloche11b}. When taking the ``infall'' signature into account we find that the cores suggested to be gravitationally bound in Cha III from the dust continuum survey are, indeed, infalling and that three additional cores are likely to become prestellar. 

\citet{smith12} modelled the line emission of irregular cores embedded in filaments via radiative transfer and found that the number of collapsing cores are underestimated when only taking the ``infall'' signature into account. The line profiles are extremely variable depending on the viewing angle and other effects, such as one-sided accretion onto filaments. These may result in a line profile not showing the expected blue asymmetry even if the core is indeed collapsing. The fraction of contracting cores in Cha I and III might therefore be even larger than the observational evidence has shown so far.

\subsection{Core evolutionary state}
\label{sec:core_evolution}

Observationally, we find a significant difference in the HC$_3$N/N$_2$H$^+$ abundance ratio between Cha I Centre, and Cha I South and West (see Fig.~\ref{fig:observed_ratio}b). The highest density cores belonging to Cha I Centre have on average lower HC$_3$N/N$_2$H$^+$ abundance ratios than the lower density cores in Cha I South and West. In the framework of the static and collapse models presented in Sect.~\ref{sec:chemical_models}, the HC$_3$N/N$_2$H$^+$ abundance ratio appears to be a good evolutionary tracer. 
Nevertheless, the use of abundance ratios as evolutionary clocks is strongly model dependent and requires knowledge of the initial conditions of the region being studied in order to be used as such.

The observed HC$_3$N to N$_2$H$^+$ abundance ratios in Fig.~\ref{fig:observed_ratio} imply that the cores in the central part of Cha I (Cha I Centre) are more evolved than the cores in the southern part of the cloud (Cha I South). The densest and most evolved core in Cha I, Cha1-C1 (Cha-MMS1; first hydrostatic core candidate) is also located in Cha I centre and has an HC$_3$N to N$_2$H$^+$ abundance ratio of $\sim0.3$.
It is interesting to note that all the cores in Cha I Centre show contraction motions manifested as the ``infall'' signature, while there are approximately equal numbers of cores showing the ``infall'' and opposite signatures in Cha I South (Fig.~\ref{fig:infallingcores}). Therefore, as there is no excess of ``infall'' signatures compared to the opposite signatures, the cores in Cha I South may only be transient structures or oscillating cores.

The HC$_3$N to N$_2$H$^+$ abundance ratio for the Cha III cores is found to be similar to Cha I South, although with low statistics for Cha III. This points to the fact that Cha III, as well as Cha I South and West, are less chemically evolved than Cha I Centre.
In addition, the evidence of contraction that we see for up to 25\% of cores in Cha III through the ``infall'' signature and the fact that we see no opposite signatures for any cores indicate that some of the Cha III cores may become prestellar. Therefore, Cha III is likely to form stars in the future.

\subsection{Overlapping velocity components}
\label{sec:discussion_2vel}

The \emph{Herschel} image of Cha I shown in \citet{winston12} (PACS 160 $\mu$m, SPIRE 250 $\mu$m, and SPIRE 500 $\mu$m) reveals the dust structure of the cloud in great detail. The elongated filament in Cha I Centre (Fig.~\ref{fig:2vel}b) corresponds to the Ced 110 cluster on the \emph{Herschel} image of \citet{winston12}. It appears very bright and dense, while retaining the same elongated structure as 
in the 870 $\mu$m LABOCA map of Cha I
\citep{belloche11a}. The direction as well as the location of the higher velocity emission in our maps seems to correspond to fainter, more dispersed and lower density material seen on the \emph{Herschel} image that is threading the dense, elongated filament in the perpendicular direction. Alternatively, the higher velocity component could simply be a projection effect.

As the higher velocity component in Cha~I North (Fig.~\ref{fig:2vel}a) is not related to the bipolar outflow of Cha-MMS2 (see Sect.~\ref{sec:multiple_components}) it might also be due to material in the cloud interacting with this elongated structure or a projection effect.
The dust continuum \emph{Herschel} map of Cha I North \citep[Ced 112;][]{winston12} shows that there is some fainter material close to the very northern part of the filament where the peak of the higher velocity emission is located and it seems to be oriented perpendicular to it. However, the material in Cha I North as seen on the \emph{Herschel} map is not as dense and homogeneous in direction as it appears in Cha I Centre.

\section{Summary and conclusions}
\label{sec:conclusions}

We performed spectral line observations with the APEX and Mopra telescopes toward 60 cores in Cha I and 29 cores in Cha III. The aim of this study is to explore the kinematical state of the cores, their future dynamical evolution, and to determine what the main driver of the different star formation activities in the two clouds is. Our conclusions are the following:

\begin{enumerate}

\item We find at most 5 prestellar cores ($\sim 8$\%) in Cha I and 1 in Cha III based on a virial analysis. Between 8 and 17 cores show contraction motions based on the ``infall'' signature in Cha I, bringing the total percentage of cores that might become, if not already, prestellar to 13--28\%. Similarly, between 2 and 5 cores are contracting in Cha III, and thus 10--25\% cores of the observed sample might become, if not already, prestellar in the future, which corresponds to 7--17\% of the whole core population.
 
\item Multiple velocity components are observed in Cha I toward a few cores in the northern and central parts of the cloud. The C$^{18}$O 1--0 integrated emission of the higher velocity component in the central elongated structure in Cha I (Cha I Centre) is oriented \emph{perpendicular} to the dust continuum emission. The higher velocity component might be caused by more disperse material interacting with the elongated structure along this direction, as suggested by the \emph{Herschel} continuum data.

\item We derive a C$^{18}$O median abundance relative to H$_2$ that is lower in Cha III than in Cha I by a factor of $\sim2$. The distributions of the individual core abundances in Cha III in C$^{18}$O, C$^{17}$O, HC$_3$N, and N$_2$H$^+$ lie at the lower end of the core abundances in Cha I, although with low statistics for Cha III.

\item The C$^{18}$O, C$^{17}$O, CH$_3$OH, and C$^{34}$S abundances of the Cha I cores decrease with increasing density, therefore suggesting that these molecules are affected by depletion in the inner parts of the cores.  

\item The turbulence level is similar in both clouds in various transitions. There is therefore no indication of turbulence playing a role in the different star formation activities of Cha I and III. 

\item Turbulence dissipation in the core interiors is seen in both clouds with the non-thermal dispersions of the high-density tracers N$_2$H$^+$ 1--0 and HC$_3$N 10--9 being smaller than the non-thermal dispersions of C$^{18}$O 2--1 and C$^{17}$O 2--1. C$^{34}$S 2--1 has a higher non-thermal velocity dispersion compared to N$_2$H$^+$ 1--0 in Cha I, which is consistent with C$^{34}$S being affected by depletion in the inner parts of the cores. 

\item A velocity distribution and collision time analysis adapted for cylindrical symmetry shows that the dynamical evolution of the cores in Cha I and III on a cloud scale is not likely to be affected by interactions between the cores and competitive accretion. Core interactions in Cha I North may, on the other hand, take place within $\sim$ 0.7~Myr.

\item Both contraction and static chemical models indicate that the HC$_3$N to N$_2$H$^+$ abundance ratio is a good core evolutionary indicator, but 
it cannot distinguish between transient and gravitationally bound cores. In the framework of these models, the observed HC$_3$N/N$_2$H$^+$ abundance ratios in Cha I and III suggest that the cores in Cha I South and Cha III are less chemically evolved than the cores in Cha I Centre. The Cha I South cores may just be oscillating or transient objects, whereas some of the Cha III cores may form stars in the future.

\end{enumerate}  

\noindent The measured HC$_3$N/N$_2$H$^+$ abundance ratio and the evidence of both small-scale and large-scale contraction seen towards some of the Cha III cores suggests that Cha III is less chemically evolved than Cha I Centre, but may form stars in the future.

\begin{acknowledgements}
We thank the referee for the feedback that greatly improved the quality of this paper. We also thank the APEX and Mopra staff for their support during the observations, as well as Axel Wei{\ss}, Fr{\'e}d{\'e}ric Schuller, Silvia Leurini, Arturo G{\'o}mez-Ruiz, Friedrich Wyrowski, and Keping Qiu for carrying out parts of the observations. In addition, we are very thankful to J. Harju for providing us with the Cha III C$^{18}$O 1--0 data from their published work. AET was supported for this research through a stipend from the International Max Planck Research School (IMPRS) for Astronomy and Astrophysics at the Universities of Bonn and Cologne. RTG acknowledges support from the NASA Astrophysics Theory Program through grant NNX11AC38G. The Australia Telescope Compact Array (/ Parkes radio telescope / Mopra radio telescope / Long Baseline Array) is part of the Australia Telescope National Facility which is funded by the Commonwealth of Australia for operation as a National Facility managed by CSIRO. The University of New South Wales Digital Filter Bank used for the observations with the Mopra Telescope was provided with support from the Australian Research Council.
\end{acknowledgements}

\bibliographystyle{aa} 
\bibliography{paper_survey}

\appendix

\section{Detailed physical parameters of Cha I and III cores}
\label{sec:appendix_chaI}

Tables~\ref{table:physical_parameters_chaI_2} and ~\ref{table:physical_parameters_chaIII_2} provide a detailed list of the observationally derived physical parameters of the cores for each transition separately. The method for extracting these values is described in Sects.~\ref{sec:definitions} and ~\ref{sec:results_kinematics}. For the spectra that have two velocity components, the derived properties based on both components are listed. The integrated intensities are listed only for the transitions that were used to derive column densities.

\clearpage

\begin{table*}
\hspace*{-40in}
\caption{Line parameters and physical parameters of cores in Cha I.} 
\vspace*{-1ex}
\hfill{}
\begin{tabular}{lllllllllll} 
\hline
\hline
Source  & Transition & $I_{\rm int}$\tablefootmark{a} & $V_{\rm LSR}$\tablefootmark{b} & $FWHM$\tablefootmark{c} &  $M_{\rm vir}$\tablefootmark{d} &   $\frac{M_{\rm tot}}{M_{\rm vir}}$\tablefootmark{e} &   $\frac{M_{\rm 50}}{M_{\rm vir}}$\tablefootmark{f} &  $\frac{\sigma_{\rm nth}}{\sigma_{\rm th,mean}}$\tablefootmark{g} & Gr.B.\tablefootmark{h} &  Vir.\tablefootmark{i} \\ 
        &            & \scriptsize{(K km s$^{-1}$)} & \scriptsize{(km s$^{-1}$)} &  \scriptsize{(km s$^{-1}$)} & \scriptsize{($M_{\rm \odot}$)}  & & & & & \\ \hline
Cha1-C1  & & & & & & & & \\ 
& C$_4$H 10$_{10}$--9$_{9}$ &  &  4.44$\pm$0.02 & 0.41$\pm$0.04 & 1.55 &      2.17 &  0.93 &0.91$\pm$ 0.10 &  yes  &  yes  \\
& C$_4$H 11$_{11}$--9$_{10}$ &  &  4.49$\pm$0.02 & 0.64$\pm$0.05 & 2.61 &      1.29 &  0.55 &1.44$\pm$ 0.11 &  yes  &  yes  \\
& H$_2$CO 3$_{1,2}$--2$_{1,1}$ &  &  4.33$\pm$0.03 & 0.81$\pm$0.06 & 3.65 & 0.92 & 0.40 &  1.82$\pm$0.13 &  yes &   no \\
& C$^{18}$O 2--1 &  3.17$\pm$0.05 &  4.55$\pm$0.01 & 0.59$\pm$0.01 & 2.28 &      1.47 &  0.63 &1.30$\pm$ 0.02 &  yes  &  yes  \\
& C$^{17}$O 2--1 &  1.05$\pm$0.07 &  4.40$\pm$0.01 & 0.47$\pm$0.03 & 1.73 &      1.94 &  0.83 &1.02$\pm$ 0.07 &  yes  &  yes  \\
& CH$_3$OH 2$_{0}$--1$_{0}$ A$^+$ &  1.01$\pm$0.03 &  4.39$\pm$0.01 & 0.47$\pm$0.01 & 1.73 &      1.94 &  0.83 &1.02$\pm$ 0.03 &  yes  &  yes  \\
& CH$_3$OH 2$_{1}$--1$_{1}$ E &  0.80$\pm$0.02 &  4.40$\pm$0.01 & 0.50$\pm$0.01 & 1.88 &      1.79 &  0.77 &1.10$\pm$ 0.03 &  yes  &  yes  \\
& N$_2$H$^+$ 1--0 &  1.55$\pm$0.02 &  4.40$\pm$0.01 & 0.49$\pm$0.01 & 1.81 &      1.85 &  0.79 &1.07$\pm$ 0.01 &  yes  &  yes  \\
& C$^{34}$S 2--1 &  0.35$\pm$0.02 &  4.44$\pm$0.01 & 0.56$\pm$0.04 & 2.17 &      1.55 &  0.66 &1.25$\pm$ 0.08 &  yes  &  yes  \\
& HC$_3$N 10--9 &  1.26$\pm$0.02 &  4.45$\pm$0.01 & 0.47$\pm$0.01 & 1.77 &      1.90 &  0.81 &1.04$\pm$ 0.02 &  yes  &  yes  \\
& N$_2$H$^{+}$ 3--2 &  &  4.47$\pm$0.01 & 0.53$\pm$0.01 & 1.98 &      1.70 &  0.73 &1.16$\pm$ 0.03 &  yes  &  yes  \\
Cha1-C2  & & & & & & & & \\ 
& C$_4$H 11$_{11}$--9$_{10}$ &  &  4.49$\pm$0.04 & 0.46$\pm$0.08 & 2.28 & 0.71 & 0.26 &  1.02$\pm$0.18 &  yes &   no \\
& H$_2$CO 3$_{1,2}$--2$_{1,1}$ &  &  4.27$\pm$0.04 & 0.60$\pm$0.10 & 3.10 & 0.52 & 0.19 &  1.34$\pm$0.22 &  yes &   no \\
& C$^{18}$O 2--1 &  3.30$\pm$0.06 &  4.61$\pm$0.01 & 0.78$\pm$0.01 & 4.55 & 0.36 &  0.13 &1.76$\pm$ 0.03 &  no  &   no \\
& C$^{17}$O 2--1 &  1.13$\pm$0.09&  4.39$\pm$0.02 & 0.59$\pm$0.06 & 3.01 & 0.54 & 0.19 &  1.31$\pm$0.13 &  yes &   no \\

& CH$_3$OH 2$_{0}$--1$_{0}$ A$^+$ &  1.56$\pm$0.05&  4.40$\pm$0.01 & 0.51$\pm$0.01 & 2.52 & 0.64 & 0.23 &  1.13$\pm$0.04 &  yes &   no \\
& CH$_3$OH 2$_{1}$--1$_{1}$ E &  1.23$\pm$0.05&  4.39$\pm$0.01 & 0.50$\pm$0.02 & 2.44 & 0.66 & 0.24 &  1.09$\pm$0.05 &  yes &   no \\
& N$_2$H$^+$ 1--0 &  0.60$\pm$0.04&  4.32$\pm$0.01 & 0.48$\pm$0.01 & 2.37 & 0.69 & 0.24 &  1.06$\pm$0.03 &  yes &   no \\
& C$^{34}$S 2--1 &  0.52$\pm$0.04 &  4.41$\pm$0.02 & 0.67$\pm$0.05 & 3.62 & 0.45 &  0.16 &1.50$\pm$ 0.11 &  no  &   no \\
& HC$_3$N 10--9 &  0.76$\pm$0.04&  4.42$\pm$0.01 & 0.50$\pm$0.07 & 2.51 & 0.65 & 0.23 &  1.12$\pm$0.17 &  yes &   no \\
& N$_2$H$^{+}$ 3--2 &  &  4.32$\pm$0.01 & 0.51$\pm$0.04 & 2.53 & 0.64 & 0.23 &  1.13$\pm$0.09 &  yes &   no \\
Cha1-C3  & & & & & & & & \\ 
& C$_4$H 10$_{10}$--9$_{9}$ &  &  4.40$\pm$0.04 & 0.37$\pm$0.09 & 1.61 & 0.72 & 0.28 &  0.81$\pm$0.21 &  yes &   no \\
& C$_4$H 11$_{11}$--9$_{10}$ &  &  4.52$\pm$0.05 & 0.54$\pm$0.11 & 2.37 & 0.49 &  0.19 &1.20$\pm$ 0.25 &  no  &   no \\
& C$^{18}$O 2--1 &  3.21$\pm$0.08 &  4.54$\pm$0.01 & 0.91$\pm$0.03 & 5.05 & 0.23 &  0.09 &2.05$\pm$ 0.06 &  no  &   no \\
& C$^{17}$O 2--1 &  0.65$\pm$0.08&  4.33$\pm$0.02 & 0.41$\pm$0.04 & 1.72 & 0.68 & 0.26 &  0.88$\pm$0.11 &  yes &   no \\
& CH$_3$OH 2$_{0}$--1$_{0}$ A$^+$ &  0.53$\pm$0.04 &  4.42$\pm$0.02 & 0.66$\pm$0.06 & 3.08 & 0.38 &  0.15 &1.47$\pm$ 0.13 &  no  &   no \\
& CH$_3$OH 2$_{1}$--1$_{1}$ E &  0.44$\pm$0.04 &  4.44$\pm$0.04 & 0.91$\pm$0.14 & 5.06 & 0.23 &  0.09 &2.05$\pm$ 0.33 &  no  &   no \\
& N$_2$H$^+$ 1--0 &  0.65$\pm$0.04&  4.36$\pm$0.01 & 0.53$\pm$0.01 & 2.29 & 0.51 & 0.20 &  1.17$\pm$0.04 &  yes &   no \\
& C$^{34}$S 2--1 &  0.32$\pm$0.05 &  4.33$\pm$0.04 & 0.57$\pm$0.10 & 2.58 & 0.45 &  0.18 &1.29$\pm$ 0.22 &  no  &   no \\
& HC$_3$N 10--9 &  0.32$\pm$0.04&  4.33$\pm$0.03 & 0.39$\pm$0.01 & 1.69 & 0.69 & 0.27 &  0.86$\pm$0.04 &  yes &   no \\
& N$_2$H$^{+}$ 3--2 &  &  4.38$\pm$0.03 & 0.31$\pm$0.05 & 1.36 & 0.85 & 0.33 &  0.64$\pm$0.13 &  yes &   no \\
Cha1-C4  & & & & & & & & \\ 
& C$_4$H 10$_{10}$--9$_{9}$ &  &  5.06$\pm$0.02 & 0.33$\pm$0.04 & 1.27 & 0.52 & 0.23 &  0.72$\pm$0.09 &  yes &   no \\
& C$_4$H 11$_{11}$--9$_{10}$ &  &  5.10$\pm$0.01 & 0.24$\pm$0.03 & 1.06 & 0.62 & 0.28 &  0.51$\pm$0.08 &  yes &   no \\
& H$_2$CO 3$_{1,2}$--2$_{1,1}$ &  &  4.95$\pm$0.06 & 0.42$\pm$0.12 & 1.53 & 0.43 &  0.19 &0.90$\pm$ 0.28 &  no  &   no \\
& C$^{18}$O 2--1 &  2.68$\pm$0.07 &  5.06$\pm$0.01 & 0.70$\pm$0.02 & 2.89 & 0.23 &  0.10 &1.56$\pm$ 0.04 &  no  &   no \\
& C$^{17}$O 2--1 &  0.80$\pm$0.08 &  4.92$\pm$0.03 & 0.60$\pm$0.08 & 2.35 & 0.28 &  0.12 &1.34$\pm$ 0.18 &  no  &   no \\
& CH$_3$OH 2$_{0}$--1$_{0}$ A$^+$ &  0.54$\pm$0.04 &  5.00$\pm$0.01 & 0.36$\pm$0.02 & 1.34 & 0.49 &  0.22 &0.77$\pm$ 0.06 &  no  &   no \\
& CH$_3$OH 2$_{1}$--1$_{1}$ E &  0.35$\pm$0.05 &  5.01$\pm$0.01 & 0.37$\pm$0.03 & 1.38 & 0.48 &  0.21 &0.80$\pm$ 0.07 &  no  &   no \\
& N$_2$H$^+$ 1--0 &  0.34$\pm$0.03&  5.06$\pm$0.01 & 0.23$\pm$0.01 & 1.01 & 0.66 & 0.29 &  0.45$\pm$0.02 &  yes &   no \\
& C$^{34}$S 2--1 &  0.49$\pm$0.03 &  5.04$\pm$0.01 & 0.36$\pm$0.02 & 1.37 & 0.48 &  0.21 &0.79$\pm$ 0.05 &  no  &   no \\
& HC$_3$N 10--9 &  1.23$\pm$0.04&  5.11$\pm$0.01 & 0.26$\pm$0.01 & 1.10 & 0.60 & 0.27 &  0.55$\pm$0.04 &  yes &   no \\
& N$_2$H$^{+}$ 3--2 &  &  5.03$\pm$0.02 & 0.22$\pm$0.03 & 0.98 & 0.67 & 0.30 &  0.41$\pm$0.09 &  yes &   no \\
Cha1-C5  & & & & & & & & \\ 
& C$^{18}$O 2--1 &  1.58$\pm$0.07 &  4.33$\pm$0.01 & 0.45$\pm$0.02 & 1.36 & 0.30 &  0.23 &0.98$\pm$ 0.04 &  no  &   no \\
& C$^{17}$O 2--1 &  0.39$\pm$0.06 &  4.17$\pm$0.03 & 0.34$\pm$0.05 & 1.04 & 0.39 &  0.30 &0.71$\pm$ 0.12 &  no  &   no \\
& CH$_3$OH 2$_{1}$--1$_{1}$ E &  0.11$\pm$0.04&  4.18$\pm$0.01 & 0.16$\pm$0.04 & 0.73 & 0.56 & 0.43 &  0.23$\pm$0.16 &  yes &   no \\
Cha1-C6  & & & & & & & & \\ 
& C$^{18}$O 2--1 &  1.45$\pm$0.05 &  4.41$\pm$0.02 & 1.08$\pm$0.04 & 2.98 & 0.07 &  0.11 &2.44$\pm$ 0.10 &  no  &   no \\
Cha1-C7  & & & & & & & & \\ 
& C$^{18}$O 2--1 &  4.28$\pm$0.06 &  4.00$\pm$0.01 & 0.66$\pm$0.02 & 3.60 & 0.23 &  0.06 &1.47$\pm$ 0.04 &  no  &   no \\
& -- \tablefootmark{j}            &  -- &  4.76$\pm$0.01 & 0.51$\pm$0.02 & 2.60 & 0.32 &  0.09 &1.13$\pm$ 0.05 &  no  &   no \\
& C$^{17}$O 2--1 &  1.13$\pm$0.05 &  3.85$\pm$0.02 & 0.49$\pm$0.06 & 2.49 & 0.33 &  0.09 &1.08$\pm$ 0.14 &  no  &   no \\
& -- \tablefootmark{j} &  -- &  4.57$\pm$0.03 & 0.40$\pm$0.08 & 2.01 & 0.41 &  0.11 &0.87$\pm$ 0.19 &  no  &   no \\
& CH$_3$OH 2$_{0}$--1$_{0}$ A$^+$ &  0.12$\pm$0.02 &  4.65$\pm$0.04 & 0.36$\pm$0.08 & 1.83 & 0.45 &  0.12 &0.78$\pm$ 0.19 &  no  &   no \\

\hline
\end{tabular}
\label{table:physical_parameters_chaI_2}
\vspace*{-1ex}
\end{table*}

\begin{table*}
\hspace*{-40in}
 \addtocounter{table}{-1}
\caption{continued.} 
\vspace*{-1ex}
\hfill{}
\begin{tabular}{lllllllllll} 
\hline
\hline
Source  & Transition & $I_{\rm int}$\tablefootmark{a} & $V_{\rm LSR}$\tablefootmark{b} & $FWHM$\tablefootmark{c} &  $M_{\rm vir}$\tablefootmark{d} &   $\frac{M_{\rm tot}}{M_{\rm vir}}$\tablefootmark{e} &   $\frac{M_{\rm 50}}{M_{\rm vir}}$\tablefootmark{f} &  $\frac{\sigma_{\rm nth}}{\sigma_{\rm th,mean}}$\tablefootmark{g} & Gr.B.\tablefootmark{h} &  Vir.\tablefootmark{i} \\ 
        &            & \scriptsize{(K km s$^{-1}$)} & \scriptsize{(km s$^{-1}$)} &  \scriptsize{(km s$^{-1}$)} & \scriptsize{($M_{\rm \odot}$)}  & & & & & \\ \hline
Cha1-C8  & & & & & & & & \\ 
& C$^{18}$O 2--1 &  2.66$\pm$0.08 &  4.51$\pm$0.01 & 0.91$\pm$0.03 & 1.70 & 0.09 &  0.13 &2.04$\pm$ 0.06 &  no  &   no \\
& CH$_3$OH 2$_{0}$--1$_{0}$ A$^+$ &  0.27$\pm$0.04 &  4.40$\pm$0.04 & 0.55$\pm$0.10 & 0.83 & 0.18 &  0.27 &1.23$\pm$ 0.24 &  no  &   no \\
& CH$_3$OH 2$_{1}$--1$_{1}$ E &  0.20$\pm$0.03 &  4.37$\pm$0.02 & 0.41$\pm$0.06 & 0.59 & 0.25 &  0.38 &0.89$\pm$ 0.14 &  no  &   no \\
& N$_2$H$^+$ 1--0 &  1.57$\pm$0.07 &  4.46$\pm$0.02 & 0.58$\pm$0.04 & 0.87 & 0.17 &  0.26 &1.28$\pm$ 0.09 &  no  &   no \\
& HC$_3$N 10--9 &  0.16$\pm$0.03 &  4.53$\pm$0.05 & 0.61$\pm$0.12 & 0.94 & 0.15 &  0.24 &1.36$\pm$ 0.28 &  no  &   no \\
& N$_2$H$^{+}$ 3--2 &  &  4.65$\pm$0.02 & 0.44$\pm$0.09 & 0.63 & 0.23 &  0.35 &0.96$\pm$ 0.22 &  no  &   no \\
Cha1-C9  & & & & & & & & \\ 
& H$_2$CO 3$_{1,2}$--2$_{1,1}$ &  &  4.86$\pm$0.05 & 0.40$\pm$0.11 & 1.93 & 0.35 &  0.10 &0.87$\pm$ 0.26 &  no  &   no \\
& C$^{18}$O 2--1 &  1.78$\pm$0.08 &  4.96$\pm$0.01 & 0.49$\pm$0.01 & 2.35 & 0.29 &  0.08 &1.07$\pm$ 0.03 &  no  &   no \\
& C$^{17}$O 2--1 &  0.53$\pm$0.06 &  4.88$\pm$0.01 & 0.31$\pm$0.04 & 1.53 & 0.44 &  0.13 &0.64$\pm$ 0.09 &  no  &   no \\
& CH$_3$OH 2$_{0}$--1$_{0}$ A$^+$ &  0.93$\pm$0.04 &  4.86$\pm$0.01 & 0.41$\pm$0.01 & 1.97 & 0.34 &  0.10 &0.89$\pm$ 0.03 &  no  &   no \\
& N$_2$H$^+$ 1--0 &  0.50$\pm$0.07 &  4.86$\pm$0.02 & 0.32$\pm$0.05 & 1.59 & 0.42 &  0.12 &0.68$\pm$ 0.11 &  no  &   no \\
& C$^{34}$S 2--1 &  0.10$\pm$0.04 &  4.86$\pm$0.02 & 0.35$\pm$0.04 & 1.73 & 0.39 &  0.12 &0.76$\pm$ 0.11 &  no  &   no \\
Cha1-C10  & & & & & & & & \\ 
& C$^{18}$O 2--1 &  2.29$\pm$0.06 &  4.60$\pm$0.01 & 0.57$\pm$0.01 & 1.89 & 0.16 &  0.10 &1.26$\pm$ 0.03 &  no  &   no \\
& C$^{17}$O 2--1 &  0.53$\pm$0.07 &  4.44$\pm$0.02 & 0.39$\pm$0.06 & 1.24 & 0.25 &  0.15 &0.83$\pm$ 0.13 &  no  &   no \\
& CH$_3$OH 2$_{0}$--1$_{0}$ A$^+$ &  0.70$\pm$0.04 &  4.44$\pm$0.01 & 0.36$\pm$0.02 & 1.18 & 0.26 &  0.16 &0.78$\pm$ 0.04 &  no  &   no \\
& CH$_3$OH 2$_{1}$--1$_{1}$ E &  0.57$\pm$0.04 &  4.46$\pm$0.01 & 0.35$\pm$0.02 & 1.14 & 0.27 &  0.17 &0.74$\pm$ 0.04 &  no  &   no \\
& N$_2$H$^+$ 1--0 &  1.30$\pm$0.07 &  4.44$\pm$0.01 & 0.39$\pm$0.04 & 1.23 & 0.25 &  0.15 &0.83$\pm$ 0.09 &  no  &   no \\
& C$^{34}$S 2--1 &  0.34$\pm$0.03 &  4.48$\pm$0.04 & 0.77$\pm$0.09 & 2.94 & 0.10 &  0.06 &1.74$\pm$ 0.21 &  no  &   no \\
& HC$_3$N 10--9 &  0.11$\pm$0.04 &  4.55$\pm$0.04 & 0.28$\pm$0.04 & 1.00 & 0.31 &  0.19 &0.61$\pm$ 0.10 &  no  &   no \\
Cha1-C11  & & & & & & & & \\ 
& C$_4$H 10$_{10}$--9$_{9}$ &  &  5.05$\pm$0.02 & 0.22$\pm$0.05 & 1.52 & 0.50 &  0.11 &0.45$\pm$ 0.12 &  no  &   no \\
& C$_4$H 11$_{11}$--9$_{10}$ &  &  5.09$\pm$0.01 & 0.16$\pm$0.03 & 1.38 & 0.55 & 0.12 &  0.30$\pm$0.09 &  yes &   no \\
& H$_2$CO 3$_{1,2}$--2$_{1,1}$ &  &  4.96$\pm$0.04 & 0.47$\pm$0.09 & 2.65 & 0.29 &  0.06 &1.04$\pm$ 0.21 &  no  &   no \\
& C$^{18}$O 2--1 &  1.61$\pm$0.07 &  5.15$\pm$0.01 & 0.38$\pm$0.01 & 2.12 & 0.36 &  0.08 &0.82$\pm$ 0.03 &  no  &   no \\
& C$^{17}$O 2--1 &  0.35$\pm$0.06 &  4.96$\pm$0.02 & 0.27$\pm$0.03 & 1.65 & 0.46 &  0.10 &0.55$\pm$ 0.07 &  no  &   no \\
& CH$_3$OH 2$_{0}$--1$_{0}$ A$^+$ &  0.49$\pm$0.04 &  5.08$\pm$0.01 & 0.36$\pm$0.03 & 2.04 & 0.37 &  0.08 &0.78$\pm$ 0.06 &  no  &   no \\
& CH$_3$OH 2$_{1}$--1$_{1}$ E &  0.33$\pm$0.07 &  5.06$\pm$0.01 & 0.28$\pm$0.02 & 1.70 & 0.45 &  0.10 &0.58$\pm$ 0.06 &  no  &   no \\
& N$_2$H$^+$ 1--0 &  1.21$\pm$0.07 &  5.06$\pm$0.01 & 0.29$\pm$0.02 & 1.74 & 0.44 &  0.09 &0.61$\pm$ 0.05 &  no  &   no \\
& C$^{34}$S 2--1 &  0.29$\pm$0.04 &  5.05$\pm$0.01 & 0.36$\pm$0.03 & 2.07 & 0.37 &  0.08 &0.79$\pm$ 0.08 &  no  &   no \\
& HC$_3$N 10--9 &  0.58$\pm$0.04 &  5.10$\pm$0.01 & 0.32$\pm$0.01 & 1.86 & 0.41 &  0.09 &0.68$\pm$ 0.04 &  no  &   no \\
Cha1-C12  & & & & & & & & \\ 
& C$^{18}$O 2--1 &  3.66$\pm$0.07 &  4.98$\pm$0.01 & 0.71$\pm$0.01 & 2.26 & 0.11 &  0.06 &1.59$\pm$ 0.03 &  no  &   no \\
& C$^{17}$O 2--1 &  1.20$\pm$0.09 &  4.83$\pm$0.02 & 0.49$\pm$0.06 & 1.37 & 0.18 &  0.10 &1.07$\pm$ 0.13 &  no  &   no \\
& CH$_3$OH 2$_{0}$--1$_{0}$ A$^+$ &  0.31$\pm$0.04 &  4.84$\pm$0.04 & 0.55$\pm$0.10 & 1.60 & 0.15 &  0.09 &1.22$\pm$ 0.22 &  no  &   no \\
& CH$_3$OH 2$_{1}$--1$_{1}$ E &  0.36$\pm$0.04 &  4.90$\pm$0.03 & 0.64$\pm$0.08 & 1.97 & 0.12 &  0.07 &1.44$\pm$ 0.19 &  no  &   no \\
Cha1-C13  & & & & & & & & \\ 
& C$^{18}$O 2--1 &  1.46$\pm$0.07 &  4.85$\pm$0.01 & 0.48$\pm$0.03 & 2.38 & 0.23 &  0.07 &1.06$\pm$ 0.06 &  no  &   no \\
& CH$_3$OH 2$_{0}$--1$_{0}$ A$^+$ &  0.20$\pm$0.04 &  4.73$\pm$0.01 & 0.26$\pm$0.03 & 1.41 & 0.38 &  0.11 &0.51$\pm$ 0.06 &  no  &   no \\
Cha1-C14  & & & & & & & & \\ 
& C$^{18}$O 2--1 &  2.17$\pm$0.08 &  4.56$\pm$0.02 & 0.87$\pm$0.04 & 2.93 & 0.07 &  0.06 &1.95$\pm$ 0.10 &  no  &   no \\
& CH$_3$OH 2$_{0}$--1$_{0}$ A$^+$ &  0.44$\pm$0.04 &  4.40$\pm$0.03 & 0.57$\pm$0.12 & 1.57 & 0.14 &  0.11 &1.26$\pm$ 0.29 &  no  &   no \\
& CH$_3$OH 2$_{1}$--1$_{1}$ E &  0.29$\pm$0.04 &  4.31$\pm$0.04 & 0.62$\pm$0.08 & 1.77 & 0.12 &  0.10 &1.38$\pm$ 0.19 &  no  &   no \\
& N$_2$H$^+$ 1--0 &  1.11$\pm$0.08 &  4.37$\pm$0.03 & 0.52$\pm$0.05 & 1.42 & 0.15 &  0.12 &1.15$\pm$ 0.11 &  no  &   no \\
Cha1-C15  & & & & & & & & \\ 
& C$^{18}$O 2--1 &  3.83$\pm$0.08 &  4.25$\pm$0.01 & 0.57$\pm$0.01 & 3.07 & 0.17 &  0.04 &1.27$\pm$ 0.02 &  no  &   no \\
& C$^{17}$O 2--1 &  1.19$\pm$0.10 &  4.12$\pm$0.02 & 0.59$\pm$0.05 & 3.22 & 0.16 &  0.04 &1.32$\pm$ 0.11 &  no  &   no \\
Cha1-C16  & & & & & & & & \\ 
& C$^{18}$O 2--1 &  2.38$\pm$0.09 &  4.74$\pm$0.01 & 0.87$\pm$0.04 & 3.38 & 0.06 &  0.03 &1.97$\pm$ 0.08 &  no  &   no \\
& C$^{34}$S 2--1 &  0.09$\pm$0.04 &  4.63$\pm$0.04 & 0.38$\pm$0.11 & 1.18 & 0.18 &  0.09 &0.83$\pm$ 0.26 &  no  &   no \\
Cha1-C17  & & & & & & & & \\ 
& C$^{18}$O 2--1 &  2.00$\pm$0.07 &  5.29$\pm$0.01 & 0.51$\pm$0.01 & 1.93 & 0.15 &  0.06 &1.12$\pm$ 0.04 &  no  &   no \\
& C$^{17}$O 2--1 &  0.58$\pm$0.07 &  5.11$\pm$0.02 & 0.41$\pm$0.06 & 1.52 & 0.19 &  0.07 &0.88$\pm$ 0.13 &  no  &   no \\
& CH$_3$OH 2$_{0}$--1$_{0}$ A$^+$ &  0.46$\pm$0.04 &  5.18$\pm$0.01 & 0.44$\pm$0.03 & 1.66 & 0.18 &  0.07 &0.96$\pm$ 0.07 &  no  &   no \\
& CH$_3$OH 2$_{1}$--1$_{1}$ E &  0.38$\pm$0.07 &  5.19$\pm$0.02 & 0.42$\pm$0.05 & 1.59 & 0.18 &  0.07 &0.92$\pm$ 0.11 &  no  &   no \\
& C$^{34}$S 2--1 &  0.41$\pm$0.04 &  5.32$\pm$0.01 & 0.40$\pm$0.04 & 1.52 & 0.19 &  0.07 &0.87$\pm$ 0.09 &  no  &   no \\
& HC$_3$N 10--9 &  0.10$\pm$0.04 &  5.45$\pm$0.02 & 0.18$\pm$0.07 & 0.97 & 0.30 &  0.12 &0.36$\pm$ 0.20 &  no  &   no \\
\hline
\end{tabular}
\label{table:physical_parameters_chaI_2}
\vspace*{-1ex}
\end{table*}

\begin{table*}
\hspace*{-40in}
 \addtocounter{table}{-1}
\caption{continued.} 
\vspace*{-1ex}
\hfill{}
\begin{tabular}{llllllllllll} 
\hline
\hline
Source  & Transition & $I_{\rm int}$\tablefootmark{a} & $V_{\rm LSR}$\tablefootmark{b} & $FWHM$\tablefootmark{c} &  $M_{\rm vir}$\tablefootmark{d} &   $\frac{M_{\rm tot}}{M_{\rm vir}}$\tablefootmark{e} &   $\frac{M_{\rm 50}}{M_{\rm vir}}$\tablefootmark{f} &  $\frac{\sigma_{\rm nth}}{\sigma_{\rm th,mean}}$\tablefootmark{g} & Gr.B.\tablefootmark{h} &  Vir.\tablefootmark{i} \\ 
        &            & \scriptsize{(K km s$^{-1}$)} & \scriptsize{(km s$^{-1}$)} &  \scriptsize{(km s$^{-1}$)} & \scriptsize{($M_{\rm \odot}$)}  & & & & & \\ \hline
Cha1-C18  & & & & & & & & \\ 
& C$^{18}$O 2--1 &  2.03$\pm$0.08 &  5.03$\pm$0.01 & 0.47$\pm$0.02 & 2.37 & 0.21 &  0.05 &1.02$\pm$ 0.04 &  no  &   no \\
& C$^{17}$O 2--1 &  0.60$\pm$0.08 &  4.89$\pm$0.02 & 0.30$\pm$0.04 & 1.59 & 0.31 &  0.07 &0.61$\pm$ 0.10 &  no  &   no \\
& CH$_3$OH 2$_{0}$--1$_{0}$ A$^+$ &  0.29$\pm$0.04 &  4.91$\pm$0.02 & 0.36$\pm$0.04 & 1.87 & 0.26 &  0.06 &0.78$\pm$ 0.10 &  no  &   no \\
& C$^{34}$S 2--1 &  0.15$\pm$0.04 &  4.95$\pm$0.04 & 0.39$\pm$0.15 & 2.03 & 0.24 &  0.06 &0.86$\pm$ 0.35 &  no  &   no \\
& HC$_3$N 10--9 &  0.09$\pm$0.03 &  5.05$\pm$0.02 & 0.18$\pm$0.02 & 1.31 & 0.37 &  0.09 &0.36$\pm$ 0.06 &  no  &   no \\
Cha1-C19  & & & & & & & & \\ 
& C$_4$H 10$_{10}$--9$_{9}$ &  &  4.93$\pm$0.01 & 0.25$\pm$0.03 & 1.08 & 0.26 &  0.09 &0.53$\pm$ 0.07 &  no  &   no \\
& C$_4$H 11$_{11}$--9$_{10}$ &  &  4.96$\pm$0.01 & 0.29$\pm$0.03 & 1.18 & 0.24 &  0.08 &0.62$\pm$ 0.07 &  no  &   no \\
& H$_2$CO 3$_{1,2}$--2$_{1,1}$ &  &  4.82$\pm$0.07 & 0.50$\pm$0.12 & 1.87 & 0.15 &  0.05 &1.10$\pm$ 0.28 &  no  &   no \\
& C$^{18}$O 2--1 &  1.99$\pm$0.08 &  4.95$\pm$0.01 & 0.44$\pm$0.02 & 1.63 & 0.17 &  0.06 &0.96$\pm$ 0.04 &  no  &   no \\
& C$^{17}$O 2--1 &  0.54$\pm$0.05 &  4.77$\pm$0.02 & 0.40$\pm$0.05 & 1.47 & 0.19 &  0.07 &0.86$\pm$ 0.11 &  no  &   no \\
& CH$_3$OH 2$_{0}$--1$_{0}$ A$^+$ &  0.47$\pm$0.04 &  4.86$\pm$0.01 & 0.34$\pm$0.03 & 1.30 & 0.22 &  0.08 &0.73$\pm$ 0.06 &  no  &   no \\
& CH$_3$OH 2$_{1}$--1$_{1}$ E &  0.33$\pm$0.04 &  4.88$\pm$0.02 & 0.34$\pm$0.04 & 1.27 & 0.22 &  0.08 &0.71$\pm$ 0.10 &  no  &   no \\
& N$_2$H$^+$ 1--0 &  1.40$\pm$0.10 &  4.91$\pm$0.01 & 0.31$\pm$0.02 & 1.19 & 0.23 &  0.08 &0.64$\pm$ 0.05 &  no  &   no \\
& C$^{34}$S 2--1 &  0.33$\pm$0.04 &  4.88$\pm$0.01 & 0.32$\pm$0.03 & 1.24 & 0.23 &  0.08 &0.68$\pm$ 0.06 &  no  &   no \\
& HC$_3$N 10--9 &  1.39$\pm$0.04 &  4.98$\pm$0.01 & 0.21$\pm$0.01 & 1.01 & 0.28 &  0.10 &0.44$\pm$ 0.03 &  no  &   no \\
Cha1-C20  & & & & & & & & \\ 
& C$^{18}$O 2--1 &  1.18$\pm$0.08 &  4.35$\pm$0.01 & 0.52$\pm$0.03 & 2.03 & 0.13 &  0.05 &1.16$\pm$ 0.06 &  no  &   no \\
& CH$_3$OH 2$_{0}$--1$_{0}$ A$^+$ &  0.24$\pm$0.05 &  4.29$\pm$0.02 & 0.34$\pm$0.05 & 1.31 & 0.20 &  0.07 &0.72$\pm$ 0.12 &  no  &   no \\
& CH$_3$OH 2$_{1}$--1$_{1}$ E &  0.14$\pm$0.04 &  4.36$\pm$0.04 & 0.51$\pm$0.07 & 1.96 & 0.14 &  0.05 &1.12$\pm$ 0.18 &  no  &   no \\
Cha1-C21  & & & & & & & & \\ 
& H$_2$CO 3$_{1,2}$--2$_{1,1}$ &  &  4.90$\pm$0.05 & 0.45$\pm$0.13 & 0.86 & 0.12 &  0.13 &0.98$\pm$ 0.32 &  no  &   no \\
& C$^{18}$O 2--1 &  1.96$\pm$0.08 &  5.03$\pm$0.01 & 0.62$\pm$0.03 & 1.28 & 0.08 &  0.09 &1.38$\pm$ 0.06 &  no  &   no \\
& C$^{17}$O 2--1 &  0.66$\pm$0.05 &  4.84$\pm$0.04 & 0.52$\pm$0.09 & 1.01 & 0.10 &  0.11 &1.14$\pm$ 0.20 &  no  &   no \\
& CH$_3$OH 2$_{0}$--1$_{0}$ A$^+$ &  0.28$\pm$0.05 &  5.01$\pm$0.02 & 0.33$\pm$0.04 & 0.65 & 0.16 &  0.17 &0.69$\pm$ 0.09 &  no  &   no \\
& CH$_3$OH 2$_{1}$--1$_{1}$ E &  0.32$\pm$0.05 &  4.96$\pm$0.02 & 0.39$\pm$0.05 & 0.75 & 0.14 &  0.15 &0.84$\pm$ 0.11 &  no  &   no \\
& N$_2$H$^+$ 1--0 &  1.43$\pm$0.08 &  5.05$\pm$0.01 & 0.25$\pm$0.01 & 0.54 & 0.19 &  0.21 &0.48$\pm$ 0.04 &  no  &   no \\
& C$^{34}$S 2--1 &  0.16$\pm$0.04 &  5.07$\pm$0.02 & 0.37$\pm$0.06 & 0.73 & 0.14 &  0.16 &0.81$\pm$ 0.13 &  no  &   no \\
& HC$_3$N 10--9 &  0.43$\pm$0.04 &  5.08$\pm$0.01 & 0.18$\pm$0.01 & 0.50 & 0.21 &  0.23 &0.36$\pm$ 0.00 &  no  &   no \\
Cha1-C22  & & & & & & & & \\ 
& C$^{18}$O 2--1 &  0.98$\pm$0.09 &  4.93$\pm$0.01 & 0.35$\pm$0.02 & 2.00 & 0.27 &  0.06 &0.73$\pm$ 0.05 &  no  &   no \\
& CH$_3$OH 2$_{0}$--1$_{0}$ A$^+$ &  0.21$\pm$0.04 &  4.84$\pm$0.01 & 0.33$\pm$0.04 & 1.92 & 0.28 &  0.06 &0.69$\pm$ 0.09 &  no  &   no \\
& CH$_3$OH 2$_{1}$--1$_{1}$ E &  0.10$\pm$0.04 &  4.81$\pm$0.01 & 0.19$\pm$0.03 & 1.45 & 0.37 &  0.08 &0.34$\pm$ 0.07 &  no  &   no \\
Cha1-C23  & & & & & & & & \\ 
& C$^{18}$O 2--1 &  2.14$\pm$0.08 &  4.97$\pm$0.01 & 0.78$\pm$0.03 & 2.33 & 0.06 &  0.04 &1.74$\pm$ 0.07 &  no  &   no \\
& CH$_3$OH 2$_{0}$--1$_{0}$ A$^+$ &  0.25$\pm$0.04 &  4.80$\pm$0.05 & 0.64$\pm$0.09 & 1.74 & 0.08 &  0.05 &1.43$\pm$ 0.20 &  no  &   no \\
& CH$_3$OH 2$_{1}$--1$_{1}$ E &  0.24$\pm$0.05 &  4.86$\pm$0.04 & 0.57$\pm$0.11 & 1.49 & 0.09 &  0.06 &1.26$\pm$ 0.26 &  no  &   no \\
& C$^{34}$S 2--1 &  0.19$\pm$0.04 &  5.12$\pm$0.02 & 0.20$\pm$0.06 & 0.67 & 0.21 &  0.13 &0.41$\pm$ 0.15 &  no  &   no \\
Cha1-C24  & & & & & & & & \\ 
& C$^{18}$O 2--1 &  1.44$\pm$0.07 &  4.53$\pm$0.01 & 0.45$\pm$0.03 & 1.26 & 0.13 &  0.06 &0.98$\pm$ 0.06 &  no  &   no \\
& CH$_3$OH 2$_{0}$--1$_{0}$ A$^+$ &  0.18$\pm$0.04 &  4.39$\pm$0.03 & 0.30$\pm$0.06 & 0.90 & 0.18 &  0.09 &0.63$\pm$ 0.16 &  no  &   no \\
Cha1-C25  & & & & & & & & \\ 
& C$^{18}$O 2--1 &  0.99$\pm$0.07 &  4.34$\pm$0.02 & 0.56$\pm$0.06 & 1.72 & 0.11 &  0.07 &1.25$\pm$ 0.14 &  no  &   no \\
& CH$_3$OH 2$_{1}$--1$_{1}$ E &  0.08$\pm$0.04 &  4.50$\pm$0.03 & 0.29$\pm$0.07 & 0.91 & 0.20 &  0.14 &0.59$\pm$ 0.18 &  no  &   no \\
Cha1-C26  & & & & & & & & \\ 
& C$_4$H 11$_{11}$--9$_{10}$ &  &  5.25$\pm$0.03 & 0.26$\pm$0.07 & 1.48 & 0.30 &  0.07 &0.54$\pm$ 0.18 &  no  &   no \\
& C$^{18}$O 2--1 &  1.60$\pm$0.10 &  5.22$\pm$0.01 & 0.39$\pm$0.01 & 1.94 & 0.23 &  0.05 &0.83$\pm$ 0.04 &  no  &   no \\
& C$^{17}$O 2--1 &  0.84$\pm$0.10 &  5.06$\pm$0.02 & 0.33$\pm$0.05 & 1.70 & 0.26 &  0.06 &0.69$\pm$ 0.12 &  no  &   no \\
& CH$_3$OH 2$_{0}$--1$_{0}$ A$^+$ &  0.24$\pm$0.05 &  5.14$\pm$0.02 & 0.36$\pm$0.06 & 1.82 & 0.24 &  0.06 &0.76$\pm$ 0.14 &  no  &   no \\
& CH$_3$OH 2$_{1}$--1$_{1}$ E &  0.21$\pm$0.06 &  5.11$\pm$0.02 & 0.34$\pm$0.04 & 1.74 & 0.25 &  0.06 &0.72$\pm$ 0.11 &  no  &   no \\
& C$^{34}$S 2--1 &  0.23$\pm$0.05 &  5.11$\pm$0.03 & 0.36$\pm$0.06 & 1.85 & 0.24 &  0.06 &0.78$\pm$ 0.14 &  no  &   no \\
& HC$_3$N 10--9 &  0.18$\pm$0.04 &  5.30$\pm$0.02 & 0.21$\pm$0.02 & 1.35 & 0.32 &  0.07 &0.42$\pm$ 0.06 &  no  &   no \\
Cha1-C27  & & & & & & & & \\ 
& C$^{18}$O 2--1 &  3.05$\pm$0.08 &  4.40$\pm$0.01 & 0.80$\pm$0.02 & 3.13 & 0.07 &  0.04 &1.80$\pm$ 0.05 &  no  &   no \\
& C$^{17}$O 2--1 &  0.76$\pm$0.06 &  4.24$\pm$0.04 & 0.67$\pm$0.11 & 2.40 & 0.09 &  0.06 &1.50$\pm$ 0.26 &  no  &   no \\
\hline
\end{tabular}
\label{table:physical_parameters_chaI_2}
\vspace*{-1ex}
\end{table*}

\begin{table*}
\hspace*{-40in}
 \addtocounter{table}{-1}
\caption{continued.} 
\vspace*{-1ex}
\hfill{}
\begin{tabular}{llllllllllll} 
\hline
\hline
Source  & Transition & $I_{\rm int}$\tablefootmark{a} & $V_{\rm LSR}$\tablefootmark{b} & $FWHM$\tablefootmark{c} &  $M_{\rm vir}$\tablefootmark{d} &   $\frac{M_{\rm tot}}{M_{\rm vir}}$\tablefootmark{e} &   $\frac{M_{\rm 50}}{M_{\rm vir}}$\tablefootmark{f} &  $\frac{\sigma_{\rm nth}}{\sigma_{\rm th,mean}}$\tablefootmark{g} & Gr.B.\tablefootmark{h} &  Vir.\tablefootmark{i} \\ 
        &            & \scriptsize{(K km s$^{-1}$)} & \scriptsize{(km s$^{-1}$)} &  \scriptsize{(km s$^{-1}$)} & \scriptsize{($M_{\rm \odot}$)}  & & & & & \\ \hline
Cha1-C28  & & & & & & & & \\ 
& C$^{18}$O 2--1 &  1.34$\pm$0.09 &  4.76$\pm$0.03 & 0.89$\pm$0.07 & 3.25 & 0.05 &  0.03 &2.00$\pm$ 0.17 &  no  &   no \\
Cha1-C29  & & & & & & & & \\ 
& H$_2$CO 3$_{1,2}$--2$_{1,1}$ &  &  4.43$\pm$0.03 & 0.50$\pm$0.07 & 1.43 & 0.12 &  0.09 &1.10$\pm$ 0.16 &  no  &   no \\
& C$^{18}$O 2--1 &  2.88$\pm$0.12 &  4.68$\pm$0.02 & 0.88$\pm$0.04 & 3.14 & 0.05 &  0.04 &1.97$\pm$ 0.09 &  no  &   no \\
& C$^{17}$O 2--1 &  0.56$\pm$0.06 &  4.45$\pm$0.04 & 0.69$\pm$0.10 & 2.16 & 0.08 &  0.06 &1.54$\pm$ 0.24 &  no  &   no \\
& CH$_3$OH 2$_{0}$--1$_{0}$ A$^+$ &  0.79$\pm$0.05 &  4.52$\pm$0.01 & 0.60$\pm$0.04 & 1.79 & 0.10 &  0.07 &1.33$\pm$ 0.09 &  no  &   no \\
& CH$_3$OH 2$_{1}$--1$_{1}$ E &  0.50$\pm$0.05 &  4.54$\pm$0.02 & 0.57$\pm$0.04 & 1.69 & 0.10 &  0.08 &1.28$\pm$ 0.10 &  no  &   no \\
& N$_2$H$^+$ 1--0 &  1.08$\pm$0.09 &  4.42$\pm$0.01 & 0.57$\pm$0.05 & 1.67 & 0.10 &  0.08 &1.26$\pm$ 0.11 &  no  &   no \\
& C$^{34}$S 2--1 &  0.49$\pm$0.04 &  4.52$\pm$0.03 & 0.76$\pm$0.09 & 2.54 & 0.07 &  0.05 &1.72$\pm$ 0.21 &  no  &   no \\
& HC$_3$N 10--9 &  0.30$\pm$0.04 &  4.46$\pm$0.04 & 0.68$\pm$0.13 & 2.13 & 0.08 &  0.06 &1.52$\pm$ 0.30 &  no  &   no \\
Cha1-C30  & & & & & & & & \\ 
& C$^{18}$O 2--1 &  0.74$\pm$0.13 &  4.78$\pm$0.01 & 0.21$\pm$0.07 & 1.41 & 0.32 &  0.07 &0.38$\pm$ 0.21 &  no  &   no \\
& CH$_3$OH 2$_{0}$--1$_{0}$ A$^+$ &  0.23$\pm$0.05 &  4.63$\pm$0.01 & 0.20$\pm$0.03 & 1.39 & 0.33 &  0.07 &0.35$\pm$ 0.07 &  no  &   no \\
& CH$_3$OH 2$_{1}$--1$_{1}$ E &  0.14$\pm$0.05 &  4.67$\pm$0.01 & 0.22$\pm$0.03 & 1.46 & 0.31 &  0.07 &0.43$\pm$ 0.07 &  no  &   no \\
& N$_2$H$^+$ 1--0 &  0.91$\pm$0.10 &  4.64$\pm$0.01 & 0.23$\pm$0.02 & 1.48 & 0.31 &  0.07 &0.45$\pm$ 0.05 &  no  &   no \\
Cha1-C31  & & & & & & & & \\ 
& C$^{18}$O 2--1 &  0.69$\pm$0.10 &  4.75$\pm$0.01 & 0.28$\pm$0.03 & 1.22 & 0.23 &  0.08 &0.57$\pm$ 0.07 &  no  &   no \\
& CH$_3$OH 2$_{0}$--1$_{0}$ A$^+$ &  0.16$\pm$0.04 &  4.65$\pm$0.01 & 0.20$\pm$0.03 & 1.04 & 0.27 &  0.10 &0.36$\pm$ 0.07 &  no  &   no \\
& CH$_3$OH 2$_{1}$--1$_{1}$ E &  0.04$\pm$0.04 &  4.62$\pm$0.01 & 0.16$\pm$0.05 & 0.97 & 0.29 &  0.10 &0.24$\pm$ 0.16 &  no  &   no \\
& N$_2$H$^+$ 1--0 &  1.08$\pm$0.09 &  4.64$\pm$0.01 & 0.19$\pm$0.02 & 1.02 & 0.27 &  0.10 &0.33$\pm$ 0.05 &  no  &   no \\
Cha1-C32  & & & & & & & & \\ 
& C$^{18}$O 2--1 &  1.71$\pm$0.12 &  4.56$\pm$0.03 & 0.69$\pm$0.08 & 3.79 & 0.10 &  0.03 &1.54$\pm$ 0.19 &  no  &   no \\
& CH$_3$OH 2$_{1}$--1$_{1}$ E &  0.12$\pm$0.04 &  4.37$\pm$0.04 & 0.34$\pm$0.08 & 1.71 & 0.23 &  0.06 &0.72$\pm$ 0.20 &  no  &   no \\
Cha1-C33  & & & & & & & & \\ 
& C$_4$H 10$_{10}$--9$_{9}$ &  &  5.30$\pm$0.02 & 0.30$\pm$0.05 & 0.99 & 0.18 &  0.10 &0.64$\pm$ 0.11 &  no  &   no \\
& C$_4$H 11$_{11}$--9$_{10}$ &  &  5.38$\pm$0.02 & 0.26$\pm$0.04 & 0.93 & 0.20 &  0.11 &0.56$\pm$ 0.10 &  no  &   no \\
& C$^{18}$O 2--1 &  1.15$\pm$0.13 &  5.32$\pm$0.02 & 0.45$\pm$0.04 & 1.40 & 0.13 &  0.07 &0.99$\pm$ 0.10 &  no  &   no \\
& CH$_3$OH 2$_{0}$--1$_{0}$ A$^+$ &  0.34$\pm$0.04 &  5.22$\pm$0.02 & 0.36$\pm$0.04 & 1.12 & 0.16 &  0.09 &0.77$\pm$ 0.10 &  no  &   no \\
& CH$_3$OH 2$_{1}$--1$_{1}$ E &  0.15$\pm$0.06 &  5.19$\pm$0.03 & 0.41$\pm$0.07 & 1.26 & 0.15 &  0.08 &0.88$\pm$ 0.16 &  no  &   no \\
& N$_2$H$^+$ 1--0 &  1.89$\pm$0.08 &  5.22$\pm$0.01 & 0.31$\pm$0.02 & 1.00 & 0.18 &  0.10 &0.65$\pm$ 0.04 &  no  &   no \\
& C$^{34}$S 2--1 &  0.31$\pm$0.05 &  5.22$\pm$0.01 & 0.30$\pm$0.03 & 1.01 & 0.18 &  0.10 &0.65$\pm$ 0.08 &  no  &   no \\
& HC$_3$N 10--9 &  0.99$\pm$0.04 &  5.32$\pm$0.01 & 0.29$\pm$0.03 & 0.98 & 0.19 &  0.10 &0.62$\pm$ 0.06 &  no  &   no \\
Cha1-C34  & & & & & & & & \\ 
& C$_4$H 10$_{10}$--9$_{9}$ &  &  5.25$\pm$0.02 & 0.27$\pm$0.05 & 1.68 & 0.30 &  0.06 &0.57$\pm$ 0.12 &  no  &   no \\
& C$_4$H 11$_{11}$--9$_{10}$ &  &  5.33$\pm$0.02 & 0.27$\pm$0.04 & 1.69 & 0.30 &  0.06 &0.57$\pm$ 0.10 &  no  &   no \\
& C$^{18}$O 2--1 &  1.29$\pm$0.10 &  5.29$\pm$0.01 & 0.33$\pm$0.03 & 1.86 & 0.27 &  0.06 &0.68$\pm$ 0.06 &  no  &   no \\
& C$^{17}$O 2--1 &  0.31$\pm$0.07 &  5.09$\pm$0.03 & 0.35$\pm$0.07 & 1.99 & 0.25 &  0.05 &0.75$\pm$ 0.18 &  no  &   no \\
& CH$_3$OH 2$_{0}$--1$_{0}$ A$^+$ &  0.34$\pm$0.05 &  5.18$\pm$0.01 & 0.36$\pm$0.03 & 2.01 & 0.25 &  0.05 &0.76$\pm$ 0.07 &  no  &   no \\
& CH$_3$OH 2$_{1}$--1$_{1}$ E &  0.31$\pm$0.07 &  5.19$\pm$0.02 & 0.35$\pm$0.04 & 1.97 & 0.26 &  0.05 &0.74$\pm$ 0.11 &  no  &   no \\
& N$_2$H$^+$ 1--0 &  0.70$\pm$0.08 &  5.18$\pm$0.01 & 0.30$\pm$0.03 & 1.77 & 0.28 &  0.06 &0.63$\pm$ 0.07 &  no  &   no \\
& C$^{34}$S 2--1 &  0.28$\pm$0.04 &  5.20$\pm$0.01 & 0.29$\pm$0.03 & 1.77 & 0.28 &  0.06 &0.63$\pm$ 0.08 &  no  &   no \\
& HC$_3$N 10--9 &  0.38$\pm$0.05 &  5.28$\pm$0.01 & 0.23$\pm$0.03 & 1.54 & 0.33 &  0.07 &0.46$\pm$ 0.08 &  no  &   no \\
Cha1-C35  & & & & & & & & \\ 
& C$_4$H 10$_{10}$--9$_{9}$ &  &  4.21$\pm$0.04 & 0.49$\pm$0.08 & 1.12 & 0.11 &  0.07 &1.09$\pm$ 0.18 &  no  &   no \\
& C$^{18}$O 2--1 &  1.94$\pm$0.13 &  4.31$\pm$0.01 & 0.54$\pm$0.04 & 1.24 & 0.10 &  0.06 &1.19$\pm$ 0.09 &  no  &   no \\
& C$^{17}$O 2--1 &  0.39$\pm$0.06 &  4.13$\pm$0.03 & 0.48$\pm$0.07 & 1.09 & 0.11 &  0.07 &1.06$\pm$ 0.17 &  no  &   no \\
& CH$_3$OH 2$_{0}$--1$_{0}$ A$^+$ &  0.29$\pm$0.04 &  4.01$\pm$0.03 & 0.35$\pm$0.10 & 0.81 & 0.15 &  0.10 &0.76$\pm$ 0.25 &  no  &   no \\
& N$_2$H$^+$ 1--0 &  0.59$\pm$0.08 &  4.13$\pm$0.02 & 0.34$\pm$0.03 & 0.78 & 0.16 &  0.10 &0.72$\pm$ 0.08 &  no  &   no \\
& C$^{34}$S 2--1 &  0.37$\pm$0.04 &  4.22$\pm$0.02 & 0.47$\pm$0.05 & 1.07 & 0.12 &  0.07 &1.04$\pm$ 0.12 &  no  &   no \\
& HC$_3$N 10--9 &  0.65$\pm$0.04 &  4.23$\pm$0.01 & 0.33$\pm$0.03 & 0.77 & 0.16 &  0.10 &0.72$\pm$ 0.07 &  no  &   no \\
Cha1-C36  & & & & & & & & \\ 
& C$^{18}$O 2--1 &  1.35$\pm$0.14 &  4.29$\pm$0.01 & 0.35$\pm$0.03 & 0.95 & 0.15 &  0.10 &0.76$\pm$ 0.07 &  no  &   no \\
& C$^{17}$O 2--1 &  0.28$\pm$0.06 &  4.12$\pm$0.03 & 0.37$\pm$0.08 & 0.98 & 0.15 &  0.10 &0.78$\pm$ 0.19 &  no  &   no \\
& CH$_3$OH 2$_{0}$--1$_{0}$ A$^+$ &  0.30$\pm$0.05 &  4.18$\pm$0.01 & 0.32$\pm$0.03 & 0.87 & 0.17 &  0.11 &0.66$\pm$ 0.09 &  no  &   no \\
& CH$_3$OH 2$_{1}$--1$_{1}$ E &  0.20$\pm$0.05 &  4.14$\pm$0.03 & 0.36$\pm$0.07 & 0.96 & 0.15 &  0.10 &0.76$\pm$ 0.16 &  no  &   no \\
Cha1-C37  & & & & & & & & \\ 
& C$^{18}$O 2--1 &  1.52$\pm$0.10 &  3.88$\pm$0.02 & 0.53$\pm$0.04 & 1.67 & 0.12 &  0.05 &1.16$\pm$ 0.10 &  no  &   no \\
\hline
\end{tabular}
\label{table:physical_parameters_chaI_2}
\vspace*{-1ex}
\end{table*}

\begin{table*}
\hspace*{-40in}
 \addtocounter{table}{-1}
\caption{continued.} 
\vspace*{-1ex}
\hfill{}
\begin{tabular}{llllllllllll} 
\hline
\hline
Source  & Transition & $I_{\rm int}$\tablefootmark{a} & $V_{\rm LSR}$\tablefootmark{b} & $FWHM$\tablefootmark{c} &  $M_{\rm vir}$\tablefootmark{d} &   $\frac{M_{\rm tot}}{M_{\rm vir}}$\tablefootmark{e} &   $\frac{M_{\rm 50}}{M_{\rm vir}}$\tablefootmark{f} &  $\frac{\sigma_{\rm nth}}{\sigma_{\rm th,mean}}$\tablefootmark{g} & Gr.B.\tablefootmark{h} &  Vir.\tablefootmark{i} \\ 
        &            & \scriptsize{(K km s$^{-1}$)} & \scriptsize{(km s$^{-1}$)} &  \scriptsize{(km s$^{-1}$)} & \scriptsize{($M_{\rm \odot}$)}  & & & & & \\ \hline
Cha1-C38  & & & & & & & & \\ 
& C$_4$H 11$_{11}$--9$_{10}$ &  &  5.32$\pm$0.03 & 0.25$\pm$0.05 & 1.26 & 0.25 &  0.08 &0.51$\pm$ 0.12 &  no  &   no \\
& C$^{18}$O 2--1 &  1.51$\pm$0.14 &  5.31$\pm$0.01 & 0.36$\pm$0.03 & 1.60 & 0.19 &  0.06 &0.78$\pm$ 0.07 &  no  &   no \\
& C$^{17}$O 2--1 &  0.48$\pm$0.09 &  5.15$\pm$0.02 & 0.27$\pm$0.03 & 1.30 & 0.24 &  0.07 &0.55$\pm$ 0.08 &  no  &   no \\
& CH$_3$OH 2$_{0}$--1$_{0}$ A$^+$ &  0.31$\pm$0.05 &  5.19$\pm$0.01 & 0.27$\pm$0.03 & 1.30 & 0.24 &  0.07 &0.55$\pm$ 0.07 &  no  &   no \\
& C$^{34}$S 2--1 &  0.20$\pm$0.04 &  5.21$\pm$0.03 & 0.37$\pm$0.07 & 1.67 & 0.19 &  0.06 &0.82$\pm$ 0.17 &  no  &   no \\
& HC$_3$N 10--9 &  0.16$\pm$0.04 &  5.32$\pm$0.02 & 0.18$\pm$0.08 & 1.13 & 0.28 &  0.09 &0.36$\pm$ 0.22 &  no  &   no \\
Cha1-C39  & & & & & & & & \\ 
& C$^{18}$O 2--1 &  2.75$\pm$0.13 &  4.52$\pm$0.01 & 0.49$\pm$0.02 & 1.86 & 0.13 &  0.05 &1.09$\pm$ 0.05 &  no  &   no \\
& C$^{17}$O 2--1 &  0.68$\pm$0.06 &  4.31$\pm$0.02 & 0.56$\pm$0.06 & 2.16 & 0.11 &  0.04 &1.24$\pm$ 0.15 &  no  &   no \\
Cha1-C40  & & & & & & & & \\ 
& C$^{18}$O 2--1 &  0.99$\pm$0.07 &  4.75$\pm$0.01 & 0.42$\pm$0.03 & 1.54 & 0.15 &  0.06 &0.90$\pm$ 0.07 &  no  &   no \\
& CH$_3$OH 2$_{0}$--1$_{0}$ A$^+$ &  0.38$\pm$0.04 &  4.62$\pm$0.01 & 0.25$\pm$0.02 & 1.05 & 0.21 &  0.09 &0.49$\pm$ 0.06 &  no  &   no \\
& CH$_3$OH 2$_{1}$--1$_{1}$ E &  0.29$\pm$0.04 &  4.63$\pm$0.01 & 0.22$\pm$0.02 & 0.99 & 0.23 &  0.10 &0.42$\pm$ 0.06 &  no  &   no \\
& N$_2$H$^+$ 1--0 &  0.67$\pm$0.08 &  4.61$\pm$0.01 & 0.19$\pm$0.02 & 0.93 & 0.24 &  0.10 &0.32$\pm$ 0.05 &  no  &   no \\
& C$^{34}$S 2--1 &  0.07$\pm$0.03 &  4.63$\pm$0.05 & 0.41$\pm$0.14 & 1.52 & 0.15 &  0.06 &0.89$\pm$ 0.33 &  no  &   no \\
& HC$_3$N 10--9 &  0.06$\pm$0.03 &  4.68$\pm$0.01 & 0.18$\pm$0.01 & 0.96 & 0.23 &  0.10 &0.36$\pm$ 0.02 &  no  &   no \\
Cha1-C41  & & & & & & & & \\ 
& C$^{18}$O 2--1 &  1.83$\pm$0.16 &  5.23$\pm$0.01 & 0.42$\pm$0.03 & 1.04 & 0.12 &  0.08 &0.90$\pm$ 0.07 &  no  &   no \\
& C$^{17}$O 2--1 &  0.65$\pm$0.06 &  5.11$\pm$0.01 & 0.31$\pm$0.04 & 0.81 & 0.16 &  0.10 &0.65$\pm$ 0.10 &  no  &   no \\
& CH$_3$OH 2$_{0}$--1$_{0}$ A$^+$ &  0.31$\pm$0.04 &  5.14$\pm$0.01 & 0.28$\pm$0.03 & 0.77 & 0.17 &  0.11 &0.58$\pm$ 0.08 &  no  &   no \\
& CH$_3$OH 2$_{1}$--1$_{1}$ E &  0.31$\pm$0.06 &  5.17$\pm$0.01 & 0.26$\pm$0.03 & 0.73 & 0.17 &  0.11 &0.53$\pm$ 0.07 &  no  &   no \\
& C$^{34}$S 2--1 &  0.27$\pm$0.04 &  5.13$\pm$0.02 & 0.37$\pm$0.05 & 0.94 & 0.14 &  0.09 &0.80$\pm$ 0.11 &  no  &   no \\
Cha1-C42  & & & & & & & & \\ 
& C$^{18}$O 2--1 &  1.22$\pm$0.07 &  4.52$\pm$0.02 & 0.85$\pm$0.06 & 2.11 & 0.05 &  0.03 &1.90$\pm$ 0.13 &  no  &   no \\
& CH$_3$OH 2$_{0}$--1$_{0}$ A$^+$ &  0.15$\pm$0.04 &  4.42$\pm$0.05 & 0.53$\pm$0.09 & 1.08 & 0.09 &  0.06 &1.17$\pm$ 0.22 &  no  &   no \\
& N$_2$H$^+$ 1--0 &  0.51$\pm$0.08 &  4.17$\pm$0.03 & 0.41$\pm$0.07 & 0.82 & 0.12 &  0.08 &0.89$\pm$ 0.17 &  no  &   no \\
Cha1-C43  & & & & & & & & \\ 
& C$^{18}$O 2--1 &  1.31$\pm$0.12 &  4.70$\pm$0.01 & 0.37$\pm$0.02 & 1.70 & 0.19 &  0.05 &0.78$\pm$ 0.05 &  no  &   no \\
& C$^{17}$O 2--1 &  0.49$\pm$0.06 &  4.47$\pm$0.01 & 0.27$\pm$0.04 & 1.37 & 0.23 &  0.06 &0.55$\pm$ 0.10 &  no  &   no \\
& CH$_3$OH 2$_{0}$--1$_{0}$ A$^+$ &  0.27$\pm$0.04 &  4.51$\pm$0.02 & 0.33$\pm$0.04 & 1.57 & 0.20 &  0.05 &0.70$\pm$ 0.09 &  no  &   no \\
& CH$_3$OH 2$_{1}$--1$_{1}$ E &  0.13$\pm$0.04 &  4.53$\pm$0.02 & 0.20$\pm$0.04 & 1.20 & 0.26 &  0.07 &0.37$\pm$ 0.11 &  no  &   no \\
Cha1-C44  & & & & & & & & \\ 
& C$^{18}$O 2--1 &  1.09$\pm$0.09 &  4.59$\pm$0.02 & 0.68$\pm$0.04 & 2.28 & 0.07 &  0.03 &1.51$\pm$ 0.10 &  no  &   no \\
Cha1-C45  & & & & & & & & \\ 
& C$^{18}$O 2--1 &  0.82$\pm$0.08 &  4.34$\pm$0.01 & 0.36$\pm$0.03 & 0.89 & 0.14 &  0.08 &0.78$\pm$ 0.06 &  no  &   no \\
& CH$_3$OH 2$_{0}$--1$_{0}$ A$^+$ &  0.22$\pm$0.04 &  4.22$\pm$0.03 & 0.42$\pm$0.07 & 1.02 & 0.12 &  0.07 &0.92$\pm$ 0.16 &  no  &   no \\
Cha1-C46  & & & & & & & & \\ 
& C$^{18}$O 2--1 &  1.19$\pm$0.09 &  4.48$\pm$0.02 & 0.52$\pm$0.04 & 0.84 & 0.07 &  0.06 &1.15$\pm$ 0.10 &  no  &   no \\
Cha1-C47  & & & & & & & & \\ 
& C$^{18}$O 2--1 &  1.26$\pm$0.10 &  4.61$\pm$0.02 & 0.62$\pm$0.05 & 2.57 & 0.09 &  0.04 &1.37$\pm$ 0.12 &  no  &   no \\
Cha1-C48  & & & & & & & & \\ 
& C$^{18}$O 2--1 &  1.50$\pm$0.09 &  5.07$\pm$0.02 & 0.85$\pm$0.05 & 2.11 & 0.04 &  0.03 &1.92$\pm$ 0.11 &  no  &   no \\
Cha1-C49  & & & & & & & & \\ 
& C$^{18}$O 2--1 &  1.70$\pm$0.12 &  5.04$\pm$0.01 & 0.45$\pm$0.04 & 0.83 & 0.10 &  0.09 &0.98$\pm$ 0.09 &  no  &   no \\
& C$^{17}$O 2--1 &  0.47$\pm$0.05 &  4.83$\pm$0.01 & 0.29$\pm$0.04 & 0.57 & 0.14 &  0.12 &0.59$\pm$ 0.10 &  no  &   no \\
& CH$_3$OH 2$_{0}$--1$_{0}$ A$^+$ &  0.13$\pm$0.05 &  4.80$\pm$0.03 & 0.21$\pm$0.07 & 0.49 & 0.17 &  0.15 &0.40$\pm$ 0.19 &  no  &   no \\
Cha1-C50  & & & & & & & & \\ 
& C$^{18}$O 2--1 &  1.00$\pm$0.06 &  4.46$\pm$0.01 & 0.31$\pm$0.04 & 1.35 & 0.18 &  0.06 &0.65$\pm$ 0.09 &  no  &   no \\
& -- \tablefootmark{j}    &  -- &  5.18$\pm$0.02 & 0.65$\pm$0.05 & 2.92 & 0.08 &  0.03 &1.44$\pm$ 0.11 &  no  &   no \\
& CH$_3$OH 2$_{0}$--1$_{0}$ A$^+$ &  0.59$\pm$0.03 &  5.11$\pm$0.03 & 0.66$\pm$0.07 & 2.98 & 0.08 &  0.03 &1.46$\pm$ 0.17 &  no  &   no \\
& CH$_3$OH 2$_{1}$--1$_{1}$ E &  0.43$\pm$0.05 &  5.17$\pm$0.03 & 0.48$\pm$0.06 & 2.00 & 0.12 &  0.04 &1.05$\pm$ 0.14 &  no  &   no \\
Cha1-C51  & & & & & & & & \\ 
& C$^{18}$O 2--1 &  3.05$\pm$0.12 &  4.86$\pm$0.03 & 1.15$\pm$0.06 & 3.56 & 0.02 &  0.02 &2.60$\pm$ 0.13 &  no  &   no \\
& CH$_3$OH 2$_{0}$--1$_{0}$ A$^+$ &  0.59$\pm$0.05 &  4.83$\pm$0.06 & 1.23$\pm$0.11 & 4.01 & 0.02 &  0.02 &2.78$\pm$ 0.26 &  no  &   no \\
\hline
\end{tabular}
\hfill{}
\label{table:physical_parameters_chaI_2}
\vspace*{-1ex}
\end{table*}

\begin{table*}
\hspace*{-40in}
 \addtocounter{table}{-1}
\caption{continued.} 
\vspace*{-1ex}
\hfill{}
\begin{tabular}{llllllllllll} 
\hline
\hline
Source  & Transition & $I_{\rm int}$\tablefootmark{a} & $V_{\rm LSR}$\tablefootmark{b} & $FWHM$\tablefootmark{c} &  $M_{\rm vir}$\tablefootmark{d} &   $\frac{M_{\rm tot}}{M_{\rm vir}}$\tablefootmark{e} &   $\frac{M_{\rm 50}}{M_{\rm vir}}$\tablefootmark{f} &  $\frac{\sigma_{\rm nth}}{\sigma_{\rm th,mean}}$\tablefootmark{g} & Gr.B.\tablefootmark{h} &  Vir.\tablefootmark{i} \\ 
        &            & \scriptsize{(K km s$^{-1}$)} & \scriptsize{(km s$^{-1}$)} &  \scriptsize{(km s$^{-1}$)} & \scriptsize{($M_{\rm \odot}$)}  & & & & & \\ \hline
Cha1-C52  & & & & & & & & \\ 
& C$^{18}$O 2--1 &  1.67$\pm$0.09 &  4.44$\pm$0.02 & 0.59$\pm$0.04 & 1.39 & 0.07 &  0.05 &1.31$\pm$ 0.10 &  no  &   no \\
& -- \tablefootmark{j}         &  -- &  5.48$\pm$0.02 & 0.50$\pm$0.04 & 1.14 & 0.09 &  0.06 &1.11$\pm$ 0.09 &  no  &   no \\
& CH$_3$OH 2$_{0}$--1$_{0}$ A$^+$ &  0.33$\pm$0.05 &  4.52$\pm$0.03 & 0.47$\pm$0.06 & 1.05 & 0.10 &  0.06 &1.03$\pm$ 0.15 &  no  &   no \\
& CH$_3$OH 2$_{1}$--1$_{1}$ E &  0.26$\pm$0.04 &  4.49$\pm$0.03 & 0.44$\pm$0.07 & 0.98 & 0.10 &  0.07 &0.96$\pm$ 0.16 &  no  &   no \\
& C$^{34}$S 2--1 &  0.15$\pm$0.02 &  4.29$\pm$0.03 & 0.33$\pm$0.07 & 0.77 & 0.13 &  0.08 &0.71$\pm$ 0.16 &  no  &   no \\
& -- \tablefootmark{j}           &  -- &  5.33$\pm$0.06 & 0.47$\pm$0.10 & 1.07 & 0.09 &  0.06 &1.05$\pm$ 0.24 &  no  &   no \\
Cha1-C53  & & & & & & & & \\ 
& C$^{18}$O 2--1 &  2.01$\pm$0.12 &  4.54$\pm$0.01 & 0.47$\pm$0.03 & 1.49 & 0.10 &  0.06 &1.03$\pm$ 0.06 &  no  &   no \\
& C$^{17}$O 2--1 &  0.48$\pm$0.06 &  4.40$\pm$0.03 & 0.45$\pm$0.05 & 1.42 & 0.11 &  0.06 &0.98$\pm$ 0.12 &  no  &   no \\
Cha1-C54  & & & & & & & & \\ 
& C$^{18}$O 2--1 &  0.71$\pm$0.06 &  4.51$\pm$0.03 & 0.51$\pm$0.09 & 1.02 & 0.07 &  0.06 &1.13$\pm$ 0.21 &  no  &   no \\
&  -- \tablefootmark{j} &  -- &  5.33$\pm$0.01 & 0.49$\pm$0.03 & 0.97 & 0.08 &  0.06 &1.08$\pm$ 0.06 &  no  &   no \\
& C$^{17}$O 2--1 &  0.44$\pm$0.04 &  5.13$\pm$0.03 & 0.50$\pm$0.09 & 0.99 & 0.08 &  0.06 &1.11$\pm$ 0.21 &  no  &   no \\
& CH$_3$OH 2$_{0}$--1$_{0}$ A$^+$ &  0.27$\pm$0.05 &  5.10$\pm$0.05 & 0.55$\pm$0.12 & 1.10 & 0.07 &  0.06 &1.21$\pm$ 0.28 &  no  &   no \\
& C$^{34}$S 2--1 &  0.18$\pm$0.03 &  5.22$\pm$0.06 & 0.36$\pm$0.14 & 0.72 & 0.11 &  0.08 &0.79$\pm$ 0.33 &  no  &   no \\
Cha1-C55  & & & & & & & & \\ 
& C$^{18}$O 2--1 &  2.21$\pm$0.09 &  4.67$\pm$0.02 & 0.99$\pm$0.04 & 2.76 & 0.03 &  0.03 &2.23$\pm$ 0.10 &  no  &   no \\
Cha1-C56  & & & & & & & & \\ 
& C$^{18}$O 2--1 &  0.97$\pm$0.10 &  4.58$\pm$0.01 & 0.32$\pm$0.03 & 0.62 & 0.12 &  0.09 &0.67$\pm$ 0.08 &  no  &   no \\
& CH$_3$OH 2$_{0}$--1$_{0}$ A$^+$ &  0.15$\pm$0.06 &  4.49$\pm$0.02 & 0.29$\pm$0.06 & 0.58 & 0.13 &  0.10 &0.59$\pm$ 0.14 &  no  &   no \\
& CH$_3$OH 2$_{1}$--1$_{1}$ E &  0.18$\pm$0.06 &  4.44$\pm$0.03 & 0.23$\pm$0.06 & 0.51 & 0.14 &  0.11 &0.44$\pm$ 0.16 &  no  &   no \\
Cha1-C57  & & & & & & & & \\ 
& C$^{18}$O 2--1 &  1.10$\pm$0.12 &  5.08$\pm$0.01 & 0.30$\pm$0.02 & 0.98 & 0.15 &  0.12 &0.62$\pm$ 0.05 &  no  &   no \\
& C$^{17}$O 2--1 &  0.38$\pm$0.06 &  4.90$\pm$0.02 & 0.27$\pm$0.00 & 0.92 & 0.16 &  0.13 &0.55$\pm$ 0.01 &  no  &   no \\
Cha1-C58  & & & & & & & & \\ 
& C$^{18}$O 2--1 &  1.62$\pm$0.14 &  5.31$\pm$0.01 & 0.36$\pm$0.02 & 0.91 & 0.12 &  0.10 &0.77$\pm$ 0.06 &  no  &   no \\
& C$^{17}$O 2--1 &  0.36$\pm$0.07 &  5.14$\pm$0.02 & 0.31$\pm$0.07 & 0.81 & 0.14 &  0.11 &0.65$\pm$ 0.16 &  no  &   no \\
Cha1-C59  & & & & & & & & \\ 
& C$^{18}$O 2--1 &  2.12$\pm$0.14 &  5.24$\pm$0.02 & 0.63$\pm$0.04 & 1.18 & 0.05 &  0.04 &1.41$\pm$ 0.09 &  no  &   no \\
& C$^{17}$O 2--1 &  0.61$\pm$0.06 &  5.02$\pm$0.03 & 0.69$\pm$0.07 & 1.32 & 0.05 &  0.04 &1.53$\pm$ 0.16 &  no  &   no \\
Cha1-C60  & & & & & & & & \\ 
& C$^{18}$O 2--1 &  2.00$\pm$0.12 &  4.47$\pm$0.01 & 0.48$\pm$0.03 & 1.79 & 0.10 &  0.04 &1.07$\pm$ 0.06 &  no  &   no \\
& C$^{17}$O 2--1 &  0.46$\pm$0.07 &  4.27$\pm$0.02 & 0.37$\pm$0.05 & 1.37 & 0.14 &  0.05 &0.80$\pm$ 0.13 &  no  &   no \\
& CH$_3$OH 2$_{0}$--1$_{0}$ A$^+$ &  0.03$\pm$0.05 &  4.42$\pm$0.03 & 0.26$\pm$0.09 & 1.07 & 0.17 &  0.07 &0.53$\pm$ 0.22 &  no  &   no \\
\hline
\end{tabular}
\hfill{}
\label{table:physical_parameters_chaI_2}
\vspace*{-1ex}
\tablefoot{
\tablefoottext{a}{Integrated intensities used to calculate the column densities in Sect.~\ref{sec:observed_abundances}. The integrated intensities are computed in $T_{\rm mb}$ scale. }
\tablefoottext{b}{Systemic velocities derived from Gaussian or hyperfine-structure fits to the spectra.}
\tablefoottext{c}{Linewidths derived from Gaussian or hyperfine-structure fits to the spectra.}
\tablefoottext{d}{Virial mass. }
\tablefoottext{e}{Ratio of the total mass to the virial mass. }
\tablefoottext{f}{Ratio of the mass within a 50$^{\arcsec}$ aperture to the virial mass.}
\tablefoottext{g}{Ratio of non-thermal to mean thermal velocity dispersions computed for a kinetic temperature of 10 K.}
\tablefoottext{h}{Gravitational boundedness of the core as $M_{\rm tot}/M_{\rm vir} > 0.5$. }
\tablefoottext{i}{Core is virialized if $M_{\rm tot}/M_{\rm vir} \ge 1$.}
\tablefoottext{j}{The dashed lines indicate the kinematic results of the same transition as the previous one listed, but at a different velocity. The integrated intensities are only listed for the velocity components used in the analysis (i.e., the strongest emission peak).}
}
\end{table*}

\begin{table*}
\caption{Line parameters and physical parameters of cores in Cha III.} 
\vspace*{-1ex}
\hfill{}
\begin{tabular}{@{\extracolsep{-3pt}}lllllllllll} 
\hline
\hline
Source  & Transition & $I_{\rm int}$\tablefootmark{a} & $V_{\rm LSR}$\tablefootmark{b} & $FWHM$\tablefootmark{c} &  $M_{\rm vir}$\tablefootmark{d} &   $\frac{M_{\rm tot}}{M_{\rm vir}}$\tablefootmark{e} &   $\frac{M_{\rm 50}}{M_{\rm vir}}$\tablefootmark{f} &  $\frac{\sigma_{\rm nth}}{\sigma_{\rm th,mean}}$\tablefootmark{g} & Gr.B.\tablefootmark{h} &  Vir.\tablefootmark{i} \\ 
        &            & \scriptsize{(K km s$^{-1}$)} & \scriptsize{(km s$^{-1}$)} &  \scriptsize{(km s$^{-1}$)} & \scriptsize{($M_{\rm \odot}$)}  & & & & & \\ \hline
Cha3-C1  & & & & & & & & \\ 
& C$^{18}$O 2--1 &  1.89$\pm$0.07 &  1.23$\pm$0.01 & 0.49$\pm$0.02 & 3.10 & 0.50 &  0.09 &1.07$\pm$ 0.04 &  no  &   no \\
& C$^{17}$O 2--1 &  0.49$\pm$0.04&  1.21$\pm$0.01 & 0.36$\pm$0.03 & 2.32 & 0.67 & 0.13 &  0.77$\pm$0.07 &  yes &   no \\
& CH$_3$OH 2$_{0}$--1$_{0}$ A$^+$ &  0.42$\pm$0.04&  1.26$\pm$0.01 & 0.45$\pm$0.03 & 2.86 & 0.54 & 0.10 &  0.99$\pm$0.08 &  yes &   no \\
& CH$_3$OH 2$_{1}$--1$_{1}$ E &  0.31$\pm$0.03&  1.29$\pm$0.02 & 0.46$\pm$0.04 & 2.90 & 0.54 & 0.10 &  1.00$\pm$0.10 &  yes &   no \\
& H$^{13}$CO$^+$ 1--0 &   &   1.24$\pm$0.03 & 0.57$\pm$0.07 & 3.75 & 0.41 & 0.08 &  1.26$\pm$0.16 &  no  &   no \\
& N$_2$H$^+$ 1--0 &  0.61$\pm$0.09&  1.24$\pm$0.03 & 0.34$\pm$0.07 & 2.21 & 0.70 & 0.13 &  0.72$\pm$0.17 &  yes &   no \\
& HC$_3$N 10--9 &  0.17$\pm$0.03&  1.46$\pm$0.04 & 0.28$\pm$0.04 & 1.99 & 0.78 & 0.15 &  0.61$\pm$0.10 &  yes &   no \\
Cha3-C2  & & & & & & & & \\ 
& C$^{18}$O 2--1 &  1.61$\pm$0.07 &  1.24$\pm$0.01 & 0.57$\pm$0.03 & 2.84 & 0.20 &  0.08 &1.28$\pm$ 0.06 &  no  &   no \\
& CH$_3$OH 2$_{0}$--1$_{0}$ A$^+$ &  0.18$\pm$0.04 &  1.19$\pm$0.05 & 0.45$\pm$0.11 & 2.13 & 0.27 &  0.11 &0.99$\pm$ 0.25 &  no  &   no \\
& H$^{13}$CO$^+$ 1--0 &   &   1.30$\pm$0.05 & 0.52$\pm$0.10 & 2.50 & 0.23 & 0.09 &  1.15$\pm$0.22 &  no  &   no \\
Cha3-C3  & & & & & & & & \\ 
& C$^{18}$O 2--1 &  1.04$\pm$0.07 &  1.45$\pm$0.01 & 0.46$\pm$0.03 & 1.20 & 0.17 &  0.09 &1.00$\pm$ 0.08 &  no  &   no \\
& CH$_3$OH 2$_{0}$--1$_{0}$ A$^+$ &  0.13$\pm$0.05 &  1.45$\pm$0.06 & 0.44$\pm$0.09 & 1.15 & 0.17 &  0.10 &0.96$\pm$ 0.22 &  no  &   no \\
Cha3-C4  & & & & & & & & \\ 
& C$^{18}$O 2--1 &  1.15$\pm$0.08 &  1.42$\pm$0.02 & 0.60$\pm$0.05 & 2.84 & 0.15 &  0.05 &1.34$\pm$ 0.11 &  no  &   no \\
& C$^{17}$O 2--1 &  0.22$\pm$0.03 &  1.40$\pm$0.04 & 0.49$\pm$0.08 & 2.18 & 0.20 &  0.07 &1.07$\pm$ 0.19 &  no  &   no \\
Cha3-C5  & & & & & & & & \\ 
& C$^{18}$O 2--1 &  1.17$\pm$0.07 &  1.69$\pm$0.01 & 0.47$\pm$0.03 & 1.96 & 0.19 &  0.07 &1.02$\pm$ 0.07 &  no  &   no \\
& CH$_3$OH 2$_{0}$--1$_{0}$ A$^+$ &  0.19$\pm$0.04 &  1.69$\pm$0.03 & 0.40$\pm$0.09 & 1.70 & 0.22 &  0.08 &0.87$\pm$ 0.21 &  no  &   no \\
& CH$_3$OH 2$_{1}$--1$_{1}$ E &  0.22$\pm$0.04 &  1.73$\pm$0.06 & 0.58$\pm$0.13 & 2.58 & 0.14 &  0.05 &1.30$\pm$ 0.31 &  no  &   no \\
Cha3-C6  & & & & & & & & \\ 
& C$^{18}$O 2--1 &  1.06$\pm$0.06 &  1.49$\pm$0.01 & 0.44$\pm$0.03 & 1.66 & 0.18 &  0.08 &0.96$\pm$ 0.07 &  no  &   no \\
Cha3-C7  & & & & & & & & \\ 
& C$^{18}$O 2--1 &  0.67$\pm$0.06 &  1.62$\pm$0.01 & 0.39$\pm$0.04 & 1.26 & 0.18 &  0.08 &0.84$\pm$ 0.09 &  no  &   no \\
& CH$_3$OH 2$_{0}$--1$_{0}$ A$^+$ &  0.09$\pm$0.04 &  1.64$\pm$0.05 & 0.36$\pm$0.10 & 1.18 & 0.20 &  0.09 &0.77$\pm$ 0.23 &  no  &   no \\
Cha3-C8  & & & & & & & & \\ 
& C$^{18}$O 2--1 &  0.87$\pm$0.06 &  1.23$\pm$0.01 & 0.34$\pm$0.02 & 1.51 & 0.25 &  0.07 &0.72$\pm$ 0.05 &  no  &   no \\
& CH$_3$OH 2$_{0}$--1$_{0}$ A$^+$ &  0.39$\pm$0.04 &  1.19$\pm$0.01 & 0.29$\pm$0.02 & 1.36 & 0.28 &  0.08 &0.61$\pm$ 0.05 &  no  &   no \\
& CH$_3$OH 2$_{1}$--1$_{1}$ E &  0.26$\pm$0.04 &  1.17$\pm$0.01 & 0.27$\pm$0.03 & 1.30 & 0.29 &  0.09 &0.55$\pm$ 0.07 &  no  &   no \\
Cha3-C9  & & & & & & & & \\ 
& C$^{18}$O 2--1 &  0.90$\pm$0.06 &  1.37$\pm$0.01 & 0.25$\pm$0.01 & 1.13 & 0.28 &  0.10 &0.50$\pm$ 0.04 &  no  &   no \\
& CH$_3$OH 2$_{0}$--1$_{0}$ A$^+$ &  0.26$\pm$0.03 &  1.36$\pm$0.02 & 0.28$\pm$0.05 & 1.19 & 0.27 &  0.09 &0.57$\pm$ 0.12 &  no  &   no \\
& CH$_3$OH 2$_{1}$--1$_{1}$ E &  0.21$\pm$0.03 &  1.38$\pm$0.02 & 0.35$\pm$0.05 & 1.40 & 0.23 &  0.08 &0.74$\pm$ 0.13 &  no  &   no \\
Cha3-C10  & & & & & & & & \\ 
& C$^{18}$O 2--1 &  1.06$\pm$0.05 &  1.51$\pm$0.01 & 0.47$\pm$0.03 & 1.14 & 0.14 &  0.07 &1.03$\pm$ 0.06 &  no  &   no \\
Cha3-C11  & & & & & & & & \\ 
& C$^{18}$O 2--1 &  -- &  1.16$\pm$0.06 & 0.63$\pm$0.12 & 2.15 & 0.09 &  0.05 &1.41$\pm$ 0.28 &  no  &   no \\
& --      &  0.79$\pm$0.07 &  2.05$\pm$0.01 & 0.21$\pm$0.01 & 0.83 & 0.24 &  0.12 &0.40$\pm$ 0.04 &  no  &   no \\

Cha3-C12  & & & & & & & & \\ 
& C$^{18}$O 2--1 &  1.01$\pm$0.06 &  1.32$\pm$0.01 & 0.37$\pm$0.02 & 1.56 & 0.20 &  0.06 &0.79$\pm$ 0.05 &  no  &   no \\
& CH$_3$OH 2$_{0}$--1$_{0}$ A$^+$ &  0.25$\pm$0.04 &  1.42$\pm$0.05 & 0.62$\pm$0.15 & 2.82 & 0.11 &  0.04 &1.39$\pm$ 0.35 &  no  &   no \\
& CH$_3$OH 2$_{1}$--1$_{1}$ E &  0.13$\pm$0.04 &  1.31$\pm$0.05 & 0.43$\pm$0.11 & 1.80 & 0.17 &  0.06 &0.94$\pm$ 0.26 &  no  &   no \\
Cha3-C13  & & & & & & & & \\ 
& C$^{18}$O 2--1 &  0.88$\pm$0.05 &  1.55$\pm$0.01 & 0.31$\pm$0.02 & 1.10 & 0.20 &  0.09 &0.65$\pm$ 0.05 &  no  &   no \\
& CH$_3$OH 2$_{0}$--1$_{0}$ A$^+$ &  0.26$\pm$0.03 &  1.57$\pm$0.03 & 0.46$\pm$0.08 & 1.57 & 0.14 &  0.06 &1.01$\pm$ 0.19 &  no  &   no \\
& CH$_3$OH 2$_{1}$--1$_{1}$ E &  0.16$\pm$0.03 &  1.50$\pm$0.02 & 0.22$\pm$0.06 & 0.91 & 0.25 &  0.11 &0.41$\pm$ 0.17 &  no  &   no \\
& C$^{34}$S 2--1 &  0.09$\pm$0.03 &  1.57$\pm$0.03 & 0.24$\pm$0.05 & 0.98 & 0.23 &  0.10 &0.51$\pm$ 0.12 &  no  &   no \\
& HC$_3$N 10--9 &  0.08$\pm$0.03 &  1.55$\pm$0.03 & 0.18$\pm$0.34 & 0.88 & 0.25 &  0.11 &0.36$\pm$ 0.90 &  no  &   no \\
Cha3-C14  & & & & & & & & \\ 
& C$^{18}$O 2--1 &  1.07$\pm$0.06 &  1.75$\pm$0.01 & 0.43$\pm$0.02 & 1.11 & 0.15 &  0.07 &0.93$\pm$ 0.06 &  no  &   no \\
& CH$_3$OH 2$_{0}$--1$_{0}$ A$^+$ &  0.26$\pm$0.05 &  1.81$\pm$0.03 & 0.43$\pm$0.07 & 1.12 & 0.15 &  0.07 &0.93$\pm$ 0.17 &  no  &   no \\
\hline
\end{tabular}
\hfill{}
\label{table:physical_parameters_chaIII_2}
\vspace*{-1ex}
\end{table*}

\begin{table*}
\hspace*{-40in}
 \addtocounter{table}{-1}
\caption{continued.} 
\vspace*{-1ex}
\hfill{}
\begin{tabular}{lllllllllll} 
\hline
\hline
Source  & Transition & $I_{\rm int}$\tablefootmark{a} & $V_{\rm LSR}$\tablefootmark{b} & $FWHM$\tablefootmark{c} &  $M_{\rm vir}$\tablefootmark{d} &   $\frac{M_{\rm tot}}{M_{\rm vir}}$\tablefootmark{e} &   $\frac{M_{\rm 50}}{M_{\rm vir}}$\tablefootmark{f} &  $\frac{\sigma_{\rm nth}}{\sigma_{\rm th,mean}}$\tablefootmark{g} & Gr.B.\tablefootmark{h} &  Vir.\tablefootmark{i} \\ 
        &            & \scriptsize{(K km s$^{-1}$)} & \scriptsize{(km s$^{-1}$)} &  \scriptsize{(km s$^{-1}$)} & \scriptsize{($M_{\rm \odot}$)}  & & & & & \\ \hline
Cha3-C15  & & & & & & & & \\ 
& C$^{18}$O 2--1 &  0.89$\pm$0.06 &  1.42$\pm$0.01 & 0.46$\pm$0.03 & 1.59 & 0.14 &  0.06 &1.01$\pm$ 0.07 &  no  &   no \\
& CH$_3$OH 2$_{0}$--1$_{0}$ A$^+$ &  0.25$\pm$0.04 &  1.47$\pm$0.02 & 0.28$\pm$0.05 & 1.06 & 0.20 &  0.08 &0.58$\pm$ 0.11 &  no  &   no \\
& CH$_3$OH 2$_{1}$--1$_{1}$ E &  0.17$\pm$0.03 &  1.42$\pm$0.03 & 0.36$\pm$0.07 & 1.27 & 0.17 &  0.07 &0.78$\pm$ 0.18 &  no  &   no \\
& H$^{13}$CO$^+$ 1--0 &   &   1.44$\pm$0.04 & 0.54$\pm$0.08 & 1.90 & 0.11 & 0.05 &  1.18$\pm$0.18 &  no  &   no \\
& N$_2$H$^+$ 1--0 &  0.30$\pm$0.09 &  1.45$\pm$0.01 & 0.21$\pm$0.04 & 0.90 & 0.24 &  0.10 &0.38$\pm$ 0.11 &  no  &   no \\
& HC$_3$N 10--9 &  0.41$\pm$0.03 &  1.51$\pm$0.01 & 0.23$\pm$0.01 & 0.98 & 0.22 &  0.09 &0.48$\pm$ 0.04 &  no  &   no \\
Cha3-C16  & & & & & & & & \\ 
& C$^{18}$O 2--1 &  0.93$\pm$0.06 &  1.29$\pm$0.01 & 0.49$\pm$0.04 & 1.68 & 0.12 &  0.05 &1.07$\pm$ 0.09 &  no  &   no \\
& CH$_3$OH 2$_{0}$--1$_{0}$ A$^+$ &  0.13$\pm$0.04 &  1.27$\pm$0.05 & 0.36$\pm$0.12 & 1.27 & 0.16 &  0.07 &0.79$\pm$ 0.29 &  no  &   no \\
Cha3-C17  & & & & & & & & \\ 
& C$^{18}$O 2--1 &  1.24$\pm$0.06 &  1.52$\pm$0.02 & 0.65$\pm$0.04 & 0.97 & 0.06 &  0.05 &1.46$\pm$ 0.10 &  no  &   no \\
Cha3-C18  & & & & & & & & \\ 
& C$^{18}$O 2--1 &  1.63$\pm$0.06 &  1.39$\pm$0.01 & 0.69$\pm$0.03 & 1.50 & 0.06 &  0.08 &1.54$\pm$ 0.07 &  no  &   no \\
& CH$_3$OH 2$_{0}$--1$_{0}$ A$^+$ &  0.36$\pm$0.04 &  1.34$\pm$0.02 & 0.46$\pm$0.05 & 0.90 & 0.10 &  0.14 &1.00$\pm$ 0.11 &  no  &   no \\
& CH$_3$OH 2$_{1}$--1$_{1}$ E &  0.31$\pm$0.04 &  1.37$\pm$0.03 & 0.53$\pm$0.08 & 1.07 & 0.09 &  0.12 &1.19$\pm$ 0.19 &  no  &   no \\
& C$^{34}$S 2--1 &  0.13$\pm$0.04 &  1.48$\pm$0.05 & 0.44$\pm$0.09 & 0.88 & 0.10 &  0.14 &0.98$\pm$ 0.21 &  no  &   no \\
Cha3-C19  & & & & & & & & \\ 
& C$^{18}$O 2--1 &  1.14$\pm$0.06 &  1.29$\pm$0.01 & 0.36$\pm$0.02 & 0.94 & 0.14 &  0.10 &0.78$\pm$ 0.05 &  no  &   no \\
& CH$_3$OH 2$_{0}$--1$_{0}$ A$^+$ &  0.31$\pm$0.04 &  1.24$\pm$0.01 & 0.26$\pm$0.04 & 0.73 & 0.17 &  0.12 &0.51$\pm$ 0.11 &  no  &   no \\
& CH$_3$OH 2$_{1}$--1$_{1}$ E &  0.16$\pm$0.04 &  1.24$\pm$0.01 & 0.23$\pm$0.02 & 0.69 & 0.18 &  0.13 &0.44$\pm$ 0.06 &  no  &   no \\
& H$^{13}$CO$^+$ 1--0 &   &   1.28$\pm$0.04 & 0.32$\pm$0.09 & 0.84 & 0.15 & 0.11 &  0.68$\pm$0.23 &  no  &   no \\
Cha3-C20  & & & & & & & & \\ 
& C$^{18}$O 2--1 &  1.26$\pm$0.06 &  1.35$\pm$0.01 & 0.41$\pm$0.02 & 0.90 & 0.12 &  0.09 &0.88$\pm$ 0.05 &  no  &   no \\
& CH$_3$OH 2$_{0}$--1$_{0}$ A$^+$ &  0.26$\pm$0.04 &  1.29$\pm$0.02 & 0.34$\pm$0.05 & 0.78 & 0.14 &  0.11 &0.73$\pm$ 0.12 &  no  &   no \\
& CH$_3$OH 2$_{1}$--1$_{1}$ E &  0.25$\pm$0.04 &  1.30$\pm$0.04 & 0.52$\pm$0.08 & 1.17 & 0.09 &  0.07 &1.14$\pm$ 0.20 &  no  &   no \\
& H$^{13}$CO$^+$ 1--0 &   &   1.38$\pm$0.05 & 0.34$\pm$0.12 & 0.76 & 0.14 & 0.11 &  0.71$\pm$0.31 &  no  &   no \\
Cha3-C21  & & & & & & & & \\ 
& C$^{18}$O 2--1 &  --    &  0.86$\pm$0.08 & 0.73$\pm$0.27 & 1.96 & 0.06 &  0.03 &1.65$\pm$ 0.62 &  no  &   no \\
& --            &  0.58$\pm$0.05 &  2.04$\pm$0.04 & 0.72$\pm$0.09 & 1.90 & 0.06 &  0.03 &1.61$\pm$ 0.21 &  no  &   no \\
Cha3-C22  & & & & & & & & \\ 
& C$^{18}$O 2--1 &  0.84$\pm$0.06 &  1.53$\pm$0.01 & 0.32$\pm$0.02 & 0.82 & 0.14 &  0.09 &0.68$\pm$ 0.06 &  no  &   no \\
Cha3-C23  & & & & & & & & \\ 
& C$^{18}$O 2--1 &  0.48$\pm$0.06 &  1.89$\pm$0.01 & 0.27$\pm$0.03 & 1.14 & 0.23 &  0.06 &0.55$\pm$ 0.07 &  no  &   no \\
Cha3-C24  & & & & & & & & \\ 
& C$^{18}$O 2--1 &  0.92$\pm$0.06 &  1.76$\pm$0.02 & 0.54$\pm$0.03 & 2.20 & 0.10 &  0.04 &1.20$\pm$ 0.08 &  no  &   no \\
Cha3-C25  & & & & & & & & \\ 
& C$^{18}$O 2--1 &  0.67$\pm$0.06 &  1.18$\pm$0.01 & 0.29$\pm$0.02 & 0.62 & 0.13 &  0.08 &0.59$\pm$ 0.05 &  no  &   no \\
Cha3-C26  & & & & & & & & \\ 
& C$^{18}$O 2--1 &  0.90$\pm$0.06 &  1.35$\pm$0.01 & 0.34$\pm$0.02 & 0.40 & 0.10 &  0.14 &0.72$\pm$ 0.05 &  no  &   no \\
Cha3-C27  & & & & & & & & \\ 
& C$^{18}$O 2--1 &  0.36$\pm$0.04 &  1.65$\pm$0.05 & 0.64$\pm$0.12 & 1.35 & 0.05 &  0.04 &1.43$\pm$ 0.29 &  no  &   no \\
Cha3-C28  & & & & & & & & \\ 
& C$^{18}$O 2--1 &  0.77$\pm$0.06 &  1.25$\pm$0.01 & 0.33$\pm$0.03 & 0.56 & 0.10 &  0.07 &0.70$\pm$ 0.06 &  no  &   no \\
Cha3-C29  & & & & & & & & \\ 
& C$^{18}$O 2--1 &  0.75$\pm$0.06 &  1.55$\pm$0.02 & 0.48$\pm$0.04 & 1.00 & 0.08 &  0.06 &1.05$\pm$ 0.09 &  no  &   no \\
\hline
\end{tabular}
\hfill{}
\label{table:physical_parameters_chaIII_2}
\vspace*{-1ex}
\tablefoot{Same as Table~\ref{table:physical_parameters_chaI_2}.
}
\end{table*}

\clearpage

\section{Critical densities}
\label{sec:crit_dens}

For the calculation of critical densities we use the Einstein A-coefficients, the collisional rate coefficients, and energies from the Leiden Atomic and Molecular Database (LAMDA\footnote{http://home.strw.leidenuniv.nl/~moldata/}).

The critical density is the density at which the rate of spontaneous emission 
is equal to the rate of collisional de-excitation,
\begin{equation}
n_{cr} = \frac{A_{ul}}{\gamma_{ul}},
\label{eq:ncrit}
\end{equation}  
where $A_{ul}$ is the Einstein coefficient for spontaneous emission (s$^{-1}$), $C_{ul} = n_{H_2}\gamma_{ul}$ the collision coefficient (collision de-excitation rate, s$^{-1}$), and $\gamma_{ul}$ the collision rate (cm$^3$ s$^{-1}$). The critical densities we derive are listed in Table~\ref{table:crit_dens} for various molecular transitions. 

\begin{table}[!h]
\caption{Critical densities.}
\begin{tabular}{lllll}
\hline\hline
Transition  & $A_{ul}$\tablefootmark{a}   & $\gamma_{ul}$\tablefootmark{b}  & $T$\tablefootmark{c}   & $n_{\rm cr}$\tablefootmark{d}  \\
            & \scriptsize{(s$^{-1}$)} & \scriptsize{(cm$^{3}$~s$^{-1}$)} & \scriptsize{(K)} & \scriptsize{(cm$^{-3}$)}  \\
\hline
C$^{18}$O 2--1               &  6.01$\times$10$^{-7}$ &   7.15$\times$10$^{-11}$ &       10 &       8.4$\times$10$^3$ \\ 
C$^{17}$O 2--1               &  6.43$\times$10$^{-7}$ &   7.15$\times$10$^{-11}$ &       10 &       9.0$\times$10$^3$ \\ 
CH$_3$OH 2$_{0}$--1$_{0}$ A$^+$ &   3.41$\times$10$^{-6}$ &   1.10$\times$10$^{-10}$ &       10 &       3.1$\times$10$^4$ \\ 
CH$_3$OH 2$_{1}$--1$_{1}$ E &   2.49$\times$10$^{-6}$ &   7.50$\times$10$^{-11}$ &       10 &       3.3$\times$10$^4$ \\ 
H$^{13}$CO$^+$ 1--0           &   3.85$\times$10$^{-5}$ &   2.60$\times$10$^{-10}$ &       10 &       1.5$\times$10$^5$ \\ 
N$_2$H$^+$ 1--0              &  3.90 $\times$10$^{-5}$&   1.60$\times$10$^{-10}$ &       10 &       2.4$\times$10$^5$ \\ 
HNC 1--0                    &  2.69$\times$10$^{-5}$ &   9.71$\times$10$^{-11}$ &       10 &       2.8$\times$10$^5$ \\ 
C$^{34}$S 2--1                &   1.60$\times$10$^{-5}$ &   5.30$\times$10$^{-11}$ &       20 &       3.0$\times$10$^5$ \\ 
CS 2--1                      &   1.68$\times$10$^{-5}$ &   5.30$\times$10$^{-11}$ &       20 &      3.2$\times$10$^5$ \\ 
HC$_3$N 10--9                &  5.81$\times$10$^{-5}$ &   9.70$\times$10$^{-11}$ &       10 &       6.0$\times$10$^5$ \\ 
\hline
\end{tabular}
\label{table:crit_dens}
\vspace*{-1ex}
\tablefoot{
\tablefoottext{a}{Einstein coefficient for the molecular transition with quantum numbers $u$ to $l$.}
\tablefoottext{b}{Collision rate.}
\tablefoottext{c}{Temperature corresponding to the listed collision rate.}
\tablefoottext{d}{Critical density computed with Equation~\ref{eq:ncrit}}.
}
\end{table}

\clearpage

\section{Optical depths }
\subsection{C$^{18}$O 2--1 opacity}
\label{sec:opacity_c18o}

C$^{18}$O 2--1 was observed toward all 60 cores in Cha I and C$^{17}$O 2--1 toward 32. We estimate the optical depth of the C$^{18}$O 2--1 transition for this sample of 32 cores. We assume an isotopic ratio 
[$^{18}$O]/[$^{17}$O] of $\sim$ 4.11, as was found for the nearby 
(140~pc) low-mass star forming cloud $\rho$ Ophiuchus 
\citep{wouterloot05}, to derive the opacity of C$^{18}$O 2--1 with the relation 
\begin{equation}
\frac{I_{\rm C^{18}O}}{I_{\rm C^{17}O}} = \frac{1-e^{-\tau_{\rm C^{18}O}}}{1-e^{-\tau_{\rm C^{17}O}}},
\end{equation}
where I$_{\rm C^{18}O}$ and I$_{\rm C^{17}O}$  are the integrated intensities of the two transitions, $\tau_{\rm C^{18}O}$ and $\tau_{\rm C^{17}O}$ their opacities, and $\tau_{\rm C^{17}O}$=$\tau_{\rm C^{18}O}$/4.11. Table~\ref{table:opacity_c18o} lists the opacity estimates for the sample of 32 cores in Cha I. Approximately half of them are somewhat optically thick ($4.2 > \tau_{\rm C^{18}O} \ge 1$) in C$^{18}$O 2--1. 

We can only calculate the C$^{18}$O 2--1 opacities for the two cores in Cha III that were also observed in C$^{17}$O 2--1, Cha3-C1 and Cha3-C4. Cha3-C1 is borderline optically thick/thin with an opacity of $\sim1$, while Cha3-C4 is optically thin (Table~\ref{table:opacity_c18o_chaIII}). They are both optically thin in C$^{17}$O 2--1.

\begin{table}[!h]
\caption{C$^{18}$O 2--1 and C$^{17}$O 2--1 opacities of the Cha I cores.}
\begin{tabular}{llll}
\hline\hline \\[-1.5ex]
Core  & $\frac{I_{\rm c18o}}{I_{\rm c17o}}$\tablefootmark{a}   & $\tau_{\rm c18o}$\tablefootmark{b} & $\tau_{\rm c17o}$\tablefootmark{c}  \\
  \hline
Cha1-C1 &       2.37$\pm$0.12 &      1.76 &      0.43\\
Cha1-C2 &       2.37$\pm$0.18 &      1.77 &      0.43\\
Cha1-C3 &       3.26$\pm$0.27 &      0.66 &      0.16\\
Cha1-C4 &       2.34$\pm$0.20 &      1.81 &      0.44\\
Cha1-C5 &       3.29$\pm$0.48 &      0.63 &      0.15\\
Cha1-C7 &       3.80$\pm$0.17 &      0.21 &      0.05\\
Cha1-C9 &       3.03$\pm$0.47 &      0.89 &      0.22\\
Cha1-C10 &       2.95$\pm$0.33 &      0.97 &      0.24\\
Cha1-C11 &       3.97$\pm$0.91 &      0.09 &      0.02\\
Cha1-C12 &       2.40$\pm$0.17 &      1.73 &      0.42\\
Cha1-C15 &       2.46$\pm$0.21 &      1.63 &      0.40\\
Cha1-C17 &       2.12$\pm$0.24 &      2.26 &      0.55\\
Cha1-C18 &       2.71$\pm$0.38 &      1.27 &      0.31\\
Cha1-C19 &       3.14$\pm$0.40 &      0.77 &      0.19\\
Cha1-C21 &       2.18$\pm$0.21 &      2.11 &      0.51\\
Cha1-C26 &       1.76$\pm$0.25 &      3.24 &      0.79\\
Cha1-C27 &       2.97$\pm$0.24 &      0.96 &      0.23\\
Cha1-C29 &       4.42$\pm$0.55 &      0.00 &      0.00\\
Cha1-C35 &       4.17$\pm$0.78 &      0.00 &      0.00\\
Cha1-C36 &       2.71$\pm$0.47 &      1.27 &      0.31\\
Cha1-C38 &       2.84$\pm$0.66 &      1.10 &      0.27\\
Cha1-C39 &       4.08$\pm$0.64 &      0.01 &      0.00\\
Cha1-C41 &       2.20$\pm$0.29 &      2.09 &      0.51\\
Cha1-C43 &       1.54$\pm$0.25 &      4.20 &      1.02\\
Cha1-C49 &       2.64$\pm$0.40 &      1.36 &      0.33\\
Cha1-C53 &       4.38$\pm$0.66 &      0.00 &      0.00\\
Cha1-C54 &       4.79$\pm$0.93 &      0.00 &      0.00\\
Cha1-C57 &       1.88$\pm$0.31 &      2.86 &      0.70\\
Cha1-C58 &       2.74$\pm$0.51 &      1.23 &      0.30\\
Cha1-C59 &       5.31$\pm$1.36 &      0.00 &      0.00\\
Cha1-C60 &       3.09$\pm$0.41 &      0.81 &      0.20\\

\hline
\end{tabular}
\label{table:opacity_c18o}
\vspace*{-1ex}
\tablefoot{
\tablefoottext{a}{Integrated intensity ratio of the two isotopologues.}
\tablefoottext{b}{C$^{18}$O 2--1 opacity.}
\tablefoottext{c}{C$^{17}$O 2--1 opacity.}
}
\end{table}

\begin{table}
\caption{C$^{18}$O 2--1 and C$^{17}$O 2--1 opacities of the Cha III cores.}
\begin{tabular}{llllll}
\hline\hline \\[-1.5ex]
Core  & $\frac{I_{\rm c18o}}{I_{\rm c17o}}$\tablefootmark{a}   & $\tau_{\rm c18o}$\tablefootmark{b}  & $\tau_{\rm c17o}$\tablefootmark{c} \\
\hline
Cha3-C1 &       2.83$\pm$0.21 &      1.11 &      0.27\\
Cha3-C4 &       3.85$\pm$0.54 &      0.17 &      0.04\\

\hline
\end{tabular}
\label{table:opacity_c18o_chaIII}
\vspace*{-1ex}
\tablefoot{
\tablefoottext{a}{Integrated intensity ratio of the two isotopologues.}
\tablefoottext{b}{C$^{18}$O 2--1 opacity.}
\tablefoottext{c}{C$^{17}$O 2--1 opacity.}
}
\end{table}

\subsection{CS 2--1 opacity}
\label{sec:opacity_cs}

We assume an isotopic ratio of [$^{32}$S]/[$^{34}$S] $\sim 22$ \citep{frerking80} to calculate the opacity of the CS 2--1 line in Cha I. All 57 CS 2--1 spectra feature detections and 23 out of the 57 are detected in C$^{34}$S 2--1. We follow the same method as above to estimate the CS 2--1 opacities toward the 23 cores that are detected in both transitions. The results are given in Table~\ref{table:opacity_cs}. 

For the remaining 34 spectra we calculate upper limits for the CS 2--1 opacity by using the ratio of the CS 2--1 integrated intensity to the value of $3 \times rms$ of the C$^{34}$S 2--1 spectra. The opacity upper limits are listed in Table~\ref{table:opacity_cs_uplim}. All 23 cores that have detections in C$^{34}$S 2--1 are optically thick in CS 2--1, with opacities ranging from $\sim 2$ -- 10.  

Two cores in Cha III are detected in C$^{34}$S 2--1 and their CS 2--1 opacity is listed in Table~\ref{table:opacity_cs_chaIII}. Both have CS 2--1 opacities greater than one, while they are optically thin in C$^{34}$S 2--1. CS 2--1 is detected toward all remaining cores, for which we calculate upper limits (Table~\ref{table:opacity_cs_uplim_chaIII}) in the same way as for the Cha I cores.

\begin{table}[!h]
\caption{CS 2--1 and C$^{34}$S 2--1 opacities of the Cha I cores.}
\begin{tabular}{llll}
\hline\hline \\[-1.5ex]
Core  & $\frac{I_{\rm cs}}{I_{\rm c34s}}$\tablefootmark{a}   & $\tau_{\rm cs}$\tablefootmark{b}   & $\tau_{\rm c34s}$\tablefootmark{c} \\
\hline
Cha1-C1 &       6.63$\pm$0.55 &      3.49 &      0.16\\
Cha1-C2 &       4.74$\pm$0.45 &      5.17 &      0.24\\
Cha1-C3 &       5.90$\pm$0.87 &      4.00 &      0.18\\
Cha1-C4 &       2.71$\pm$0.27 &     10.11 &      0.46\\
Cha1-C9 &       9.24$\pm$5.91 &      2.23 &      0.10\\
Cha1-C10 &       5.44$\pm$0.85 &      4.40 &      0.20\\
Cha1-C11 &       6.43$\pm$1.13 &      3.60 &      0.16\\
Cha1-C16 &       2.94$\pm$0.35 &      9.17 &      0.42\\
Cha1-C17 &       2.94$\pm$0.35 &      9.17 &      0.42\\
Cha1-C18 &       7.47$\pm$3.04 &      3.00 &      0.14\\
Cha1-C19 &       5.35$\pm$0.77 &      4.49 &      0.20\\
Cha1-C21 &       6.87$\pm$1.93 &      3.31 &      0.15\\
Cha1-C23 &       5.15$\pm$1.41 &      4.71 &      0.21\\
Cha1-C26 &       4.27$\pm$1.00 &      5.86 &      0.27\\
Cha1-C29 &       4.73$\pm$0.54 &      5.20 &      0.24\\
Cha1-C33 &       5.05$\pm$0.89 &      4.80 &      0.22\\
Cha1-C34 &       3.08$\pm$0.52 &      8.66 &      0.39\\
Cha1-C35 &       5.52$\pm$0.75 &      4.34 &      0.20\\
Cha1-C38 &       5.07$\pm$1.56 &      4.80 &      0.22\\
Cha1-C40 &       5.12$\pm$3.22 &      4.74 &      0.22\\
Cha1-C41 &       5.06$\pm$1.10 &      4.80 &      0.22\\
Cha1-C52 &       8.77$\pm$2.39 &      2.40 &      0.11\\
Cha1-C54 &       8.57$\pm$3.38 &      2.49 &      0.11\\

\hline
\end{tabular}
\label{table:opacity_cs}
\vspace*{-1ex}
\tablefoot{
\tablefoottext{a}{Integrated intensity ratio of the two isotopologues.}
\tablefoottext{b}{CS 2--1 opacity.}
\tablefoottext{c}{C$^{34}$S 2--1 opacity.}

}
\end{table}

\begin{table}
\caption{CS 2--1 upper limit opacities of the Cha I cores.}
\begin{tabular}{lllll}
\hline\hline \\[-1.5ex]
Core  & $\frac{I_{\rm cs}}{3 \times \sigma_{\rm I,c34s}}$\tablefootmark{a}   & $\tau_{\rm cs}$\tablefootmark{b}  \\
  \hline
Cha1-C5 &       3.56$\pm$0.29 &      7\\
Cha1-C6 &       4.69$\pm$0.36 &      5\\
Cha1-C7 &       7.73$\pm$0.39 &      3\\
Cha1-C8 &       9.54$\pm$0.35 &      2\\
Cha1-C12 &       6.43$\pm$0.34 &      4\\
Cha1-C13 &       2.40$\pm$0.35 &     12\\
Cha1-C14 &       9.58$\pm$0.43 &      2\\
Cha1-C15 &       3.34$\pm$0.28 &      8\\
Cha1-C16 &       1.77$\pm$0.33 &     18\\
Cha1-C20 &       1.76$\pm$0.42 &     18\\
Cha1-C22 &       3.94$\pm$0.29 &      6\\
Cha1-C24 &       0.92$\pm$0.37 &    > 1000 \tablefootmark{c}\\
Cha1-C25 &       2.06$\pm$0.35 &     15\\
Cha1-C27 &       2.18$\pm$0.38 &     14\\
Cha1-C28 &       2.25$\pm$0.36 &     13\\
Cha1-C30 &       1.36$\pm$0.37 &     29\\
Cha1-C31 &       3.58$\pm$0.36 &      7\\
Cha1-C32 &       2.90$\pm$0.36 &      9\\
Cha1-C36 &       3.40$\pm$0.34 &      8\\
Cha1-C37 &       3.45$\pm$0.39 &      8\\
Cha1-C39 &       6.18$\pm$0.37 &      4\\
Cha1-C42 &       5.34$\pm$0.30 &      5\\
Cha1-C43 &       1.18$\pm$0.38 &     41\\
Cha1-C44 &       1.81$\pm$0.32 &     18\\
Cha1-C45 &       2.04$\pm$0.30 &     15\\
Cha1-C46 &       2.91$\pm$0.35 &      9\\
Cha1-C47 &       2.27$\pm$0.40 &     13\\
Cha1-C48 &       3.95$\pm$0.37 &      6\\
Cha1-C49 &       6.48$\pm$0.31 &      4\\
Cha1-C50 &       8.82$\pm$0.34 &      2\\
Cha1-C51 &       1.56$\pm$0.39 &     23\\
Cha1-C53 &       4.31$\pm$0.34 &      6\\
Cha1-C55 &       1.97$\pm$0.32 &     16\\
Cha1-C56 &       2.40$\pm$0.35 &     12\\

\hline
\end{tabular}
\label{table:opacity_cs_uplim}
\vspace*{-1ex}
\tablefoot{
\tablefoottext{a}{Ratio of CS 2--1 integrated intensity to $3\sigma$ of the C$^{34}$S 2--1 non-detections.}
\tablefoottext{b}{CS 2--1 upper limit opacity.}
\tablefoottext{c}{For integrated intensity ratios $< 1$ the estimated upper limit opacity is $> 1000$.}
}
\end{table}

\begin{table}
\caption{CS 2--1 and C$^{34}$S 2--1 opacities of the Cha III cores.}
\begin{tabular}{llll}
\hline\hline \\[-1.5ex]
Core  & $\frac{I_{\rm cs}}{I_{\rm c34s}}$\tablefootmark{a}   & $\tau_{\rm cs}$\tablefootmark{b}  & $\tau_{\rm c34s}$\tablefootmark{c} \\
\hline
Cha3-C13 &       7.26$\pm$2.23 &      3.11 &      0.14\\
Cha3-C18 &       6.91$\pm$1.66 &      3.31 &      0.15\\

\hline
\end{tabular}
\label{table:opacity_cs_chaIII}
\vspace*{-1ex}
\tablefoot{
\tablefoottext{a}{Integrated intensity ratio of the two isotopologues.}
\tablefoottext{b}{CS 2--1 opacity.}
\tablefoottext{c}{C$^{34}$S 2--1 opacity.}

}
\end{table}

\begin{table}
\caption{CS 2--1 upper limit opacities of the Cha III cores.}
\begin{tabular}{lllll}
\hline\hline \\[-1.5ex]
Core  & $\frac{I_{\rm cs}}{3 \times \sigma_{\rm I,c34s}}$\tablefootmark{a}   & $\tau_{\rm cs}$\tablefootmark{b}  \\
  \hline
Cha3-C1 &      13.76$\pm$0.36 &      1\\
Cha3-C2 &       7.26$\pm$0.32 &      3\\
Cha3-C3 &       2.71$\pm$0.36 &     10\\
Cha3-C4 &       4.30$\pm$0.33 &      6\\
Cha3-C5 &       5.28$\pm$0.38 &      5\\
Cha3-C6 &       2.94$\pm$0.34 &      9\\
Cha3-C7 &       4.22$\pm$0.49 &      6\\
Cha3-C8 &       4.08$\pm$0.30 &      6\\
Cha3-C9 &       4.64$\pm$0.37 &      5\\
Cha3-C10 &       4.19$\pm$0.44 &      6\\
Cha3-C11 &       4.21$\pm$0.59 &      6\\
Cha3-C12 &       5.54$\pm$0.48 &      4\\
Cha3-C14 &       7.40$\pm$0.66 &      3\\
Cha3-C15 &       6.57$\pm$0.40 &      4\\
Cha3-C16 &       7.19$\pm$0.43 &      3\\
Cha3-C17 &       5.57$\pm$0.43 &      4\\
Cha3-C19 &       5.90$\pm$0.36 &      4\\
Cha3-C20 &       6.92$\pm$0.33 &      3\\

\hline
\end{tabular}
\label{table:opacity_cs_uplim_chaIII}
\vspace*{-1ex}
\tablefoot{
\tablefoottext{a}{Ratio of CS 2--1 integrated intensity to $3\sigma$ of the C$^{34}$S 2--1 non-detections.}
\tablefoottext{b}{CS 2--1 upper limit opacity.}
}
\end{table}

\clearpage

\section{Density conversions}
\label{sec:density_conversions}

For the conversion of hydrogen densities to free-particle densities and vice versa we use the mean molecular weight per free-particle, $\mu \sim 2.37$ \citep[see Appendix A.1 in][]{kauffmann08}. The cosmic mass ratio of hydrogen to the total mass of matter is $\sim 0.71$, while that of helium is $\sim0.27$ \citep{cox00}. The total hydrogen density, $n_{\rm H}$, is equivalent to
\begin{equation}
n_{\rm H} \sim n_{\rm fp} \times \mu \times 0.71,
\end{equation} 
which gives
\begin{equation}
n_{\rm H} \sim n_{\rm fp} \times 1.683,
\end{equation} 
with $n_{\rm fp}$ the free-particle density.

\section{Spectra of Cha I and III cores}
\label{sec:appendix_spectra}

The spectra of the transitions used to search for signatures of contraction 
motions in the sources listed in Table~\ref{table:lineshapes_table} are 
displayed in Fig.~\ref{fig:appendix_spectra}. 

\begin{figure*}[]
%\centerline{\includegraphics[width=0.70\paperwidth,angle=0]{Figs/all_spectra1.eps}}
\centerline{\includegraphics[width=0.70\paperwidth,angle=0]{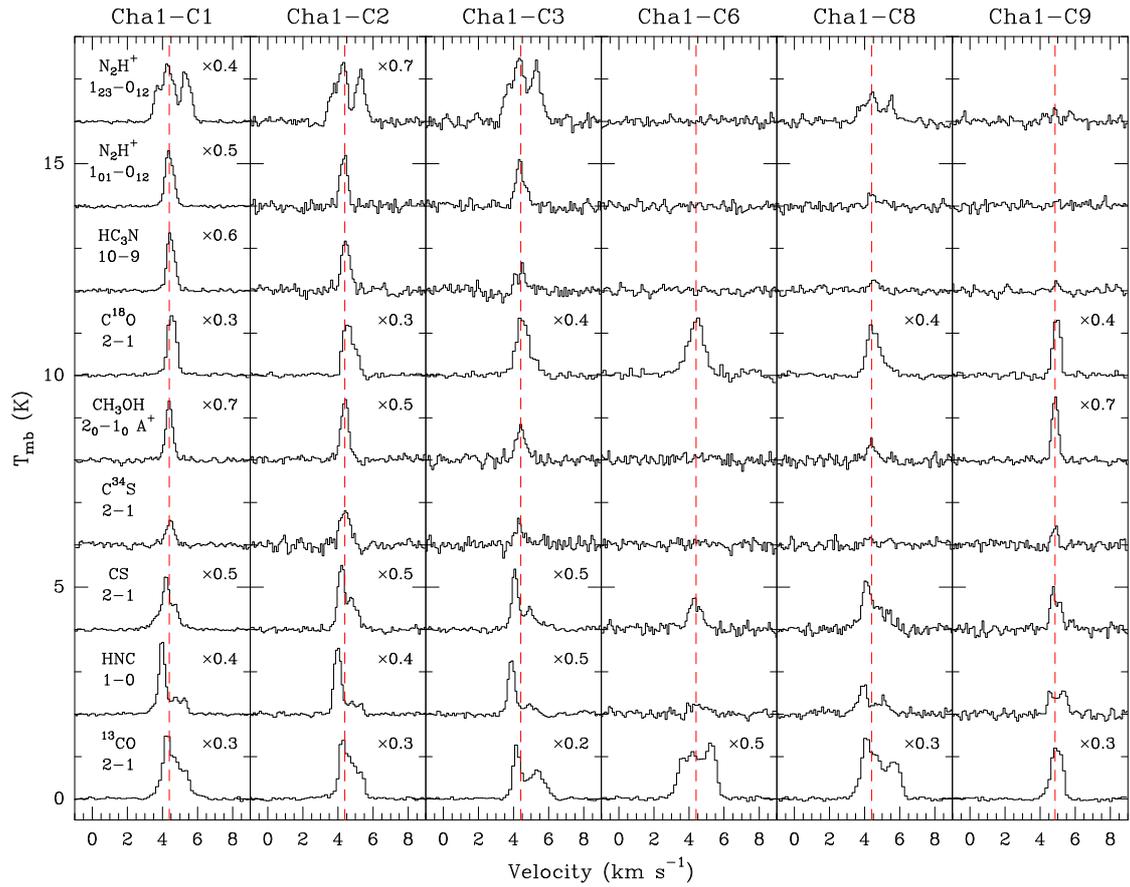}}
\caption[]{Spectra obtained with APEX and the Mopra telescope toward the Cha~I
and III sources listed in Table~\ref{table:lineshapes_table}, displayed in 
main-beam temperature scale. Spectra that were rescaled to fit in the figure 
have their scaling factor indicated on the right. The vertical
dashed line marks the systemic velocity as derived from a Gaussian fit to
the CH$_3$OH~$2_0$--$1_0$~A$^+$ transition, except for sources Cha1-C6, 
Cha1-C24, Cha1-C27, Cha1-C35, and Cha3-C10 for which the C$^{18}$O~2--1 
transition was used.}
\label{fig:appendix_spectra}
\end{figure*}

\begin{figure*}[]
%\centerline{\includegraphics[width=0.70\paperwidth,angle=0]{Figs/all_spectra2.eps}}
\centerline{\includegraphics[width=0.70\paperwidth,angle=0]{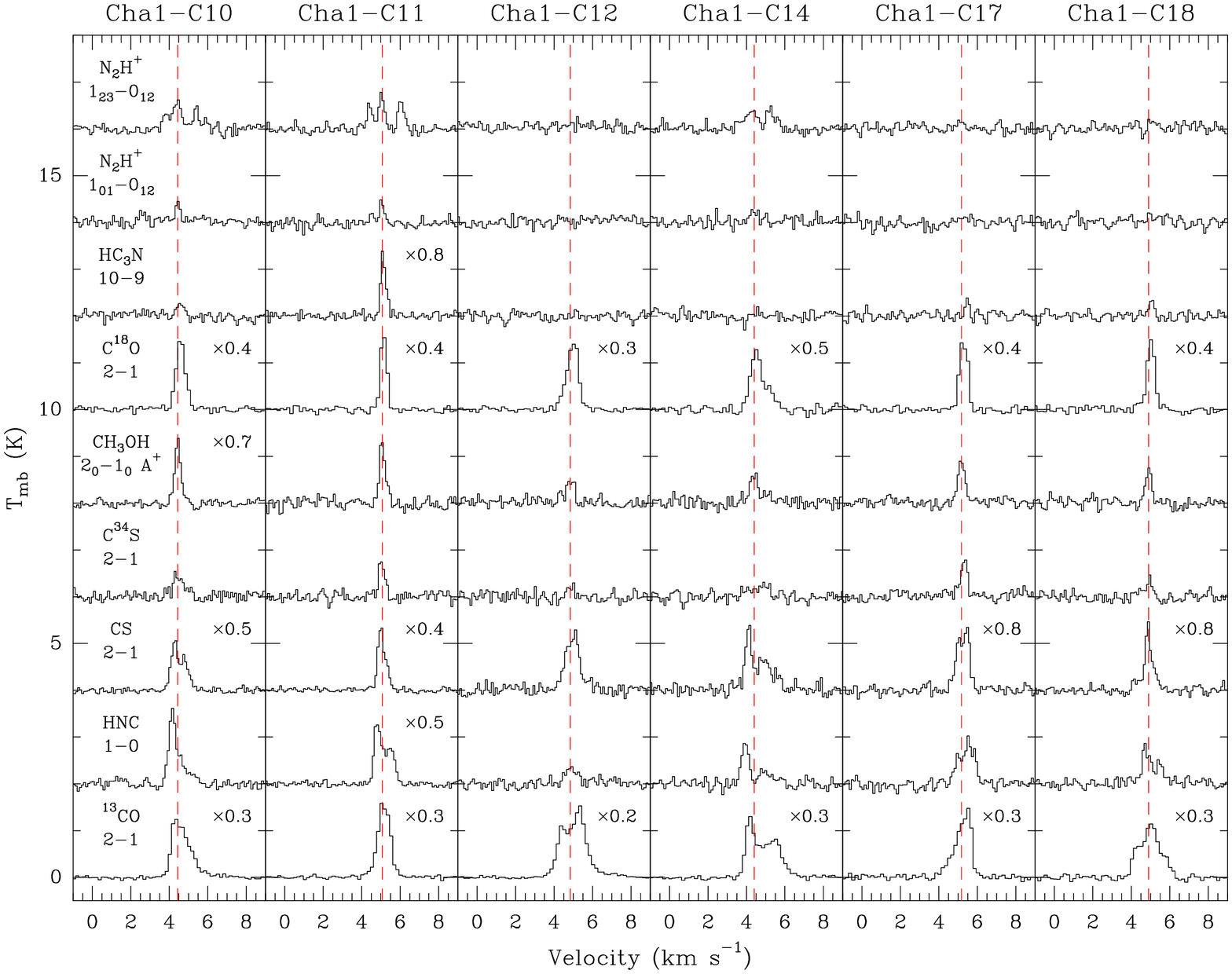}}
\vspace*{2ex}
%\centerline{\includegraphics[width=0.70\paperwidth,angle=0]{Figs/all_spectra3.eps}}
\centerline{\includegraphics[width=0.70\paperwidth,angle=0]{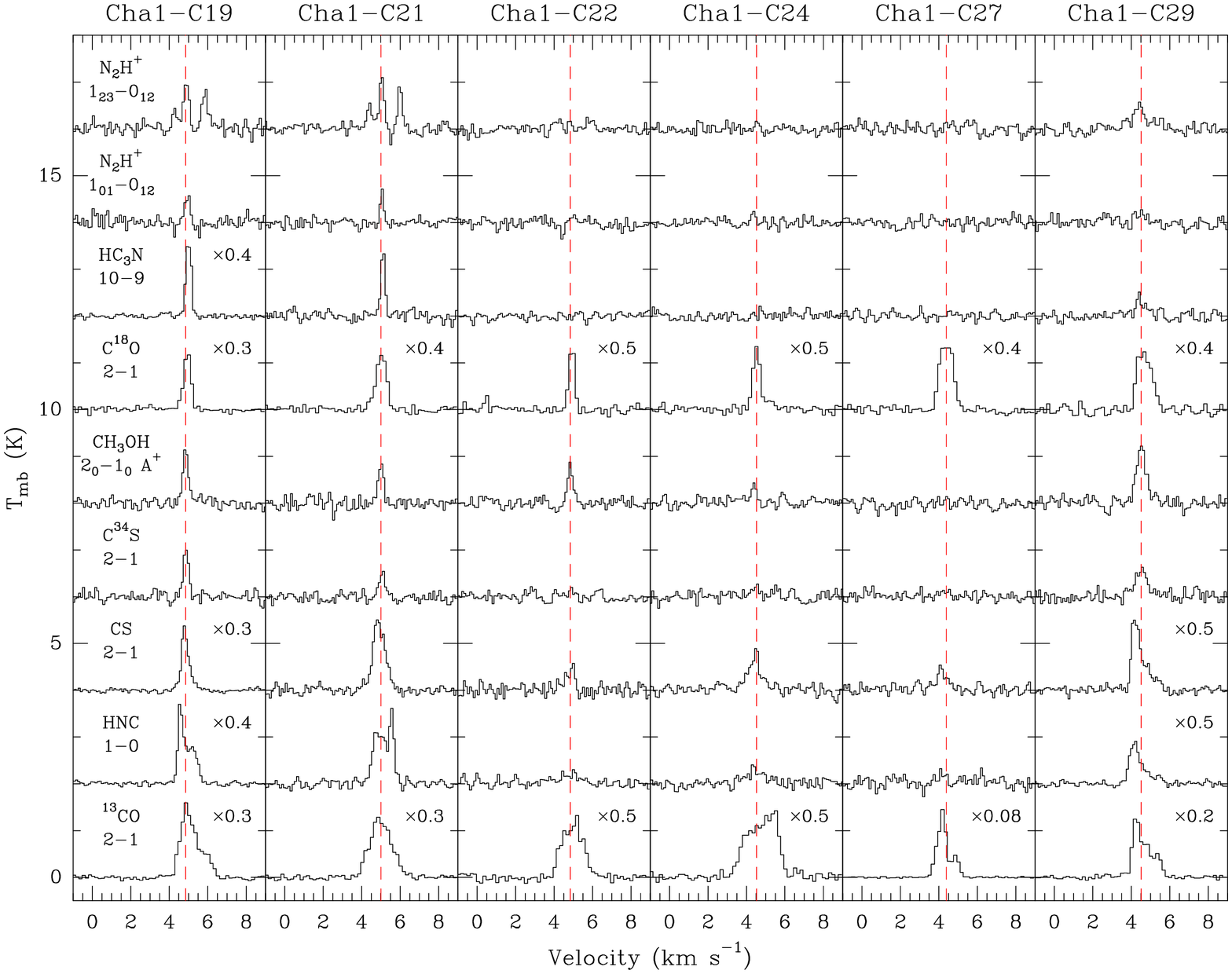}}
\addtocounter{figure}{-1}
\caption[]{continued.}
\end{figure*}

\begin{figure*}[]
%\centerline{\includegraphics[width=0.70\paperwidth,angle=0]{Figs/all_spectra4.eps}}
\centerline{\includegraphics[width=0.70\paperwidth,angle=0]{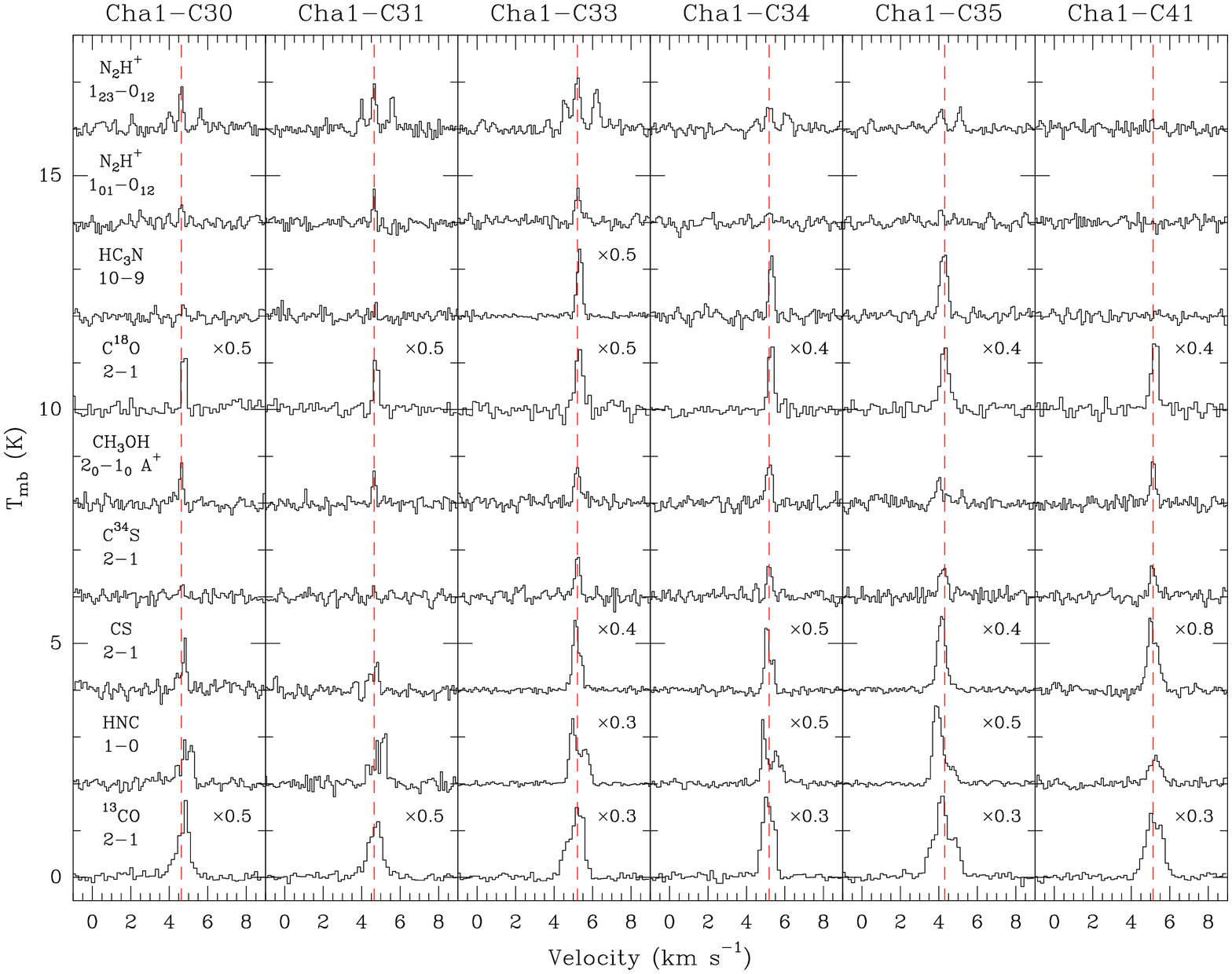}}
\vspace*{2ex}
%\centerline{\includegraphics[width=0.70\paperwidth,angle=0]{Figs/all_spectra5.eps}}
\centerline{\includegraphics[width=0.70\paperwidth,angle=0]{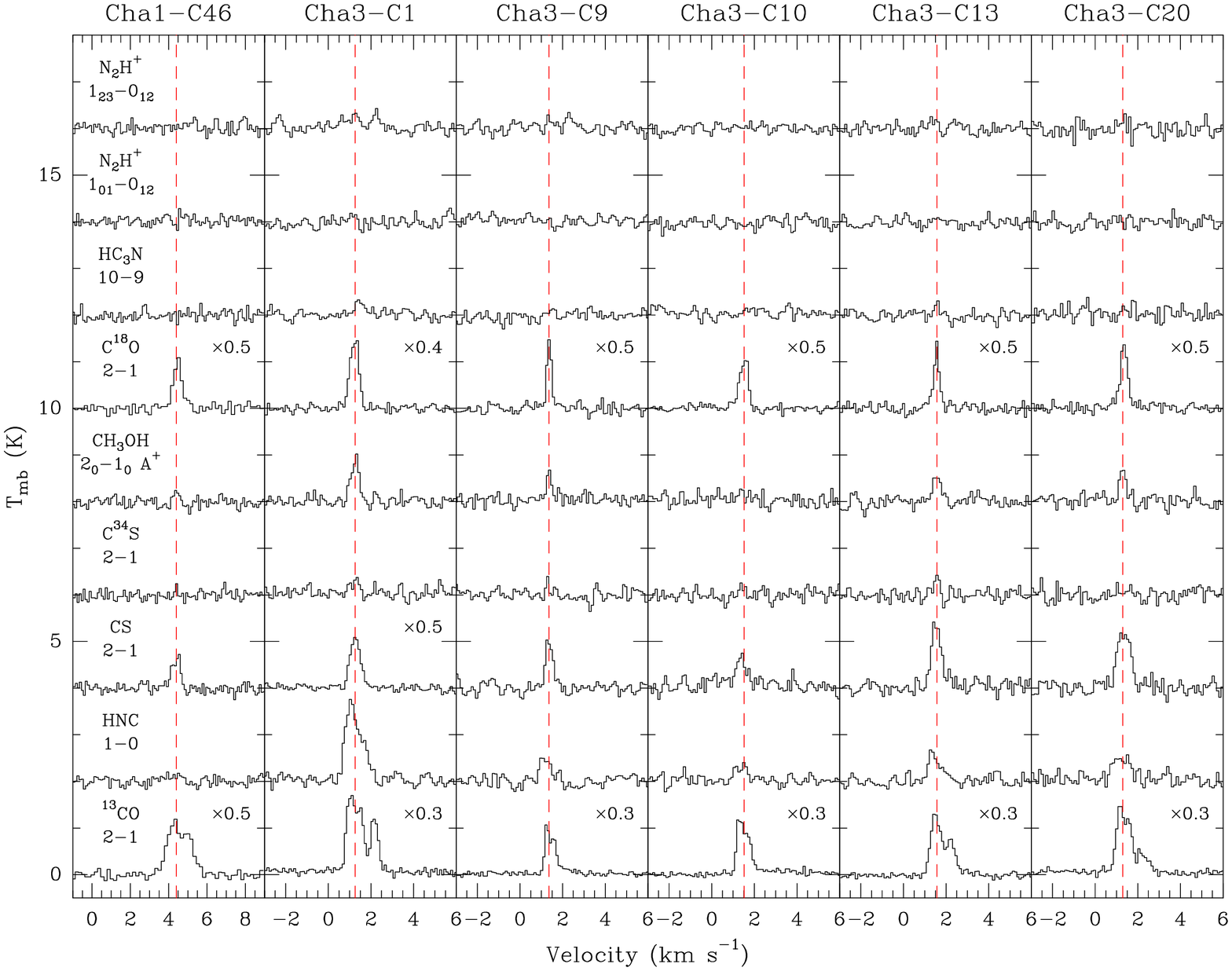}}
\addtocounter{figure}{-1}
\caption[]{continued.}
\end{figure*}

\end{document}